\documentclass[a4paper,11pt]{article}
\pdfoutput=1 

\usepackage{jheppub} 

\usepackage[T1]{fontenc} 
\usepackage{slashed}
\usepackage{subfigure}
\usepackage[dvipsnames]{xcolor}
\usepackage{enumitem}
\usepackage{bm,comment}
\usepackage{ulem}
\usepackage{booktabs}

\newcommand{\eq}[1]{Eq.~\eqref{eq:#1}}
\newcommand{\eqs}[2]{Eqs.~\eqref{eq:#1} and \eqref{eq:#2}}

\renewcommand{\sec}[1]{Sec.~\ref{sec:#1}}

\newcommand{\appx}[1]{App.~\ref{app:#1}}

\newcommand{\tab}[1]{Table~\ref{tab:#1}}
\newcommand{\nn}{\nonumber}

\usepackage{placeins}

\usepackage{amsmath}
\usepackage{tikz}
\usetikzlibrary{shapes.geometric, arrows.meta, positioning, fit, backgrounds}

\title{Centauric 1-Jettiness in DIS and Universal Power Corrections}

\author[a,b]{Andrew Dotson,}
\author[a,c]{June-Haak Ee,}
\author[a,d]{Christopher Lee,}
\author[e]{Yiannis Makris,}
\author[a,f]{and John Terry}

\affiliation[a]{Theoretical Division, Los Alamos National Laboratory, Los Alamos, New Mexico 87545, USA}
\affiliation[b]{Department of Physics, New Mexico State University, Las Cruces, NM, 88003, USA}
\affiliation[c]{Center for Theoretical Physics – a Leinweber Institute, Massachusetts Institute of Technology, Cambridge, MA 02139, USA}
\affiliation[d]{Physics Division, Los Alamos National Laboratory, Los Alamos, New Mexico 87545, USA}
\affiliation[e]{INFN Sezione di Pavia, via Bassi 6, I-27100 Pavia, Italy}
\affiliation[f]{Physics Division, Argonne National Laboratory, Lemont, Illinois 60439, USA}

\emailAdd{adots004@nmsu.edu}
\emailAdd{jhee@mit.edu}
\emailAdd{clee@lanl.gov}
\emailAdd{terryj@anl.gov}

\abstract{
We introduce the \emph{Centauric 1-jettiness}, $\tau_1^C$, a generalized event shape for Deep Inelastic Scattering (DIS) with adjustable beam and jet reference vectors and thus beam and jet regions. We demonstrate that a specific choice of weights allows this observable to exactly reproduce the geometric boundaries of the Centauro jet algorithm in the Breit frame.
Within the framework of Soft-Collinear Effective Theory (SCET), we derive a factorized cross section in the small-$\tau_1^C$ region in terms of known perturbative ingredients. This allows the resummation of large logarithms to N$^3$LL accuracy, which we then match to fixed-order NLO QCD ($\mathcal{O}(\alpha_s^2)$) predictions from \texttt{NLOJet++}.
We establish that the soft measurement reduces to a rescaled hemisphere measurement, placing Centauric 1-jettiness in the same universality class, for the leading non-perturbative corrections, as DIS thrust and jet mass.
As a consequence, the leading non-perturbative shift depends on the same universal first-moment-shift parameter $\Omega_1$ and scales exactly as $1/R$ with the jet radius, thanks to the boost invariance of the Centauro algorithm along the photon axis in the Breit frame, a scaling that we test using \textsc{Pythia} simulations.
These results open new strategies for determining the strong coupling from DIS event shapes, with $R$ providing a handle to break the degeneracy between $\alpha_s$ and the universal non-perturbative shift parameter $\Omega_1$.
}

\begin{document} 
\preprint{
\vspace{-0.75cm}

\setcounter{tocdepth}{2}
\begin{flushright} 
LA-UR-26-21204
\\
MIT-CTP/5994
\end{flushright} \vspace*{-0.75cm}
}
\maketitle
\flushbottom

\section{Introduction}\label{sec:intro}
Fifty years after the prediction and discovery of hadronic jets in QCD
\cite{Brandt:1964sa,Farhi:1977sg},
the study of jets and hadronic final states in $e^+e^-$ and $ep$ collisions
continues to provide deep insight into the dynamics of the strong interaction.
Measurements of the structure of hadronic radiation have played a central role
in establishing the partonic picture of QCD and have evolved into precision
tools for probing the strong coupling, parton distribution functions, and
non-perturbative effects associated with hadronization.
Two complementary approaches to observe and measure jet-like structure have been particularly powerful in this effort: jet algorithms, which identify localized energy flow in the final state, and event shapes, which provide global measures of the radiation pattern \cite{Dasgupta:2003iq}.
In this introduction, we review the status of event-shape measurements in DIS and motivate a new twist on existing event-shape observables, \emph{Centauric 1-jettiness} ($\tau_1^C$), whose jet region is defined through a geometric clustering boundary, providing a tunable jet-radius parameter $R$ within an otherwise standard global event-shape framework.

Our focus is on the event-shape observable known as DIS 1-jettiness
\cite{Stewart:2010tn,Kang:2012zr,Kang:2013nha} or DIS thrust \cite{Antonelli:1999kx}, which partitions final-state
radiation in Deep Inelastic Scattering (DIS) into two regions, one centered around the proton beam direction, and a second (``jet'' or ``current'') region around an additional jet or spray of hadrons produced by the energetic photon. In the Breit frame, for some versions of thrust/1-jettiness ($\tau_Q$ in \cite{Antonelli:1999kx} or $\tau_1^b$ in \cite{Kang:2013nha}), these regions are back-to-back hemispheres.
In the construction studied here, we modify the definition of these regions by associating them with a jet clustering algorithm, specifically the Centauro algorithm \cite{Arratia:2020ssx}, which is optimized for $ep$ collisions to maintain boost invariance along the beam direction while being able to properly reconstruct jets close to the ``backward'' or photon direction.
This procedure introduces an explicit jet-radius parameter $R$ into the definition of the observable, providing a new degree of freedom that can be used to probe both perturbative radiation and non-perturbative effects in DIS final states. 
We will find that the boost-invariance properties of the Centauro algorithm lead to a cleanly predictable dependence of the leading non-perturbative effects on the jet radius $R$.
Before motivating this construction in detail, we first review the current status of event shapes as tools for precision studies of the strong coupling in QCD.

Hadronic event shapes are global measures of the structure of a hadronic final state \cite{Dasgupta:2003iq}, most studied in $e^+e^-$ collisions, but also in DIS and $pp$ collisions.
A two-jet event shape in $e^+e^-$ collisions, for instance, is one that goes to zero in the limit of two back-to-back pencil-like jets in the CM frame, with larger values indicating broadening of these jets or the radiation of additional jets.
Thrust \cite{Brandt:1964sa,Farhi:1977sg}, $C$-parameter \cite{Gardi:2003iv,Donoghue:1979vi}, heavy jet mass \cite{Chien:2010kc}, jet broadening \cite{Rakow:1981qn,Catani:1992jc}, and angularities \cite{Berger:2003iw} are well studied examples.
Some of these are the basis of some of the most precise determinations of the strong coupling, being found amenable to high-order resummed and fixed-order perturbative computation in QCD and with understandable and provably universal properties of the leading non-perturbative corrections to them
\cite{Abbate:2010xh,Hoang:2014wka,Chien:2010kc,Gehrmann:2012sc,Hoang:2015hka,Kardos:2018kqj,Workman:2022ynf,ParticleDataGroup:2024cfk}.
Now that such determinations are approaching the percent level of uncertainty, however, competitive with the global world average in the PDG itself, a swell of renewed energy and enthusiasm has emerged in the field to understand in detail all aspects of the theoretical predictions and the determinations of $\alpha_s$ from experimental data---including methods of resummation, leading and subleading non-perturbative power corrections, and even revisiting the determinations of experimental uncertainties from some of the experiments sourcing the relevant data
\cite{Luisoni:2020efy,Caola:2021kzt,Caola:2022vea,Nason:2023asn,Bell:2023dqs,Benitez-Rathgeb:2024ylc,Benitez:2024nav,Benitez:2025vsp,Nason:2025qbx,Electron-PositronAlliance:2025hze}.
We can hope that consensus about the uncertainties coming from these sources will enable a more complete compilation in PDG of $\alpha_s$ determinations from $e^+e^-$ event shapes, and lead to achievement of $\lesssim 1\%$ level of precision \cite{ParticleDataGroup:2024cfk}.

DIS events provide a very different environment in which to test much of the same theoretical technology, although with additional dependence on proton PDFs, and different kinematic handles and experimental systematics.
From the perspective of future precision programs, DIS will be central at the Electron-Ion Collider \cite{AbdulKhalek:2021gbh} in the coming decade, and continued analyses of HERA data, e.g. \cite{H1:2024aze}, already provide rich testing grounds for modern theoretical methods. 
Hadronic event shapes in DIS have been studied since the turn of the millennium \cite{Antonelli:1999kx,Dasgupta:2001eq,Dasgupta:2003iq}, 
with early HERA measurements and phenomenology motivating continued interest in precision final-state observables. In this context, DIS 1-jettiness---a special case of the more general and popular $N$-jettiness framework \cite{Stewart:2010tn}---has received sustained attention over the past decade or so \cite{Kang:2012zr,Kang:2013nha,Kang:2014qba,Kang:2013lga,Kang:2015swk,Kang:2015moa,AbdulKhalek:2021gbh,Cao:2024ota}
for its amenability to high perturbative accuracy and for the controlled treatment of universal non-perturbative effects, closely paralleling the classic $e^+e^-$ event-shape program.

A version of DIS 1-jettiness that has attracted considerable theoretical and experimental attention is $\tau_1^b$ in the notation of Refs.~\cite{Kang:2013nha,Kang:2014qba}, which is equivalent to DIS thrust $\tau_Q$ introduced in \cite{Antonelli:1999kx}. Defined in the Breit frame, $\tau_1^b$ partitions radiation into two hemispheres.
Its structure closely parallels thrust in $e^+e^-$ annihilation
\cite{Brandt:1964sa,Farhi:1977sg}: both are global observables (hence free of
non-global logarithms) and are therefore well suited to high-accuracy
resummation \cite{Dasgupta:2002dc}.
Historically, other variants such as $\tau_1$ and $\tau_1^a$ have also been developed with alternative axis choices and measurements \cite{Kang:2012zr,Kang:2013nha,Jouttenus:2011wh}, and fixed-order and resummed predictions have been advanced steadily over time \cite{Kang:2013lga,Kang:2014qba,Kang:2015swk,Kang:2015moa,AbdulKhalek:2021gbh,Chu:2022jgs,Cao:2024ota}.
On the fixed-order side, state-of-the-art calculations and implementations relevant for DIS phenomenology rely on established partonic tools and matching strategies \cite{Nagy:2001xb,Nagy:2003tz}.

At small $\tau_1$ the relevant physics is naturally organized in Soft-Collinear Effective Theory (SCET)
\cite{Bauer:2000ew,Bauer:2000yr,Bauer:2001ct,Bauer:2001yt,Bauer:2002nz},
where factorization theorems enable systematic resummation and power expansions.
In the dijet (i.e. one beam jet and one ``current'' jet) region $\Lambda_{\rm QCD}/Q \ll \tau_1 \ll 1$, the leading non-perturbative effect for a broad class of event shapes takes the form of a universal shift of the perturbative spectrum \cite{Dokshitzer:1995zt,Dokshitzer:1995qm,Lee:2006fn,Mateu:2012nk}.
Modern precision treatments include methods to remove renormalon ambiguities and define the leading power-correction parameter in a short-distance scheme; one widely used option is the $R$-gap scheme \cite{Hoang:2008fs,Jain:2008gb,Hoang:2008yj}, which we apply here following the treatment for $\tau_1^b$ \cite{Ee:2025scz}.
These developments have enabled DIS event-shape predictions to approach the theoretical sophistication achieved in $e^+e^-$ studies.

Experimentally, DIS event-shape measurements at HERA \cite{H1:1997hbl,H1:1999wfh,H1:2005zsk,ZEUS:1997nib,ZEUS:2002tyf} provide an essential testing ground, and the recent H1 measurement of $\tau_1^b$ (triple-differential in 
$\tau_1^b$, $Q^2$, and $y$) \cite{H1:2024aze,H1:2024nde} has opened the door to precision phenomenology in combination with modern resummation and power-correction technology, in addition to forecasted experimental performance for 1-jettiness measurements at EIC \cite{AbdulKhalek:2021gbh}.
These recent H1 measurements \cite{H1:2024aze,H1:2024nde} have obtained experimental uncertainties that are only a few percent, while the measured distributions exhibit visible sensitivity to the treatment of fixed order corrections, logarithmic resummation, and hadronization effects. This presents the exciting potential to make DIS event shapes a high-precision laboratory for QCD phenomenology, including percent-level probing of $\alpha_s$, if the effects of perturbative radiation, resummation, and non-perturbative power corrections can be understood simultaneously in a rigorous theoretical framework.

At the same time, a conceptual distinction persists between global event shapes, which partition phase space through minimization procedures, and jet observables defined through sequential clustering algorithms with an explicit jet radius. 
The possibility to define jets and jet regions by minimizing various definitions of $N$-jettiness with suitably defined measures to control the sizes and shapes of jet regions has been pursued since $N$-jettiness was itself defined 
\cite{Stewart:2010tn,Thaler:2010tr,Thaler:2011gf,Jouttenus:2011wh,Jouttenus:2013hs}. 
These ideas were unified, refined, and made concrete with the XCone jet algorithm \cite{Stewart:2015waa}, which provides an explicit bridge between jet finding and event-shape-like partitions through a single $N$-jettiness minimization with a tunable measure.

Conceptually, our construction below can be related to the XCone framework: both define jet regions through $N$-jettiness measures crafted to produce jets with the desired characteristics in a given collision system. The specific XCone measures of \cite{Stewart:2015waa}, however, were designed to capture high-$p_T$ jets in $pp$ collisions, whereas the Centauric measure used here is tailored to $ep$ collisions, where the relevant jets are characterized by large rapidity in the Breit frame. Our measure could in principle be viewed as a Breit-frame-adapted variant within the XCone framework, but here we work directly with the Centauro jet algorithm, which provides a natural geometric realization of this measure for the DIS kinematics. We also restrict the partition to a single jet region with the beam region taken as its complement, rather than identifying a separate exclusive jet around the beam.

Among the possibilities for jet-clustering-based event shapes in DIS, the Centauro jet algorithm \cite{Arratia:2020ssx} is particularly well suited, as it properly reconstructs jets produced in the photon or ``current'' direction away from the proton beam and preserves longitudinal invariance along the beam direction, which naive applications of $pp$ or $e^+e^-$-based $k_T$-clustering algorithms do not do. A key property of the algorithm is that the geometric jet boundary depends only on the rapidity of particles in the Breit frame, where the algorithm is naturally defined. 
As a result, the clustering within the jet is boost-invariant along the beam direction. The clustering of soft radiation, in particular, is determined purely by rapidity, which will allow us to derive an exact scaling relation between the leading non-perturbative corrections and the jet radius $R$. 
Recent fixed-order studies have explored DIS 1-jettiness observables in which the beam and jet regions are defined using standard jet algorithms, including anti-$k_T$ and Centauro, and have demonstrated nontrivial dependence on the jet radius in such constructions \cite{Chu:2022jgs}.

In this work, we build on these developments by introducing Centauric 1-jettiness $\tau_1^C$, an event-shape observable that incorporates an explicit jet-radius dependence while preserving the factorization and resummation properties of global DIS 1-jettiness. Our motivation is to ask whether the high-precision predictability and universal non-perturbative structure that make event shapes powerful tools for QCD studies can be preserved when the jet region is defined geometrically, and whether the resulting tunable jet radius can serve as a new handle for disentangling the partially degenerate dependence on $\alpha_s$ and on the non-perturbative parameter $\Omega_1$.

Using SCET factorization and renormalization-group evolution (with standard ingredients such as beam/jet functions and their operator structure \cite{Stewart:2009yx,Jain:2011iu,Gaunt:2014xxa}), we derive a factorized form for the differential cross section for Centauric 1-jettiness in the small-$\tau_1^C$ region and resum the associated large logarithms at high perturbative accuracy. We demonstrate that the leading non-perturbative effect can be captured by a universal shift governed by the same non-perturbative parameter $\Omega_1$ that appears in DIS event-shape analyses, with a coefficient that scales as $1/R$. We test this prediction using hadron-level \textsc{Pythia} simulations across a range of $Q^2$ and $x_B$, with $\Omega_1$ extracted both through a global fit and an independent Symbolic Regression (SR) analysis, providing a new lever arm for isolating and constraining non-perturbative physics, and find that these simulations follow the predicted $R$-dependence of the shift of the $\tau_1^C$ distributions quite well. Centauric 1-jettiness thus unifies a specific jet geometry definition with precision resummed predictions for global event shapes in DIS, providing a new tool for HERA analyses and for the upcoming Electron-Ion Collider.

The remainder of this paper is organized as follows. In Sec.~\ref{sec:Obs}, we introduce the Centauric 1-jettiness observable and establish its relation to the Centauro jet algorithm, demonstrating how a suitable choice of reference-vector weights reproduces the geometric jet boundary. In Sec.~\ref{sec:theory-formalism}, we develop the theoretical framework within Soft-Collinear Effective Theory, deriving a factorization theorem for the small-$\tau_1^C$ region and establishing the universality structure of the leading non-perturbative corrections. In Sec.~\ref{sec:fixed-order}, we discuss the matching to fixed-order perturbation theory and present the numerical implementation used to obtain stable perturbative predictions when transitioning from the resummation to fixed-order region. 
In Sec.~\ref{sec:omega1}, we determine the leading non-perturbative shift through a global fit to hadron-level \textsc{Pythia} simulations and study its dependence on the jet radius, demonstrating the predicted $1/R$ scaling. In Sec.~\ref{sec:SR}, we present an independent extraction of the same shift using symbolic regression, providing a methodologically distinct cross-check of the result.
In Sec.~\ref{sec:results}, we present phenomenological predictions and comparisons with hadron-level simulations, illustrating how varying the jet radius modifies the sensitivity to non-perturbative effects. Finally, in Sec.~\ref{sec:conc}, we summarize our conclusions and discusses the implications of Centauric 1-jettiness for future precision studies of QCD event shapes in DIS.

\section{Centauric 1-Jettiness and Centauro Jets}\label{sec:Obs}
In this section we establish the framework for \emph{Centauric 1-jettiness} by connecting DIS 1-jettiness event-shape definitions to the Centauro jet clustering algorithm. The goal is to construct an observable whose partition of radiation reproduces the geometric jet boundary of the Centauro algorithm while maintaining the theoretical advantages of the 1-jettiness framework. 
This construction is central to the analysis that follows, allowing the observable to be described within the standard SCET factorization framework while introducing the jet radius $R$ as a new lever arm for studying both perturbative radiation and non-perturbative effects.

We begin in Sec.~\ref{subsec:Kin} by reviewing the DIS kinematics in the Breit frame which are used throughout the paper. In Sec.~\ref{subsec:Def} we then recall the standard definitions of DIS 1-jettiness and introduce a generalized form in which both the directions and magnitudes of the reference vectors are allowed to vary. This leads to the definition of \emph{Centauric 1-jettiness}, characterized by adjustable scales $\omega_B$ and $\omega_J$. Next, in Sec.~\ref{subsec:centauro-jets}, we review the Centauro jet algorithm in the Breit frame and analyze its geometric distance measure in the massless limit. Using this analysis, we derive the condition on $\omega_B$ and $\omega_J$ required for the minimization in Centauric 1-jettiness to reproduce exactly the Centauro jet boundary. With this choice, the partition of radiation into beam and jet regions defined by the event shape coincides with the clustering boundary of the jet algorithm. This correspondence will play a central role in Sec.~\ref{sec:theory-formalism}, where it enables the factorization of the cross section and leads to the prediction of a universal, $R$-dependent non-perturbative power correction.

\subsection{Kinematics}\label{subsec:Kin}
The Centauric 1-jettiness $\tau_1^C$ that we define below is a Lorentz 
scalar and can be evaluated in any frame, whereas the Centauro jet 
algorithm is defined in the Breit frame and its clustering output is 
invariant only under longitudinal ($\hat z$) boosts along the photon axis in this frame. The equivalence 
between the $\tau_1^C$ partition and the Centauro jet boundary therefore holds in the Breit frame and in 
any frame related to it by a $\hat z$-boost.

We consider the DIS process,
\begin{align}
\label{eq:DIS}
\ell\left(k\right) + p\left(P\right) \to \ell\left(k'\right) + X\left(p_X\right)\,,
\end{align}
in which an incoming lepton ($\ell$) scatters off a proton ($p$), producing an outgoing lepton ($\ell'$) together with a hadronic final-state ($X$).
Here, $k$ and $k'$ are the initial and final lepton momenta, $P$ is the proton momentum, and $p_X=\sum_{i\in X} p_i$ is the total momentum of the hadronic final state $X$. The momentum transfer is mediated by an off-shell photon with 
\begin{align}
q = k - k'\,.
\end{align}
For later reference, we define the DIS invariant variables
\begin{align}
    s = (P+k)^2\,,
    \qquad
    Q^2 = -q^2\,,
    \qquad
    x_B = \frac{Q^2}{2 P\cdot q}\,,
    \qquad
    y = \frac{P\cdot q}{P\cdot k}\,,
\end{align}
where $s$ is the lepton-proton center-of-mass (CM) energy squared, $Q^2>0$ is the virtuality of the exchanged photon, $x_B$ is the Bjorken scaling variable, and $y$ denotes the inelasticity.

To describe the kinematics in the Breit frame we introduce the standard light-cone basis vectors
\begin{align}
    n^\mu = \left(1, 0, 0, 1\right)\,,
    \qquad
    \bar{n}^\mu = \left(1, 0, 0, -1\right)\,,
\end{align}
and an arbitrary four-vector can be decomposed as
\begin{align}
\label{eq:light-cone-coordinates}
r^\mu = r^+ \frac{\bar{n}^\mu}{2}
+r^- \frac{n^\mu}{2}
+r_\perp^\mu
=
(r^+, r^-, r_\perp)\,,
\end{align}
with the light-cone components
\begin{align}
\label{eq:light-front-coordinate-convention}
r^+ = {n}\cdot r\,,
\quad
r^- = \bar{n}\cdot r\,,
\quad
r_\perp^\mu = r^\mu 
-r^+ \frac{\bar{n}^\mu}{2}
-r^- \frac{n^\mu}{2}\,,
\end{align}
and $n\cdot \bar{n} = 2$.
Working in the approximation where proton mass corrections are neglected (suppressed by $\Lambda_{\rm QCD}^2/Q^2$), the proton and photon momenta in the Breit frame take the simple form 
\begin{align}
\label{eq:P-q-mu-in-breit}
    P^\mu 
    \stackrel{\text{Breit}}{=} 
    \frac{Q}{x_B} \frac{\bar{n}^\mu}{2}=\left(\frac{Q}{x_B}, 0, 0_\perp\right)\,,
    \qquad
    q^\mu 
    \stackrel{\text{Breit}}{=} 
    Q\, \frac{n^\mu-\bar{n}^\mu}{2} = 
    \left(-Q, Q, 0_\perp\right)\,.
\end{align}
In this frame, the struck quark reverses its longitudinal momentum, and the resulting jet emerges approximately along the $+z$ direction, providing a clean geometrical separation between beam and jet regions.
Throughout this paper, we use italic font to denote four-vectors and bold italic font to denote three-vectors (e.g., $p_i$ for the four-momentum and $\bm{p}_i$ for the three-momentum).

\subsection{Definition of the Centauric 1-jettiness}\label{subsec:Def}

The standard DIS 1-jettiness is defined as 
\cite{Kang:2013nha}
\begin{align}
\label{eq:original-1-jettiness}
\tau_1 = \frac{2}{Q^2} \sum_{i \in X} \min \left\{ q_B \cdot p_i,\; q_J \cdot p_i \right\}\,,
\end{align}
where the sum runs over all final-state hadrons in $X$. Here, $q_B^\mu$ and $q_J^\mu$ are reference four-vectors that determine how radiation is partitioned between the beam and jet regions: the minimization assigns each hadron to either the beam region or the jet region.
We denote the resulting beam and jet regions as $\mathcal{H}_B$ and $\mathcal{H}_J$ and define their total momenta as
\begin{align}
\label{eq:physical_beam_jet_momenta}
p_B^\mu = \sum_{i\in \mathcal{H}_B}p_i^\mu,
\qquad
p_J^\mu = \sum_{i\in \mathcal{H}_J}p_i^\mu.
\end{align}

The standard DIS 1-jettiness variants $\tau_1^a$ and $\tau_1^b$ correspond to the following choices of reference vectors~\cite{Kang:2013nha}:
\begin{itemize}
\item $\tau_1^a$: $q_B^\mu = x P^\mu$, with $q_J^\mu$ aligned with the 
physical jet axis determined by a jet algorithm.
\item $\tau_1^b$: $q_B^\mu = x P^\mu$ and $q_J^\mu = q^\mu + x P^\mu$, 
which in the Breit frame correspond to back-to-back hemispheres.
\end{itemize}
These variants differ in their sensitivity to initial-state radiation, and therefore lead to distinct factorization structures in SCET \cite{Kang:2013nha}. 
For $\tau_1^b$, the Breit-frame reference vectors take the explicit form
\begin{align}
\label{eq:reference-vector-tau1b}
q_B^\mu \big|_{\tau_1^b}
\stackrel{\text{Breit}}{=} 
Q \frac{\bar{n}^\mu}{2}\,, 
\qquad
q_J^\mu \big|_{\tau_1^b}
\stackrel{\text{Breit}}{=} 
Q \frac{{n}^\mu}{2}\,,
\end{align}
so that the minimization in Eq.~\eqref{eq:original-1-jettiness} partitions all particles into two hemispheres separated by the plane $z=0$ in the Breit frame. It is the same as DIS thrust $\tau_Q$ \cite{Antonelli:1999kx,Kang:2013nha}.

\begin{figure}
    \centering
    \includegraphics[width=0.6\linewidth]{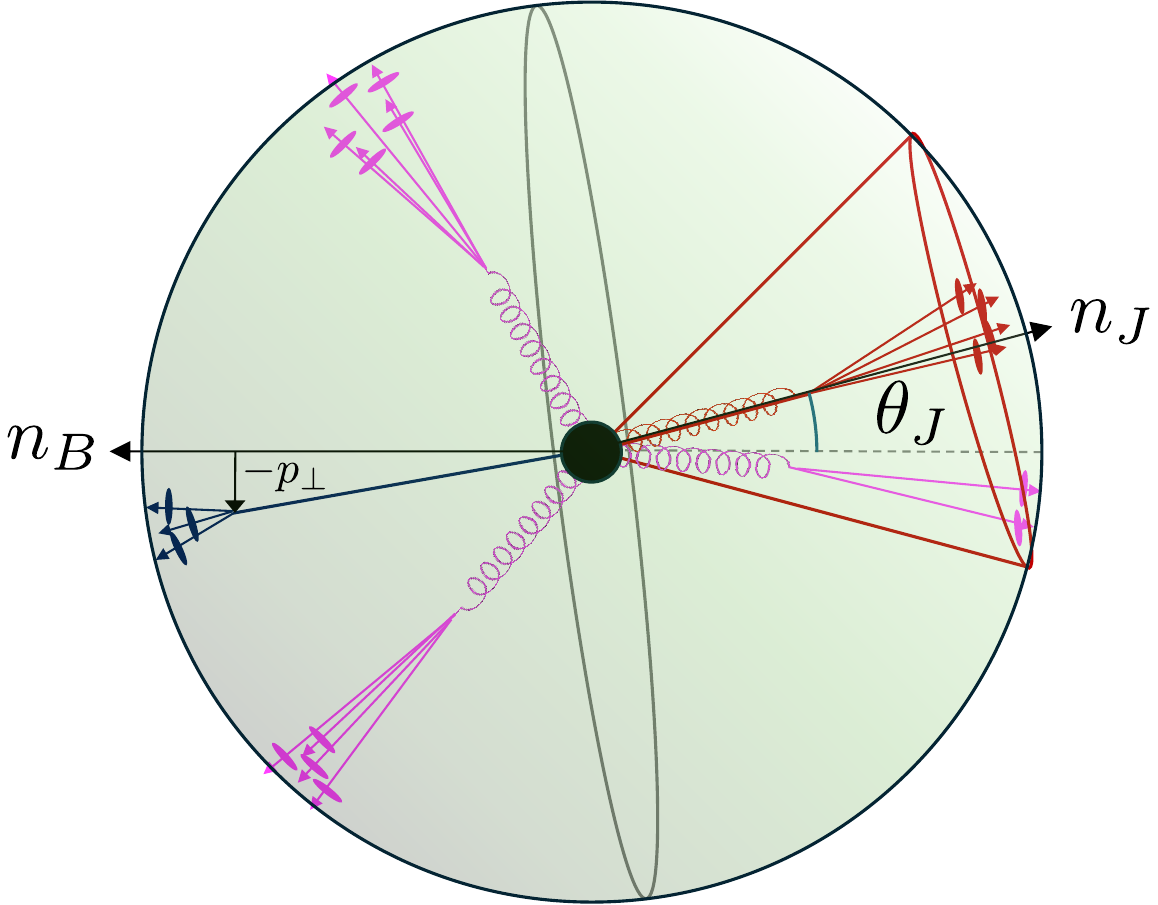}
    \caption{Representation of the Centauric 1-jettiness. The green sphere represents the celestial sphere. The black circle represents the point at which the photon strikes the quark. The red, blue, and magenta colors represent the contributions of collinear, anti-collinear, and soft radiation, respectively. The lines piercing the celestial sphere represent final-state hadrons. The green circle represents the plane that separates the beam and jet hemispheres in DIS 1-jettiness for $\tau_1^a$. The red cone marks the jet boundary defined by the Centauric 1-jettiness. Note that while the physical jet axis $n_J$ defines the reference axis for the measurement, the effective geometric center of the jet region $\mathcal{H}_J$ determined by the minimization may deviate slightly from $n_J$ due to the asymmetric nature of the Centauro jet algorithm.}
    \label{fig:1-Jet}
\end{figure}

For $\tau_1^a$, the physical jet axis $n_J^\mu = (1, \bm{n}_J)$ with $\bm{n}_J^2=1$ is used. Here we have
\begin{align}
q_J^\mu \big|_{\tau_1^a} \stackrel{\text{Breit}}{=} Q \frac{n_J^\mu}{2}\,,
\end{align}
where only the leading-power contribution is retained. 
The resulting minimization still divides the particles into two hemispheres, but the separating plane is slightly tilted toward the physical jet direction, as illustrated in Fig.~\ref{fig:1-Jet}.
In this work, we generalize DIS 1-jettiness $\tau_1^a$ by allowing both the directions and magnitudes of the reference vectors to vary. In a general frame, we can write
\begin{align}
\label{eq:reference-general}
q_B^\mu 
= 
\omega_B\,\frac{n_B^\mu}{2}\,, \qquad
q_J^\mu
= 
\omega_J\,\frac{n_J^\mu}{2}\,,
\end{align}
where $n_B^\mu = (1, \bm{n}_B)$ (with $\bm{n}_B^2=1$) denotes the light-like vector associated with the beam remnant direction and $n_J^\mu$ is aligned with the physical jet axis.
Critically, $\omega_B$ and $\omega_J$ are adjustable energy scales. The introduction of these adjustable scales is the key generalization underlying our construction: by choosing $\omega_B \neq \omega_J$, we obtain an asymmetric partition of hadrons (Fig.~\ref{fig:1-Jet}) that, as we show in Sec.~\ref{subsec:centauro-jets}, can be tuned to reproduce the Centauro jet boundary exactly.

To motivate this choice, it is useful to consider that an asymmetric partition of phase space can in general be obtained in two conceptually different ways. First, one may vary the normalization of the reference vectors in the 1-jettiness measure, which shifts the geometric boundary determined by the minimization. Second, one may instead define the regions using a jet clustering algorithm, such as the Centauro jet algorithm, and then construct an event shape based on the resulting partition.

In the clustering-based approach, the partition of particles in the final state is
defined directly through the jet algorithm rather than through the minimization in
DIS 1-jettiness. In this case, we define the total momenta in the jet and beam
regions as
\begin{align}
\label{eq:physical_beam_jet_momenta_Cent}
\left(p_{J}^{\rm Cent}\right)^\mu = \sum_{i\in J}p_i^\mu\,,
\qquad
\left(p_{B}^{\rm Cent}\right)^\mu = \sum_{i \notin J}p_i^\mu\,,
\end{align}
where $i \in J$ represents particles that are clustered into the jet by the
Centauro algorithm and $i \notin J$ represents particles outside the jet. One may
then define a DIS event shape using this partition by projecting the total momenta
in each region onto the reference vectors,
\begin{align}
\label{eq:original-1-jettiness-Centauro}
\tau_1^{\rm Cent} = \frac{2}{Q^2}
\left(
q_B \cdot p_B^{\rm Cent}
+
q_J \cdot p_J^{\rm Cent}
\right).
\end{align}
In this definition, the region assignment is determined entirely by the jet
algorithm rather than by the minimization in Eq.~\eqref{eq:original-1-jettiness},
and the observable therefore depends explicitly on the jet boundary defined by the
clustering procedure.

Because the boundary between the beam and jet regions is determined by a jet
algorithm rather than by a global minimization measure, soft radiation near the jet
boundary can influence the region assignment in a non-global way. As a result, the
observable defined in Eq.~\eqref{eq:original-1-jettiness-Centauro} does not in general
correspond to a purely global event-shape measurement and can be sensitive to
non-global logarithms \cite{Dasgupta:2001sh,Dasgupta:2002dc}. In practice, these effects are expected to be reduced for
sufficiently large jet radii $R$, but they nevertheless complicate the factorization
structure and resummation.

To avoid this issue, we instead construct a generalized DIS 1-jettiness observable
in which the normalization of the reference vectors is chosen such that the
minimization reproduces the Centauro jet boundary exactly. In addition, the weights
are chosen such that particles lying on the jet boundary contribute equally to the
observable whether they are assigned to the beam or jet region. This ensures that
the measurement is continuous across the boundary and depends only on the total
radiation in each region, thereby restoring the global structure required for
factorization and avoiding sensitivity to non-global logarithms. This construction
leads to the definition of the \emph{Centauric 1-jettiness},
\begin{align}
\label{eq:gen-1-jettiness}
\tau_1^C =
\frac{1}{Q^2}
\sum_{i \in X} \min \left\{
\omega_B\, n_B\cdot p_i, 
\omega_J\, n_J \cdot p_i
\right\},
\end{align}
where we inserted Eq.~\eqref{eq:reference-general} into Eq.~\eqref{eq:original-1-jettiness}, and where $\omega_B$ and $\omega_J$ here will have to satisfy a condition so that the minimization in \eq{gen-1-jettiness} reconstructs the Centauro jet boundary: we will determine this condition in the next subsection, see \eq{wJ-wB-cond}.

\subsection{Centauro Jets}\label{subsec:centauro-jets}
The Centauro jet algorithm of Ref.~\cite{Arratia:2020ssx} defines jets in the Breit frame using a distance measure that depends only on the angular separation between particles and the jet axis, leading to a cone-like geometric boundary characterized by the jet radius $R$. We show below that, in the massless limit, there exists a choice of reference-vector weights for which the minimization in the generalized 1-jettiness observable reproduces exactly this geometric jet boundary.

The two main features of the Centauro jet algorithm are that it defines jet clustering that is invariant under boosts along the $z$ axis from the Breit frame, similar to the longitudinally boost-invariant anti-$k_T$ algorithm at hadron colliders, while at the same time it can reconstruct jets produced along the positive $z$ direction, as in spherically invariant jet algorithms for $e^+e^-$ collisions (see also, recently, \cite{vanBeekveld:2026rez}). These properties are achieved by defining the distance measures
\begin{align}
\label{eq:distance-measures}
d_{ij} = \min\left(z_i^{2p}, z_{j}^{2p}\right) \frac{(\bar{\eta}_i - \bar{\eta}_j)^2 + 2\bar{\eta}_i \bar{\eta}_j (1-\cos\Delta \phi_{ij})}{R^2}\,,
\qquad
d_{iB} = z_i^{2p}\,,
\end{align}
where $z_i = \bar{n}\cdot p_i / Q$ denotes the momentum fraction of the final-state hadron with momentum $p_i$ relative to the struck-quark momentum, and $\Delta\phi_{ij} = \phi_i - \phi_j$ with $\phi_i$ being the azimuthal angle of $p_i$ in the Breit frame, and $p$ is an integer taken to be $0$ for the Cambridge--Aachen (C--A) algorithm or $-1$ for the anti-$k_T$ algorithm. The quantity $\bar{\eta}_i$ is defined as
\begin{align}
\label{eq:eta_i-defined}
\bar{\eta}_i \stackrel{\text{Breit}}{=}
\frac{2p_i^\perp}{\bar{n}\cdot p_i},
\end{align}
where $p_i^\perp$ is the magnitude of the transverse momentum of $p_i$ with respect to the photon direction. 

In the massless limit and for the C--A choice $p=0$, the prefactor $\min\left(z_i^{2p}, z_{j}^{2p}\right)=1$ and $d_{ij}$ depends only on the angular separations between the particles. In this case, $\bar{\eta}_i$ can be expressed in terms of the polar angle $\theta_i$ of $p_i$ as 
\begin{align}
\label{eq:eta_definition}
\bar{\eta}_i = 2\sqrt{\frac{{n} \cdot p_i}{\bar{n}\cdot p_i}}
\stackrel{\text{massless}}{\to}
2e^{-y_i}
=
2\tan(\theta_i/2),
\end{align}
where $y_i$ is the rapidity with respect to the $+z$-axis, and we have used
\begin{align}
\bar{n}\cdot p_i &= E_i(1+\cos\theta_i) = 2E_i\cos^2(\theta_i/2),\nonumber\\
n\cdot p_i &= E_i(1-\cos\theta_i) = 2E_i\sin^2(\theta_i/2),
\end{align}
valid for massless $p_i$.
The polar angle is defined in the Breit frame and satisfies
$0\le \theta_i <\pi$, with $\theta_i\sim 0$ corresponding to the photon direction.  
Inserting Eq.~\eqref{eq:eta_definition} into Eq.~\eqref{eq:distance-measures} yields
\begin{align}
d_{ij} = \frac{\tan^2(\theta_i/2) + \tan^2(\theta_j/2) - 2\cos\Delta\phi_{ij}
\tan(\theta_i/2)\tan(\theta_j/2)}
{R^2/4},
\end{align}
which depends only on the polar and azimuthal angles of the two particles. 

Let us now consider the distance between a final-state hadron with momentum $p_i$ and a given jet. We assume that a jet axis has been identified, with polar and azimuthal 
angles $0\le\theta_J< \pi$ and $\phi_J=0$ in the Breit frame. 
The corresponding Euclidean unit vector can be written as
\begin{align}
\bm{n}_J = \left(
\sin\theta_J, 0, \cos\theta_J
\right).
\end{align}
The general case with arbitrary azimuthal angle $\phi_J$ can be obtained by a rotation around the $z$-axis, making use of the azimuthal symmetry of the initial proton-photon system.

The distance between a particle with momentum $p_i$ and the jet axis is given by $d_{iJ}$, which takes the form
\begin{align}
\label{eq:diJ-i-to-Jet}
d_{iJ} = \frac{\tan^2(\theta_i/2) + \tan^2(\theta_J/2) - 2\cos\phi_{i}
\tan(\theta_i/2)\tan(\theta_J/2)}
{R^2/4},
\end{align}
where $\theta_i$ and $\phi_i$ are the polar and azimuthal angles of $p_i$ in Breit frame. 
A particle is clustered into the jet if $d_{iJ}\le 1$, while $d_{iJ}> 1$ places it outside the jet. Thus, the geometric jet boundary is determined by the locus of points satisfying $d_{iJ}=1$. 
The Centauro 1-jettiness $\tau_1^C$ of Eq.~\eqref{eq:gen-1-jettiness} instead partitions the final-state particles through its minimization operator: each particle is assigned to the beam or the jet region according to which of the two weighted projections, $\omega_B\, n_B\cdot p_i$ or $\omega_J\, n_J\cdot p_i$, is smaller. The boundary separating the two regions is therefore the locus of momenta satisfying $\omega_B\, n_B\cdot p_i = \omega_J\, n_J\cdot p_i$.
In Appendix~\ref{sec:equivalence-derivation} we show that, in the massless limit, there exists a choice of $\omega_B$ and $\omega_J$ for which the minimization in the generalized 1-jettiness observable reproduces exactly the geometric jet boundary defined by the Centauro algorithm. The condition is
\begin{align}
\label{eq:wJ-wB-cond}
\frac{\omega_J}{\omega_B}
=
\frac{4}{R^2\cos^2(\theta_J/2)}\,.
\end{align}
This determines only the ratio $\omega_J/\omega_B$, leaving an overall normalization to be specified. In Sec.~\ref{sec:fixed-order}, we will explore the impact of these choices on the observable.

A brief remark on mis-clustering: due to the sequential nature of the Centauro algorithm, hadrons lying within the geometric boundary ($d_{iJ}\lesssim 1$) may not always be included in the final jet, particularly at large $R$ where wide-angle radiation can initially lie outside the jet but subsequently radiate emissions falling within it. 
In the resummation region, however, these contributions arise from soft emissions near the jet boundary. The mis-clustering of such emissions result in numerically small effects on $\tau_1^C$, due to the symmetric treatment of soft radiation at the boundary, i.e.~whether soft radiation near the boundary has been mis-clustered or not, that soft radiation will contribute the same to a global observable like $\tau_1^C$ regardless of which side of the boundary it falls. (See also discussions in, e.g., Refs.~\cite{Ellis:2010rw,Kelley:2012kj}.)

\section{Factorization and the Leading Non-perturbative Shift} \label{sec:theory-formalism}
In this section we develop the theoretical framework used to describe the Centauric 1-jettiness observable introduced in Sec.~\ref{sec:Obs}. In the resummation region, $\tau_1^C \ll 1$, the final state is dominated by a single energetic jet recoiling against the proton beam, and additional radiation is constrained to be either collinear to the jet or beam directions, or soft. This hierarchy of scales motivates an analysis within SCET, in which the cross section factorizes into contributions associated with the hard scattering, collinear radiation along the jet and beam directions, and soft wide-angle emissions. The resulting factorization theorem enables the resummation of logarithms of the 1-jettiness variable and provides a systematic framework for describing the leading non-perturbative effects. In particular, we will show that the soft measurement can be mapped onto a rescaled hemisphere measurement, placing Centauric 1-jettiness in the same universality class as thrust-like event shapes and implying that the leading non-perturbative correction takes the form of a shift that scales with the jet radius as $1/R$. Considerations related to the fixed-order calculation and its matching to the resummed prediction are deferred to Sec.~\ref{sec:fixed-order}.

We begin in Sec.~\ref{subsec:PC} by identifying the kinematic regions relevant for the $\tau_1^C$ distribution and the corresponding power counting, and by introducing the effective hard scale $Q_R$ that governs the onset of the resummation region. In Sec.~\ref{subsec:overview} we then outline the general structure of the resummed prediction, including the separation between singular and non-singular contributions and the treatment of non-perturbative effects through a shape function. Next, in Sec.~\ref{subsec:fac} we derive the SCET factorization theorem relevant for Centauric 1-jettiness, identifying the Lorentz invariants that govern the measurement and through which the jet-radius dependence enters the cross section. 
In Sec.~\ref{subsec:profile}, we introduce the profile functions used to define the renormalization scales across the peak, tail, and far-tail regions and to estimate perturbative uncertainties. 
Finally, in Sec.~\ref{subsec:soft} we analyze the structure of the soft function and the associated non-perturbative power corrections, demonstrating that Centauric 1-jettiness belongs to the same universality class as thrust-like event shapes and determining how the leading power correction depends on the jet radius $R$.

\subsection{Power Counting}\label{subsec:PC}
The structure of the cross section is determined by the relative size of the measurement variable $\tau_1^C$ compared to the characteristic hard scale $Q_R$. Three kinematic regions emerge--peak, tail, and far-tail--each governed by a different but smoothly connected description. The corresponding power counting is 
\begin{align}
\label{eq:power-counting-tau1c}
\begin{split}
\textrm{peak region}: & \quad 
\tau_1^C \sim \Lambda_\textrm{QCD}/Q_R \ll 1,
\\
\textrm{tail region}: & \quad 
\Lambda_\textrm{QCD}/Q_R \ll \tau_1^C \ll 1,
\\
\textrm{far-tail region}: & \quad
\tau_1^C \sim 1.
\end{split}
\end{align}
Here, $Q_R$ denotes the characteristic $R$-dependent hard scale of the observable. Unlike standard DIS observables where the hard scale is universally $Q$, the Centauric 1-jettiness introduces adjustable weights $\omega_{B,J}$ in its definition. The scale $Q_R$ will appear naturally in the factorized description
derived below as the combination of $Q$ and $\omega_{B,J}$ that
normalizes the soft-function argument. It represents the effective hard scale governing the soft and collinear radiation's phase spaces and entering the measurement of $\tau_1^C$ as the 
natural normalization of the soft momentum, directly dictating the onset of the resummation region.
For the Centauric 1-jettiness this scale takes the form
$Q_R = QR/[2\cos^2(\theta_J/2)]$, derived from 
Eq.~\eqref{eq:1-jet-invariants} below using the Centauric weights given 
in Table~\ref{tab:tau1-variables}.
Throughout this section we assume $Q\gg \Lambda_{\rm QCD}$, $R\sim 1$, and $\theta_J\ll 1$, so that the power counting in Eq.~\eqref{eq:power-counting-tau1c} remains well defined.

The hierarchy of scales established above motivates an analysis within SCET. Within this framework, renormalization-group evolution between the corresponding scales allows for the systematic resummation of logarithms of the 1-jettiness variable. We note that the perturbative expansion also generates logarithms of the jet radius $R$.
In this work we focus on the regime $R \sim 1$,
for which these logarithms remain modest and do not require additional
resummation. The resummation of $\ln R$ terms relevant in the
small-radius limit $R \ll 1$ is left for future work.

\subsection{Overview of Resummation Region}
\label{subsec:overview}
Given the power counting established in Sec.~\ref{subsec:PC}, we now outline the structure of the theoretical description in the resummation region $\tau_1^C \ll 1$. In this regime the cross section is dominated by logarithmically enhanced contributions that can be systematically resummed within the SCET framework. As discussed in Sec.~\ref{sec:Obs}, the Centauro jet algorithm can be reproduced by a suitably reweighted 1-jettiness observable, characterized by Eq.~\eqref{eq:wJ-wB-cond}. This observation implies that the dependence on the jet algorithm can be absorbed into an appropriate choice of the reference vectors $q_{B,J}$. Consequently, the general SCET formulation of 1-jettiness with arbitrary $\omega_{B,J}$ and directions $n_{B,J}$, derived in Ref.~\cite{Kang:2013nha}, applies directly to the Centauric observable studied here. 

In what follows, we follow the standard theoretical treatment developed for $\tau_1^b$ in Ref.~\cite{Ee:2025scz}, adapted as needed for $\tau_1^C$. 
After combining the predictions in the peak, tail, and far-tail regions using profile scales and a smooth matching procedure, the $\tau_1^C$ distribution can be written as
\begin{align}
\label{eq:k-int-cumulant}
\sigma\left(\tau_1^C\right)
&=
\int dk\,
\sigma_\textrm{PT}\left(\tau_1^C-\frac{k}{Q_R}\right)
\left[
e^{-2\delta(R_{\textrm{gap}},\mu_S)(d/dk)}
F
\left(
k-2\Delta(R_{\textrm{gap}},\mu_S)
\right)
\right],
\end{align}
where $\sigma_\textrm{PT}$ is the perturbative cross section and $F$ is a non-perturbative shape function. The quantities $\delta,\Delta,R_\text{gap}$ are associated with a renormalon subtraction scheme to remove ambiguities in the definitions of perturbative and non-perturbative contributions that exist in the $\overline{\text{MS}}$ scheme, as will be explained below.
Throughout this work, $\sigma$ and $\sigma_\textrm{PT}$ can represent either the differential or cumulative $\tau_1^C$ distribution.
\begin{align}\label{eq:cumulant-def}
\sigma\left(\tau_1^C\right) = \frac{d\sigma}{d\tau_1^C}\,,
\quad
\textrm{or}
\quad
\sigma\left(\tau_1^C\right) = \sigma_c\left(\tau_1^C\right) = \int_0^{\tau_1^C} d\tau \frac{d\sigma}{d\tau}\,.
\end{align}
with the differentials in $x$ and $Q$ left implicit for notational simplicity. 

The perturbative cross section is decomposed into the singular and non-singular components:
\begin{align}
\label{eq:sing-plug-ns}
\sigma_\textrm{PT}\left(\tau_1^C; \mu_H, \mu_J, \mu_B, \mu_S, \mu_\textrm{ns}\right)
=
\sigma_\textrm{PT}^\textrm{s}\left(\tau_1^C; \mu_H, \mu_J, \mu_B, \mu_S\right)
+
\sigma_\textrm{PT}^\textrm{ns}\left(\tau_1^C; \mu_\textrm{ns}\right)\,.
\end{align}
The singular contribution $\sigma_{\textrm{PT}}^\textrm{s}$ contains logarithmically enhanced terms, which are resummed at leading power using SCET. 
The non-singular contribution $\sigma_{\textrm{PT}}^\textrm{ns}$ consists of 
terms that are suppressed by powers of $\tau_1^C$ and is obtained from fixed-order QCD after subtracting the singular terms to avoid double counting.  
The scales $\mu_{H,J,B,S}$ correspond to the hard, jet, beam, and soft functions in the factorization theorem, while $\mu_\textrm{ns}$ corresponds to the non-singular part. 
Their specific choices are introduced in Sec.~\ref{subsec:profile} and
further details are given in Appendix~\ref{sec:profile}.

The shape function $F$ models non-perturbative soft radiation at the $\Lambda_{\rm QCD}$ scale. It is peaked around  $k \sim \Lambda_{\textrm{QCD}}$, and falls off faster than any power for large $k$.
The perturbative cross section $\sigma_\textrm{PT}$, computed in the $\overline{\textrm{MS}}$ scheme, contains an $\mathcal{O}(\Lambda_\textrm{QCD})$ renormalon ambiguity originating from the partonic soft function. 
This ambiguity is removed by the exponential operator acting on the shape function, which implements an order-by-order renormalon subtraction. The function $\delta(R_{\mathrm{gap}},\mu_S)$ cancels the leading renormalon, while $\Delta(R_{\mathrm{gap}},\mu_S)$ is a gap parameter associated with the physical hadronic threshold. Further details on the shape function, renormalon subtraction, and gap scheme are given in Appendix~\ref{sec:power-corrections}.

We follow the standard perturbative order counting used in Refs.~\cite{Kang:2013nha, Ee:2025scz}.
Fixed-order contributions at $\mathcal{O}(\alpha_s)$ and $\mathcal{O}(\alpha_s^2)$ are referred to as LO and NLO, respectively, since the tree-level contribution at $\mathcal{O}(\alpha_s^0)$ is proportional to $\delta\left(\tau_1^C\right)$.
The ingredients entering the resummed predictions at NLL,~NNLL, N$^3$LL accuracy are summarized in Table~\ref{tab:order}. Strictly speaking, at these ``unprimed'' orders, the truncations of the fixed-order ingredients $\{H,B,J,S\}$ should be applied to the cumulative momentum-space distribution or Laplace-space distribution, not a naive differential momentum-space distribution, to avoid missing terms at a given N$^k$LL order of resummed accuracy \cite{Almeida:2014uva}. To avoid this problem, we will work with resumming the cumulative momentum-space distribution and then differentiate it to obtain the differential spectrum, as we shall explain further below.

\begin{table}
    \centering
    \begin{tabular}{c|c|c|c|c|c|c|c}
        \hline
        \hline
        \rule{0pt}{2.8ex}
        Accuracy & $\Gamma_\textrm{cusp}[\alpha_s]$ & $\gamma_{H,B,J,S}[\alpha_s]$ & $\beta[\alpha_s]$ & $\{H,B,J,S\}[\alpha_s]$ & non-singular & $\gamma_\Delta^{\mu,R_\textrm{gap}}$ & $\delta$ \\[0.6ex]
         \hline
         \hline
          \rule{0pt}{2.8ex}
        NLL     & $\alpha_s^2$ & $\alpha_s$ & $\alpha_s^2$ & $\alpha_s^0$ & - & $\alpha_s$ & - \\[0.5ex]
        NNLL    & $\alpha_s^3$ & $\alpha_s^2$ & $\alpha_s^3$ & $\alpha_s$ & $\alpha_s$ & $\alpha_s^2$ & $\alpha_s$\\[0.5ex]
        N$^3$LL & $\alpha_s^4$ & $\alpha_s^3$ & $\alpha_s^4$ & $\alpha_s^2$ & $\alpha_s^2$ & $\alpha_s^3$ & $\alpha_s^2$ \\[0.5ex]
         \hline
         \hline
    \end{tabular}
    \caption{Ingredients included at N$^k$LL accuracy.
    $\Gamma_\textrm{cusp}$ and $\gamma_{H,B,J,S}$ denote the cusp and non-cusp anomalous dimensions of the hard, beam, jet, and soft functions, respectively. $\beta$ is the QCD beta function, $\gamma_\Delta^{\mu,R_\textrm{gap}}$ are the anomalous dimensions associated with the $R$-gap scheme, and $\delta$ denotes the corresponding renormalon-subtraction terms.}
    \label{tab:order}
\end{table}

\subsection{Factorization and Resummation}\label{subsec:fac}
Having outlined the general structure of the resummed prediction in
Sec.~\ref{subsec:overview}, we now derive the SCET factorization theorem
relevant for Centauric DIS 1-jettiness in the resummation region
$\tau_1^C \ll 1$.

In this regime, the measurement is sensitive only to the small light-cone components of collinear and anti-collinear radiation, as well as soft radiation. We therefore define the SCET expansion parameter $\lambda$ through $\tau_1^C \sim \lambda^2$ in the limit $\tau_1^C \ll 1$. For $R \sim \mathcal{O}(1)$, the relevant momentum regions obey the following power counting:
\begin{align}
\text{hard:}          &\qquad k^\mu \sim Q\, (1,\, 1,\, 1)\,, \label{eq:mode-hard} \\[4pt]
\text{collinear:}     &\qquad k^\mu \sim Q\, (\lambda^2,\, 1,\, \lambda)\,, \label{eq:mode-coll} \\[4pt]
\text{anti-collinear:}&\qquad k^\mu \sim Q\, (1,\, \lambda^2,\, \lambda)\,, \label{eq:mode-anticoll} \\[4pt]
\text{soft:}          &\qquad k^\mu \sim Q\, (\lambda^2,\, \lambda^2,\, \lambda^2)\,, \label{eq:mode-soft}
\end{align}
where the components are written in light-cone coordinates as in Eq.~\eqref{eq:light-cone-coordinates}.

At leading power in $\lambda$, the singular part of the perturbative $\tau_1^C$ spectrum factorizes into the following form, extending the framework derived in Ref.~\cite{Kang:2013nha}:
\begin{align}
\label{eq:inclusive-FT-renorm}
\frac{d\sigma^{\rm s}_{\rm PT}}{d\tau_1^C}
&=
\sum_q
\int d^2\bm{p}_\perp
\sigma_0
\int dt_J dt_B dk_S\,
\delta\left(\tau_1^C-\frac{t_J}{s_J}-\frac{t_B}{s_B}-\frac{k_S}{Q_R}\right)
\nonumber \\
&
\times
S_\textrm{PT}
\left(
k_S,
\mu\right)
J_q
\left(
t_J- (\bm{q}_\perp + \bm{p}_\perp)^2,\mu
\right)
\nonumber \\
&
\times
\left[
H_{q}(q_J,q_B,Q^2,\mu)
\mathcal{B}_q
\left(
t_B,
x_B,
\bm{p}_\perp,
\mu
\right)
+
(q\to \bar{q})
\right]\,,
\end{align}
where $H_q$, $J_q$, $\mathcal{B}_q$, and $S_{\rm PT}$ denote the hard, (inclusive) quark jet, generalized quark beam~\cite{Stewart:2009yx,Jain:2011iu}, and perturbative soft functions, respectively.
The factor $\sigma_0$ corresponds to the Born-level cross section for photon exchange:
\begin{align}
\label{eq:born-general}
\sigma_0
=
\frac{d\sigma_0}{dx dQ^2}
=
\frac{4\pi \alpha_{\rm EM}^2}{x_B^2s^2Q^2}
\frac{q_J\cdot k'\,q_B\cdot k + q_J\cdot k\, q_B\cdot k'}
{q_J\cdot q_B}
\,,
\end{align}
with $\alpha_\textrm{EM}$ being the electromagnetic coupling constant at the scale $Q$. Note that the matching coefficients yielding Eq.~\eqref{eq:born-general} are determined by contracting the leptonic and hadronic tensors using the label momenta defined by the reference vectors $n_{B,J}$, as detailed in Ref.~\cite{Kang:2013nha}.

The perturbative ingredients of these functions are universal for DIS 1-jettiness observables. Consequently, their explicit forms at $\mathcal{O}(\alpha_s^2)$ accuracy are identical to those compiled in Ref.~\cite{Ee:2025scz}. Specifically, the hard, soft, and jet functions correspond to Eqs.~(3.10), (3.16), and (3.17) of Ref.~\cite{Ee:2025scz}. The soft and jet functions are conventionally expressed in terms of the plus distributions $\mathcal{L}_m$ as
\begin{align}
\label{eq:S_J_plus}
S_\textrm{PT}(k,\mu) = \frac{1}{\mu} \sum_{m=-1} S_m(\alpha_s) \mathcal{L}_m(k/\mu), \quad
J(t,\mu) = \frac{1}{\mu^2} \sum_{m=-1} J_m(\alpha_s) \mathcal{L}_m(t/\mu^2),
\end{align}
where the plus distribution is defined by
\begin{align}\label{eq:plus_equation}
\begin{split}
\mathcal{L}_{m}(u)
=
\begin{cases}
\displaystyle
\delta(u), & \textrm{for $m=-1$,}
\\[1.5ex]
\displaystyle
\left[
\frac{\theta(u)\ln^m(u)}{u}
\right]_+, & \textrm{for $m>-1$.}
\end{cases}
\end{split}
\end{align}
The generalized beam function $\mathcal{B}_q$ is given in Eq.~(3.20) of Ref.~\cite{Ee:2025scz}, and its reduction to the ordinary beam function for $\tau_1^C$ is discussed below.

The factorization formula in Eq.~\eqref{eq:inclusive-FT-renorm} 
is governed by the following Lorentz-invariant scales
(see Table II of Ref.~\cite{Kang:2013nha}):
\begin{align}
\label{eq:1-jet-invariants}
s_J 
&=
\frac{q_B\cdot q}{q_B\cdot q_J} Q^2
\stackrel{\text{Breit}}{=} \frac{Q^3}{\omega_J\cos^2(\theta_J/2)}\,,
\nonumber \\
s_B
&=
-\frac{q_J\cdot q}{q_B\cdot q_J} Q^2
\stackrel{\text{Breit}}{=}\frac{Q^3\cos\theta_J}{\omega_B\cos^2(\theta_J/2)}\,,
\nonumber \\
Q_R 
&
=\frac{Q^2}{\sqrt{2q_B\cdot q_J}}
\stackrel{\text{Breit}}{=} \frac{Q^2}{\sqrt{\omega_B \omega_J}\cos(\theta_J/2)}\,.
\end{align}
The results after the second equalities are obtained by 
making use of Eqs.~\eqref{eq:P-q-mu-in-breit}, \eqref{eq:reference-general}, $n_B=\bar{n}$, and 
\begin{align}
n_J\cdot n_B &= 1-\bm{n}_J\cdot \bm{n}_B = 1 + \cos\theta_J = 
2\cos^2(\theta_J/2),
\end{align}
where the jet angle $\theta_J$ is measured in the Breit frame. 

Physically, $(s_B, s_J)$ relate the transverse virtualities ($t_B, t_J$) of the beam and jet sectors to the observable $\tau_1^C$, respectively, while $Q_R$ acts as the effective hard scale connecting the soft momentum $k_S$ to $\tau_1^C$. Choosing the specific values of $\omega_{B,J}$, these invariants encode the observable's dependence on the jet radius $R$ and the jet angle $\theta_J$.

While the geometric boundaries of the Centauro algorithm only fix the ratio $\omega_B/\omega_J$ (Eq.~\eqref{eq:wJ-wB-cond}), numerical evaluation requires the specific values $\omega_B$ and $\omega_J$. 
Among several possible normalization schemes, we adopt the choice that  optimizes perturbative stability and convergence of fixed-order results (referred to as Choice II below; defined in Eq.~\eqref{eq:tau1-choice2} and motivated in Sec.~\ref{sec:fixed-order}).
The resulting values for $\omega_{B,J}$, $Q_R$, and $s_{B,J}$ are summarized in Table~\ref{tab:tau1-variables}.

\begin{table}
    \centering
    \begin{tabular}{c|c|c|c|c|c}
        \hline
        \hline
        \rule{0pt}{2.8ex}
        Generic $\tau_1$ & $\omega_B$ & $\omega_J$ & $Q_R$ & $s_B$ & $s_J$  \\[0.6ex]
         \hline
         \hline
          \rule{0pt}{2.8ex}
        $\tau_1^{a,b}$ & $Q$ & $Q$ & $Q$ & $Q^2$ & $Q^2$ \\[0.5ex]
        $\tau_1^C$     & $Q\cos^2(\theta_J/2)$ & $Q\frac{4}{R^2}$ & $Q\frac{R}{2\cos^2(\theta_J/2)}$ & $Q^2\frac{\cos\theta_J}{\cos^4(\theta_J/2)}$ & $Q^2\frac{R^2}{4\cos^2(\theta_J/2)}$ \\[0.5ex]
         \hline
         \hline
    \end{tabular}
    \caption{
    Kinematic variables for different versions of 1-jettiness in the Breit frame. The definitions of the Lorentz invariants $Q_R$ and $s_{B,J}$ are given in Eq.~\eqref{eq:1-jet-invariants}. For Centauric 1-jettiness, the specific choice of $\omega_{B,J}$ follows Choice II in Eq.~\eqref{eq:tau1-choice2}, which optimizes perturbative stability. In the small jet-angle limit $\theta_J\to 0$, the exact expression $Q_R =
QR/[2\cos^2(\theta_J/2)]$ reduces to $Q_R = QR/2$.
    }
    \label{tab:tau1-variables}
\end{table}

\begin{figure}
    \centering
    \includegraphics[width=0.7\linewidth]{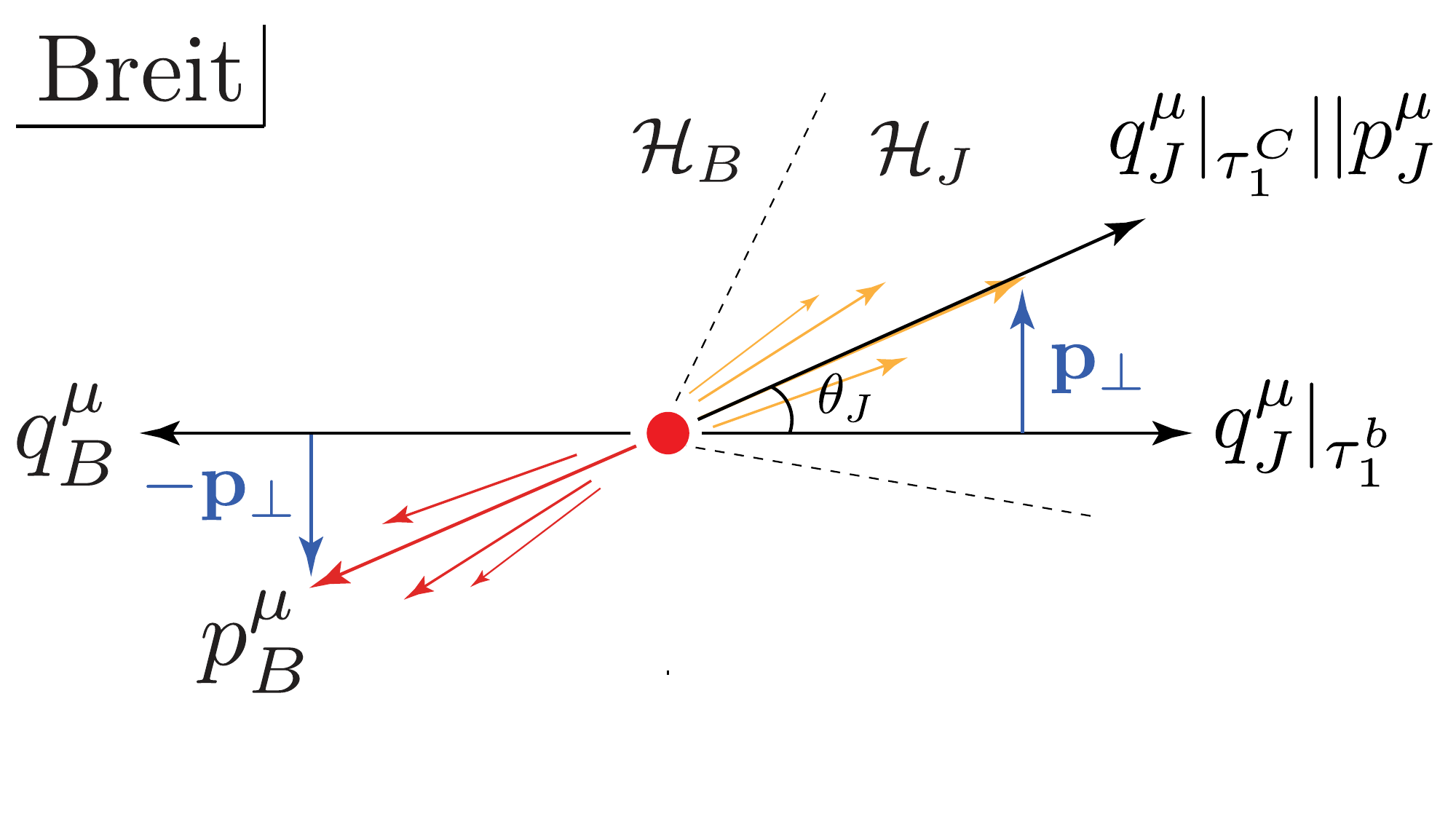}
    \vspace{-1em}
    \caption{Schematic of the Breit frame kinematics. The reference vectors $q_{B,J}$ define the measurement axes, while $p_{B,J}$ denote the physical momenta of the beam and jet. 
    $q_J^\mu|_{\tau_1^C}$ is the jet reference vector of $\tau_1^C$,
    while $q_J^\mu|_{\tau_1^b}$ is the jet reference vector of $\tau_1^b$.
    The angle $\theta_J$ represents the deviation of the jet axis from the canonical $z$-direction (see Sec.~\ref{subsec:centauro-jets}).}
    \label{fig:qJ-two-versions}
\end{figure}

A crucial feature of Centauric 1-jettiness is the dynamic alignment of the jet reference vector $q_J$ with the physical jet axis $p_J$. In the Breit frame, where the virtual photon is purely longitudinal ($\bm{q}_\perp = 0$), the transverse recoil in Eq.~\eqref{eq:inclusive-FT-renorm} simplifies based on this alignment. Unlike the $\tau_1^b$ case, where the jet axis is fixed opposite to the beam (leading to a coupled transverse recoil between the beam and jet sectors), 
the Centauric 1-jettiness selects $q_J$ to lie along the physical jet axis $p_J$. As in $\tau_1^a$~\cite{Kang:2013nha}, this alignment absorbs the jet-sector recoil into the reference vector itself.
Consequently, at leading power ($\tau_1^C \ll 1$), the relative transverse momentum between the physical jet and its reference vector vanishes:
\begin{align}
(\bm{q}_\perp+\bm{p}_\perp)^2 \to 0
\qquad\text{(for aligned }q_J\parallel p_J\text{ at leading power)}.
\end{align}
This simplification decouples the jet and beam measurements, allowing the cross section to be expressed as a simpler convolution. 
The corresponding geometric configurations for these reference vectors are illustrated in Fig.~\ref{fig:qJ-two-versions}.

With this simplification, Eq.~\eqref{eq:inclusive-FT-renorm} reduces to
\begin{align}
\label{eq:FT_aligned}
\frac{d\sigma^{\rm s}_{\rm PT}}{d\tau_1^C}
&=
\sum_q
\int d^2\bm{p}_\perp\,
\sigma_0
\int dt_J\,dt_B\,dk_S\,
\delta\left(\tau_1^C-\frac{t_J}{s_J}-\frac{t_B}{s_B}-\frac{k_S}{Q_R}\right)
\nonumber\\
&\times
S_{\rm PT}(k_S,\mu)\,
J_q(t_J,\mu)\,
\left[
H_q(q_J,q_B,Q^2,\mu)\,
\mathcal{B}_q(t_B,x_B,\bm{p}_\perp,\mu)
+(q\to\bar q)
\right].
\phantom{XX}
\end{align}
In Eq.~\eqref{eq:FT_aligned}, the dependence on $\bm{p}_\perp$ is explicit only in the generalized quark beam function $\mathcal{B}_q$, while the remaining factors depend on the jet direction (and thus implicitly on $\theta_J$ and $\bm{p}_\perp$) through $q_{B,J}$ and the invariants $Q_R$, $s_{B,J}$. For the kinematic regime of interest, $\theta_J\ll 1$, the residual $\bm{p}_\perp$-dependence of the hard, soft, and Born factors is power suppressed. At leading power one may therefore drop this dependence except for that of $\mathcal{B}_q$, such that the transverse integral reduces the generalized quark beam function to the ordinary quark beam function, \cite{Gaunt:2014xxa}
\begin{align}\label{eq:integrated-beam}
{B}_q(t_B,x_B,\mu)
=
\int d^2 \bm{p}_\perp \,
\mathcal{B}_q(t_B,x_B,\bm{p}_\perp,\mu)\,.
\end{align}
This approximation is formally justified because the power corrections scale as $\theta_J^2$ and $\bm{p}_\perp^2/Q^2$. Since we work in the regime where the jet is aligned along the $z$-axis (small $\theta_J$) and recoil is small (small $\tau_1^C$), these power corrections are subleading in the SCET power counting ($\lambda^2$).

\begin{figure}
    \centering
    \includegraphics[width=0.49\linewidth]{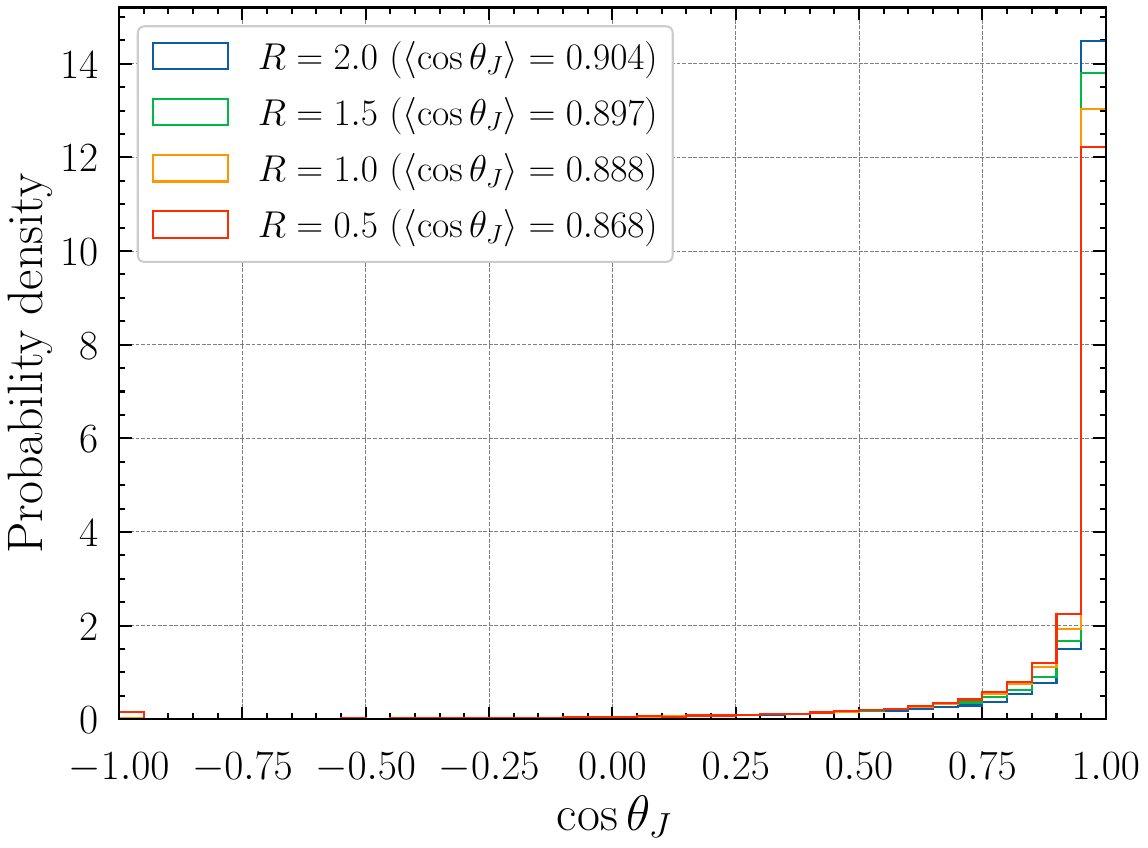}
    \vspace{-1em}
    \caption{The normalized distributions of $\cos\theta_J$ for various $R$ obtained from \textsc{Pythia} simulations at $\sqrt{s}=319~\mathrm{GeV}$, $Q=50~\mathrm{GeV}$ and $x_B=0.05$. The strong peak near $\cos\theta_J=1$ supports the approximation $\theta_J \ll 1$ used in the factorization theorem. The mean value of $\cos\theta_J$ for each $R$ is given by $\langle \cos\theta_J\rangle$.}
    \label{fig:costhJ}
\end{figure}

To empirically validate the approximation $\theta_J \ll 1$, we analyze the distribution of the jet axis polar angle using \textsc{Pythia} simulations. Figure~\ref{fig:costhJ} displays the normalized distributions of $\cos\theta_J$ for various jet radii $R$. The distributions are strongly peaked near $\cos\theta_J \approx 1$, confirming that the primary jet axis remains well-aligned with the $+z$-direction in the Breit frame. Quantitatively, the mean values of $\cos\theta_J$ lie in the range $[0.86, 0.90]$ for the $R$ values considered, providing strong support for the validity of the power expansion. 

The ordinary quark beam function in Eq.~\eqref{eq:integrated-beam} is factorized into the PDF $f_j$ and its radiative kernel $\mathcal{I}_{ij}$ as follows:
\begin{equation}
B_q(t_B,x_B,\mu)
=
\sum_j \int_x^1 \frac{dz}{z} 
\mathcal{I}_{qj}(t_B,z,\mu)\,
f_j(x_B/z,\mu)\,,
\end{equation}
where the index $j$ sums over the parton flavors in the proton. In perturbation theory, the kernel $\mathcal{I}_{ij}$ is expressed as a series of the plus distributions, and their explicit forms can be obtained from Refs.~\cite{Stewart:2010qs,Gaunt:2014xga}.
Following the convention in Ref.~\cite{Ee:2025scz}, we write the beam function as 
\begin{align}\label{eq:Bi-plus}
{B}_q(t,x,\mu)
=
\frac{1}{\mu^2}
\sum_{m=-1} 
{B}_{q,m}(x,\mu;\alpha_s)
\mathcal{L}_m(t/\mu^2),
\end{align}
where
\begin{equation}\label{eq:Bim-Jijm}
{B}_{q,m}(x,\mu;\alpha_s)
=
\sum_{j}
\int_x^1 \frac{dz}{z}
\mathcal{I}_{qj,m}(z;\alpha_s)
f_j(x/z,\mu).
\end{equation}

In the limit $\theta_J\to 0$, the Born prefactor $\sigma_0$ in
Eq.~\eqref{eq:FT_aligned} reduces to $\sigma_0^b$;
likewise, the hard function reduces to that of the $\tau_1^b$, as $H_q(q_J,q_B,Q^2,\mu)\to H_q(q+xP,P,Q^2,\mu)
\equiv H_q(y,Q^2,\mu) 
$ (following the convention of Ref.~\cite{Kang:2013nha}) and the Lorentz invariants reduce to $Q_R\to QR/2$, $s_B\to Q^2$, $s_J\to Q^2R^2/4$.%
\footnote{Throughout the remainder of this work, $Q_R$, $s_J$, and $s_B$ denote their leading-power values $QR/2$, $Q^2 R^2/4$, and $Q^2$, respectively, unless explicitly stated otherwise.}
Finally, the perturbative singular cross section in Eq.~\eqref{eq:FT_aligned} reduces to
\begin{align}
\label{eq:jet-mass-before-simp-2}
\frac{d\sigma^{\rm s}_{\rm PT}}{d\tau_1^C}
&=
\sigma_0^b
\int dt_J dt_B dk_S\,
\delta\left(\tau_1^C-\frac{t_J}{s_J}-\frac{t_B}{s_B}-\frac{k_S}{Q_R}\right)
S_\textrm{PT}
\left(
k_S,
\mu\right)
\nonumber \\
&
\times
\sum_q
J_q
\left(
t_J,\mu
\right)
\left[
H_{q}(y,Q^2,\mu)
B_q
\left(
t_B,
x_B,
\mu
\right)
+
(q\to \bar{q})
\right].
\end{align}
While this approximation is formally valid in the singular region ($\tau_1^C \ll 1$), contributions from finite $\theta_J$ are systematically accounted for through the fixed-order non-singular cross section discussed in Sec.~\ref{sec:fixed-order}. We now note that, for completeness, the anomalous dimensions required for N$^3$LL accuracy and the renormalization-group evolution kernels are summarized in Appendix~\ref{sec:N3LL-evolution}.

To provide a realistic description of the hadron-level distribution, we convolve  the partonic singular cross section with the gap-subtracted shape function as in Eq.~\eqref{eq:k-int-cumulant}. Performing the convolution and incorporating the renormalon subtraction defined in Eq.~\eqref{eq:shape-function-after-R-gap} yields the explicit form:
\begin{align}
\label{eq:singular-after-resummation-renorm-subtraction}
\sigma^\textrm{s}\left(\tau_1^C\right)
&=
\int dk\,
\sigma^\textrm{s}_\textrm{PT}\left(\tau_1^C-\frac{k}{Q_R}\right)
\left[
e^{-2\delta(R_{\textrm{gap}},\mu_S)(d/dk)}
F
\left(
k-2\Delta(R_{\textrm{gap}},\mu_S)
\right)
\right]
\nonumber \\
&=
\sigma_0^b
\frac{e^{\mathcal{K}-\gamma_\textrm{E}\Omega}}
{\Gamma(1+\Omega)}
\left(\frac{Q}{\mu_H}\right)^{\eta_H}
\left(\frac{\xi\left(\tau_1^C\right) s_B}{Q_R\mu_B^2}\right)^{\eta_{B}}
\left(\frac{\xi\left(\tau_1^C\right) s_J}{Q_R\mu_J^2}\right)^{\eta_{J}}
\left(\frac{\xi\left(\tau_1^C\right)}{\mu_S}\right)^{2\eta_S}
\nonumber \\
&
\times
\sum_q
\bigg[
H_{q}(y,Q^2,\mu_H)
\sum_{ \substack{ m_1,m_2, \\ m_3=-1 } }
{J}_{m_1}\left(\frac{\xi\left(\tau_1^C\right) s_J}{Q_R\mu_J^2}\right)
B_{q,m_2}\left(x_B,\mu_B,\frac{\xi\left(\tau_1^C\right) s_B}{Q_R\mu_B^2}\right)
S_{m_3}\left(\frac{\xi\left(\tau_1^C\right)}{\mu_S}\right)
\nonumber \\
&
\times
\sum_{\ell_1=-1}^{m_1+m_2+1}
\sum_{\ell_2=-1}^{\ell_1+m_3+1}
\sum_{\ell_3=-1}^{\ell_2+1}
V_{\ell_1}^{m_1m_2} 
V_{\ell_2}^{\ell_1 m_3}
V_{\ell_3}^{\ell_2}(\Omega)
I_{\ell_3}^{\Omega}
\left(\tau_1^C\right)
+
(q\leftrightarrow \bar{q})
\bigg]\,.
\phantom{XX}
\end{align}
Here, we introduce the rescaling variable $\xi\left(\tau_1^C\right) = Q_R\tau_1^C -2\Delta(R_{\textrm{gap}},\mu_S)$, 
which absorbs the gap subtraction into the natural soft scale and simplifies the convolution.
The function $I_{\ell_3}^{\Omega}\left(\tau_1^C\right)$ is 
\begin{align}
I_{\ell_3}^{\Omega}\left(\tau_1^C\right)
=
\xi\left(\tau_1^C\right)
\int_0^1 du\,
G_{\ell_3}^{\Omega}
\left(u\right)
\left\{
\exp\left[{\frac{2\delta(R_{\textrm{gap}},\mu_S)}
{\xi\left(\tau_1^C\right)}\frac{d}{du}}\right]
F
\left[
\xi\left(\tau_1^C\right)(1-u)
\right]
\right\},
\end{align}
where $G_\ell^\Omega(u)$ is given in Eq.~\eqref{eq:def-of-G}. 
For consistent renormalon subtraction, both the fixed-order SCET functions and the subtraction series in $\delta(R_{\textrm{gap}},\mu_S)$ are expanded in $\alpha_s$ and truncated at the same perturbative order (see Refs.~\cite{Abbate:2010xh,Bell:2023dqs}). 
Further theoretical details, including the treatment of non-perturbative power corrections and renormalon subtractions, are discussed in Appendix~\ref{app:Theory-Formalism}. 

Finally, the differential $\tau_1^C$ singular distribution is obtained by differentiating the cumulative cross section with respect to $\tau_1^C$. 
In the course of this process, the $\tau_1^C$-dependent scales, $\mu_i\left(\tau_1^C\right)$,  
are kept identical in the numerator below to avoid spurious contributions \cite{Abbate:2010xh}:
\begin{align}\label{eq:cumulant-to-differential}
\frac{d\sigma}{d\tau_1^C}
= \lim_{\epsilon\to 0}
\frac{\sigma_c(\tau_1^C +\epsilon, \mu_i\left(\tau_1^C\right))-\sigma_c(\tau_1^C -\epsilon, \mu_i\left(\tau_1^C\right))}{2\epsilon}.
\end{align}
Computing the differential distribution in this way by starting with the resummed cumulative distribution ensures full logarithmic accuracy at a given order, with ingredients truncated according to \tab{order} \cite{Almeida:2014uva}.

\subsection{Profile Functions}\label{subsec:profile}
The resummed cross section for the Centauric 1-jettiness distribution is governed by the renormalization scales of the hard, beam, jet, and soft functions ($\mu_H, \mu_B, \mu_J$, and $\mu_S$). To ensure the validity of the resummation and the perturbative stability of the matching across the entire spectrum, these scales must be specified as functions of the observable, $\mu_i\left(\tau_1^C\right)$.

The primary role of these profile functions is to satisfy the canonical scaling constraints required to avoid large logarithms in different kinematic regimes and ensure a smooth transition between the regimes:
\begin{enumerate}
\item \textbf{Peak region} ($\tau_1^C \sim \Lambda_{\rm QCD}/Q_R$): The scales maintain a hierarchy $\mu_H \gg \mu_{J,B} \gg \mu_S \sim \Lambda_{\rm QCD}$. The soft scale freezes out at a non-perturbative scale $\mu_0 > \Lambda_{\rm QCD}$, where the cross section is described by a convolution with a non-perturbative shape function.
\item \textbf{Tail region} ($\Lambda_{\rm QCD}/Q_R \ll \tau_1^C \ll 1$): The hierarchical separation $\mu_H \gg \mu_{B,J} \gg \mu_S$ persists.
In this region, the soft scale enters the perturbative regime, and the soft radiation is described by the perturbative soft function. The non-perturbative effects can be expanded via an operator product expansion (OPE), where their leading contribution is captured by a universal shift parameter, $\Omega_1$ (defined in Sec.~\ref{subsec:soft}).
The natural profile scales that minimize the fixed-order logarithms in Eq.~\eqref{eq:singular-after-resummation-renorm-subtraction} are\footnote{In Choice II (Table~\ref{tab:tau1-variables}), the $R$-dependence is absorbed entirely into $\omega_J$; consequently $\mu_B \sim Q\sqrt{\tau_1^C}$ contains no factor of $R$.}:
\begin{align}
\mu_H \sim Q, \quad \mu_B \sim Q\sqrt{\tau_1^C}, \quad \mu_J \sim \frac{QR}{2}\sqrt{\tau_1^C}, \quad \mu_S \sim \frac{QR}{2}\tau_1^C.
\label{eq:profile-scale-r-dependent}
\end{align}
The factor of $R/2$ in $\mu_J$ and $\mu_S$ reflects the $R$-dependent hard scale $Q_R = QR/2$ entering the jet and soft sectors of the factorization theorem.

\item \textbf{Far-tail region} ($\tau_1^C \sim 1$): Resummation is no longer required as the logarithms become small. To match onto standard fixed-order QCD and ensure proper cancellation between the singular and non-singular cross sections, all scales must converge to the common hard scale, $\mu_H = \mu_B = \mu_J = \mu_S \sim Q$.

\end{enumerate}

While these canonical scalings are dictated by the underlying theory, the profile functions that smoothly interpolate between the three regions are not uniquely determined. Since the resummed cross section retains a residual dependence on these functional forms at any finite perturbative order, we use variations of the profiles to estimate the perturbative uncertainty associated with the resummation and matching procedure.

In this work, the specific functional forms of the profile functions follow the construction established in Ref.~\cite{Ee:2025scz}, which we have adapted to incorporate the explicit jet-radius ($R$) dependence of the Centauro jet definition. 
We define the profile functions as:
\begin{align}
\label{eq:profile_function}
\begin{split}
\mu_H &= \mu, \\
\mu_{B,J}\left(\tau_1^C\right) &= \left[1+e_{B,J}\,g\left(\tau_1^C\right)\right] \sqrt{\mu\, \mu_\textrm{run}\left(\tau_1^C,\mu\right)}, \\
\mu_{S}\left(\tau_1^C\right) &= \left[1+e_{S}\,g\left(\tau_1^C\right)\right] \mu_\textrm{run}\left(\tau_1^C,\mu\right),
\end{split}
\end{align}
where $e_{B,J,S}$ are parameters for scale variations. The function $g$ defined in Eq.~\eqref{eq:g_profile} ensures  
that scale variations remain frozen in both the deep non-perturbative regime (${\tau_1^C} < t_0$) and the fixed-order-dominated region (${\tau_1^C} > t_3$).
Note that we define the beam, jet, and soft scales in terms of a common running scale $\mu_{\rm run}(\tau_1^C, \mu)$ to ensure that all scales strictly merge with the hard scale $\mu$ in the far-tail region, even though the canonical scales in Eq.~\eqref{eq:profile-scale-r-dependent} carry explicit
$R$-dependence.

The $R$-dependence is instead incorporated into the transition parameters $\{\mu_0, r, t_0, t_1, t_2, t_3\}$ of the running scale, defined by
\begin{align}
\label{eq:running-scale}
\mu_\textrm{run}(\tau_1^C, \mu, \{t_i\}) = \mu \times
\begin{cases}
\mu_0/\mu, & 0 \le \tau_1^C < t_0 \\
\zeta(\tau_1^C, \{t_0,\mu_0, 0\}, \{t_1, rt_1, r\}), & t_0 \le \tau_1^C < t_1 \\
r \tau_1^C, & t_1 \le \tau_1^C < t_2 \\
\zeta(\tau_1^C, \{t_2,rt_2, r\}, \{t_3, 1, 0\}), & t_2 \le \tau_1^C < t_3 \\
1, & \tau_1^C \ge t_3,
\end{cases}
\end{align}
where $\zeta$ is a quadratic spline that ensures $C^1$ continuity (continuous first derivative) between segments.
The transition parameters are chosen with explicit $R$-dependence as
\begin{align}\label{eq:profile_setting}
\begin{split}
\mu &= Q, \quad \mu_0 = 1.1~\textrm{GeV}, \quad r = \frac{R}{2}, 
\\
t_0 &= \frac{2}{R}\frac{1~{\rm GeV}}{Q}, \quad
t_1 = \frac{2}{R}\frac{4~{\rm GeV}}{Q},
\\
t_2 &= \sqrt{\frac{2}{R}}\frac{12.5~\textrm{GeV}}{Q}, \quad
t_3 = \min
\left[
\sqrt{\frac{2}{R}}\frac{17.5~\textrm{GeV}}{Q},
0.9
\right].
\end{split}
\end{align}
The freeze-out scale $t_0$ is set by the condition that subleading power corrections to the OPE-based shift become non-negligible (above the percent level) compared with the leading-power contribution.
The tail region $[t_1, t_2]$ is chosen as the range over which the singular fixed-order cross section dominates over the non-singular contribution.
The transition point $t_3$ marks the onset of the far-tail regime, above which the $\tau_1^C$ distribution is described entirely by fixed-order QCD.
The slope $r$ in the tail interval is chosen to match the natural scaling $\mu_S \sim (QR/2)\tau_1^C$ given in Eq.~\eqref{eq:profile-scale-r-dependent}.

\begin{figure}[t]
    \centering
    \vspace{-1em}
    \includegraphics[width=0.8\linewidth]{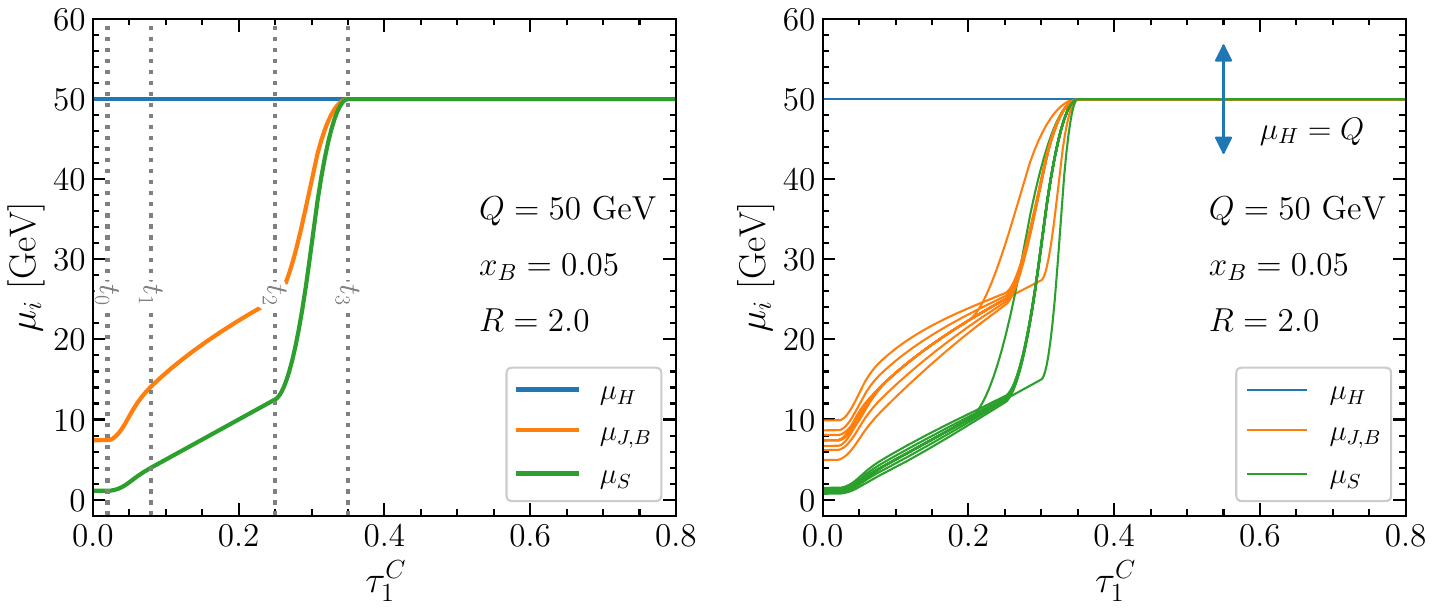}
    \includegraphics[width=0.8\linewidth]{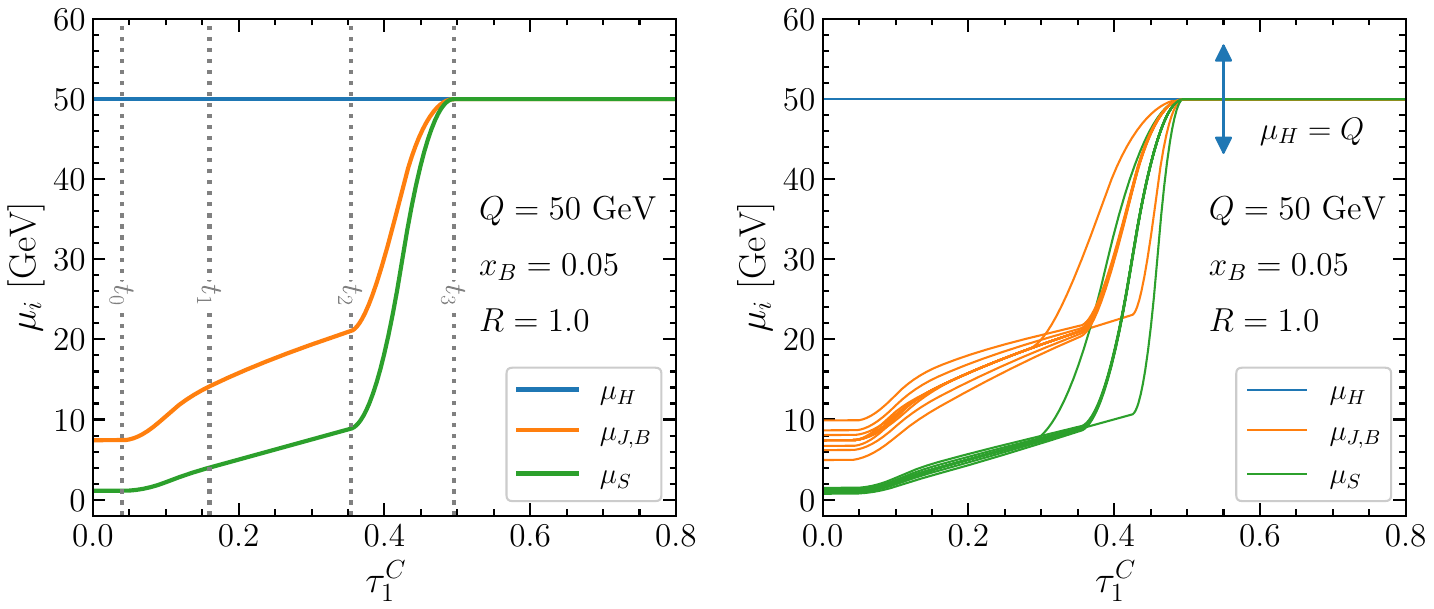}
    \vspace{-1em}
\caption{
    The profile functions $\mu_H, \mu_{J,B}, \mu_S$ at $(Q,x_B) = (50~\textrm{GeV}, 0.05)$ with $R=2.0$ (top) and with $R=1.0$ (bottom).
    Left panels show the central profiles and transition points $t_i$.
    Right panels illustrate the envelope of scale variations defined in Eq.~\eqref{eq:singular_scale_variations}.
}
    \label{fig:profile_functions}
\end{figure}

Figure~\ref{fig:profile_functions} shows representative profile functions for two jet radii together with the scale variations used to estimate theoretical uncertainties. Further details on the spline function $\zeta$, the systematic variations of the profile function, and the treatment of the non-singular scale $\mu_{\rm ns}$ are provided in Appendix~\ref{sec:profile}.

\subsection{Universality and the \texorpdfstring{$R$}{R}-Dependent Leading Non-perturbative Shift}
\label{subsec:soft}
Having established the factorized description of the singular cross section in
Sec.~\ref{subsec:fac}, we now analyze the soft function and the structure of the
leading non-perturbative corrections for DIS 1-jettiness observables involving
Centauro jet boundaries. The goal of this subsection is to determine the form of
the leading non-perturbative power correction and its dependence on the jet radius
$R$. We proceed in two steps. First, in Sec.~\ref{subsubsec:Cent-soft} we consider DIS
1-jettiness defined directly using the Centauro jet algorithm and show that the
measurement function in the soft function depends only on rapidity at leading power, implying that the leading
non-perturbative corrections are governed by the same universal matrix element
$\Omega_1$ that appears in thrust-like event shapes, with coefficients that scale
as $R$ and $1/R$. We then show in Sec.~\ref{subsubsec:Centauric-soft} that the
Centauric DIS 1-jettiness observable can be mapped onto the hemisphere soft
function through a simple rescaling of the reference directions. This mapping
determines the perturbative soft function and fixes the final form of the
non-perturbative shift in the $\tau_1^C$ distribution, which is found to scale as
$\Omega_1/Q_R \sim 1/R$.

\subsubsection{DIS 1-jettiness with a Centauro jet algorithm}\label{subsubsec:Cent-soft}
We begin by considering the soft function associated with the DIS 1-jettiness
observable when the jet region is defined by the Centauro jet clustering
algorithm without additional weighting factors. The soft function can be
written in terms of soft Wilson lines and energy-flow operators as the vacuum
matrix element
\begin{align}
S(k_J, k_B, n_J \cdot n_B,\mu)
& =
\frac{1}{N_c}
\operatorname{Tr}\,
\Bigg\langle
[Y_{n}^\dagger(0) Y_{\bar n}(0)] \delta\left(
k_J - \int dr\,dy\,d\phi\,
f^J_{\rm alg}(r,y,\phi)\,
\hat{\mathcal{E}}_T(r,y,\phi)
\right)
\nonumber\\
&\times
\delta\left(
k_B - \int dr\,dy\,d\phi\,
f^B_{\rm alg}(r,y,\phi)\,
\hat{\mathcal{E}}_T(r,y,\phi)
\right)
[Y_{\bar n}^\dagger(0) Y_n(0)]
\Bigg\rangle ,
\end{align}
where $f^{\kappa=J,B}_{\rm alg}$ encodes the measurement in the jet and beam
regions, and $\hat{\mathcal{E}}_T$ denotes the transverse velocity operator
\cite{Mateu:2012nk}. Acting on a soft final state $|X_s\rangle$,
this operator measures the transverse energy flow,
\begin{align}
\hat{\mathcal{E}}_{T}(r,y,\phi)|X_s\rangle
&=
\sum_{i\in X_s}
m_{T i}\,
\delta(r-r_i)\delta(y-y_i)\delta(\phi-\phi_i)
|X_s\rangle ,
\end{align}
where $r_i=p_{T i}/m_{T i}$ and $m_{T i}$ is the transverse mass of particle
$i$.

The soft Wilson lines are
\begin{align}
Y_n(x)
&=
\mathcal{P}
\exp\Bigl[
i g \int_0^\infty ds\, n\cdot A_s(x+s n)
\Bigr],
\\
Y_{\bar n}^\dagger(x)
&=
\mathcal{P}
\exp\Bigl[
i g \int_{-\infty}^0 ds\, \bar n\cdot A_s(x+s\bar n)
\Bigr].
\end{align}
Here $Y_n$ is the outgoing Wilson line along the jet direction, while
$Y_{\bar n}^\dagger$ is the incoming Wilson line along the beam direction.
Equivalently, $Y_{\bar n}$ denotes the corresponding Wilson line in the
conjugate amplitude. The mixed incoming--outgoing structure distinguishes the
DIS soft function from the $e^+e^-$ dijet case, which contains two outgoing
Wilson lines, and from Drell-Yan in $pp$ collisions, which contains two incoming Wilson
lines and one outgoing Wilson line \cite{Chay:2004zn,Arnesen:2005nk,Kang:2015moa}.

For dijet production in an $e^+e^-$ collision, the corresponding soft function
is
\begin{align}
S^{e^+ e^-}(k_J, k_B, n_J \cdot n_B,\mu)
& =
\frac{1}{N_c}
\operatorname{Tr}\,
\Bigg\langle
[Y_{n}^{\dagger}(0) Y_{\bar n}^{\dagger}(0)] \delta\!\left(
k_n - \int dr\,dy\,d\phi\,
f^n_{\rm alg}(r,y,\phi)\,
\hat{\mathcal{E}}_T(r,y,\phi)
\right)
\nonumber\\
\times &
\delta\!\left(
k_{\bar n} - \int dr\,dy\,d\phi\,
f^{\bar n}_{\rm alg}(r,y,\phi)\,
\hat{\mathcal{E}}_T(r,y,\phi)
\right)
[Y_{\bar n}(0) Y_n(0)]
\Bigg\rangle \,.
\end{align}
This distinction matters for the universality of the non-perturbative matrix
element discussed below.

To specialize the above expressions to DIS 1-jettiness with a Centauro jet algorithm, we separate the measurement into two ingredients: the
partition of the soft emission into the jet or beam region, and the weight with which that emission contributes to the observable. 
Using Breit-frame rapidity variables, and taking the jet direction to be
aligned with the positive $z$ axis, the soft momentum satisfies
\begin{align}
    n_B\cdot k_i = m_{T i} e^{-y_i}, 
    \qquad
    n_J\cdot k_i = m_{T i} e^{y_i},
\end{align}
up to power corrections. Thus the measurement weights appearing in the
energy-flow representation are $e^{-y}$ in the beam region and $e^{y}$ in the
jet region.

We now determine the Centauro partition in the soft limit. The simple
rapidity boundary used below assumes a single energetic jet aligned with the
$+z$ direction in the Breit frame, so that the jet axis is described by
$\theta_J\simeq 0$. It also assumes the anti-$k_T$ version of the Centauro algorithm, for which soft emissions do not cluster among themselves and instead cluster directly to the hard jet or beam directions. In this limit, the Centauro distance of a soft particle to the jet is determined by the condition $\bar\eta_i < R$. Thus for a massless emission, we find from Eq.~\eqref{eq:eta_definition} that
\begin{align}
f^{B}_{\rm Cent}(r,y,\phi)
= \Theta\left(\ln\left(\frac{R}{2}\right)-y\right)\,
e^{y}\,,
\qquad
f^{J}_{\rm Cent}(r,y,\phi)
= \Theta\left(y - \ln\left(\frac{R}{2}\right)\right)\,
e^{-y}\,.
\end{align}
At leading power, both partitions depend only on the rapidity $y$, with 
no residual $r$ or $\phi$ dependence. We retain the full $(r, y, \phi)$ 
signature for notational consistency with the general energy-flow 
representation.
We have verified that replacing the anti-$k_T$ clustering with the Cambridge--Aachen algorithm leads to negligible changes in the extracted boundary and resulting measurement functions at leading power, indicating that the result is insensitive to this choice of clustering metric. 

Since the Centauro partition and measurement weights depend only on rapidity at leading power, the leading non-perturbative correction can be expressed in terms of the universal matrix element
\begin{align}
\label{eq:Omega_1-matr}
\Omega_1
=
\frac{1}{N_c}\,
\mathrm{Tr}\,
\langle [Y_n^\dagger Y_{\bar n}]
\,\hat{\mathcal{E}}_T(y=0)\,
[Y_{\bar n}^\dagger Y_n] \rangle ,
\end{align}
which is the same soft matrix element governing DIS thrust and jet-mass event
shapes \cite{Mateu:2012nk,Kang:2013nha}, all of which belong to the ``J-scheme'' universality class. It is not necessarily the same as the similar parameter in $e^+e^-$ or $pp$ event shapes due to the different directions of the Wilson lines (incoming vs. outgoing). To at least order $\alpha_s^2$ they are known to be the same perturbatively \cite{Kang:2015moa}, but not necessarily non-perturbatively. The rough values extracted for $\Omega_1$ for DIS thrust in \cite{Ee:2025scz} suggest it may be a bit larger than that in $e^+e^-$ event shapes. A relation between $\Omega_1$ in DIS thrust vs. the parameter governing the leading shift in jet mass in $pp\to\text 1$ jet at leading order in an expansion in small $R$ does exist non-perturbatively \cite{Stewart:2014nna}.

In this case the non-perturbative corrections to the beam and jet regions take
the form
\begin{align}
\label{eq:JBshifts}
\Omega^J_{\rm Cent} = C^J_{\rm Cent}(R)\,\Omega_1,
\qquad
\Omega^B_{\rm Cent} = C^B_{\rm Cent}(R)\,\Omega_1,
\end{align}
with coefficients
\begin{align}
\label{eq:JBcoeffs}
C_{\rm Cent}^J = \frac{R}{2},
\qquad
C_{\rm Cent}^B = \frac{2}{R}.
\end{align}
The coefficients $C^J_{\rm Cent}$ and $C^B_{\rm Cent}$ are determined by the
location of the jet boundary in rapidity and therefore scale linearly and
inversely with the jet radius $R$, respectively.
These coefficients are exact in $R$---no expansion for small or large $R$ has been made to obtain these values, as is needed for some other algorithms (\appx{soft-universality}) or for observables like jet mass in $pp\to 1$ jet \cite{Stewart:2014nna}.

A comment about the ``J-scheme'' mentioned after \eq{Omega_1-matr}: the leading-power soft measurement in $\tau_1^C$ therefore depends only on
rapidity, with no residual transverse-velocity dependence. In the
terminology of Ref.~\cite{Mateu:2012nk}, this is
the defining property of the \emph{J-scheme} universality class, in
which all dijet event shapes share a common non-perturbative shift
governed by the single matrix element $\Omega_1$ of
Eq.~\eqref{eq:Omega_1-matr}. A notable feature of the
Centauro construction is that the rapidity-based clustering metric
realizes this scheme automatically, with no scheme substitution
required.

Results for some other jet algorithms are given in Appendix~\ref{app:soft-universality}. Since the Centauric DIS 1-jettiness observable introduced in Sec.~\ref{subsec:Def} reproduces the same jet boundary through a global minimization measure, it belongs to the same universality class. We now determine its perturbative soft function by mapping it onto the hemisphere soft function.

\subsubsection{Centauric DIS 1-Jettiness}\label{subsubsec:Centauric-soft}
We now consider the Centauric DIS 1-jettiness observable, in which the  measurement is modified by the inclusion of weights $\omega_{B,J}$ that alter the soft function while preserving the underlying Centauro jet partition.

The soft function for the Centauric observable therefore differs from that of the unweighted Centauro case only through these measurement weights. We now specialize the measurement functions to this observable and determine the resulting soft function. We find that the restriction on the phase space is governed by
\begin{align}
f^J_{\rm Cent}(r,y,\phi)
& =
\Theta\left(
\frac{R}{2}\, n_B\cdot k_i
-
\frac{2}{R}\, n_J\cdot k_i
\right)\,,
\nn \\
f^B_{\rm Cent}(r,y,\phi) & = \Theta\left(
\frac{2}{R}\, n_J\cdot k_i-\frac{R}{2}\, n_B\cdot k_i
\right)\,,
\label{eq:centauro-partition}
\end{align}
where, in contrast to the unweighted case in Sec.~\ref{subsubsec:Cent-soft}, the same notation $f^{J,B}_{\rm Cent}$ now denotes the partition function alone, with the measurement weights $\omega_{B,J}$ already absorbed.

Introducing rescaled reference directions
\begin{align}
n_J' = \frac{n_J}{R/2},
\qquad
n_B' = \frac{n_B}{2/R},
\end{align}
this partition can be written as
\begin{align}
\Theta_{\rm alg}(k_i)
=
\Theta\left(
n_B'\cdot k_i - n_J'\cdot k_i
\right).
\end{align}
This form corresponds precisely to the hemisphere decomposition familiar from
thrust-like event shapes. Consequently, the Centauric soft function can be
mapped onto the standard hemisphere soft function after a simple rescaling of
the momentum variables,
\begin{align}
S(k_J,k_B,n_J,n_B,\mu)
=
S_{\rm hemi}\left(
\frac{k_J}{R/2},\,\frac{k_B}{2/R},\,\mu
\right).
\end{align}
This mapping allows the perturbative soft function entering the Centauric
factorization theorem to be obtained directly from the known hemisphere soft
function, whose two-loop expression is available in the literature and forms
the basis of our N$^3$LL resummed predictions.

Using the weights in Table~\ref{tab:tau1-variables}, and taking $\theta_J \rightarrow 0$, we find that the non-perturbative shift from the soft function takes on the form
\begin{align}
\label{eq:delta-tau-main}
\boxed{    \Delta\tau = \frac{4 (\Omega_1 - \Delta)}{Q R} + \mathcal{O}\left(\frac{\Lambda_{\rm QCD}^2}{Q^2}\right) } \,,
\end{align}
in the tail region of $\tau_1^C$, $\Lambda_\textrm{QCD}/Q_R \ll \tau_1^C \ll 1$. Our choice of the weights used in Table~\ref{tab:tau1-variables} will be motivated in Sec.~\ref{sec:fixed-order}. \textbf{The exactness of the $R$-dependence in \eq{delta-tau-main}, and more generally \eqs{JBshifts}{JBcoeffs}, with no expansion needed either in small or large $R$, is the central result of our paper.}

\section{Fixed-Order Calculation and Matching}
\label{sec:fixed-order}
In this section, we construct the fixed-order ingredients required to match the prediction for the resummed singular cross section in \eq{singular-after-resummation-renorm-subtraction} onto the full QCD cross section, accurate for both small and large $\tau_1^C$, for the Centauric 1-jettiness observable. A central complication is that the Centauric 1-jettiness observable depends explicitly on both the identification of a primary jet axis and on $R$-dependent measurement weights. In the resummation region this distinction is immaterial up to power corrections, but in the fixed-order calculation for larger $\tau_1^C$, the presence of widely separated energetic particles in the final state makes the choice of jet-selection and measurement procedure essential. This section therefore develops a fixed-order implementation of the observable that preserves the geometric connection to the Centauro jet algorithm while maintaining perturbative stability, and uses it to extract the non-singular contribution needed for the fixed-order matching. We emphasize that these complications do not significantly affect the resummation region where we propose to focus phenomenological effort on extracting $\alpha_s$ and $\Omega_1$ from measurements of Centauric 1-jettiness; however, we deal with them nevertheless in order to present results that will be easier to compare to experimental measurements for all values of $\tau_1^C$. The reader not interested in these details relevant for large $\tau_1^C$ may skip to \sec{omega1} to continue with our discussion of non-perturbative corrections in the resummation region.

This section is organized as follows. In Sec.~\ref{subsec:setup}, we define the non-singular contribution and describe how it is obtained from fixed-order QCD calculations. In Sec.~\ref{subsec:instability}, we examine the perturbative stability of the observable at fixed order and show how certain choices of measurement weights can lead to discontinuities in the distribution, motivating the hybrid definition adopted in this work. In Sec.~\ref{subsec:implementation}, we describe the numerical implementation used to compute the fixed-order cross sections using \texttt{NLOJet++}. Finally, in Sec.~\ref{subsec:non-singular_results}, we present the resulting non-singular cross sections and demonstrate how they combine with the singular contributions to produce a prediction that is valid across the full $\tau_1^C$ spectrum.

\subsection{Calculation Setup}\label{subsec:setup}
We determine the non-singular contributions using fixed-order cross sections computed up to NLO [$\mathcal{O}(\alpha_s^2)$] accuracy. By combining the resummed singular contribution with the non-singular contribution, as introduced in Eq.~\eqref{eq:sing-plug-ns}, we obtain the complete N$^3$LL + NLO prediction, accurate across the full $\tau_1^C$ spectrum.

The non-singular cross section is defined by subtracting the fixed-order expansion of the singular contribution, $d\sigma^\textrm{s, FO}_\textrm{PT}$, from the full fixed-order QCD cross section, $d\sigma^\textrm{FO}_\textrm{PT}$:
\begin{align}
\label{eq:def-of-non-singular}
\frac{d\sigma^\textrm{ns}_\textrm{PT}}{d\tau_1^C}
=
\frac{d\sigma^\textrm{FO}_\textrm{PT}}{d\tau_1^C}
-
\frac{d\sigma^\textrm{s, FO}_\textrm{PT}}{d\tau_1^C}.
\end{align}
Here, all terms are evaluated at a fixed order in $\alpha_s$ (without resummation).
Consistent with the approximation used in the resummation derivation, we evaluate $d\sigma^\textrm{s, FO}_\textrm{PT}/d\tau_1^C$ in the limit $\theta_J \to 0$. In contrast, the full fixed-order computation $d\sigma^\textrm{FO}_\textrm{PT}/d\tau_1^C$ retains the exact $\theta_J$ dependence determined by the Centauro algorithm. 
Consequently, the non-singular term defined in Eq.~\eqref{eq:def-of-non-singular} systematically captures the finite-$\theta_J$ corrections. Although these corrections are formally power-suppressed in the SCET counting in the small $\tau_1^C$ region, they become numerically important beyond the tail region, where they constitute a sizable fraction of the total cross section.

Unlike the previously-studied $\tau_1^a$ and $\tau_1^b$ observables, for which analytic non-singular results are available at $\mathcal{O}(\alpha_s)$ in Ref.~\cite{Chu:2022jgs} and Ref.~\cite{Kang:2014qba}, respectively, the Centauric 1-jettiness involves explicit use of the jet algorithm and $R$-dependent weights that make a purely analytic treatment difficult.\footnote{Note that in Ref.~\cite{Chu:2022jgs}, while the jet axis was chosen for general $R$ using Centauro jet algorithm, the weighting factors $\omega_{B,J}$ were set to $Q$, which makes the observable consistent with the $\tau_1^a$ introduced in \cite{Kang:2013nha}.} For this reason, we compute the full QCD distribution $d\sigma^\textrm{FO}_\textrm{PT}$ numerically using \texttt{NLOJet++} \cite{Nagy:2001xb, Nagy:2003tz}, which implements the Catani-Seymour dipole subtraction method \cite{Catani:1996vz}. We perform this calculation at both $\mathcal{O}(\alpha_s)$ and $\mathcal{O}(\alpha_s^2)$ in order to extract the non-singular contributions entering the N$^3$LL+NLO prediction.

The numerical results from \texttt{NLOJet++} are provided as histograms for the differential cross section, making it natural to work directly with the differential cross section. Since \texttt{NLOJet++} includes only virtual photon exchange, the non-singular contributions in our predictions are restricted to this channel, while contributions from $Z^0$-boson exchange are included in the singular terms through the hard function in the SCET factorization formula \cite{Kang:2013nha}.
This treatment is justified not by a parametric suppression of the $Z^0$ contribution, but by where it enters our prediction: the $Z^0$-exchange is retained in full through the hard function in the singular terms, and is absent only from the non-singular remainder, which is itself a higher-order correction to the singular cross section. 
In the kinematic range considered here, $Q \sim M_Z/2$, the momentum transfer is spacelike and remains far from the $Z^0$ resonance, so the $\gamma^*$/$Z^0$ propagator ratio $Q^2/(Q^2+M_Z^2) \approx 0.2$ is moderate, and the $\gamma^*$-$Z^0$ interference and pure $Z^0$ contributions amount to at most a few percent of the neutral-current cross section. 
Since these effects are dropped \emph{only} in the non-singular remainder, their net impact on our predictions is well below the percent level, comfortably within our perturbative accuracy. This ensures that our results remain robust for DIS phenomenology at current and future facilities, such as the EIC. It would be appropriate, for higher precision phenomenology in the future, to include the $Z^0$-boson contributions, e.g. using NNLOJET \cite{NNLOJET:2025rno}.

To ensure numerical accuracy and validate our setup, we have checked that our numerical results recover the known analytic predictions for $\tau_1^a$ in the appropriate limit ($\theta_J \to 0$, $R=2$).

\subsection{Observable Definition and Perturbative Stability}
\label{subsec:instability}
In order to construct the non-singular contribution at fixed order, the Centauric
1-jettiness observable must be evaluated on partonic final states with a finite
number of particles. In this regime, the observable depends not only on the
definition of the jet axis but also on the choice of measurement weights
$\omega_{B,J}$ that determine how radiation is partitioned between the beam and
jet regions. While different choices of weights that satisfy the Centauro
boundary condition are equivalent up to power corrections in the resummation
region, their behavior at fixed order can differ significantly and can affect the
perturbative stability of the cross section. In this subsection we examine this
behavior and show that certain choices of weights can lead to discontinuities in
the fixed-order distribution at larger $\tau_1^C$, motivating the hybrid definition adopted in this
work, which preserves the geometric connection to the Centauro jet boundary while
ensuring a stable perturbative prediction suitable for matching.

\begin{figure}[t]
    \centering
    \includegraphics[width=0.49\linewidth]{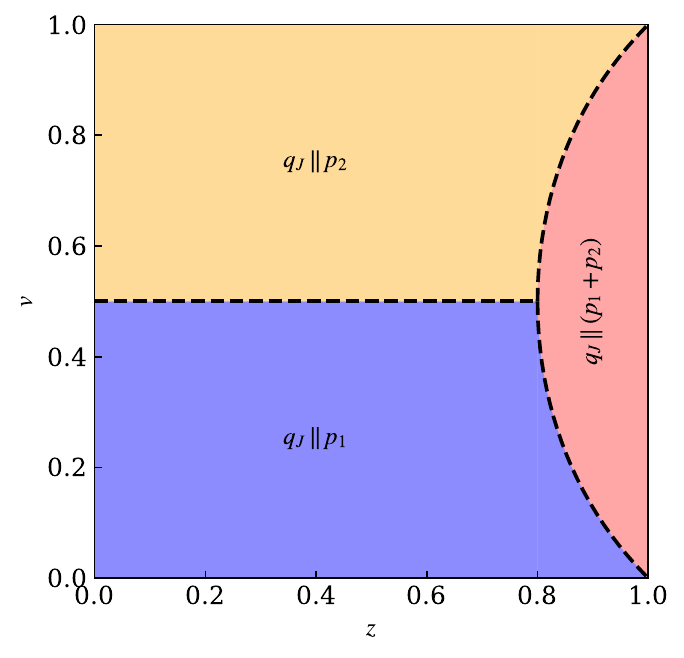}
\vspace{-1em}
    \caption{Phase space for the LO contribution (two-parton final state) illustrating the domains where the jet axis aligns with $p_1$, $p_2$, or $p_1+p_2$. The LO phase space is parameterized by $z$ and $v$, as detailed in Appendix~\ref{app:two-particle-kinematics}. We note that the lower bound of $z$ is in practice given by the momentum fraction of the incoming parton, $x$. The boundaries represent the transition points where the algorithm switches the jet definition.}
    \label{fig:phase_space_LO}
\end{figure}
\begin{figure} 
\centering \vspace{-1em} 
\includegraphics[width=0.89\linewidth]{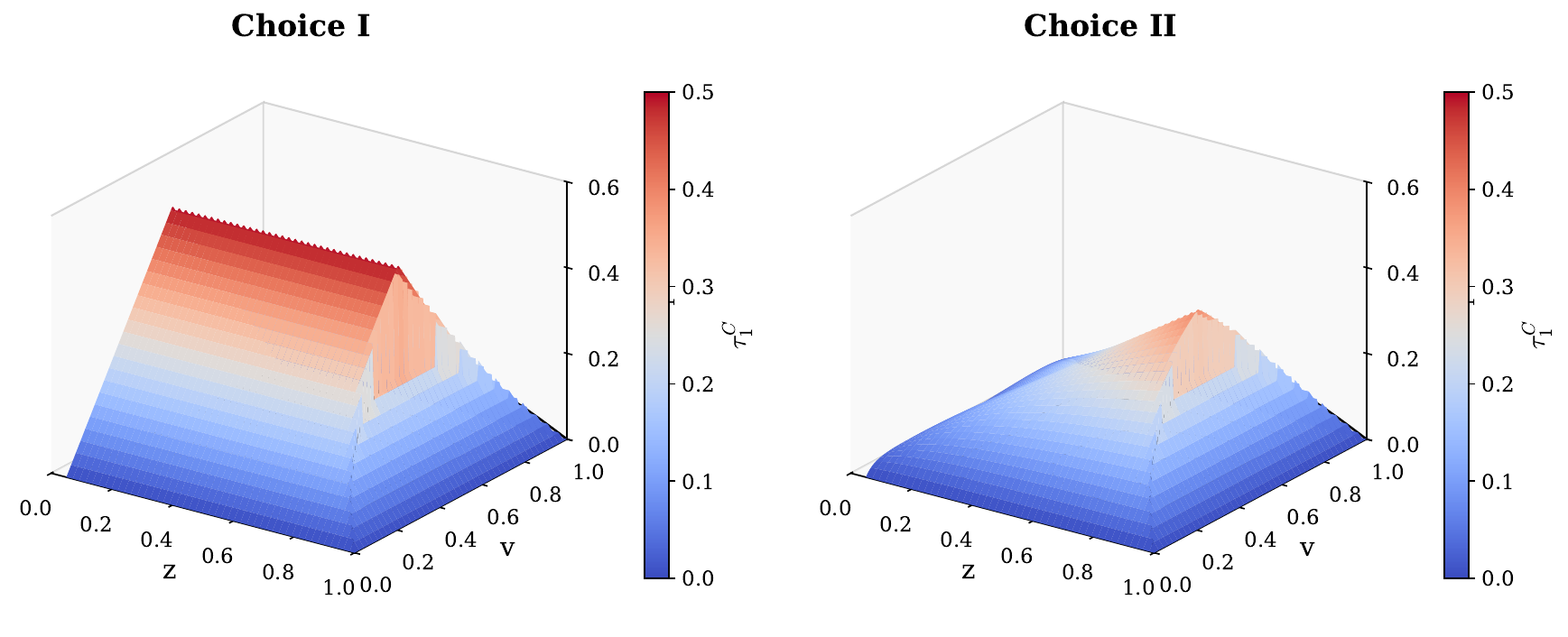} 
\vspace{-1em} 
\caption{The $\mathcal{O}(\alpha_s)$ fixed-order $\tau_1^C$ distributions for $R = 2.0$ as functions of $z$ and $v$. The left panel corresponds to Choice I, while the right panel corresponds to Choice II. In the left panel, $\tau_1^C$ becomes independent of $z$ along the boundary $v = 0.5$, whereas in the right panel $\tau_1^C$ varies smoothly across this line. Note that the available phase space vanishes at $\tau_1^C = 0.5$. In particular, with Choice I, there is a whole ridge of events that will pile up at $\tau_1^C$ before the distribution falls to zero, while with Choice II, there is only one point in phase space with $\tau_1^C=0.5$, causing a smoother vanishing of the distribution.
The kinks observed in Fig.~\ref{fig:LO-fixed-tau1a} can be attributed to the singular structures in these surface plots at the boundary $q_J \parallel (p_1 + p_2)$.
} 
\label{fig:LO-fixed-tau1a-two-vers}
\end{figure}
\begin{figure} 
\centering \vspace{-1em} 
\begin{subfigure} 
\centering 
\includegraphics[width=0.49\linewidth]{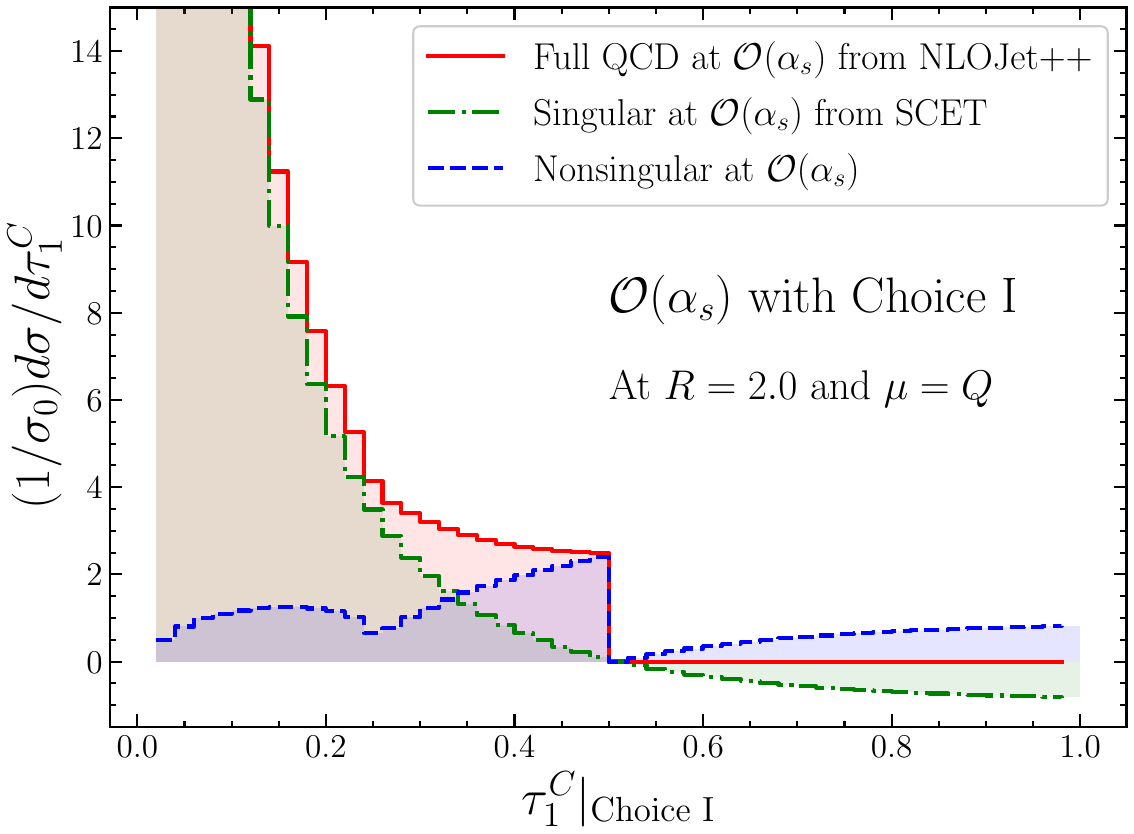} 
\end{subfigure} 
\begin{subfigure} 
\centering 
\includegraphics[width=0.49\linewidth]{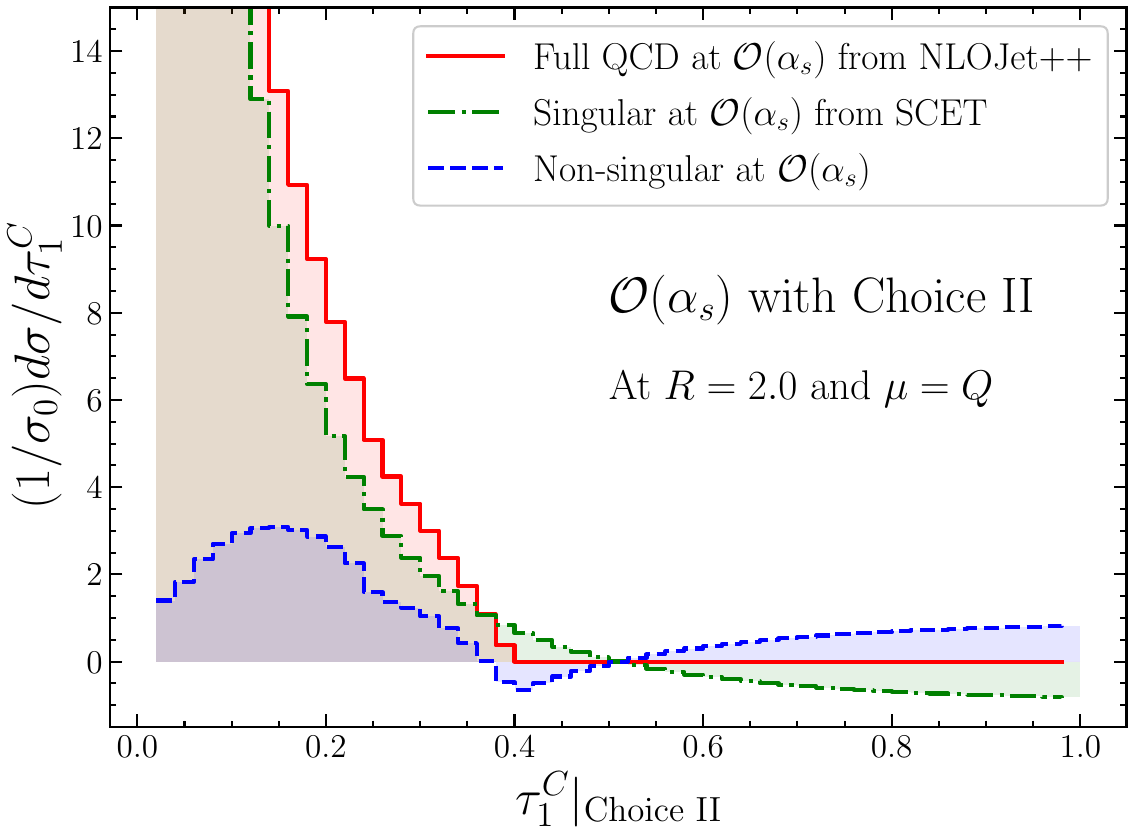} 
\end{subfigure} 
\begin{subfigure} 
\centering 
\includegraphics[width=0.49\linewidth]{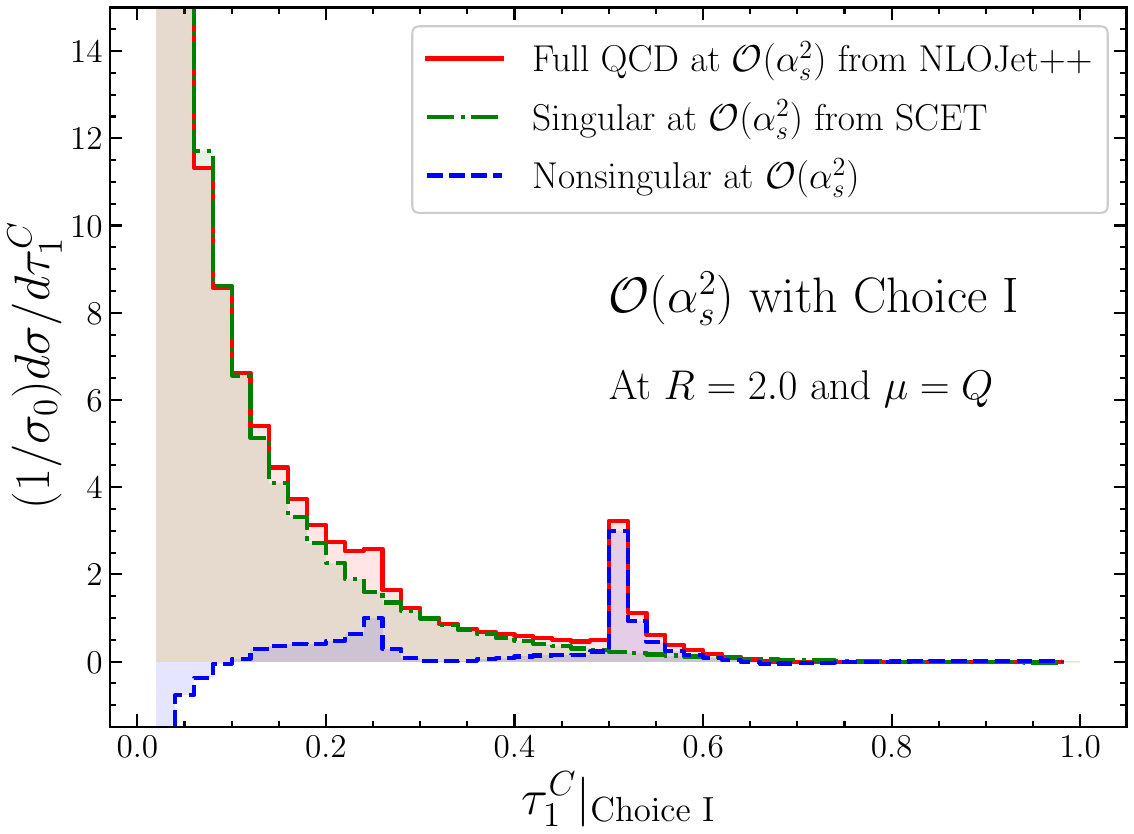} 
\end{subfigure} 
\begin{subfigure} 
\centering 
\includegraphics[width=0.49\linewidth]{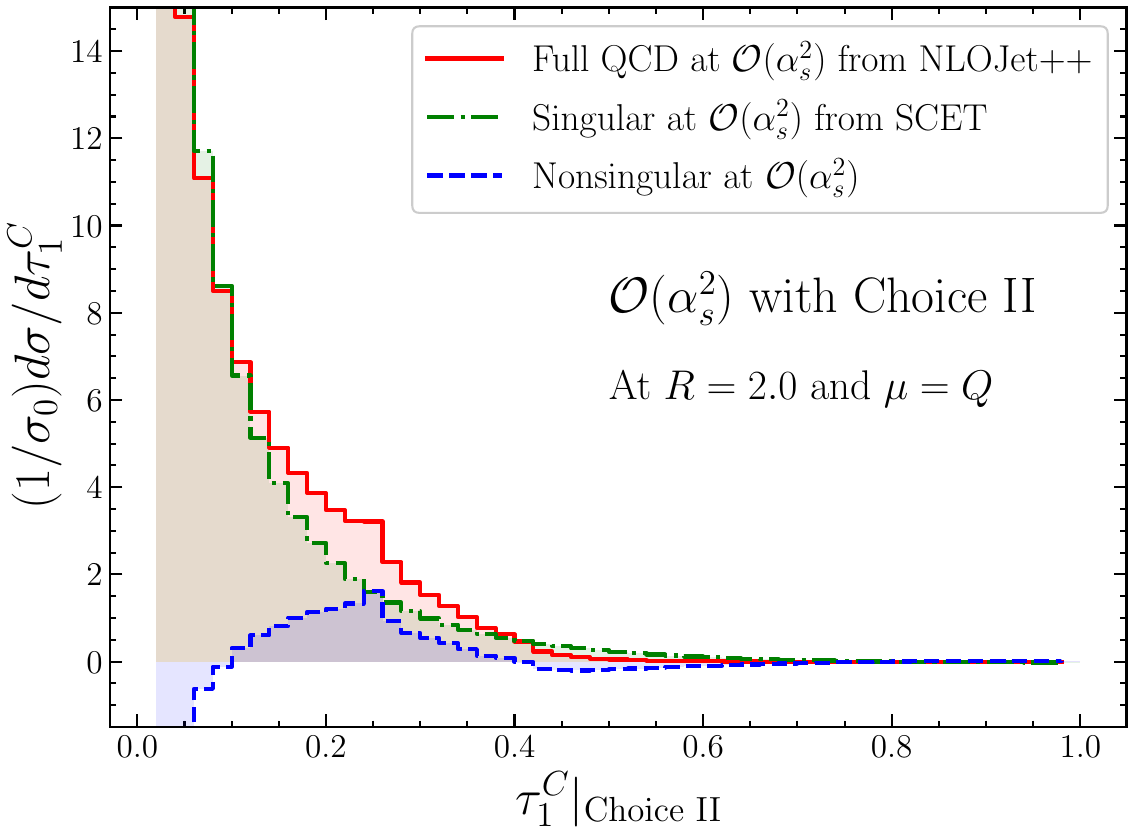} 
\end{subfigure} 
\vspace{-1em} 
\caption{The $\mathcal{O}(\alpha_s)$ (upper panels) and $\mathcal{O}(\alpha_s^2)$ (lower panels) fixed-order $\tau_1^C$ distributions for $R=2.0$. The left panels show the distributions with $\omega_{J,B}$ given in Eq.~\eqref{eq:tau1-choice1} (Choice I), while the right panels show the distributions with $\omega_{J,B}$ given in Eq.~\eqref{eq:tau1-choice2} (Choice II). 
The left panels clearly exhibit the step-like discontinuities at $\mathcal{O}(\alpha_s)$ and the resulting oscillatory instabilities at $\mathcal{O}(\alpha_s^2)$.
The right panels show smooth behavior without such perturbative artifacts.} \label{fig:LO-fixed-tau1a} 
\end{figure}

To study this dependence, we consider two choices of weights that satisfy the Centauro boundary condition:
\begin{align}
\label{eq:tau1-choice1}
\text{\textbf{Choice I}:} \quad
\omega_B = Q\,, 
\qquad
\omega_J = Q\frac{4}{R^2\cos^2(\theta_J/2)}\,,
\end{align}
and
\begin{align}
\label{eq:tau1-choice2}
\text{\textbf{Choice II}:} \quad
\omega_B = Q\cos^2(\theta_J/2)\,,
\qquad
\omega_J = Q\frac{4}{R^2}\,.
\end{align}
Both choices define valid versions of the Centauric 1-jettiness observable and differ only by power corrections in the resummation region.
However, they behave differently when applied to fixed-order final states.\footnote{Retaining the $R$ dependence in $\omega_J$ ensures that the maximum value of $\tau_1^C$ reaches 1 regardless of $R$.}

Choice I provides a natural starting point as the jet-definition parameters, $R$ and $\theta_J$, are directly encoded in the jet-region weighting factor $\omega_J$, while $\omega_B$ remains fixed at $Q$. However, this definition encounters significant issues in fixed-order perturbative calculations. While the distribution would be continuous for high-multiplicity final states as in experiments, low-multiplicity configurations at LO [$\mathcal{O}(\alpha_s)$] and NLO [$\mathcal{O}(\alpha_s^2)$] are highly sensitive to the discrete nature of the jet algorithm. We clarify the origin of this issue below.

At LO, the Centauro algorithm must select a jet axis $n_J$ from three discrete possibilities: $p_1$, $p_2$, or $p_1+p_2$. 
Figure~\ref{fig:phase_space_LO} illustrates the phase space for the selected jet axis at LO. The LO phase space is characterized by the two parameters, $v$ and $z$, as detailed in Appendix~\ref{app:two-particle-kinematics}. As shown in the figure, the boundaries between these regions represent phase-space points where the algorithm switches the axis definition.

The left panel of Fig.~\ref{fig:LO-fixed-tau1a-two-vers} (Choice I) shows the value of $\tau_1^C$ in the LO phase space: $\tau_1^C$ becomes degenerate along the boundary $v=0.5$, with all events accumulating at $\tau_1^C = 0.5$, in contrast to the smooth variation seen in the right panel (Choice II). 
The resulting step-like discontinuity in the differential cross section is visible in the left panels of Fig.~\ref{fig:LO-fixed-tau1a}.

While the NLO contribution introduces counter-discontinuities, the cancellation is incomplete at $\mathcal{O}(\alpha_s^2)$, leading to oscillatory behavior and poor perturbative convergence. This situation is reminiscent of the $C$-parameter distribution in $e^+e^-$ collisions \cite{Hoang:2014wka}, where the $\mathcal{O}(\alpha_s)$ distribution is discontinuous. While such artifacts become negligible when sufficient higher-order corrections (e.g., NNLO; $\mathcal{O}(\alpha_s^3)$) are included, they present an inconvenient perturbative instability for our current NLO-matched accuracy. There is nothing intrinsically wrong or unphysical about an observable with discontinuities or kinks at true kinematic boundaries, but if we have a choice, we prefer to seek a version that avoids them.

\begin{figure}[t]
    \centering
    \includegraphics[width=0.65\linewidth]{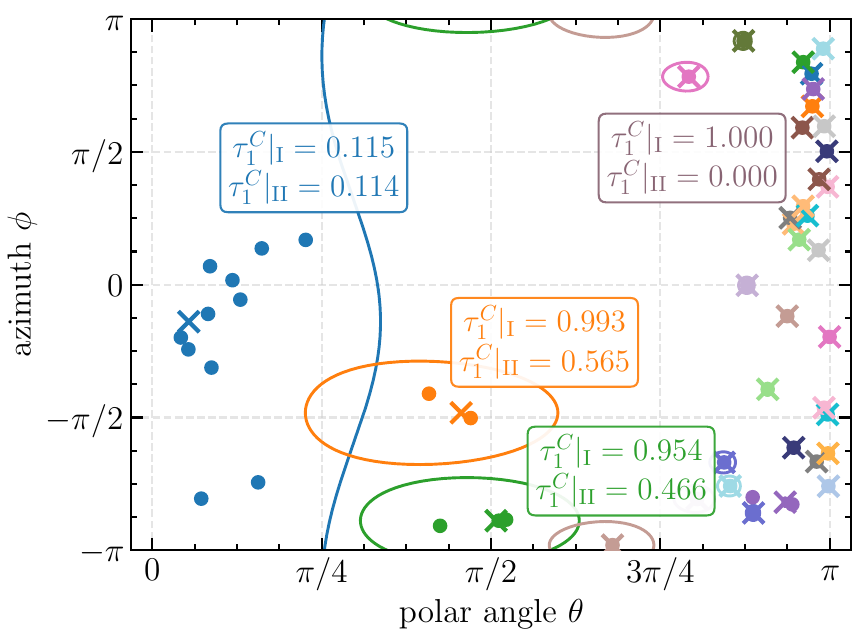}
    \caption{The Centauro jet algorithm applied to a single $ep\to e+X$ event in the
    Breit frame, shown in the $(\theta,\phi)$ plane, where $\theta$ is the polar angle
    and $\phi$ is the azimuth. Crosses ($\times$) mark the axes of the candidate jets
    found by the algorithm, and the solid curves are the corresponding Centauro
    boundaries for a jet radius $R=1.0$, the locus $d_{iJ}=1$ of Eq.~\eqref{eq:diJ-i-to-Jet} (equivalently, the 1-jettiness boundary fixed by Eq.~\eqref{eq:wJ-wB-cond}); final-state particles are
    colored by their candidate-jet assignment. For four representative candidates we
    quote $\tau_1^C|_{\mathrm I}$ and $\tau_1^C|_{\mathrm{II}}$ evaluated with the Choice~I and Choice~II weights of Eqs.~\eqref{eq:tau1-choice1} and \eqref{eq:tau1-choice2}.
    The primary jet is the candidate that minimizes $\tau_1^C|_{\mathrm I}$---here the jet candidate at small $\theta$, with $\tau_1^C|_{\mathrm I}=0.115$. The candidate near the beam direction ($\theta\to\pi$) illustrates the failure of Choice~II for jet selection: its $\tau_1^C|_{\mathrm I}=1.000$ correctly disfavors it as a jet, yet $\tau_1^C|_{\mathrm{II}}=0.000$ would make it the spurious minimum under Choice~II.}
    \label{fig:jet_algorithm}
\end{figure}

Using Choice II for the measurement removes this degeneracy because the weights depend continuously on $\theta_J$, ensuring that $\tau_1^C$ varies smoothly across the phase-space boundaries. However, Choice II introduces its own drawback as a primary-jet selection criterion.
For a given event, the Centauro algorithm may identify multiple candidate jets, as illustrated in Fig.~\ref{fig:jet_algorithm}: a single event yields a few multi-particle jets together with many low-multiplicity candidates toward the beam region ($\theta\to \pi$).
To select the primary jet in a theoretically consistent way, we could adopt a thrust-like minimization procedure: for each candidate jet $k$, we compute a trial value $\tau_1^{C,\,\rm trial}$ and select the candidate that minimizes it as the primary jet.

Choice II, however, is not well-suited for identifying the jet axis through this minimization. Because $\omega_B = Q\cos^2(\theta_J/2)$ vanishes as $\theta_J \to \pi$, the trial value $\tau_1^C$ can be made arbitrarily small by choosing a candidate axis $n_J^{\rm trial}$ pointing nearly along the beam direction. 
In Fig.~\ref{fig:jet_algorithm}, the candidate nearest the beam ($\theta\sim \pi$) has $\tau_1^C|_{\mathrm{II}}=0.000$, far below its Choice~I value $\tau_1^C|_{\mathrm I}=1.000$, so a Choice~II minimization would select this beam-aligned axis in place of the genuine struck-quark jet ($\tau_1^C|_{\mathrm I}\approx \tau_1^C|_{\mathrm{II}}\approx 0.115$). The minimization would then unphysically prefer such configurations, effectively assigning nearly all radiation to the beam region to artificially minimize $\tau_1^C$.

This tension between geometric identification and perturbative stability motivates a hybrid scheme that leverages the respective strengths of both definitions: Choice I is employed to robustly identify the primary jet axis, whereas Choice II is used for the actual measurement to ensure a stable fixed-order prediction. The procedure is summarized as follows:

\begin{enumerate}
\item \textbf{Clustering.} Identify candidate jets using the Centauro algorithm.
\item \textbf{Selection.} For each candidate, compute a trial value 
$\tau_1^{C, \rm trial}$ using Choice I weights and select the candidate 
that minimizes it as the primary jet, thereby fixing the axis $n_J$.
\item \textbf{Measurement.} Compute the physical $\tau_1^C$ using this $n_J$ 
and the Choice II weights.
\end{enumerate}
This approach secures the correct geometric identification of the struck-quark jet while preserving the perturbative stability required for high-precision theoretical predictions.

It is worth noting that this framework represents a direct generalization of the standard DIS 1-jettiness $\tau_1^a$. In the hemisphere limit $R=2$ and for $\theta_J \to 0$, the measurement weights in Choice II reduce to $\omega_B = \omega_J = Q$, so that the beam and jet regions are separated by the exact hemisphere, and the observable reduces to $\tau_1^a$.

\subsection{Numerical Implementation}
\label{subsec:implementation}
In this subsection, we describe the numerical implementation used to compute the fixed-order cross sections for the Centauric 1-jettiness observable using \texttt{NLOJet++}. As discussed in the previous subsection, a stable fixed-order definition of $\tau_1^C$ requires a hybrid procedure in which the jet axis is selected using Choice I while the observable is evaluated using Choice II. The purpose of this subsection is to translate this hybrid definition into an explicit algorithm that can be applied to low-multiplicity partonic final states in a manner that preserves infrared safety and is compatible with the dipole subtraction method used in fixed-order calculations. We therefore outline the clustering, jet-selection, and measurement steps used in our implementation, and discuss the conditions required to ensure numerical stability of the fixed-order subtraction. This implementation provides the fixed-order cross sections used to determine the non-singular contribution and to construct the matched prediction described in this section.

At $\mathcal{O}(\alpha_s)$, the final state consists of two partons ($p_1, p_2$), and at $\mathcal{O}(\alpha_s^2)$, up to three partons ($p_1, p_2, p_3$). Following the hybrid scheme introduced in Sec.~\ref{subsec:instability}, the procedure is as follows:
\begin{itemize}
    \item \textbf{Step 1: Clustering.} Run the Centauro jet algorithm (with $p=0$) on the final state partons.
    \begin{itemize}
        \item For 2 partons: Compute $d_{12}$. If $d_{12} < 1$, the candidate jet is the combined system $p_1+p_2$. If $d_{12} \ge 1$, we have two candidate jets, $p_1$ and $p_2$.
        \item For 3 partons: Compute all $d_{ij}$. If $\min(d_{ij}) < 1$, cluster the closest pair (e.g., $p_i+p_j$) and treat the system as a 2-body state, then repeat the logic above. If all $d_{ij} \ge 1$, we have three candidate jets.
    \end{itemize}
    
    \item \textbf{Step 2: Selection (Choice I).} For each candidate jet identified in Step 1, define a trial axis $n_J^{\text{trial}}$ aligned with the candidate's momentum. Compute the trial 1-jettiness value using the selection weights (Choice I):
    \begin{align}
        \tau_{1}^{\text{trial}} =\frac{1}{Q} \sum_{i} \min \left\{ 
        n_B\cdot p_i, \; \frac{4 n_J^{\text{trial}}\cdot p_i}{R^2\cos^2(\theta_J^{\text{trial}}/2)} 
        \right\}.
    \end{align}
    The candidate that minimizes $\tau_{1}^{\text{trial}}$ is selected as the primary jet, fixing the physical axis $n_J$.
    
    \item \textbf{Step 3: Measurement (Choice II).} Using the chosen jet axis $n_J$, compute the final observable $\tau_1^C$ using the measurement weights (Choice II) from Eq.~\eqref{eq:tau1-choice2}.
\end{itemize}

A comment on numerical stability is in order. \texttt{NLOJet++} relies on the cancellation of infrared poles between real and virtual corrections using the dipole subtraction method. This cancellation requires that the observable definition be infrared safe.
For axis-dependent observables like $\tau_1^C$, the jet axis must not flip discontinuously in the soft/collinear limit, as this would cause a mismatch between the real and virtual contributions. 
While $\tau_1^b$ (with fixed axes) is trivially stable, $\tau_1^C$ requires care. Our hybrid scheme ensures robustness: the minimization in Step 2 effectively acts like a thrust-axis finding, which is known to be stable. Furthermore, the use of Choice II in Step 3 ensures that even if the axis choice were to exhibit minor instabilities, the observable value would remain continuous, thereby preserving the validity of the fixed-order subtraction.

\subsection{Non-Singular Contribution to the Cross Sections}
\label{subsec:non-singular_results}
\begin{figure}
    \centering
    \includegraphics[width=\linewidth]{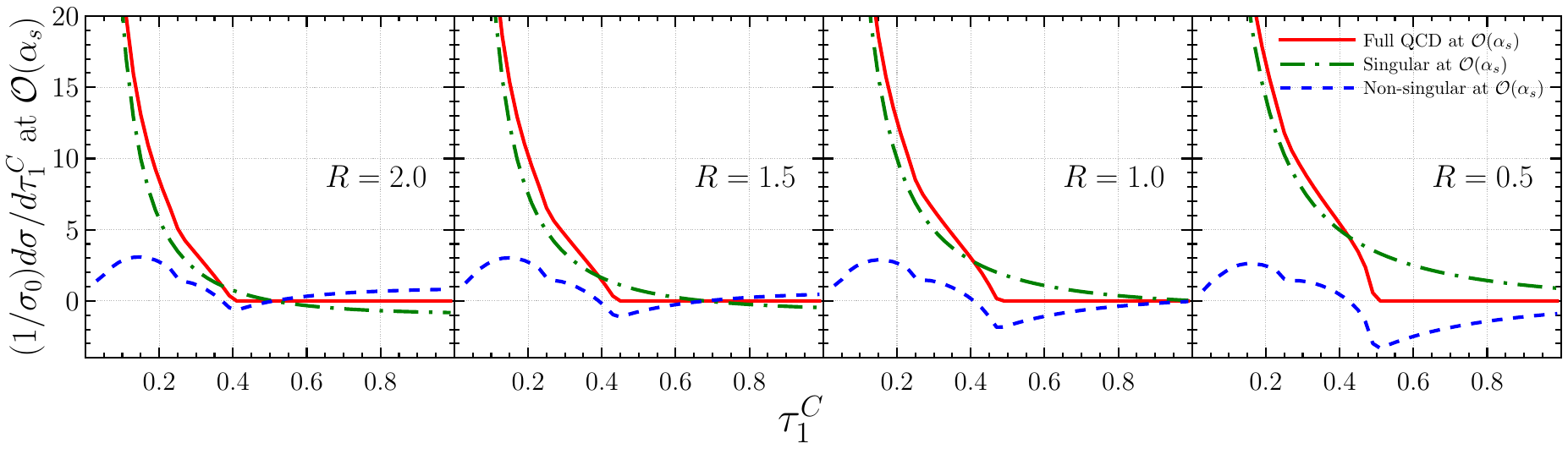}
    \includegraphics[width=\linewidth]{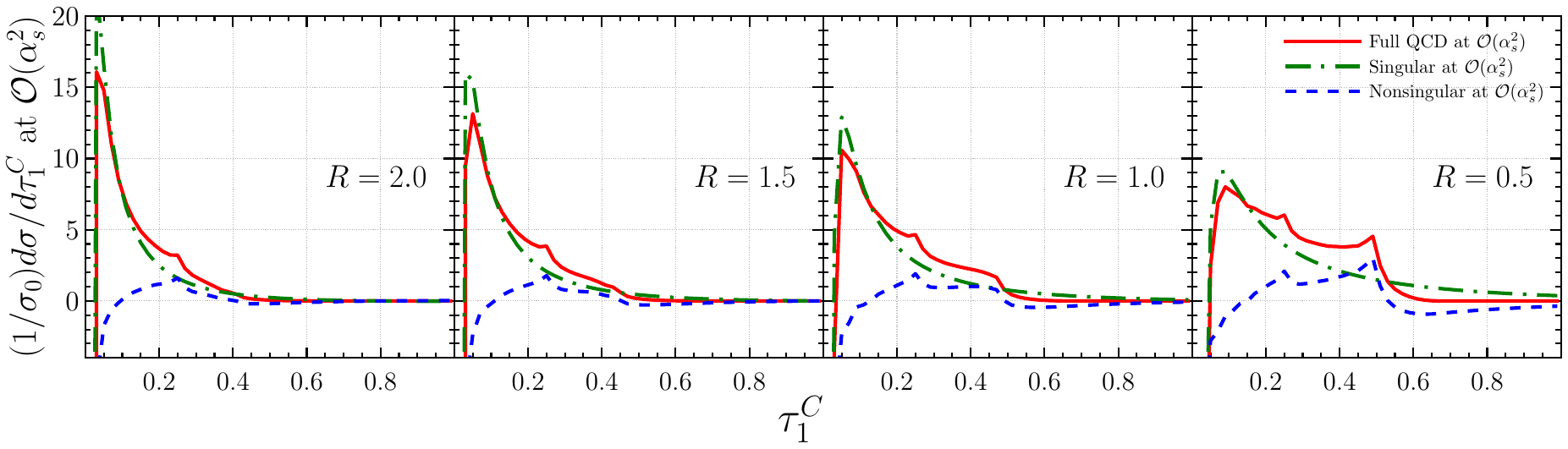}
    \caption{
    Decomposition of the fixed-order cross section components at $\mathcal{O}(\alpha_s)$ (upper) and $\mathcal{O}(\alpha_s^2)$ (lower) for various jet radii $R \in \{2.0, 1.5, 1.0, 0.5\}$ at $\mu=Q$. The results are evaluated at $\sqrt{s}=319~\textrm{GeV}$, $Q=50~\textrm{GeV}$, and $x_B=0.05$. The red solid lines denote the full QCD results from \texttt{NLOJet++}, the green dot-dashed lines represent the singular terms derived from the SCET factorization, and the blue dashed lines indicate the extracted non-singular contributions.
    }
    \label{fig:ns_LO}
\end{figure}

In this subsection, we present the non-singular cross sections obtained from the fixed-order calculations described in the previous subsections and examine their role in the matched prediction for the Centauric 1-jettiness distribution. The singular contributions derived from the factorization theorem reproduce the leading singular structure of the full QCD cross section in the small-$\tau_1^C$ region, while the non-singular terms provide the remaining finite contributions that become important in the transition and far-tail regions. To demonstrate that the subtraction defined in Eq.~\eqref{eq:def-of-non-singular} correctly isolates these contributions, we show in Fig.~\ref{fig:ns_LO} the components of the matching at $\mathcal{O}(\alpha_s)$ and $\mathcal{O}(\alpha_s^2)$ for several representative jet radii. In each case, we compare the full fixed-order result, the fixed-order expansion of the singular contribution, and their difference, which defines the non-singular cross section. 
We then study the perturbative uncertainties of the non-singular contribution and describe how it is incorporated into the final matched prediction used in Sec.~\ref{sec:results} for the comparison with \textsc{Pythia}.

Crucially, in the small-$\tau_1^C$ limit (the peak region), the singular contribution correctly captures the leading non-integrable singularities (scaling as $1/\tau_1^C$ and $\ln^k\tau_1^C/\tau_1^C$) of the full QCD cross section. Consequently, the remaining non-singular residual is free of such divergences. While it may still contain integrable singularities (e.g., logarithmic terms scaling as $\ln \tau_1^C$), the plots demonstrate that the non-singular term remains numerically subdominant compared to the singular peak in the $\tau_1^C \to 0$ limit.
Conversely, in the tail and the far-tail regions, the singular approximation deviates from the full QCD result. In these regimes, the non-singular contribution becomes equally important as the singular, providing the necessary corrections to accurately describe hard, wide-angle radiation not captured by the dijet configurations. 

We consider five distinct scale choices: the central scale $\mu_{\rm ns} = Q$, variations by factors of two ($\mu_{\rm ns} = \{Q/2, 2Q\}$), and two additional variations where $\mu_{\rm ns}$ tracks the profile scales used in the singular region, namely $\mu_J$ and $\mu_{JS} \equiv (\mu_J + \mu_S)/2$. This choice allows for a robust estimation of uncertainties arising from the hard scale and the transition toward the resummation regime.

\begin{figure}
    \centering
    \includegraphics[width=\linewidth]{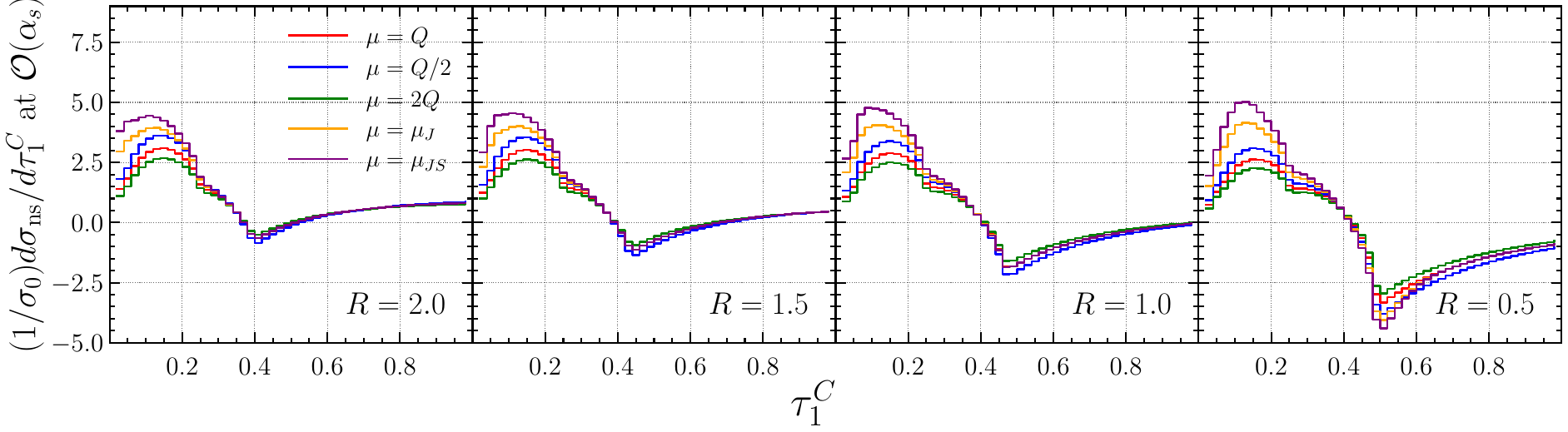}
    \includegraphics[width=\linewidth]{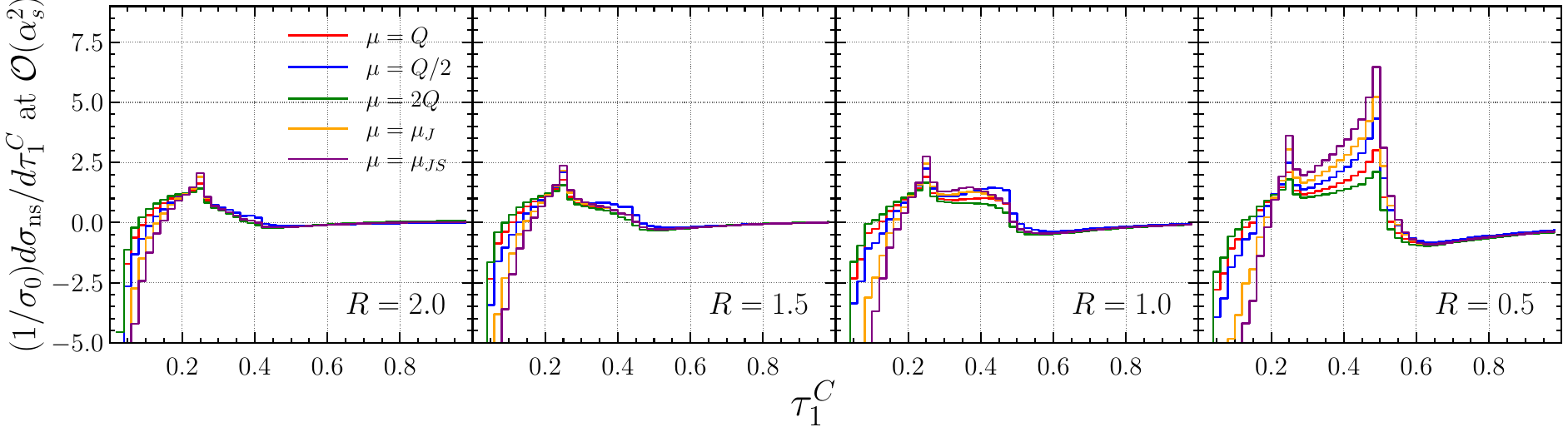}
    \caption{Scale variations of the non-singular contributions at $\mathcal{O}(\alpha_s)$ (upper panels) and $\mathcal{O}(\alpha_s^2)$ (lower panels) for $\sqrt{s}=319~\textrm{GeV}$, $Q=50~\textrm{GeV}$, and $x_B=0.05$. The five curves in each panel correspond to the scale choices discussed in the text.}
    \label{fig:ns_LO_NLO_scale_var}
\end{figure}

Figure~\ref{fig:ns_LO_NLO_scale_var} illustrates these scale variations across different jet radii $R$. 
We observe that the $\mathcal{O}(\alpha_s^2)$ non-singular contributions effectively smooth out the unphysical cusps present in the $\mathcal{O}(\alpha_s)$ distributions. These artifacts originate from either the abrupt switching of the jet axis (from $p_i$ to $p_1+p_2$) or the restricted phase-space boundaries inherent in low-multiplicity final states. The inclusion of the $\mathcal{O}(\alpha_s^2)$ contributions leads to a smoother transition and improved perturbative convergence. The resulting scale envelope is taken as the perturbative theoretical uncertainty of the non-singular component in our final predictions.

Since the \texttt{NLOJet++} results are obtained as binned differential cross sections for $\tau_1^C$, it is natural to implement the non-perturbative corrections directly at the differential level. Specifically, we convolve the non-singular cross sections with the shape function and perform the renormalon subtraction in the differential distribution, in contrast to the singular contributions where these operations are applied to the cumulative distribution (see Eq.~\eqref{eq:cumulant-to-differential}).

In the peak region ($\tau_1^C \sim \Lambda_\textrm{QCD}/Q_R$), convolving the non-singular distribution with the shape function is formally subleading, as this amounts to a non-perturbative correction to a term that is already power-suppressed. However, in the transition region where the singular and non-singular cross sections are of comparable magnitude, it is crucial that non-perturbative corrections be implemented consistently to preserve the perturbative cancellation between the two components.
Therefore, we apply the same shape function and renormalon subtraction scheme to the non-singular term as used for the singular term. This ensures that the renormalon subtraction $\delta(R_{\textrm{gap}},\mu_S)$ and the non-perturbative shift $\Omega_1(R_{\textrm{gap}},\mu_S)$ are treated on the same footing across the entire kinematic range of $\tau_1^C$, following the approach in Refs.~\cite{Abbate:2010xh,Hoang:2014wka,Bell:2018gce}. Detailed studies of multi-jet non-perturbative power corrections are beyond the scope of this work and are left for future investigation.

The behavior of the fixed-order distributions in the large-$\tau_1^C$ region can be understood from the kinematic endpoint implied by low-multiplicity final states. 
For a configuration with $n+1$ final-state partons, consider a symmetric arrangement in which all
partons carry equal large light-cone components, equal small light-cone components, and
their transverse momenta are distributed uniformly in azimuth. In the small-$R$ limit, this
configuration provides a simple estimate of the maximal value of the Centauric observable,
\begin{align}
\label{eq:tau1C-max-relation}
\tau_{1,\max}^C \;\xrightarrow[R\to 0]{}\; \frac{n}{n+1}\,.
\end{align}
Thus, at $\mathcal{O}(\alpha_s)$, where the final state contains two partons, the endpoint tends
to $\tau_{1,\max}^C \to 1/2$, while at $\mathcal{O}(\alpha_s^2)$, where configurations with three
partons first appear, the corresponding endpoint tends to $\tau_{1,\max}^C \to 2/3$.

This expectation is borne out by the fixed-order results. In the small-$R$ limit, the
$\mathcal{O}(\alpha_s)$ distributions in Fig.~\ref{fig:ns_LO} (upper panels) fall rapidly to zero near
$\tau_1^C \simeq 0.5$, reflecting the two-parton kinematic boundary. Likewise, the
$\mathcal{O}(\alpha_s^2)$ distributions in Fig.~\ref{fig:ns_LO} (lower panels) extend further and die off near
$\tau_1^C \simeq 2/3$, consistent with the opening of the three-parton phase space. These
endpoints are therefore a simple kinematic consequence of the partonic multiplicity and
provide a useful check on the fixed-order implementation of the observable.

\section{Extraction of the Non-perturbative Shift}\label{sec:omega1}
In this section, we determine the leading non-perturbative parameter $\Omega_1$, which governs the universal power correction to the Centauric 1-jettiness distribution.
The factorization theorem, combined with the geometric mapping of the Centauro jet boundary, predicts that the dominant hadronization effect in the tail region, $\Lambda_{\rm QCD}/Q_R \ll \tau_1^C \ll 1$, takes the form of a universal shift (cf. \cite{Dokshitzer:1995zt,Dokshitzer:1997ew}) of the spectrum (see Appendix~\ref{sec:power-corrections} for the derivation),
\begin{align}
\label{eq:ope-in-main-text}
\tau_1^C \longrightarrow
\tau_1^C - \frac{4}{R}\frac{\Omega_1-\Delta}{Q}\,,
\end{align}
where $\Omega_1$ is the first moment of the universal soft shape function defined in the $J$-scheme, and the characteristic $1/R$ scaling reflects the geometry of the Centauro jet boundary. 
The parameter $\Delta$ denotes the gap parameter in the $R$-gap scheme, which is introduced to ensure a renormalon-free definition of the non-perturbative shift. For our analysis, we adopt the value $\Delta(R_\Delta, \mu_\Delta) = 0.05$~GeV at the reference scales $R_\Delta = \mu_\Delta = 2$~GeV, following the detailed $R$-gap formulation provided in the appendix.
To test this leading-power prediction, we use the hybrid scheme for jet identification developed in Sec.~\ref{sec:fixed-order} and extract the effective \textsc{Pythia}-level parameter,
\begin{equation}
\label{eq:relation_omega_pythia_R_gap}
\Omega_1^\textrm{Pythia}\equiv\Omega_1-\Delta,
\end{equation}
using Monte Carlo simulations generated with \textsc{Pythia}~\cite{Bierlich:2022pfr}. We treat \textsc{Pythia} as a controlled environment where hadronization effects can be systematically isolated while keeping the perturbative dynamics and jet definitions fixed. This allows us to verify the predicted $1/R$ scaling of the leading power correction across various jet radii.

The remainder of this section is organized as follows. In Sec.~\ref{subsec:procedure}, we describe the numerical method for comparing parton- and hadron-level $\tau_1^C$ distributions and define the $\chi^2$-based fitting procedure used to determine the optimal shift. In Sec.~\ref{subsec:shift-extract}, we apply this procedure to extract the per-radius shifts and determine the universal value of $\Omega_1^\textrm{Pythia}$ through both a local and a global fit, with statistical uncertainties quantified by a non-parametric bootstrap. Finally, in Sec.~\ref{sec:SR}, we present an alternative extraction using Symbolic Regression (SR), providing an independent cross-check of the leading-power shift hypothesis and its $R$-dependence.

\subsection{Extraction Procedure}
\label{subsec:procedure}
\begin{figure}[t]
  \centering
  \includegraphics[width=0.49\linewidth]{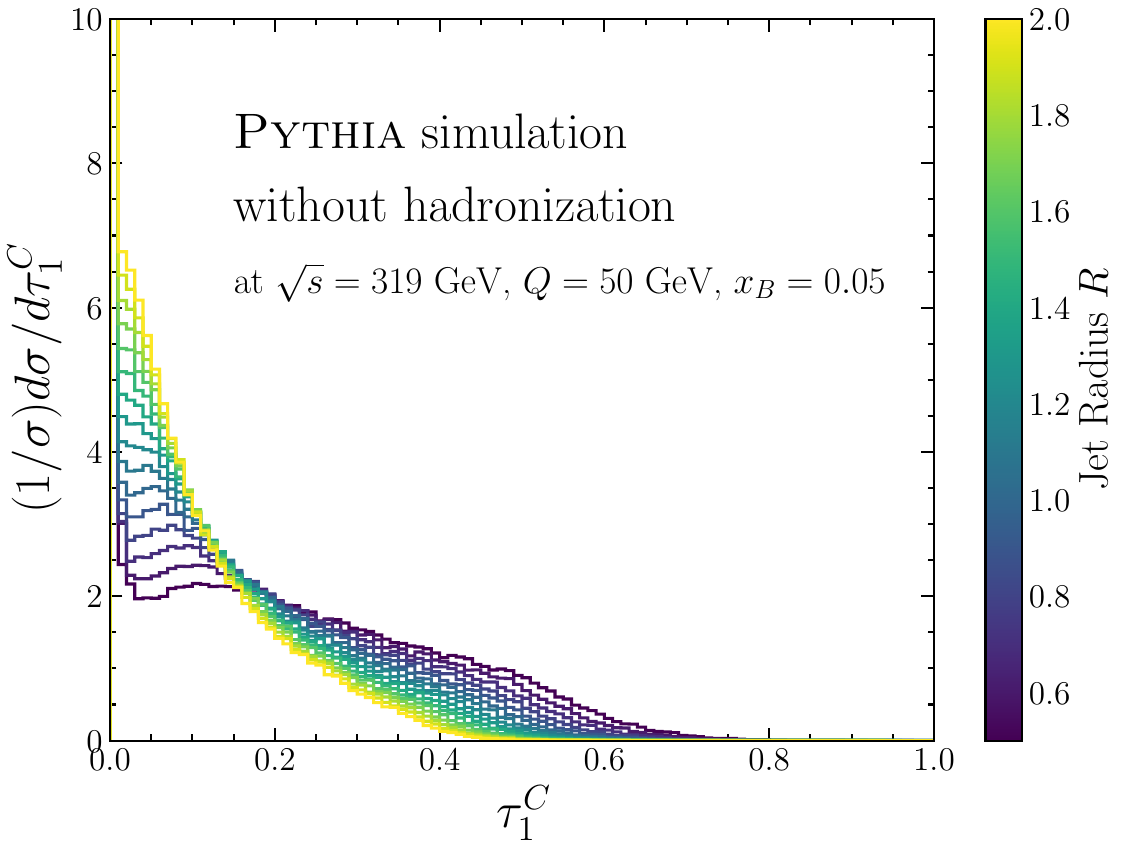}
  \includegraphics[width=0.49\linewidth]{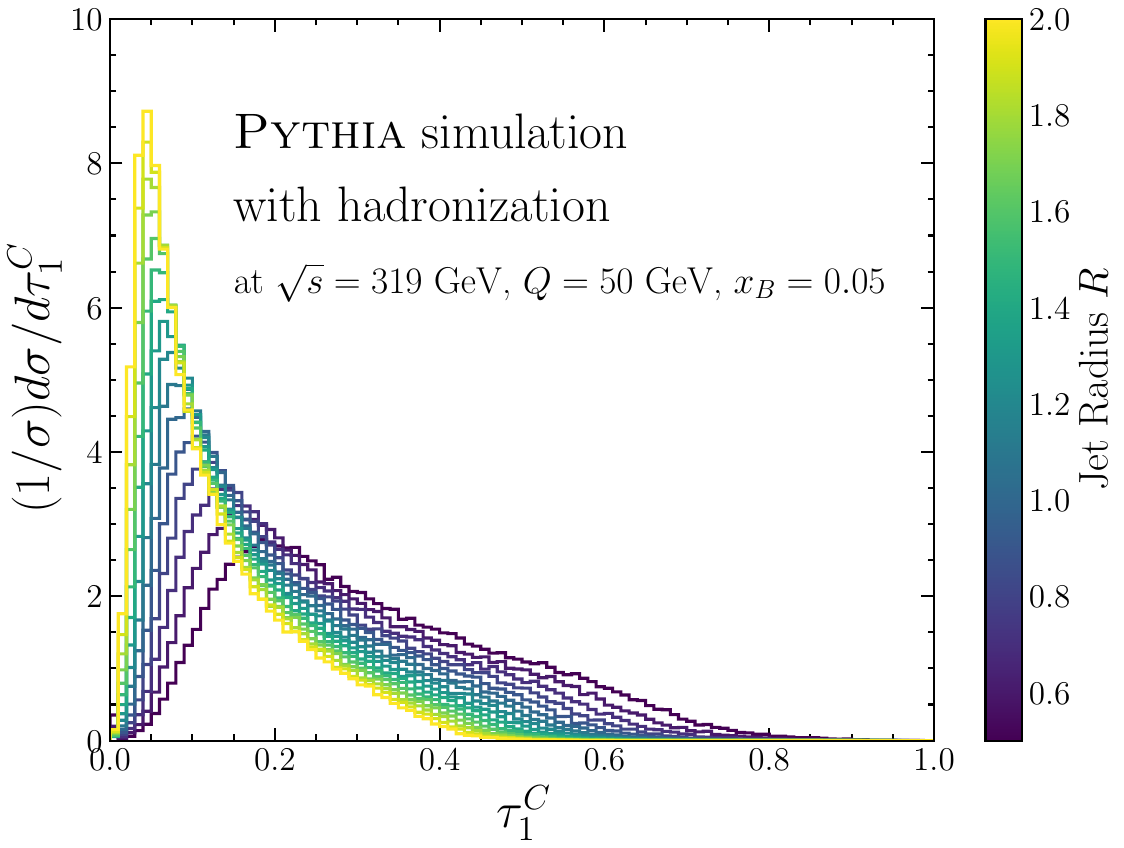}
  \caption{
    Distributions of the Centauric 1-jettiness $\tau_1^C$ obtained from \textsc{Pythia} simulations at $\sqrt{s}=319~\textrm{GeV}$, $Q=50~\textrm{GeV}$, $x_B=0.05$, without hadronization (left) and with hadronization (right), for jet radii $0.5\le R \le 2.0$. The distributions shift toward larger values of $\tau_1^C$ with increasing $R$.
  }
  \label{fig:tau_distributions-pythia}
\end{figure}

The extraction of $\Omega_1^\textrm{Pythia}$ is based on a direct comparison between parton-level and hadron-level Centauric 1-jettiness distributions generated with \textsc{Pythia}.
For fixed values of the DIS kinematic variables $(Q, x_B)$, we generate two event samples:
\begin{itemize}
  \item A \textbf{partonic sample}, obtained by disabling hadronization using the \textsc{Pythia} flag \texttt{HadronLevel:all = off}.
  \item A \textbf{hadronic sample}, obtained using the default hadronization model (Lund string model) with \texttt{HadronLevel:all = on}.
\end{itemize}
In both cases, the Centauro jet algorithm and the $\tau_1^C$ measurement are applied identically for a given jet radius $R$, following the procedure described in Secs.~\ref{subsec:instability} and \ref{subsec:implementation}. For each run, $\tau_1^C$ is computed at 16 jet radii $R \in \{0.5, 0.6, \ldots, 2.0\}$ from $N_p \sim 10^5$ parton-level and $N_h \sim 10^5$ hadron-level events per DIS kinematic point $(Q, x_B)$. Both the parton-level and hadron-level samples are histogrammed into 100 uniform bins over $\tau_1^C \in [0, 1]$ and normalized to unit area.
The resulting distributions for various $R$ are displayed in Fig.~\ref{fig:tau_distributions-pythia}.

Comparing the two sets of distributions, we find that hadronization strongly reshapes the spectrum in the peak region $\tau_1^C \sim \Lambda_{\rm QCD}/Q_R$, while in the tail region its primary effect is a rigid displacement of the distribution toward larger values of $\tau_1^C$, leaving the shape largely intact. This behavior is consistent with the expectation from the factorization theorem, in which the leading non-perturbative correction in the tail region is captured by the universal shift parameter $\Omega_1$.

Motivated by this observation, we model the leading non-perturbative effect in \textsc{Pythia} as a rigid horizontal shift of the partonic distribution:
\begin{align}
\frac{d\sigma}{d\tau_1^C}
\left(\tau_1^C\right)
\;\longrightarrow\;
\frac{d\sigma_{\rm PT}}{d\tau_1^C}
\left(\tau_1^C - \Delta\tau\right)\,.
\end{align}
A more rigorous treatment would fit the whole shape function and determine this shift as its first moment parameter, but for our illustrative purposes in this paper, the shift model (cf. \cite{Dokshitzer:1995zt,Dokshitzer:1995qm}) will suffice.
The optimal shift $\Delta\tau$ is determined by minimizing a windowed $\chi^2$ goodness-of-fit measure between the shifted partonic and hadronic distributions:
\begin{align}
\label{eq:chi2-def}
\chi^2(\Delta\tau;\, R, Q)
=
\sum_{i\,\in\,\text{window}}
\frac{\left[\,h_i - p_i(\Delta\tau)\,\right]^2}
     {\sigma_{h,i}^2 + \sigma_{p,i}^2(\Delta\tau)}\,,
\end{align}
where $h_i$ and $p_i(\Delta\tau)$ denote the hadron-level and parton-level normalized densities in bin $i$, respectively, with the latter evaluated at the shifted coordinate $\tau_i - \Delta\tau$.
The fit window is defined by $\tau_1^C\in [\tau_\textrm{min}, \tau_\textrm{max}]$. The exact bounds for these will be discussed briefly.
The per-bin statistical uncertainties $\sigma_{h,i}$ and $\sigma_{p,i}$ (not to be confused
with the cross section $\sigma$ or $\sigma_{\rm PT}$) are taken as the Poisson uncertainties of the
normalized histogram, $\sigma_i = \sqrt{n_i}/(N\cdot\text{bin width})$,
where $n_i$ is the raw count in bin $i$ and $N$ is the total event count. 
Bins where both uncertainties vanish are excluded.

The lower cutoff $\tau_{\rm min}$ is chosen to restrict the fit to the tail region where the OPE underlying Eq.~\eqref{eq:ope-in-main-text} is valid. In the tail region, subleading power corrections scale as $\mathcal{O}(\alpha_s \Lambda_{\rm QCD}/(Q_R \tau_1^C))$ and $\mathcal{O}(\Lambda_{\rm QCD}^2/(Q_R \tau_1^C)^2)$ relative to the leading term. Requiring these corrections to remain below the percent level implies that $Q_R \tau_1^C$ must be sufficiently large compared to $\Lambda_{\rm QCD}$, leading to the condition
\begin{align}
\label{eq:tau_low_condition}
Q_R {\tau_1^C} \ge Q_{\rm low}\,,
\end{align}
where $Q_{\rm low} \gg \Lambda_{\rm QCD}$ is a perturbative cutoff scale.\footnote{ Requiring the next-to-leading power corrections $\mathcal{O}(\alpha_s \Lambda_{\rm QCD}/(Q_R \tau_1^C))$ and $\mathcal{O}(\Lambda_{\rm QCD}^2/(Q_R \tau_1^C)^2)$ to remain below $\mathcal{O}(1\%)$ of the leading term implies $Q_R \tau_1^C \gtrsim 100\,\alpha_s \Lambda_{\rm QCD} \sim 10\,\Lambda_{\rm QCD}$. Taking $\alpha_s \sim 0.1$ and $\Lambda_{\rm QCD} \sim 0.3~\textrm{GeV}$ gives $Q_{\rm low} = Q_R \tau_{\rm min} \sim 3~\textrm{GeV}$.}

\begin{figure}[t]
  \centering
  \includegraphics[width=\linewidth]{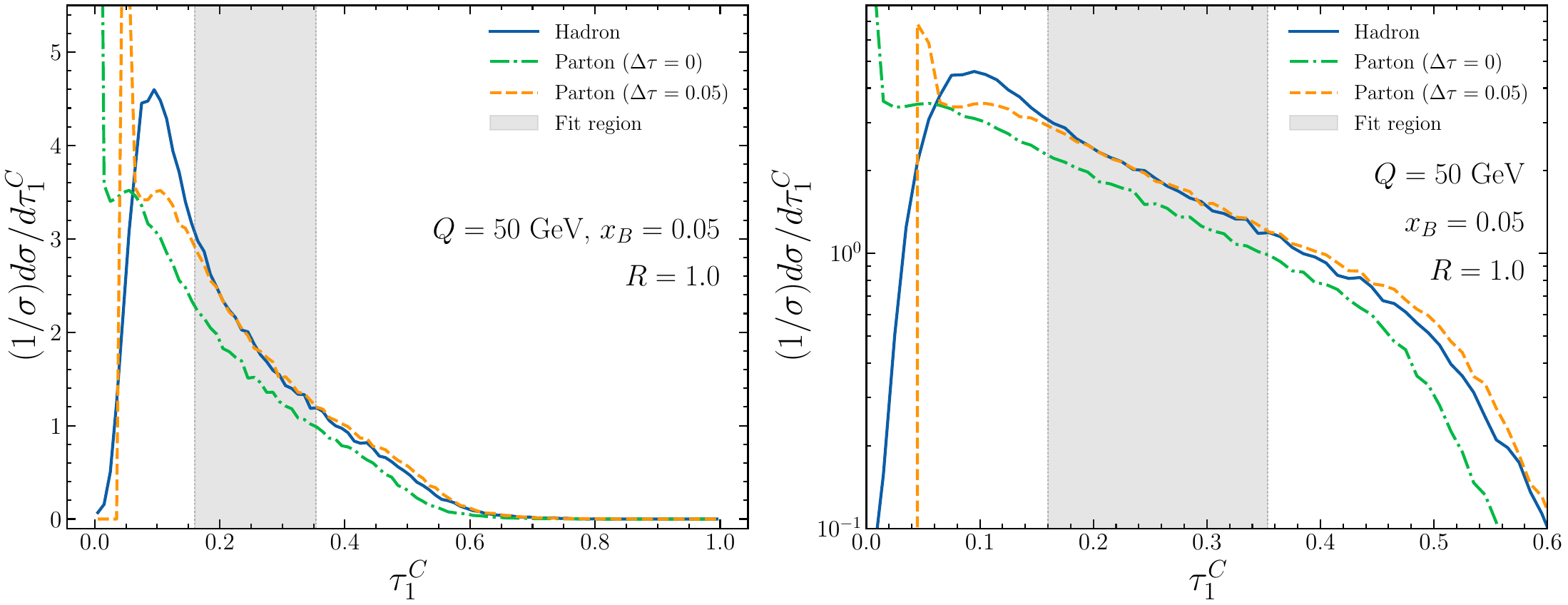}
\caption{
    Comparison of the Centauric 1-jettiness $\tau_1^C$ distributions at $R=1.0$
    obtained from \textsc{Pythia} simulations at $\sqrt{s}=319~\textrm{GeV}$, $Q=50~\textrm{GeV}$, and $x_B=0.05$.
    The blue solid and green dash-dotted curves represent the hadron-level and parton-level results, respectively.
    The orange dashed curve illustrates the parton-level distribution after applying the optimal non-perturbative shift $\Delta\tau$.
    Shaded gray bands indicate the fit interval $[t_1, t_2]$ used for the extraction.
    Within the tail region, the hadronization effects are well described by a rigid horizontal translation of the spectrum,
    largely preserving the perturbative spectral shape.
    The left and right panels display the distributions on linear and logarithmic scales, respectively.
}
  \label{fig:tau_distributions}
\end{figure}

In practice, we choose the fit window $[\tau_\textrm{min}, \tau_\textrm{max}]$ to be $[t_1, t_2]$ where $t_1$ and $t_2$ are the profile-function parameters defined in Sec.~\ref{subsec:profile}; this choice confines the fit to the resummation region where our resummed predictions are most accurate, while 
the upper cutoff $t_2$ avoids the far-tail region where \textsc{Pythia}'s 
incomplete higher-order matrix elements may distort the spectrum, and where further power corrections beyond those to dijet factorization may enter, cf. \cite{Caola:2022vea,Bell:2023dqs,Benitez-Rathgeb:2024ylc}. Only bins inside this window enter the $\chi^2$. This procedure is illustrated in Fig.~\ref{fig:tau_distributions} for a representative jet radius ($R = 1.0$). The partonic distribution shifted by the optimal $\Delta\tau$ closely tracks the hadronic distribution throughout the fitting region, confirming that the leading non-perturbative effect is well described by a horizontal shift. 

The statistical uncertainty of the optimal shift $\Delta\tau(R, Q)$ is obtained from a non-parametric bootstrap, 
which estimates how much the extracted shift would fluctuate under repeated experiments of the same finite size, without generating totally new event samples. For each replica we draw new parton- and hadron-level samples of the original sizes by resampling the events with replacement, drawn randomly from the same original set of generated events, with possible duplication of events entering the resampled set. We then rebuild the normalized histograms, and re-extract the shift by minimizing the windowed $\chi^2(\Delta\tau)$ of Eq.~\eqref{eq:chi2-def} with a parabolic sub-grid interpolation around the discrete minimum.

Repeating this $N_{\rm boot}=500$ times yields the distribution $\{\Delta\tau^{(b)}(R, Q)\}_{b=1}^{N_{\rm boot}}$; this procedure simulates the statistical uncertainty of the extracted optimal $\Delta\tau$ without repeating the computationally costly generation of new event samples. We take its mean as the central value and its standard deviation as the statistical uncertainty $\sigma_{\rm boot}(R, Q)$, used as the per-point uncertainty in all subsequent fits.

\subsection{Extraction of the Shift}\label{subsec:shift-extract}
We now extract the optimal shift $\Delta\tau(R, Q)$ for each jet radius and test the factorization-theorem prediction
\begin{align}
\Delta\tau(R, Q) = \frac{4}{R}\,\frac{\Omega_1^\textrm{Pythia}}{Q}\,.
\label{eq:shift-fit}
\end{align}
Crucially, the $1/R$ dependence in Eq.~\eqref{eq:shift-fit} is not imposed during the extraction of the individual shifts. Instead, it serves as a non-trivial test of the factorization theorem: if the leading non-perturbative effect is genuinely universal, the extracted shifts must scale linearly with $1/R$, and the slope must yield a value of $\Omega_1^\textrm{Pythia}$ that is independent of the kinematic variables $Q$ and $x_B$.

\begin{figure}[t]
  \centering
  \includegraphics[width=1.0\linewidth]{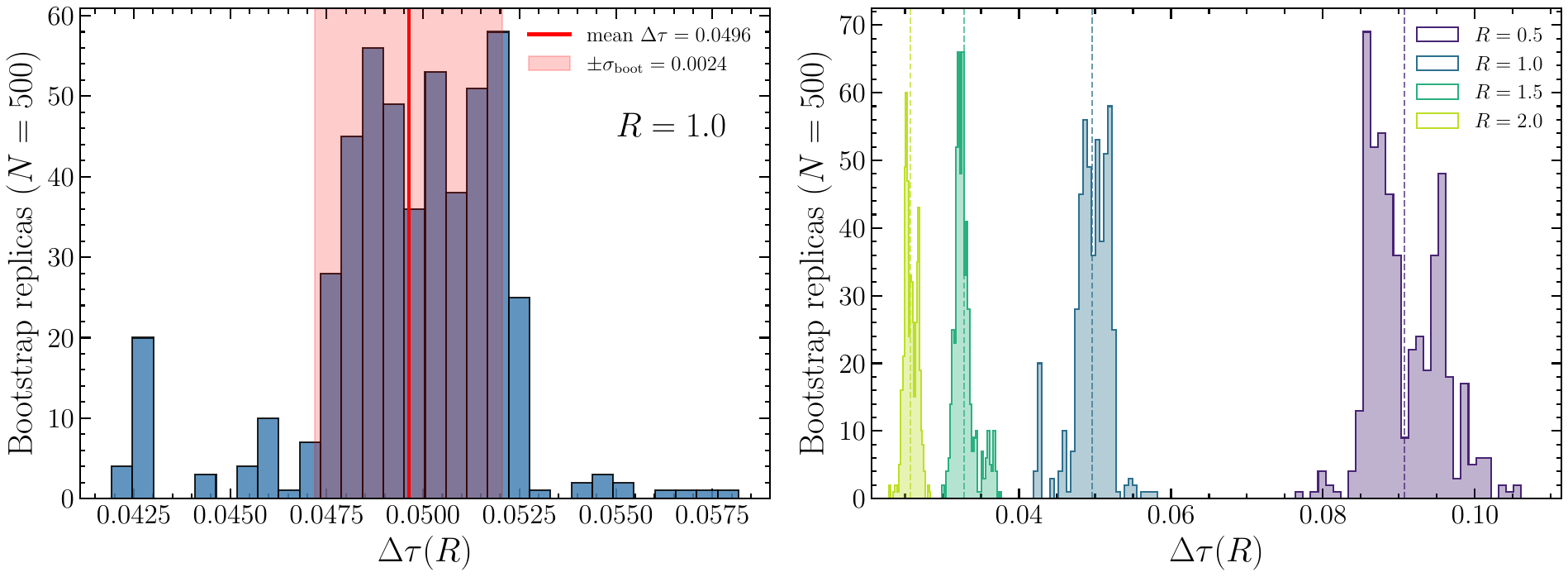}
  \vspace{-2em}
    \caption{
    Bootstrap distributions of the optimal shift $\Delta\tau(R)$ between the hadron-level and shifted parton-level $\tau_1^C$ distributions at $\sqrt{s}=319~\textrm{GeV}$, $Q=50~\textrm{GeV}$, and $x_B=0.05$, from $N_{\rm boot}=500$ replicas.
    The left panel shows the bootstrap distribution at $R=1.0$, with the red vertical line marking the mean and the shaded band marking $\pm\sigma_{\rm boot}$.
    The right panel overlays the corresponding distributions for $R\in\{0.5,1.0,1.5,2.0\}$. 
    The mean of each distribution defines the optimal shift $\Delta\tau(R)$; its standard deviation is the statistical uncertainty quoted in subsequent fits.}
  \label{fig:bootstrap_dt}
\end{figure}

For each jet radius $R$, the $\chi^2$ defined in Eq.~\eqref{eq:chi2-def} is scanned over a fine grid of candidate shifts $\Delta\tau \in [0, 0.25]$ (300 grid points). The location of the minimum determines the optimal shift $\Delta\tau(R, Q)$ for that radius, with statistical uncertainty $\sigma_{\rm boot}(R, Q)$ obtained from the bootstrap procedure described above. 
A representative example of this bootstrap procedure is shown in Fig.~\ref{fig:bootstrap_dt}. The left panel shows the bootstrap distribution of the optimal shift $\Delta\tau(R, Q)$ at $R = 1.0$, obtained from $N_{\rm boot} = 500$ resamplings of the parton- and hadron-level event sets.
The mean of this distribution defines the central value of $\Delta\tau(R, Q)$, and its standard deviation $\sigma_{\rm boot}$ is taken as the statistical uncertainty.
The right panel shows the same distributions for $R \in \{0.5, 1.0, 1.5, 2.0\}$ (color gradient indicating $R$); their positions track the expected $1/R$ scaling of the leading-power hadronization shift.

\subsubsection{Local fit}
The 16 per-radius shifts $\{\Delta\tau(R, Q) \pm \sigma_{\rm boot}(R, Q)\}_{R=0.5,\ldots,2.0}$ are fit to the leading-power model in Eq.~\eqref{eq:shift-fit} via weighted least squares, using $\sigma_{\rm boot}(R, Q)$ as the per-point uncertainty. This yields a central value of $\Omega_1^{\rm Pythia}$, whose statistical uncertainty $\sigma_{\Omega_1}^{\rm local}$ is obtained by repeating the fit on each of the $N_{\rm boot}=500$ bootstrap replicas $\{\Delta\tau^{(b)}(R, Q)\}_{b=1}^{N_{\rm boot}}$ and taking the standard deviation of the resulting distribution $\{\Omega_1^{(b)}\}_{b=1}^{N_{\rm boot}}$.

\begin{figure}[t]
  \centering
  \includegraphics[width=0.32\linewidth]{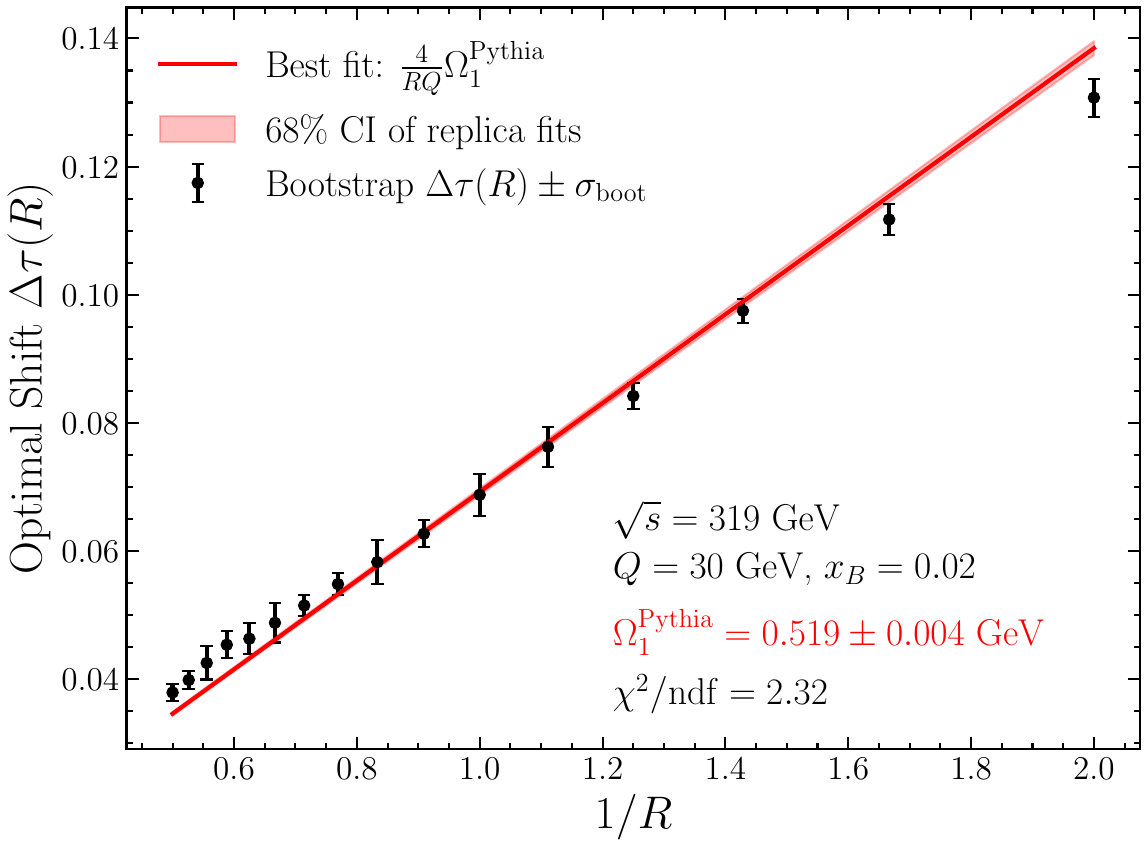}
  \includegraphics[width=0.32\linewidth]{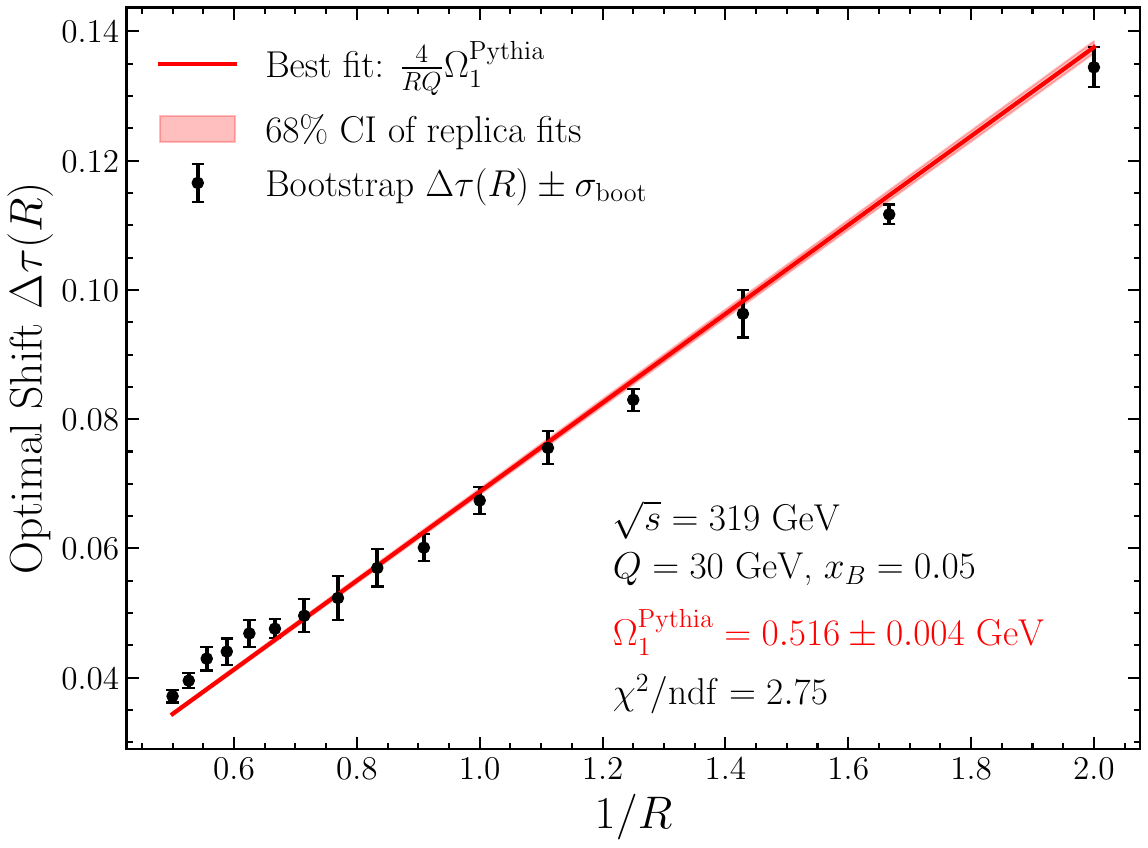}
  \includegraphics[width=0.32\linewidth]{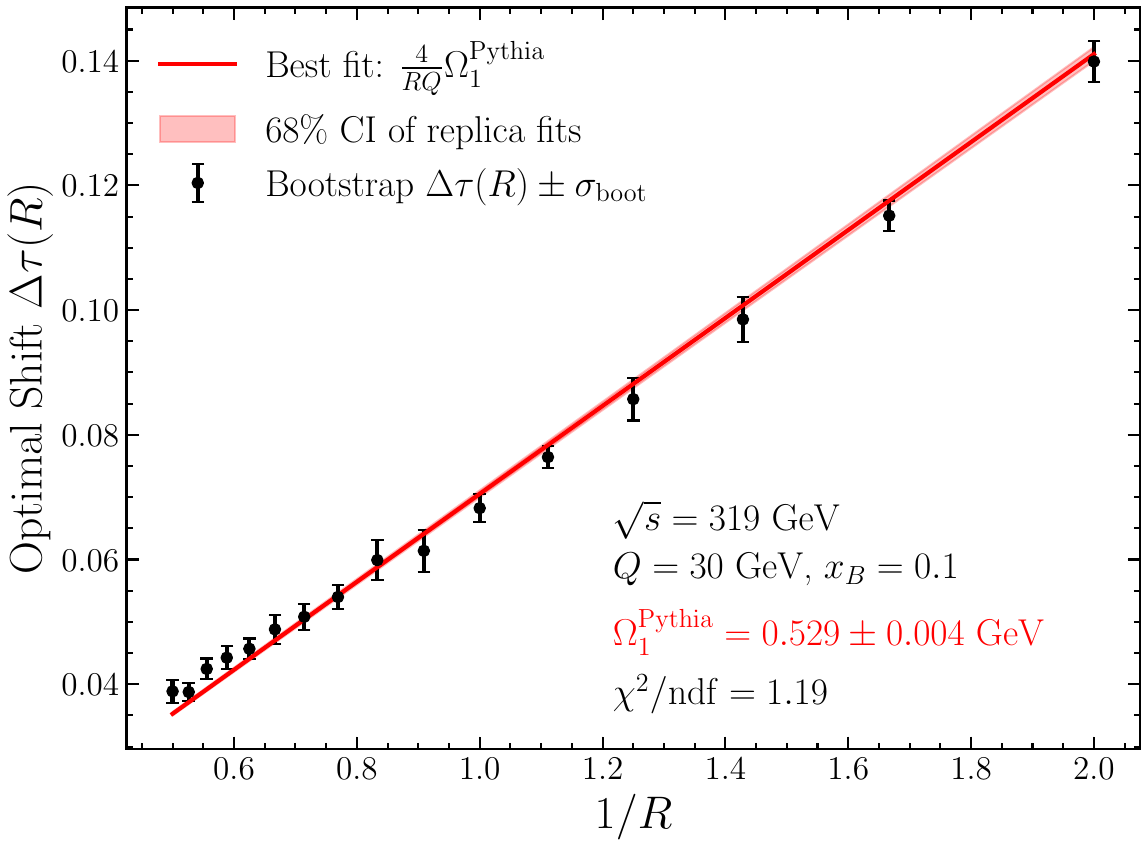}
  \includegraphics[width=0.32\linewidth]{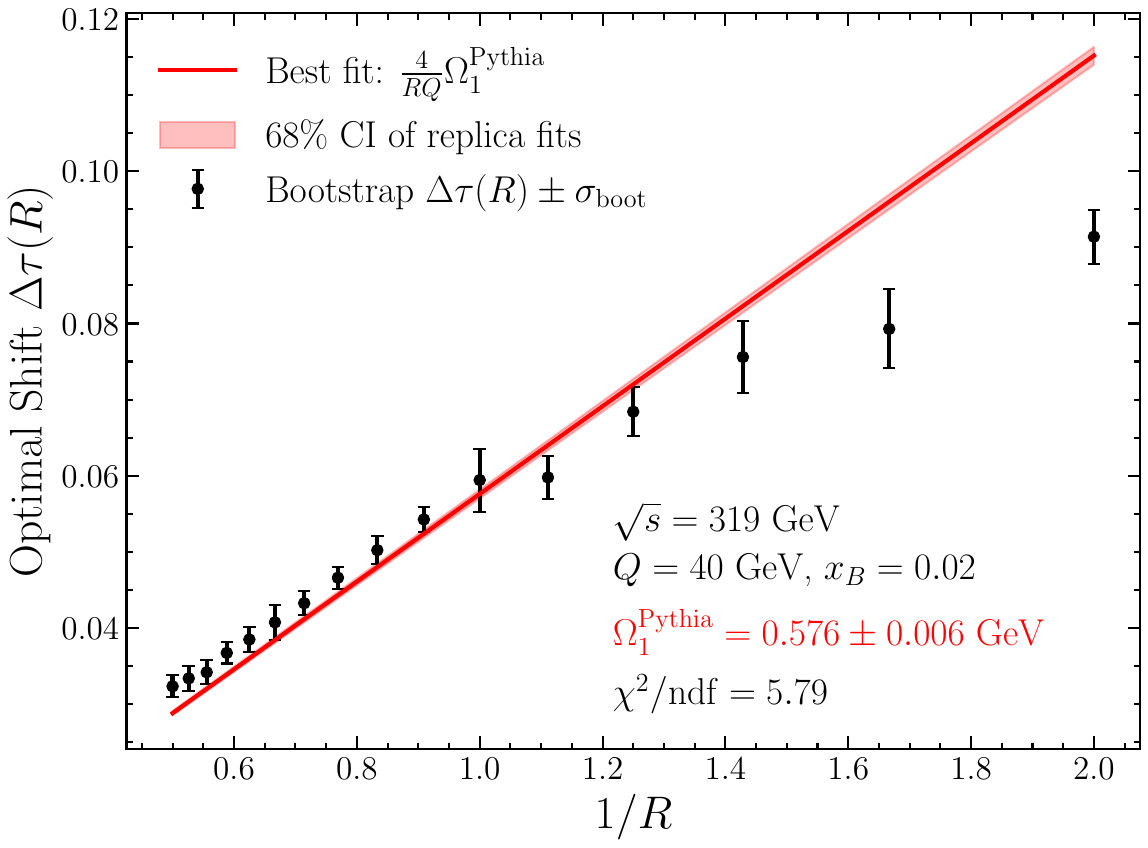}
  \includegraphics[width=0.32\linewidth]{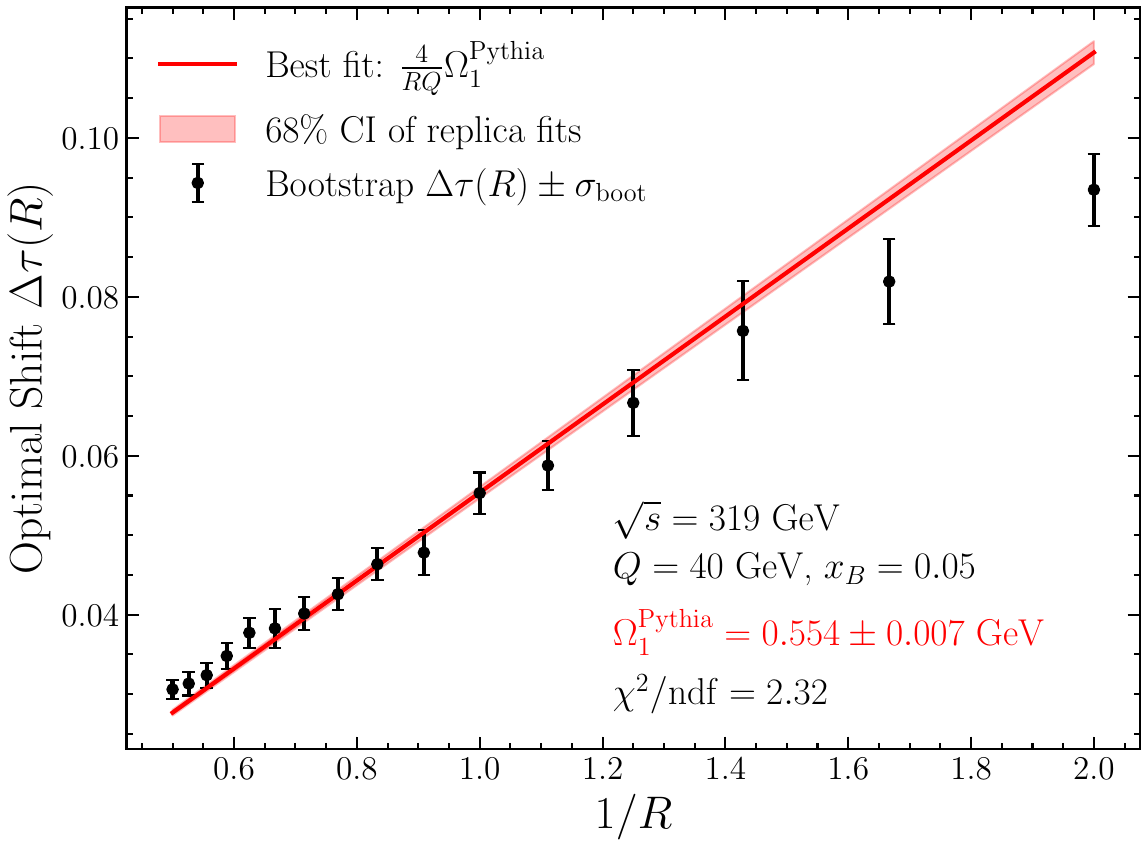}
  \includegraphics[width=0.32\linewidth]{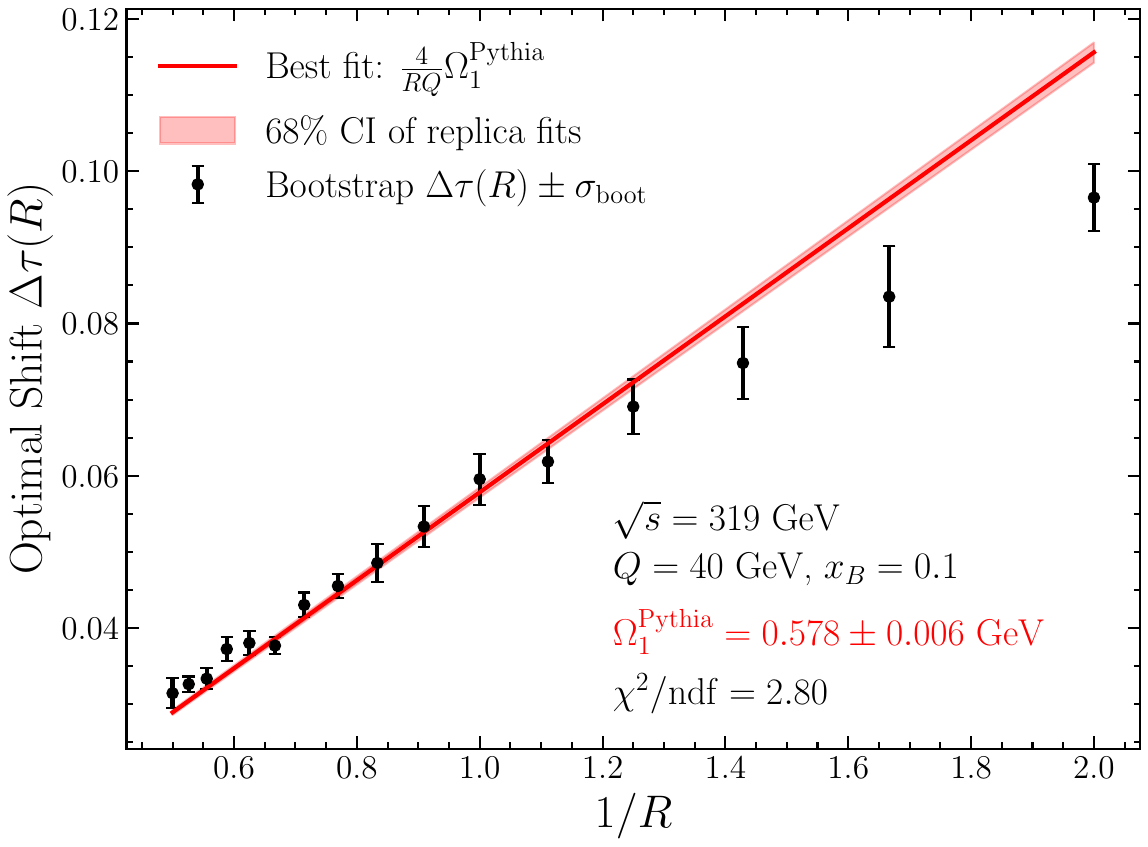}
  \includegraphics[width=0.32\linewidth]{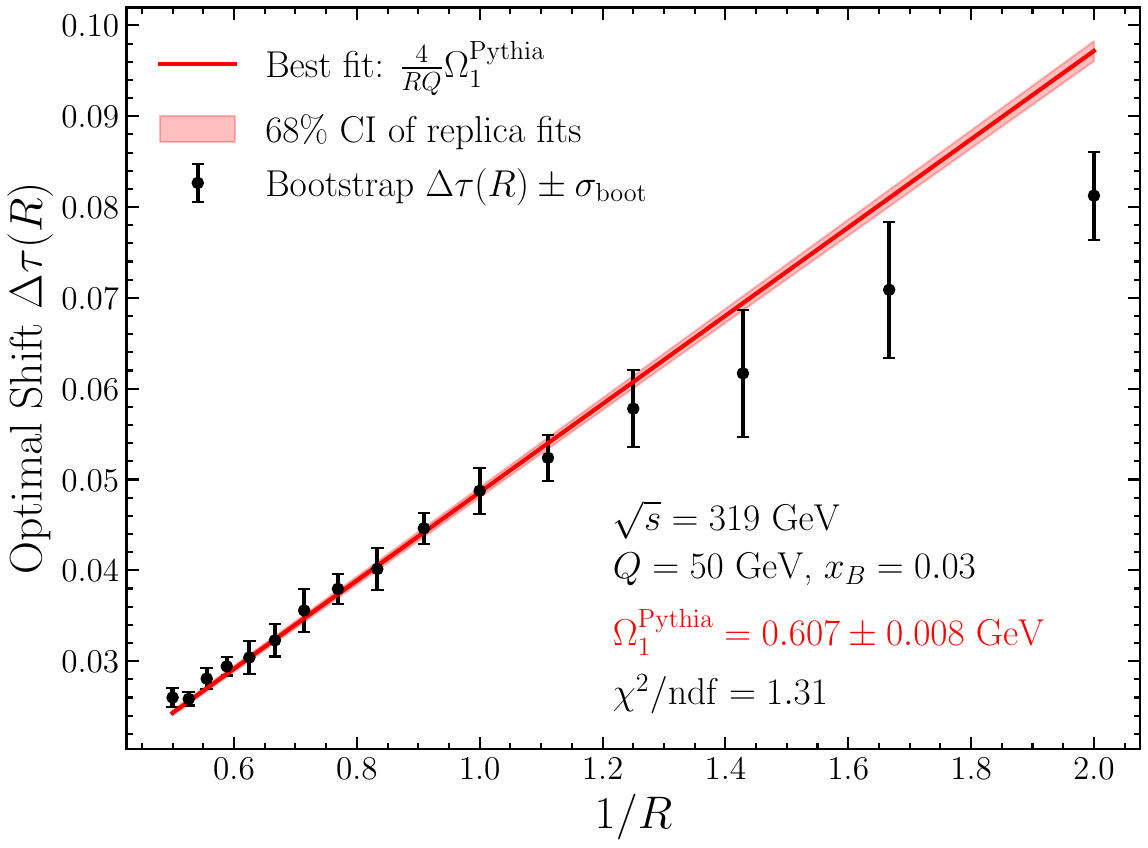}
  \includegraphics[width=0.32\linewidth]{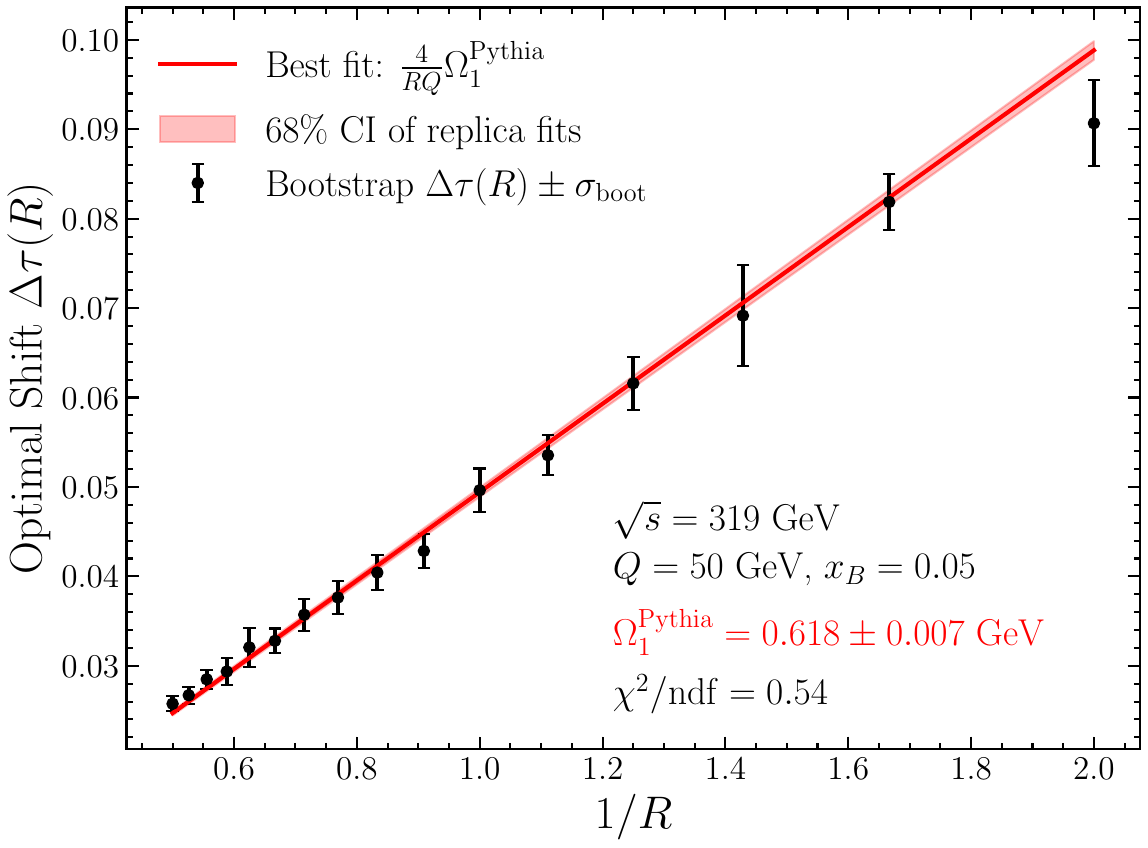}
  \includegraphics[width=0.32\linewidth]{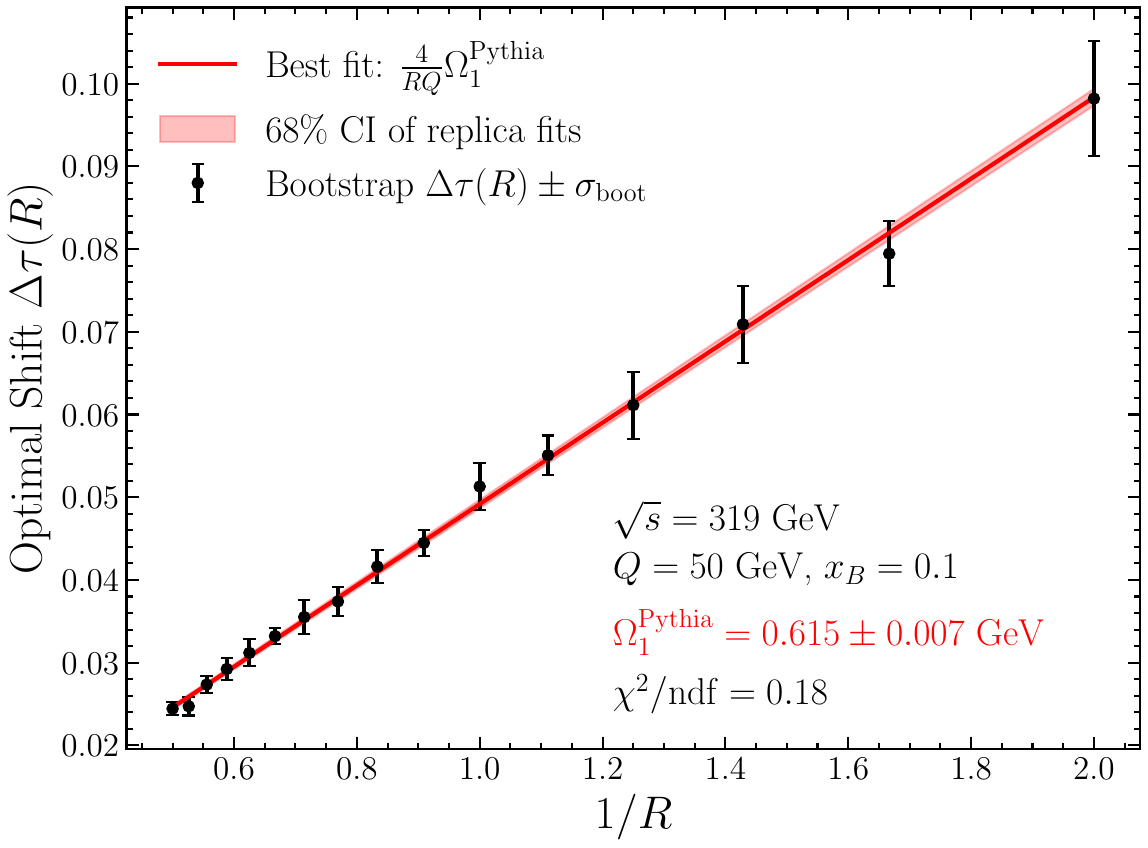}
\caption{
    Extracted optimal shifts $\Delta\tau(R, Q)$ as a function of $1/R$ for representative kinematic configurations $(Q, x_B)$, arranged in a $3 \times 3$ grid.
    Rows correspond to $Q = 30,\,40,\,50~\mathrm{GeV}$ (top to bottom),
    and columns correspond to $x_B = 0.02,\,0.05,\,0.10$ (left to right),
    except for the bottom row where $x_B = 0.03,\,0.05,\,0.10$.
    Black points show the optimal $\Delta\tau$ at each $R$ with bootstrap-resampling uncertainties. The red line is the local best-fit prediction $\Delta\tau = (4/R)(\Omega_1^{\mathrm{Pythia}}/Q)$ obtained from a weighted least-squares fit; the red band is the 68\% interval of the corresponding fit performed on each bootstrap replica.
    The approximately linear behavior observed across the various hard scales $Q$ and Bjorken-$x_B$ values supports the $1/R$ scaling of the leading-power non-perturbative shift.
}
  \label{fig:shift_vs_R}
\end{figure}

This procedure is carried out independently for each kinematic configuration. The resulting set of optimal shifts $\Delta\tau(R, Q)$ is then compared to the theoretical expectation in Eq.~\eqref{eq:shift-fit} and plotted as a function of $1/R$ in Fig.~\ref{fig:shift_vs_R} for representative configurations. 
A clear linear dependence on $1/R$ is observed across the entire range of jet radii considered, consistent with the predicted $R$-scaling of the leading power correction.
The goodness-of-fit ratio $\chi^2_{\rm local}/{\rm ndf}$ (with ${\rm ndf} = 16-1 = 15$) indicates the consistency with the leading-power $1/R$ form. 
The local fit yields $\chi^2_{\rm local}/\text{ndf}$ values ranging over $[0.2, 5.8]$ across the 9 kinematic configurations. The non-vanishing values reflect a mild deviation from the strict $1/R$ form (cf. Fig.~\ref{fig:shift_vs_R}), attributable to the subleading-power residual effects discussed below, especially at smaller $R\to 0.5$ ($1/R\to 2$) where the assumption $R\sim 1$ in our factorization theorem becomes more strained; nonetheless the linear $1/R$ scaling captures the dominant trend at every kinematic point.

\subsubsection{Global fit}

To further test the universality of the non-perturbative parameter, we perform a complementary global fit as a cross-check of the local fit. Rather than first extracting individual $\Delta\tau(R, Q)$ values and then fitting their $R$ dependence, the global fit scans $\Omega_1^{\rm Pythia}$ directly by minimizing the total $\chi^2$ summed over all jet radii simultaneously. For a candidate value $\Omega_1$, we define
\begin{align}
\label{eq:chi2-global}
\chi^2_{\rm tot}(\Omega_1)
=
\sum_{R=0.5}^{2.0}
\chi^2\!\left(\Delta\tau = \frac{4\Omega_1}{RQ};\, R\right)\,,
\end{align}
where the per-radius $\chi^2$ is as given in Eq.~\eqref{eq:chi2-def}. This total $\chi^2$ is scanned over $\Omega_1 \in [0.2, 0.8]~\textrm{GeV}$ on a 1000-point grid, and the minimum defines $\Omega_1^{\rm Pythia,\,global}$. Its statistical uncertainty $\sigma_{\Omega_1}^{\rm global}$ is obtained from $N_{\rm boot} = 500$ bootstrap replicas of the joint data.

The local and global fits ask the same physical question through different weighting schemes. The local fit assumes the $1/R$ form holds exactly and propagates per-$R$ bootstrap uncertainties through a single-parameter least-squares fit; the global fit lets each $R$ contribute to a single $\Omega_1$ via the full windowed $\chi^2$, weighting each histogram bin rather than each radius. If the leading-power $1/R$ scaling is accurate, the two values must agree within their bootstrap uncertainties, providing a non-trivial internal consistency check.

To further test universality across kinematics, we combine the data from all kinematic configurations considered in this work into a grand global fit by extending the sum in Eq.~\eqref{eq:chi2-global} over all kinematic points as well as all jet radii. The kinematic grid covers $Q \in \{30, 40, 50\}~\textrm{GeV}$, with $x_B \in \{0.02, 0.05, 0.1\}$ for $Q \in \{30, 40\}~\textrm{GeV}$ and $x_B \in \{0.03, 0.05, 0.1\}$ for $Q = 50~\textrm{GeV}$, yielding 9 kinematic configurations in total. 
(For $Q = 50~\textrm{GeV}$ the value $x_B = 0.02$ falls below the kinematic limit at $\sqrt{s} = 319~\textrm{GeV}$, which requires $x_B \gtrsim 0.025$, so $x_B = 0.03$ is the lowest accessible value.)

\begin{figure}[t]
  \centering
  \includegraphics[width=0.85\linewidth]{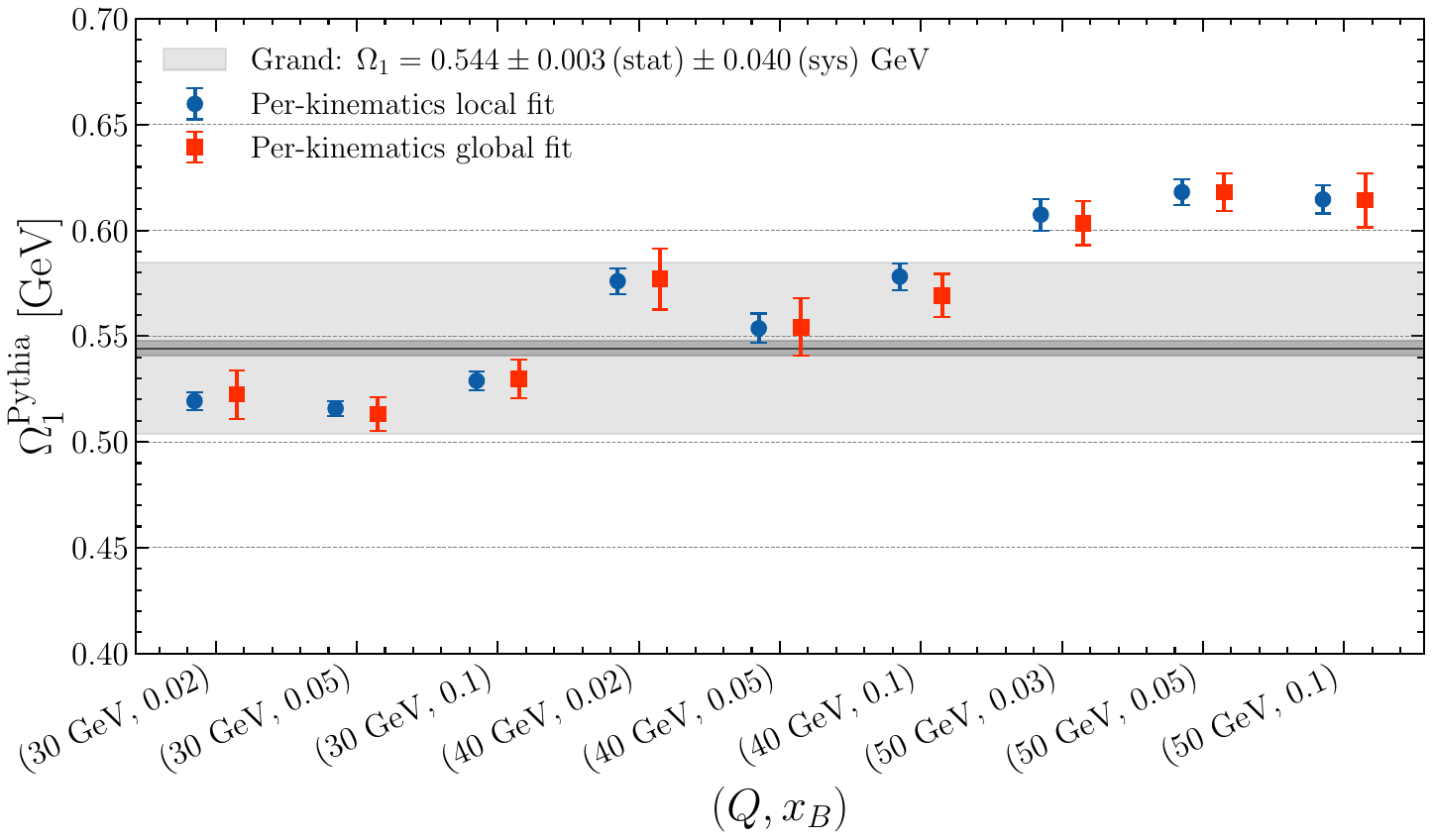}
  \caption{
    Three extractions of $\Omega_1^{\rm Pythia}$ across the 9 kinematic configurations of this study. 
    Blue circles show the per-kinematics local fit (Eq.~\eqref{eq:shift-fit} fit to the 16 per-radius shifts $\Delta\tau(R)\pm \sigma_{\rm boot}(R)$);
    red squares show the per-kinematics single-kinematics global fit (minimum of $\chi^2_{\rm tot}$ in Eq.~\eqref{eq:chi2-global} at fixed $(Q, x_B)$).
    Both error bars are bootstrap statistical uncertainties at $N_{\rm boot}=500$, and the slight horizontal offsets between the two markers are for visual clarity only. 
    The dark grey band shows the grand global fit central value with statistical uncertainty $\pm\sigma_{\rm stat}$ (Eq.~\eqref{eq:omega_1_pythia_global-stat}); the lighter outer band represents the total uncertainty $\pm\sqrt{\sigma_{\rm stat}^2 + \sigma_{\rm sys}^2}$ (Eq.~\eqref{eq:omega_1_pythia_global}), where the systematic component $\sigma_{\rm sys}$ quantifies the spread of the per-kinematics extractions (Eq.~\eqref{eq:sigma-sys}). Local and global extractions agree at every kinematic point. The residual upward drift of the central values with $Q$ is discussed in the main text.
  }
  \label{fig:omega_extraction}
\end{figure}

The result of this global analysis, together with the per-kinematics local and global fits, is summarized in
Fig.~\ref{fig:omega_extraction}.  The minima of the per-kinematics $\chi^2_{\rm tot}$ profiles are well-defined and approximately parabolic, allowing the bootstrap procedure of Sec.~\ref{subsec:procedure} to be applied directly. 
Combining all 9 kinematic configurations in the grand global fit yields a central value with statistical uncertainty
\begin{align}
\label{eq:omega_1_pythia_global-stat}
\Omega_1^{\rm Pythia}\big|_{\rm grand} \;=\; 0.544 \pm 0.003~(\text{stat})~\text{GeV},
\end{align}
where the statistical uncertainty is obtained from the joint bootstrap.
However, the per-kinematics single-global extractions exhibit a residual spread that exceeds this statistical uncertainty by an order of magnitude. We quantify this as a systematic uncertainty by taking the sample standard deviation of the 9 per-kinematics central values,
\begin{align}
\label{eq:sigma-sys}
\sigma_{\rm sys} \;=\; \mathrm{std}\big[\{\Omega_1^{\rm Pythia, (i)}\}_{i=1}^{9}\big] 
\;\simeq\; 0.040~\text{GeV}.
\end{align}
Combining the two contributions, our final extraction reads
\begin{align}
\label{eq:omega_1_pythia_global}
\Omega_1^{\rm Pythia} \;=\; 0.544 \pm 0.003~(\text{stat}) \pm 0.040~(\text{sys})~\text{GeV}.
\end{align}

The systematic uncertainty $\sigma_{\rm sys} = 0.040$~GeV in Eq.~\eqref{eq:omega_1_pythia_global} is dominated by a residual $Q$-dependence of the per-kinematics extractions, with central values rising from $\Omega_1^{\rm Pythia}\simeq 0.52~{\rm GeV}$ at $Q=30~{\rm GeV}$ to $\simeq 0.61~{\rm GeV}$ at $Q=50~{\rm GeV}$, while the variation across $x_B$ at fixed $Q$ is small. This drift may be a consequence of the hadronization treatment within \textsc{Pythia} or sub-leading power corrections, or both. We leave the investigation of this effect for a later study.

To relate this Monte Carlo-extracted value to the formal definitions in our factorization theorem, we apply the relations in Eqs.~\eqref{eq:relation_omega_pythia_R_gap} and \eqref{eq:R-dep-shift}, accounting for the renormalon-free gap parameter $\Delta$. This yields the non-perturbative inputs for the theoretical predictions,
\begin{equation}
\label{eq:R-gap-parameters}
\Omega_1(R_\Delta, \mu_\Delta) \simeq 0.59 \pm 0.04~\textrm{GeV}\,, \quad 
\Delta(R_\Delta, \mu_\Delta) = 0.05~\textrm{GeV}\,,
\end{equation}
where the uncertainty on $\Omega_1(R_\Delta, \mu_\Delta)$ is dominated by the systematic component of Eq.~\eqref{eq:omega_1_pythia_global}. These values are consistent with $\Omega_1^{\rm Pythia} = \Omega_1 - \Delta \simeq 0.54~\textrm{GeV}$ defined in Eq.~\eqref{eq:relation_omega_pythia_R_gap}.

The interpretation of this result as a leading-power non-perturbative effect is supported by several nontrivial consistency checks, while one feature--the residual
$Q$-dependence--requires further commentary:
\begin{itemize}
\item The extracted value of $\Omega_1^\textrm{Pythia}$ shows only a small spread of $\sim 2\text{--}4\%$ across the three accessible $x_B$ values at each fixed $Q$, with no systematic monotonic trend in $x_B$. This relative $x_B$-stability is in line with the universality of the leading-power soft function in the longitudinal momentum fraction.

\item Beyond the leading $1/R$ form, two mild systematic residuals are present: a slight deviation from the strict linear $1/R$-dependence within each kinematic point, and an overall $\sim 17\%$ drift of the central $\Omega_1^{\rm Pythia}$ with $Q$. Both are qualitatively consistent with subleading-power corrections beyond the leading OPE;
however, the size of the $Q$-drift substantially exceeds the natural NLP scale of $\mathcal{O}(\Lambda_{\rm QCD}^2/Q_R^2)$, suggesting that part of both residuals may also be associated with the specific implementation of the hadronization model in \textsc{Pythia}. Comparison to other Monte Carlos should be done, but lies outside the scope of this paper.  A definitive disentanglement requires applying the same procedure to experimental data, which we plan to do in a follow-up study. This size of uncertainty in $\Omega_1$, however, is not out of line with similar determinations of the $\Omega_1$ shift parameter contributing to $e^+e^-$ event shape studies, see for instance studies on thrust \cite{Bell:2023dqs, Benitez:2024nav}, and C parameter \cite{Hoang:2015hka}. 

\item The local and global fit values agree within their bootstrap uncertainties at each kinematic point, the largest deviation being 0.8$\sigma$ at ($Q=40~\textrm{GeV}$, $x_B=0.10$). 
Since the local fit assumes the $1/R$ form exactly while the global fit imposes it implicitly through Eq.~\eqref{eq:chi2-global}, this internal agreement validates both the functional form of the shift and the uncertainty estimation procedure.

\item The value $\Omega_1(R_\Delta, \mu_\Delta)\simeq 0.59$~GeV obtained in Eq.~\eqref{eq:R-gap-parameters} is comparable to the value $\Omega_1(R_\Delta, \mu_\Delta) \simeq 0.5$~GeV used in the analysis of the $\tau_1^b$ observable~\cite{Ee:2025scz}, which was estimated by comparing analytic predictions with experimental data from the H1 collaboration~\cite{H1:2024aze}.
This agreement is nontrivial, given that neither the normalization nor the $R$-dependence was imposed in our \textsc{Pythia}-based extraction. It is also somewhat larger than the analogous $\Omega_1$ for $e^+e^-$ event shapes, suggesting they may indeed be independent quantities at the non-perturbative level \cite{Kang:2015moa}.

\item The close alignment between the shifted partonic and hadronic distributions throughout the tail region (Fig.~\ref{fig:tau_distributions}) confirms the central prediction of the operator product expansion: at leading power, hadronization does not distort the shape of the perturbative spectrum but induces a universal translation whose magnitude is controlled by $\Omega_1$ and the jet radius $R$.
\end{itemize}

\section{Extraction of the Non-perturbative Shift using Symbolic Regression}
\label{sec:SR}
In this section, we take a detour to test our prediction of the universal scaling of the non-perturbative shift to Centauric 1-jettiness in a completely different way from \sec{omega1}. The reader wishing to see our final results may skip to \sec{results}.

We now use symbolic regression (SR) to probe the structure of the dominant non-perturbative correction to the Centauric 1-jettiness distribution in a model-independent way. The leading-power factorization theorem predicts that hadronization in the tail region acts as a universal shift, $\Delta\tau_1^C \sim 4\Omega_1/(Q R)$, independent of $\tau_1^C$. The extraction of Sec.~\ref{sec:omega1} confirmed that this picture describes the \textsc{Pythia} spectra well. Here we ask a sharper question: when the simulated data are given full functional freedom, do they favor any residual dependence on \(\tau_1^C\), or any dependence on $R$, $Q$, and $x_B$ beyond the leading-power form? Such effects would signal higher moments of the shape function or subleading corrections to the factorization theorem. 

Rather than imposing a functional ansatz, we extract the shift directly from the parton- and hadron-level distributions and let SR determine its analytic form. Two features distinguish the present analysis. 
First, the shift is measured locally as a function of $\tau_1^C$ through a windowed $\chi^2$ comparison between the shifted parton-level and hadron-level spectra, so that any $\tau_1^C$-dependence is read off directly from the data rather than assumed. 
Second, the extraction is performed simultaneously across nine $(Q, x_B)$ kinematic configurations, allowing SR to resolve the joint $(\tau_1^C, R, Q, x_B)$ structure of the shift and to test the universality of the leading-power $4\Omega_1/(QR)$ scaling against the data. 

This section is organized as follows. 
In Sec.~\ref{subsec:sr-shift-interpretation} we clarify the physical interpretation of a general $\tau_1^C$-dependent shift and its relation to
higher moments of the shape function. 
In Sec.~\ref{subsec:sr} we review the SR framework and our model-selection criteria. 
In Sec.~\ref{subsec:SR-training} we describe the local shift extraction and the construction of the combined multi-kinematic dataset. 
In Sec.~\ref{subsec:SR-Model} we present the resulting
SR models and compare them to the leading-power prediction.

\subsection{Interpretation of a General Shift}
\label{subsec:sr-shift-interpretation}

Before introducing the SR analysis, it is useful to clarify what is meant
by allowing a more general shift function. Starting from the leading-power
result derived in Eq.~\eqref{eq:power-correction}, expanding the
perturbative spectrum for \(k/Q_R \ll 1\) gives
\begin{align}
\sigma\left(\tau_1^C\right)
&=
\sigma_{\rm PT}\left(\tau_1^C\right)
-
\frac{2\Omega_1}{Q_R}
\sigma_{\rm PT}'\left(\tau_1^C\right)
+
\frac{\Omega_2}{Q_R^2}
\sigma_{\rm PT}''\left(\tau_1^C\right)
+
\mathcal{O}\!\left(\frac{\Lambda_{\rm QCD}^3}{Q_R^3}\right),
\label{eq:sr-ope-expanded}
\end{align}
where the moments of the shape function are defined by
\begin{align}
\Omega_n
=
\int dk\,k^n\,F(k).
\end{align}
At leading power, the first correction in
Eq.~\eqref{eq:sr-ope-expanded} is equivalent to a rigid shift of the perturbative spectrum,
\begin{align}
\sigma\left(\tau_1^C\right)
=
\sigma_{\rm PT}\!\left(
\tau_1^C-\frac{2\Omega_1}{Q_R}
\right)
+
\mathcal{O}\!\left(\frac{\Lambda_{\rm QCD}^2}{Q_R^2}\right).
\label{eq:sr-leading-shift}
\end{align}
For the Centauric observable, this corresponds to the characteristic
\(4\Omega_1/(Q R)\) scaling derived in
Sec.~\ref{subsubsec:Centauric-soft}.

The analyses performed in the following subsections are designed to test
the extent to which the hadron-level spectra are described by this
leading-power structure. In particular, we separately consider possible
deviations in the \(R\)-dependence and in the \(\tau_1^C\)-dependence of
the shift. Additional \(R\)-dependence beyond the leading
\(4\Omega_1/(Q R)\) behavior can arise from subleading corrections to the
factorization theorem, including higher-order contributions from the soft,
beam, and jet sectors. Since the Centauric measurement weights introduce
explicit \(R\)-dependence into the factorized beam and jet functions, such
effects can generate additional effective \(R\)-dependence in the extracted
shift. In the present work, we do not attempt to disentangle these
different contributions. Instead, we use SR as a model-independent probe
of possible residual structure beyond the leading-power \(1/R\) scaling.

We now turn to the interpretation of a \(\tau_1^C\)-dependent shift. To
understand how higher moments of the shape function generate effective
\(\tau_1^C\)-dependence, we compare
Eq.~\eqref{eq:sr-ope-expanded} with a generalized shifted form of the
spectrum,
\begin{align}
\sigma\left(\tau_1^C\right)
=
\sigma_{\rm PT}\!\left(
\tau_1^C-\Delta\tau_{\rm eff}\left(\tau_1^C;R, Q, x_B\right)
\right).
\end{align}
Writing
\begin{align}
\Delta\tau_{\rm eff}\left(\tau_1^C;R, Q, x_B\right)
=
\frac{2\Omega_1}{Q_R}
+
\delta\tau\left(\tau_1^C;R, Q, x_B\right),
\end{align}
where \(\delta\tau\) scales parametrically like
\(\Lambda_{\rm QCD}^2/Q_R^2\), and expanding through order
\(\Lambda_{\rm QCD}^2/Q_R^2\), we find
\begin{align}
\sigma_{\rm PT}\!\left(
\tau_1^C-\Delta\tau_{\rm eff}
\right)
&=
\sigma_{\rm PT}\left(\tau_1^C\right)
-
\Delta\tau_{\rm eff}
\sigma_{\rm PT}'\left(\tau_1^C\right)
+
\frac{2\Omega_1^2}{Q_R^2}
\sigma_{\rm PT}''\left(\tau_1^C\right)
+
\mathcal{O}\!\left(\frac{\Lambda_{\rm QCD}^3}{Q_R^3}\right).
\end{align}
Matching this expression to Eq.~\eqref{eq:sr-ope-expanded} gives
\begin{align}
\delta\tau\left(\tau_1^C;R, Q, x_B\right)
=
-
\frac{\Omega_2-2\Omega_1^2}{Q_R^2}
\frac{
\sigma_{\rm PT}''\left(\tau_1^C\right)
}{
\sigma_{\rm PT}'\left(\tau_1^C\right)
}
+
\mathcal{O}\!\left(\frac{\Lambda_{\rm QCD}^3}{Q_R^3}\right).
\label{eq:sr-effective-shift-correction}
\end{align}
Equation~\eqref{eq:sr-effective-shift-correction} demonstrates that a
\(\tau_1^C\)-dependent effective shift does not imply a breakdown of the
leading-power factorization theorem. Rather, such dependence arises
naturally when higher-order corrections are absorbed into an approximate
shifted representation of the spectrum. In this sense, a
\(\tau_1^C\)-dependent shift should be interpreted as an effective
parameterization of subleading corrections, including higher moments of the
shape function, that distort the spectrum rather than simply translating
it.
In the following, we extract this effective shift $\Delta\tau_{\rm eff}(\tau_1^C;R, Q, x_B)$ directly from the simulated spectra and use SR to determine whether the data require any $\delta\tau(\tau_1^C; Q, R, x_B)$ beyond the leading-power rigid shift.

\subsection{Overview of SR}
\label{subsec:sr}

SR is a machine-learning technique that constructs analytic expressions directly from data~\cite{Schmidt:2009ie}, and has been applied in physics~\cite{Udrescu:2019mnk, Angelis:2023bvv,Dotson:2025omi}. Unlike traditional regression, where a functional form is specified \emph{a priori} and only the numerical parameters are fitted, SR simultaneously determines both the structure of the expression and its parameters. It therefore searches over a large space of candidate analytic expressions and identifies compact functional forms that accurately reproduce the data.

In the present context, the goal is not merely to reproduce the simulated spectra, but to test whether the dominant non-perturbative correction follows the leading-power structure predicted by the factorization theorem. Concretely, we ask whether the data favor a rigid shift of the form
\(
\Delta\tau_1^C \sim 4\Omega_1/(Q R)
\),
or whether residual dependence on \(R\), \(Q\), \(x_B\), or \(\tau_1^C\) improves the description of the spectra. The advantage of SR for this purpose is that it yields explicit analytic expressions that can be compared directly with the functional forms predicted by the factorization theorem.

We extract the shift directly from the parton- and hadron-level distributions by a windowed \(\chi^2\) comparison (described in Sec.~\ref{subsec:SR-training}), and apply SR to the extracted shift. The regression target is the effective shift \(\Delta\tau_{\rm eff}(\tau_1^C;R, Q, x_B)\), and the features are \((\tau_1^C, R, Q, x_B)\).

SR is implemented using the \textsc{PySR} framework~\cite{cranmer:2023sr}, which employs an evolutionary algorithm: candidate expressions built from elementary binary operations \(\{+,-,\times,\div\}\) and unary operators, e.g. \{exp, log, etc.\}, are iteratively optimized through mutation, crossover, and parameter fitting. Each candidate is scored by its fit quality, quantified by the weighted mean-squared error (WMSE),
\begin{align}
\label{eq:MSE}
\mathrm{WMSE}
=
\frac{\sum_{i=1}^{N} w_i\left|y_i^{\rm model} - y_i^{\rm data}\right|^2}
     {\sum_{i=1}^{N} w_i}\,,
\qquad
w_i=\frac{1}{\sigma_i^2}\,.
\end{align}
and by its complexity, defined as the number of nodes in the corresponding expression tree. This trade-off favors compact analytic representations and helps suppress over-fitting.
The WMSE is normalized by $\sum_i w_i$ and therefore measures the weighted-mean squared residual in units of the per-point variance; it serves as a relative measure of fit quality across models.

For each run, the algorithm returns a set of models at increasing complexity, the ``Hall of Fame,'' representing different compromises between accuracy and simplicity. To identify representative models we construct the Pareto front Ref.~\cite{Pareto1906} of WMSE versus complexity over the full dataset---the set of models for which no alternative achieves both lower WMSE and lower complexity---and quantify the improvement in fit quality per unit complexity through the score
\begin{align}
\label{eq:sr-score}
\mathrm{score}(c_i)
=
\frac{\left|\Delta\log \mathrm{WMSE}\right|}{\Delta c}\,,
\end{align}
where \(\Delta c\) is the change in complexity between neighboring points on the Pareto front. Models at which the score is large mark the points where additional complexity yields a genuine improvement in the description of the data; we take these as the representative candidate forms and compare them with the leading-power prediction.
We did not need to set aside a held-out validation sample to monitor generalization: with only a handful of free parameters per candidate, over-fitting is controlled by the complexity penalty in the model ranking and by the requirement that the learned structures be stable across independent SR runs.

To verify that the learned structures are not artifacts of the chosen operator set, we performed the search with and without the inclusion of unary operators (log, exp). No such model was favored by the selection criterion over the forms obtained with the elementary binary operators; the dominant learned structure remained the leading-power \(4\Omega_1/(Q R)\) rigid shift. 
We additionally perform multiple independent SR runs (``replicas'') with different random seeds and retain only structures that recur across replicas, ensuring that the selected forms are robust against the stochastic optimization.

\subsection{Training Procedure}
\label{subsec:SR-training}

The dataset consists of normalized differential cross sections in \(\tau_1^C\) constructed from \textsc{Pythia}-generated DIS events at both parton and hadron level, at a center-of-mass energy \(\sqrt{s}=319~\mathrm{GeV}\). To probe the joint dependence of the shift on the hard scale and the jet radius, we combine nine kinematic configurations, with \(Q \in \{30,40,50\}~\mathrm{GeV}\) and \(x_B \in \{0.02, 0.05, 0.10\}\) for \(Q=30,40~\mathrm{GeV}\) and \(x_B \in \{0.03, 0.05, 0.10\}\) for \(Q=50~\mathrm{GeV}\), consistent with Sec.~\ref{sec:omega1}. 
For each configuration the Centauric 1-jettiness is computed event by event and histogrammed for the jet radii \(R \in \{0.5, 0.6, \ldots, 2.0\}\), with the fit restricted to the resummation-dominated window \([t_1, t_2]\) introduced in Sec.~\ref{subsec:profile}, within which the operator product expansion underlying Eq.~\eqref{eq:sr-leading-shift} is expected to hold.

The effective shift \(\Delta\tau_{\rm eff}(\tau_1^C;R, Q, x_B)\) is measured locally, as a function of \(\tau_1^C\), directly from the parton- and hadron-level spectra. For a trial displacement \(\delta\), the parton-level events are rigidly shifted, \(\tau_1^C \to \tau_1^C + \delta\), and rebinned to form the shifted parton distribution \(p_\delta\), which is compared to the hadron distribution \(h\) through a local windowed \(\chi^2\) evaluated over a small band of bins centered on each \(\tau_1^C\),
\begin{align}
\label{eq:local-chi2}
\chi^2_{\rm loc}(\delta;\tau_1^C)
=
\sum_{i \in \text{band}(\tau_1^C)}
\frac{\left[h_i - p_{\delta,i}\right]^2}
     {\sigma_{h,i}^2 + \sigma_{p_\delta,i}^2}\,,
\end{align}
where the band spans \(\pm 3\) bins around the central bin, central bins are
sampled in steps of two, and the per-bin uncertainties are the Poisson errors
of the normalized histograms. Minimizing Eq.~\eqref{eq:local-chi2} over \(\delta\) at
each central bin, with a parabolic interpolation around the discrete minimum,
yields \(\Delta\tau_{\rm eff}(\tau_1^C;R, Q, x_B)\). Because the parton-level events
are physically displaced before rebinning, the extraction conserves probability
and incorporates the Jacobian associated with a \(\tau_1^C\)-dependent shift;
within each band the displacement is treated as constant, so that the procedure
measures a locally rigid shift whose variation across bands captures the
effective \(\tau_1^C\)-dependence of
Eq.~\eqref{eq:sr-effective-shift-correction}. The statistical uncertainty on
\(\Delta\tau_{\rm eff}\) is obtained from a non-parametric bootstrap as in Sec.~\ref{subsec:procedure}: for each
\((R, Q,x_B)\) we resample the parton- and hadron-level events with replacement,
re-extract the shift on a common grid, and take the mean and standard deviation
over \(N_{\rm boot}=300\) replicas as the central value and uncertainty
\(\sigma_{\rm boot}\). The latter sets the regression weight
\(w=1/\sigma_{\rm boot}^2\) in Eq.~\eqref{eq:MSE}.

Repeating the extraction for all nine configurations and all jet radii yields the combined dataset used for the SR analysis, comprising 1767 points, each carrying the features \((\tau_1^C, R, Q, x_B)\), the target \(\Delta\tau_{\rm eff}\), and the weight \(1/\sigma_{\rm boot}^2\). Pooling the configurations is essential: at fixed \(Q\) the leading-power scaling \(1/(Q R)\) is degenerate with a pure \(1/R\) dependence, whereas varying \(Q\) allows SR to resolve the \(1/(Q R)\) structure of the leading-power shift and its normalization \(4\Omega_1\) directly from the data.

\subsection{Model Selection and Results}
\label{subsec:SR-Model}
\begin{figure}[t]
\noindent
\begin{minipage}{0.45\textwidth}
\centering
\includegraphics[width=\linewidth]{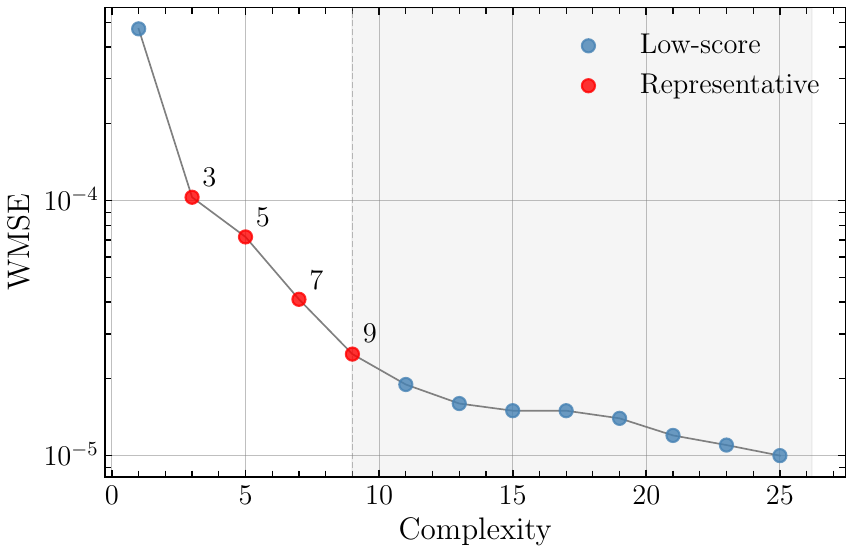}
\end{minipage}%
\hspace{2em}%
\begin{minipage}{0.40\textwidth}
\centering
\footnotesize
\setlength{\tabcolsep}{3pt}
\renewcommand{\arraystretch}{1.2}
\begin{tabular}{lccc}
\toprule
Model & $c$ & WMSE $[10^{-5}]$ & score \\
\midrule
$0.061/R$ & 3 & 10.3 & 0.758 \\
$2.203/(QR)$ & 5 & 7.21 & 0.180 \\
$(2.638 - \tau_1^C)/(QR)$ & 7 & 4.06 & 0.287 \\
$(1.725 + 0.154/\tau_1^C)/(QR)$ & 9 & 2.53 & 0.237 \\
\bottomrule
\end{tabular}
\end{minipage}%
\hfill
\caption{\textit{Left:} Pareto front of WMSE versus complexity for $\Delta\tau_{\rm eff}(\tau_1^C, R, Q, x_B)$ models in the Hall of Fame; red points mark the representative models. \textit{Right:} Representative models selected from the Pareto front by the score of Eq.~\eqref{eq:sr-score}, with numerical coefficients quoted with $Q$ in GeV. The rigid model at complexity~5 fixes the leading-power normalization, Eq.~\eqref{eq:omega1-sr}; higher-complexity models introduce $\tau_1^C$-dependence with only modest reductions in WMSE.}
\label{fig:pareto-front}
\end{figure}

\begin{figure}[t]
  \centering
  \includegraphics[width=1.0\linewidth]{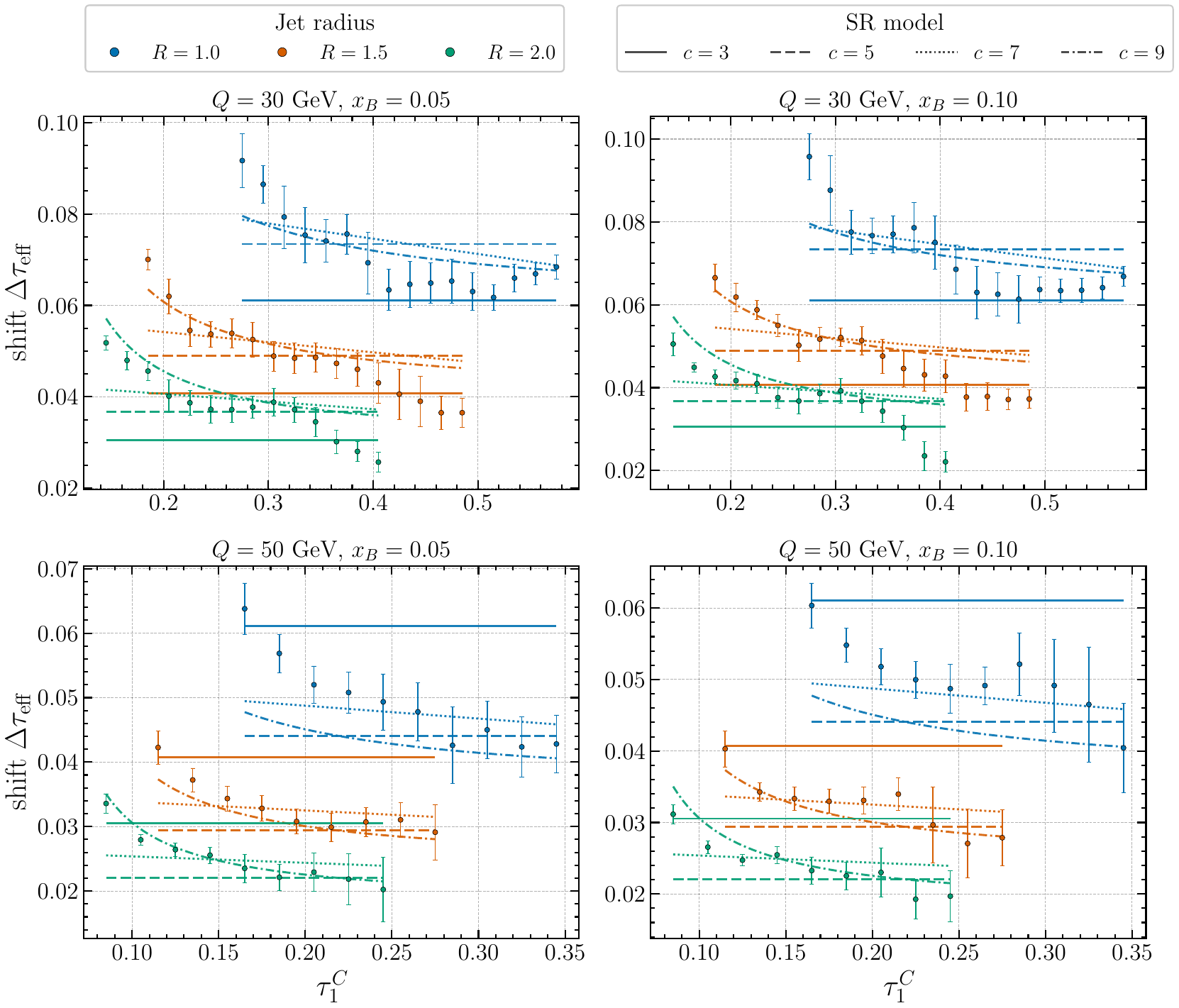}
  \caption{
  Effective shift $\Delta\tau_{\rm eff}(\tau_1^C;R,Q,x_B)$ extracted from the \textsc{Pythia} spectra (points, with bootstrap uncertainties) compared to the representative SR models of the right side of Fig.~\ref{fig:pareto-front} (curves), for selected kinematic configurations $(Q, x_B)$ and jet radii $R\in \{1.0, 1.5, 2.0\}$ (colors). The four models (complexities 3, 5, 7, 9) are distinguished by line style. The higher complexity models describe the effective shifts marginally better, but the leading-power, $\tau_1^C$-independent model (complexity 5, dashed) already describes the extracted shifts well across all configurations, with the \(\tau_1^C\)-dependent models (complexities 7, 9) providing only a mild additional tilt.
  }
  \label{fig:shift_SR_fit}
\end{figure}

Applying the procedure of Sec.~\ref{subsec:SR-training} to the combined dataset, SR returns a Hall of Fame of analytic models for the effective shift $\Delta\tau_{\rm eff}(\tau_1^C; R, Q, x_B)$ at increasing complexity. The resulting Pareto front of WMSE versus complexity is shown in Fig.~\ref{fig:pareto-front}: the WMSE drops sharply at low complexity, where SR is still identifying the leading-power structure of the shift, and then flattens rapidly beyond complexity~9, indicating diminishing returns from additional structure and the onset of overfitting. We therefore restrict our attention to models up to complexity~9; the representative models within this range, selected by the score of Eq.~\eqref{eq:sr-score}, are summarized in the right side of Fig.~\ref{fig:pareto-front} and compared to the extracted shift in Fig.~\ref{fig:shift_SR_fit}. As we discuss below, the largest single gain in fit quality occurs at the lowest complexities, where SR isolates the leading-power $1/(QR)$ scaling, with higher-complexity refinements producing only marginal further improvements. Although the complexity 3 model in Fig. \ref{fig:pareto-front} has the highest score, this is an artifact of the score quantifying relative improvement: since a complexity 1 model can only be a single variable or constant, it cannot capture any real functional dependence in the data, so even a modest increase in complexity produces a large relative gain in WMSE, inflating the score. The complexity 5, 7, and 9 models also fit the data well at low complexity, but have the added benefit of capturing non-trivial Q or $\tau_1^C$ dependence.

The dominant structure identified by the regression is the leading-power rigid shift. The largest single gain in fit quality occurs already at low complexity, where SR isolates the \(1/R\) scaling, and is completed once the hard-scale dependence is included, yielding the representative rigid model (Complexity 5), which carries no \(\tau_1^C\)- or \(x_B\)-dependence and is therefore written as a function of \((R, Q)\) alone,
\begin{align}
\label{eq:sr-rigid}
\Delta\tau_{\rm eff}^{\rm SR}(R,Q)
=
\frac{2.203}{Q R}\,,
\end{align}
with $Q$ in GeV. Identifying the coefficient with the leading-power prediction \(4\Omega_1/(QR)\) gives
\begin{align}
\label{eq:omega1-sr}
\Omega_1^{\rm SR} \simeq 0.55~\mathrm{GeV}\,,
\end{align}
in good agreement with the grand global determination of Sec.~\ref{sec:omega1}. We emphasize that neither the \(1/(QR)\) functional form nor its normalization was imposed: both emerge directly from the simulated data across the nine kinematic configurations. The absence of any \(x_B\)-dependence in the learned expression provides a data-driven confirmation of the universality of the leading-power soft function in the longitudinal momentum fraction.

Allowing for residual \(\tau_1^C\)-dependence, the leading correction (Complexity 7), which likewise carries no \(x_B\)-dependence, is
\begin{align}
\label{eq:sr-taudep}
\Delta\tau_{\rm eff}^{\rm SR}(\tau_1^C,R,Q)
=
\frac{2.638 - \tau_1^C}{Q R}\,,
\end{align}
which augments the rigid shift with a term scaling as \(-\tau_1^C/(QR)\). The leading-power coefficient is most cleanly read from the rigid model, Eq.~\eqref{eq:sr-rigid}; in Eq.~\eqref{eq:sr-taudep} it shifts upward because the linear term partially absorbs the constant, and the value in Eq.~\eqref{eq:omega1-sr} should therefore be taken from the rigid fit. As anticipated in Sec.~\ref{subsec:sr-shift-interpretation}, such a \(\tau_1^C\)-dependent term is the expected signature of subleading corrections, including higher moments of the shape function, absorbed into an effective shifted representation of the spectrum.

While the \(\tau_1^C\)-dependent model of Eq.~\eqref{eq:sr-taudep} achieves a lower WMSE than the rigid shift, the improvement is modest relative to the increased complexity, and the more complex models in the right side of Fig.~\ref{fig:pareto-front} yield only marginal further gains. The extracted \(\tau_1^C\)-dependence should accordingly be regarded as an effective parameterization of residual spectral distortions rather than a unique prediction for the structure of subleading corrections; its detailed form depends on the model-selection criterion and on the restricted kinematic window used in the fit. The dominant, robust conclusion of the SR analysis is therefore the leading-power result of Eqs.~\eqref{eq:sr-rigid}--\eqref{eq:omega1-sr}:
even without imposing the expected functional form, the simulated data favor a rigid, \(\tau_1^C\)-independent shift with the predicted \(1/(QR)\) scaling and a normalization consistent with the factorization theorem, as illustrated in Fig.~\ref{fig:shift_SR_fit}.

\section{Theoretical Predictions and Comparison with \textsc{Pythia}}
\label{sec:results}
In this section, we present the final theoretical predictions for the Centauric 1-jettiness distribution and compare them directly to hadron-level Monte Carlo results. The predictions are obtained using the factorization and resummation framework developed in Sec.~\ref{sec:theory-formalism}, the matched cross section constructed in Sec.~\ref{sec:fixed-order}, and the non-perturbative parameter $\Omega_1$ determined in Sec.~\ref{sec:omega1}. 
The purpose of this comparison is twofold. First, we test whether the predicted leading non-perturbative correction, in particular the universal $1/R$-shift, correctly describes the hadron-level spectrum. Second, we assess the perturbative convergence of the resummed predictions. We also comment on the limitations of the current treatment in the peak region.

Throughout this discussion, \textsc{Pythia} is treated as a numerical laboratory for isolating hadronization effects and testing scaling behaviors, rather than as a substitute for first-principles QCD calculations. Accordingly, our emphasis is on parametric behaviors, such as the $R$ dependence and universality of the leading power correction, rather than on a precise tuning of the peak region dynamics.

In Sec.~\ref{subsec:comparison}, we compare the resummed predictions at NLL, NNLL, and N$^3$LL accuracy to hadron-level \textsc{Pythia} results for several jet radii and study the perturbative convergence and universality of the non-perturbative shift. In Sec.~\ref{subsec:scope}, we discuss the scope and limitations of this comparison and clarify what aspects of the non-perturbative dynamics are tested by our analysis.

\subsection{Comparison with Hadron-Level Distributions}\label{subsec:comparison}
We now compare the theoretical predictions at NLL, NNLL, and N$^3$LL accuracy directly with hadron-level Centauric 1-jettiness distributions from \textsc{Pythia} at several jet radii. The goal is not precise tuning of Monte Carlo parameters, but rather to test the parametric predictions of the factorization theorem, in particular the $R$ dependence of the leading power correction.
The matched cross section is constructed using the hybrid scheme of Sec.~\ref{sec:fixed-order}, which preserves the geometric Centauro jet identification while ensuring perturbative stability.

Our predictions incorporate hadronization through convolution with the non-perturbative shape function in the $R$-gap scheme (Sec.~\ref{subsec:soft}), while \textsc{Pythia} models the same effects via Lund string fragmentation.

\begin{figure}[t]
    \centering
    \includegraphics[width=0.40\linewidth]{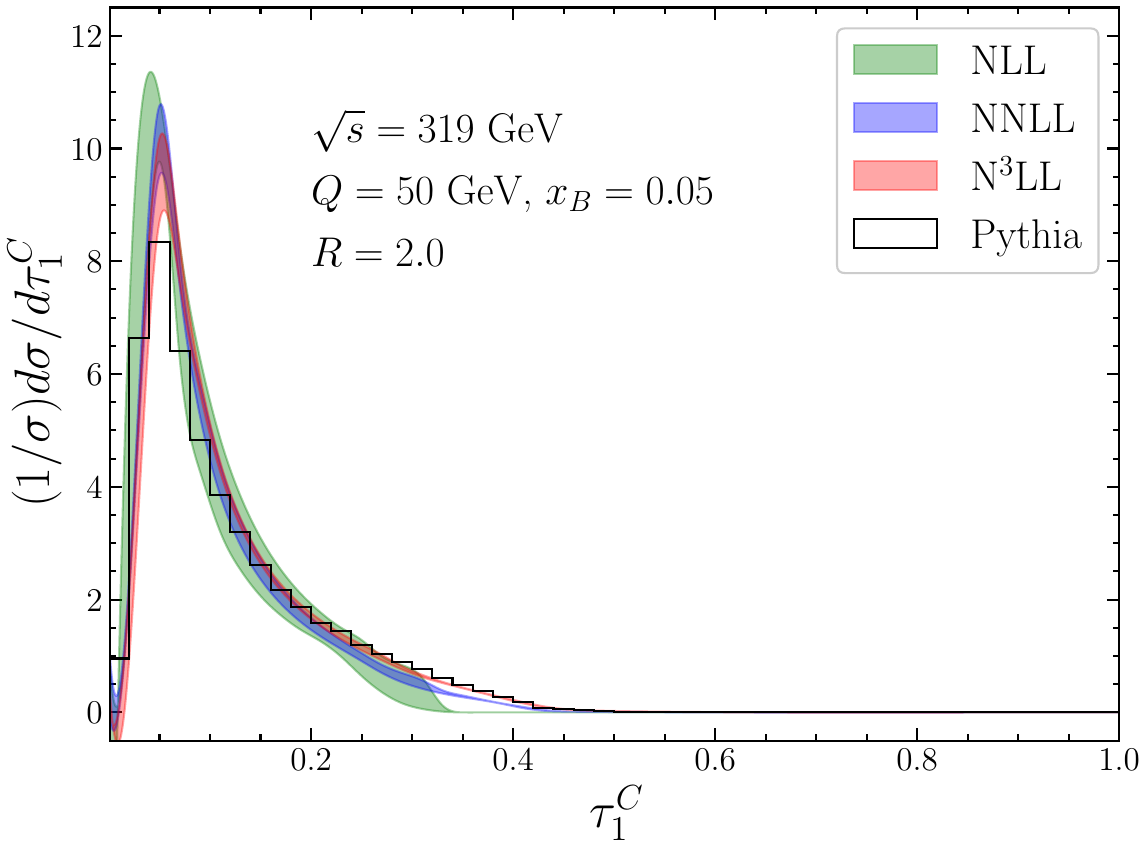}
    \includegraphics[width=0.40\linewidth]{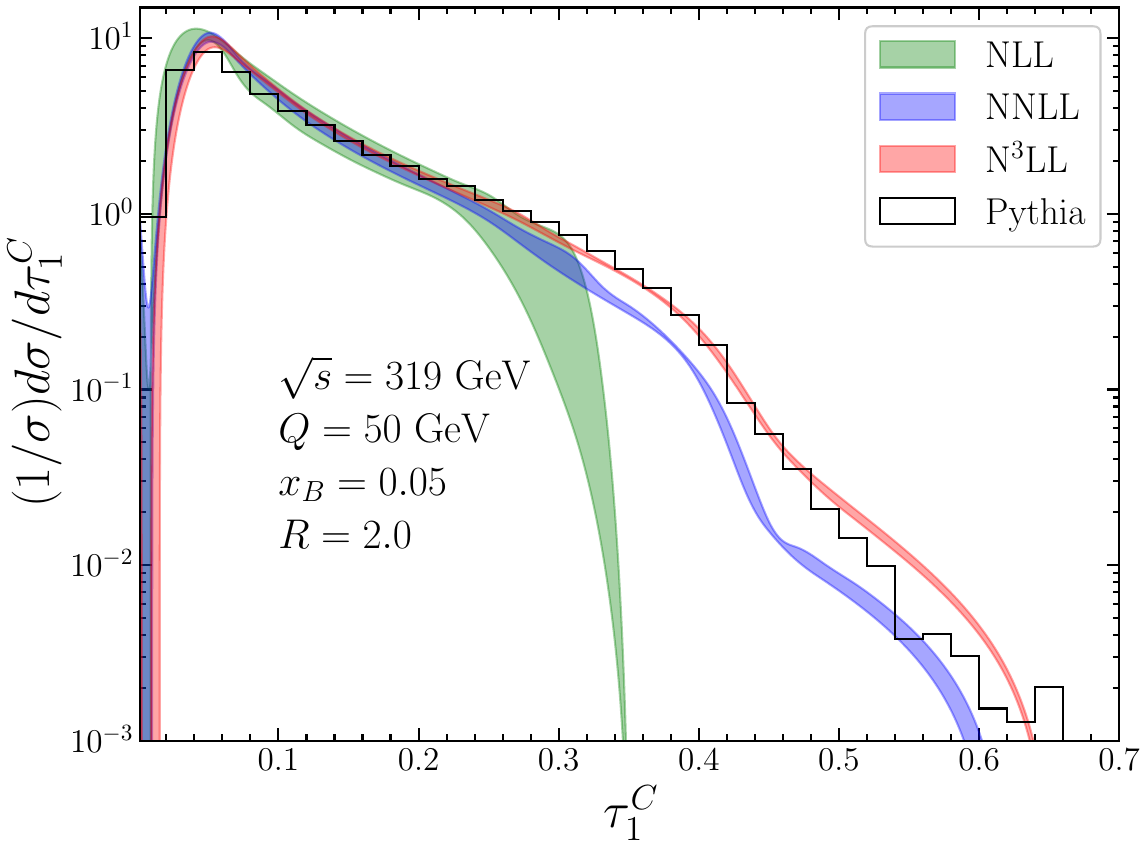}
    \includegraphics[width=0.40\linewidth]{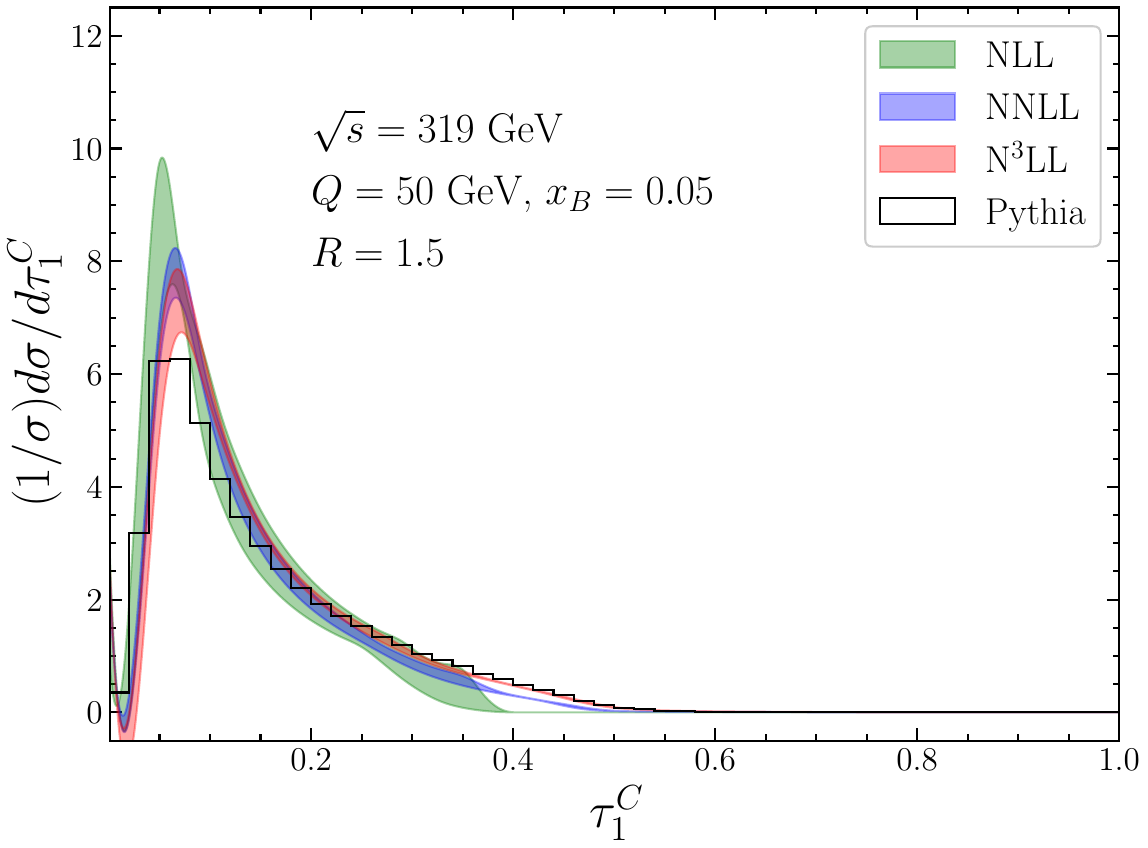}
    \includegraphics[width=0.40\linewidth]{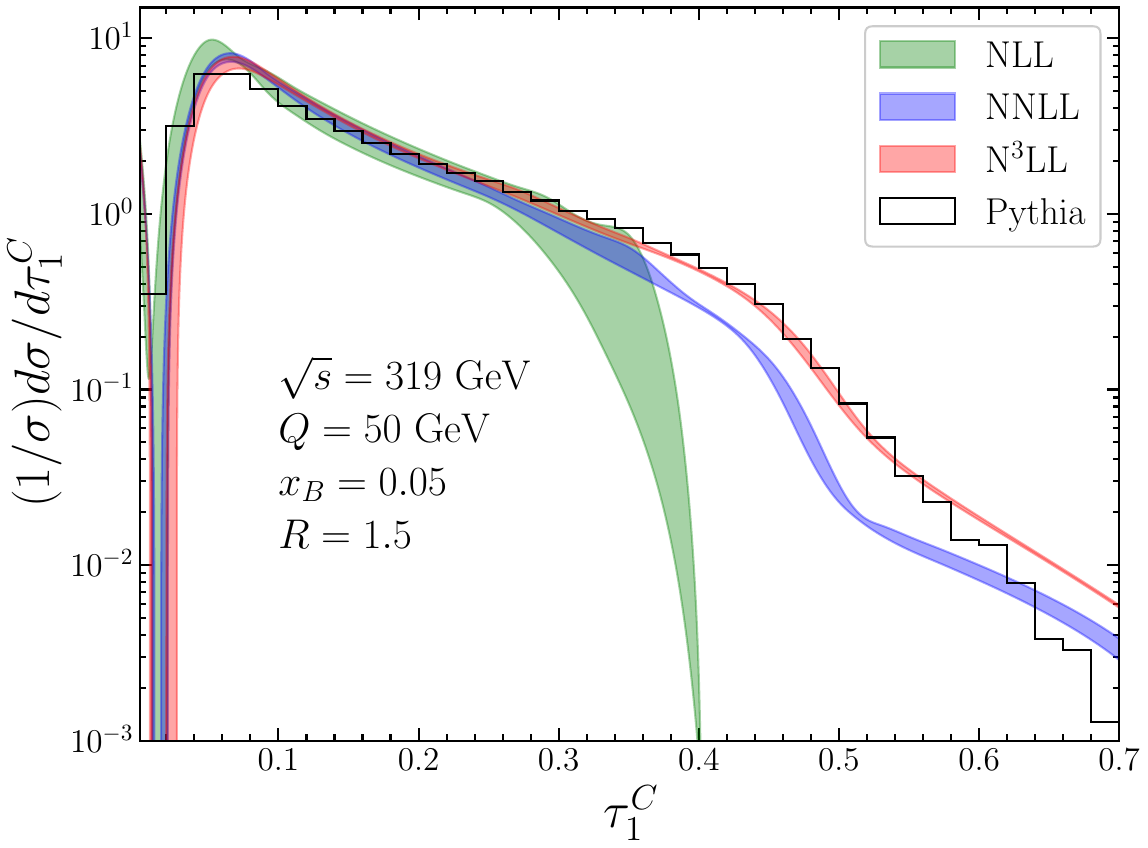}
    \includegraphics[width=0.40\linewidth]{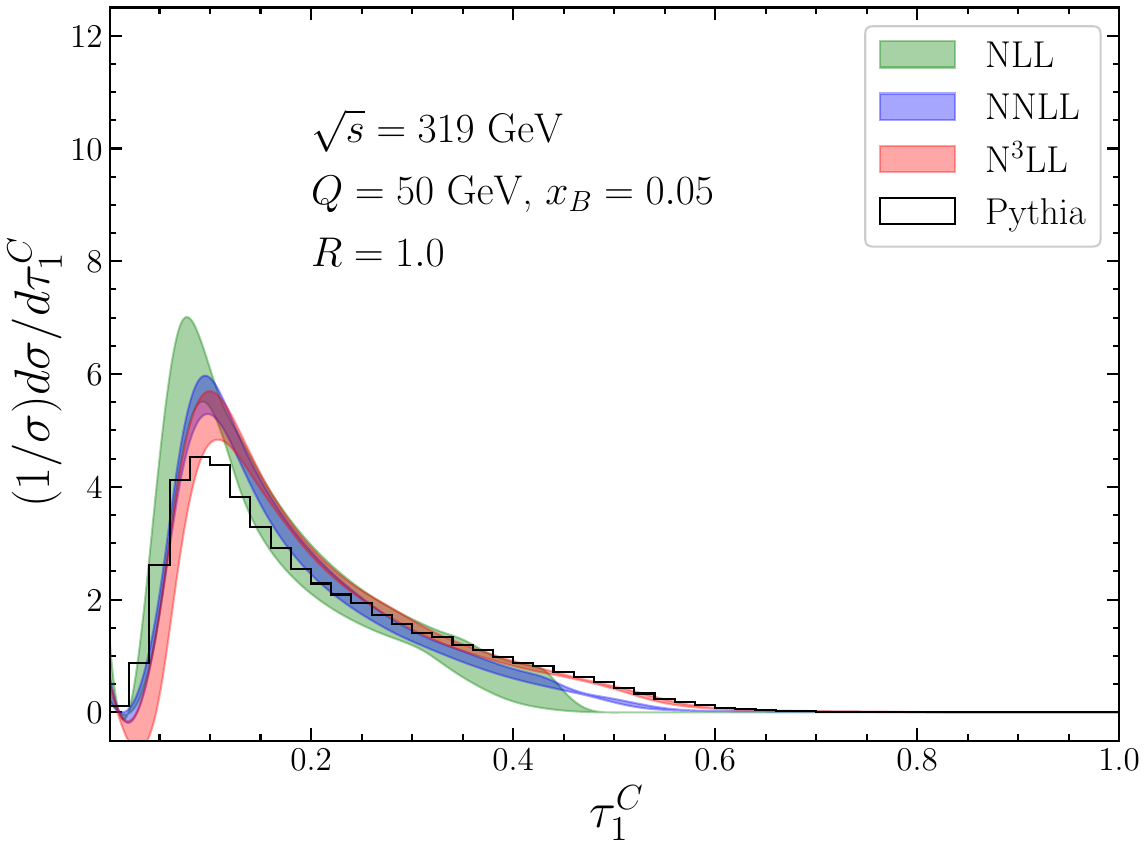}
    \includegraphics[width=0.40\linewidth]{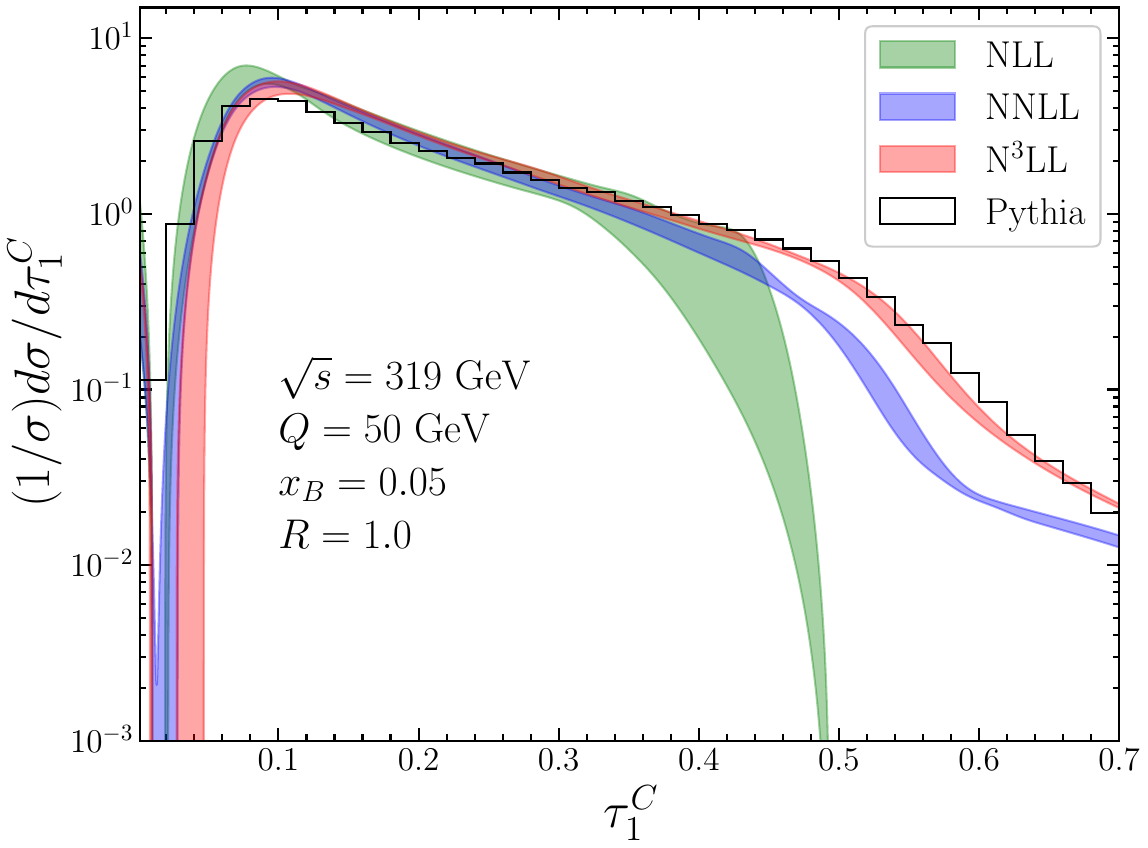}
    \includegraphics[width=0.40\linewidth]{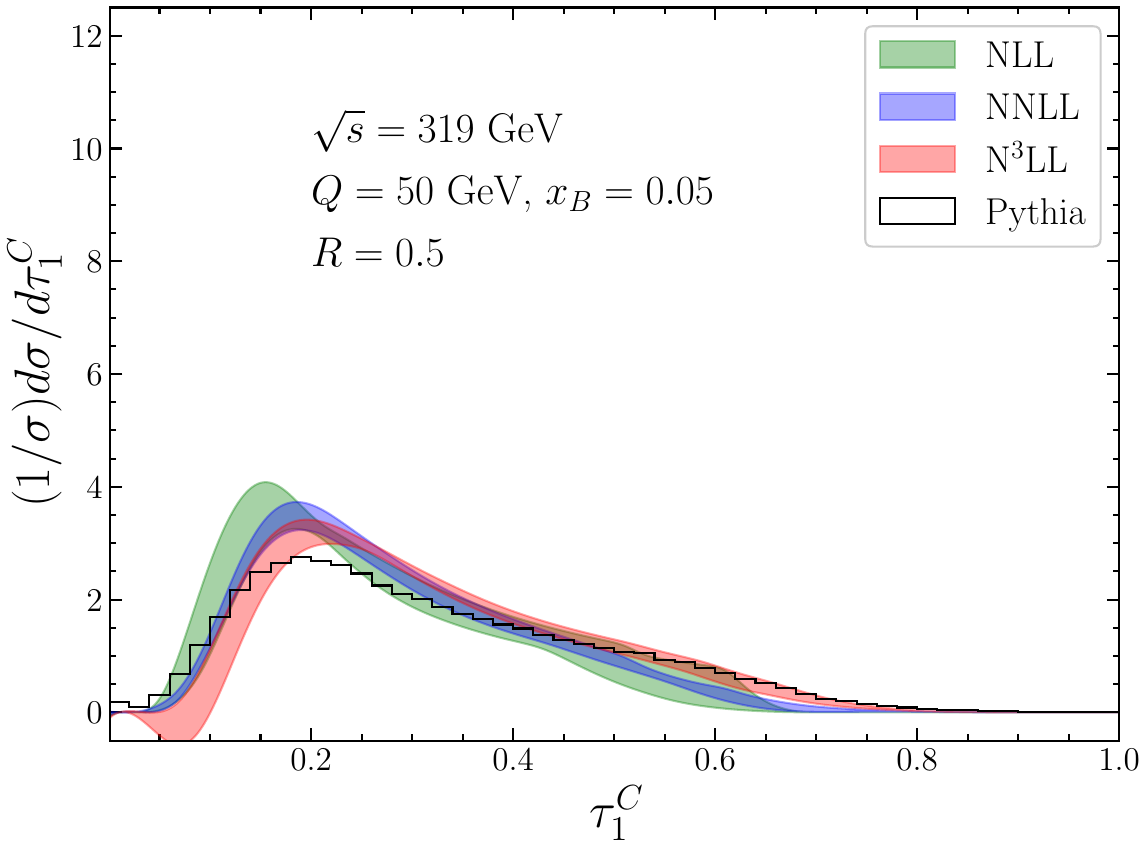}
    \includegraphics[width=0.40\linewidth]{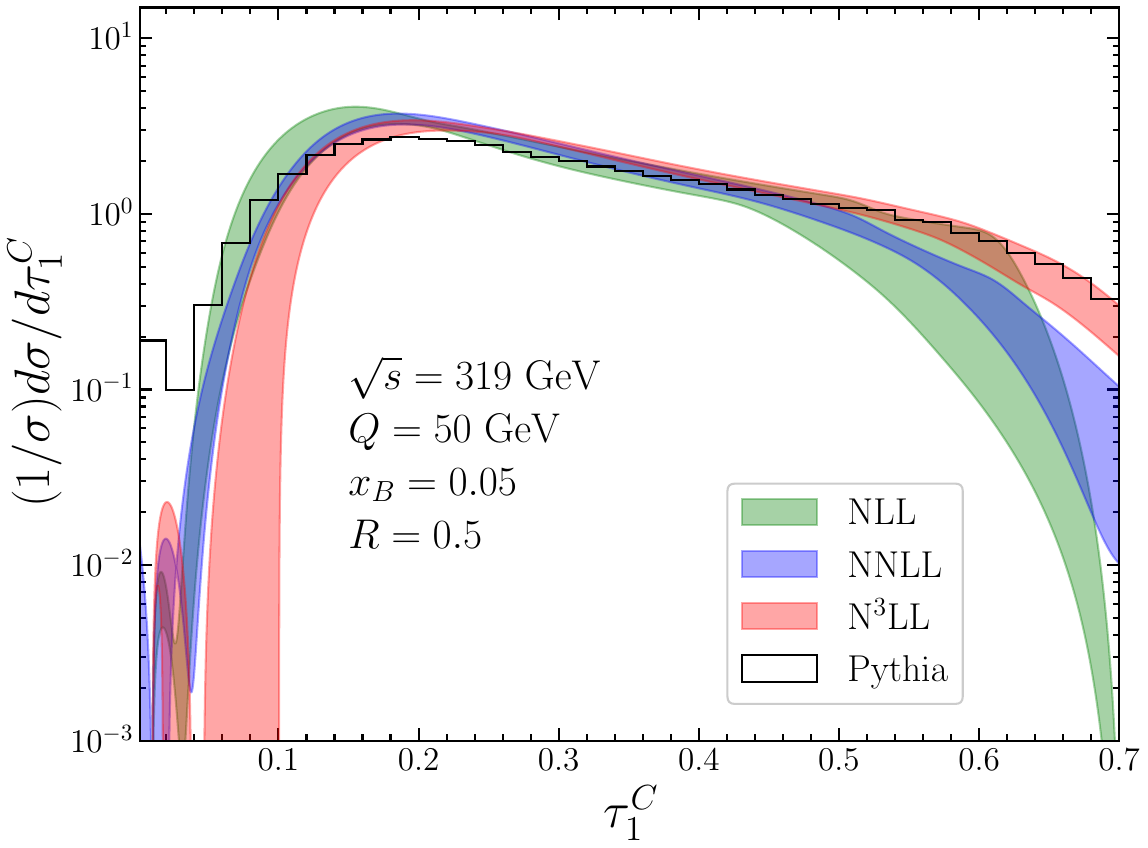}
    \vspace{-1em}
    \caption{Comparison of the resummed theoretical predictions at NLL (green), NNLL (blue), and N$^3$LL (red) accuracy with hadron-level \textsc{Pythia} results (black histogram) for jet radii $R \in \{2.0, 1.5, 1.0, 0.5\}$. The theoretical bands represent the perturbative uncertainty estimated by scale variations. The left panels show the distributions on a linear scale, highlighting the overall shape including the peak region, while the right panels employ a logarithmic scale to illustrate the behavior in the tail and far-tail regions. The perturbative convergence improves systematically at higher orders, with the N$^3$LL results providing an excellent description of the Monte Carlo data across the resummation region and the transition toward the fixed-order regime.}
    \label{fig:theory-final-hadron}
\end{figure}

Figure~\ref{fig:theory-final-hadron} displays the comparison for four representative jet radii $R \in \{0.5, 1.0, 1.5, 2.0\}$.
The distributions are evaluated at the kinematic point $(Q, x_B) = (50~\text{GeV}, 0.05)$; comparisons at other kinematic points are provided in Appendix~\ref{app:additional-kinematics} and yield qualitatively similar agreement.
The theoretical curves represent the fully resummed distributions at three orders of accuracy: 
\begin{itemize}
    \item NLL (green bands): Matched to tree-level fixed-order QCD ($\mathcal{O}(\alpha_s^0)$) corresponding to the Born-level configuration with a single final-state parton.
    \item NNLL (blue bands): Matched to LO fixed-order QCD ($\mathcal{O}(\alpha_s^1)$) incorporating two-parton final states.
    \item N$^3$LL (red bands): Matched to NLO fixed-order QCD ($\mathcal{O}(\alpha_s^2)$) including up to three-parton final states.
\end{itemize}
The theoretical uncertainty bands are estimated by varying the profile scales as described in Sec.~\ref{subsec:profile} and Appendix~\ref{sec:profile}. 
For the non-perturbative input, we employ the values determined from the global fit to the \textsc{Pythia} partonic shift in Eq.~\eqref{eq:R-gap-parameters}: 
$\Omega_1(R_\Delta, \mu_\Delta) \simeq 0.59$ GeV with the gap subtraction $\Delta(R_\Delta, \mu_\Delta) \simeq 0.05$ GeV.
This value is consistent with the independent extraction from the SR analysis of Sec.~\ref{sec:SR}.
The shape function is modeled using the basis function $f_0$ defined in Eq.~\eqref{eq:shape-definition}.

Three key observations can be made from these results:
\begin{enumerate}
\item Universality of the Shift: 
The N$^3$LL prediction (red curves) reproduces the \textsc{Pythia} distribution (black histogram) within the perturbative uncertainty throughout the tail region. This agreement holds for all considered jet radii ($R=0.5$ to $2.0$) using a single value of $\Omega_1$. 
The non-trivial nature of this agreement deserves emphasis: in Sec.~\ref{sec:omega1}, $\Omega_1$ was determined by extracting only the shift between the partonic and hadronic \textsc{Pythia} distributions and verifying its $1/(QR)$ scaling. Here, in contrast, the same value of $\Omega_1$ is used as input to the full factorization theorem to predict the entire hadron-level distribution across the resummation region. The agreement across all jet radii therefore constitutes a non-trivial test of the universality of the leading-power shape function moment, demonstrating that the leading non-perturbative correction enters as a universal shift scaled by $1/R$.

\item Perturbative Convergence: 
The uncertainty bands narrow systematically from NLL to N$^3$LL across all $R$ values, reflecting the convergence of our resummation framework. The N$^3$LL bands generally overlap with or are contained within the NNLL bands, indicating the stability of the matching procedure.
We caution, however, that the displayed bands reflect only perturbative scale variation and therefore underestimate the total theoretical uncertainty in the peak region, where the distribution shape is sensitive to the explicit form of the shape function rather than to the leading moment alone. In the tail region this distinction does not arise, since the only non-perturbative information entering the prediction is the leading moment $\Omega_1$ of the shape function, which generates the universal $1/R$ shift. Scale variation alone is also not sufficient to capture the uncertainty in the purely fixed-order predictions in the far-tail region at large $\tau_1^C$, where each new order (and thus particle multiplicity) captures new kinematic configurations not present at lower orders.

\item Peak Region Behavior: 
In the non-perturbative peak region ($\tau_1^C \lesssim \Lambda_{\rm QCD}/Q_R$), the theoretical prediction captures the general features of the \textsc{Pythia} distribution but shows some deviations. These deviations originate from two sources: on the theoretical side, the peak shape depends on higher moments $\Omega_{k>1}$ of the shape function, for which we use a one-parameter ansatz; on the \textsc{Pythia} side, the Lund string fragmentation model produces a specific pattern of higher moments that we do not aim to reproduce. The residual differences therefore reflect the limitations of comparing two distinct hadronization models in the peak region rather than a failure of the factorization theorem; as discussed in Ref.~\cite{Ee:2025scz}, incorporating higher moments provides a path for further phenomenological tuning. Studies of other Monte Carlo event generators with different hadronization models, e.g. Herwig \cite{Bellm:2015jjp}, Powheg \cite{Banfi:2023mhz}, would be in order, but lie outside the scope of the present paper.
\end{enumerate}

\subsection{Scope and Limitations}\label{subsec:scope}
The comparison presented in this section is primarily designed to validate the structural integrity of the factorization theorem, specifically the predicted $1/R$ scaling of  the leading-power non-perturbative shift and its universality. Accordingly, the agreement with hadron-level simulations should be interpreted as a verification of these leading-power features rather than a comprehensive modeling of all QCD effects across the entire kinematic range.

While our results confirm the dominance of the universal shift in the tail region, a quantitative treatment of the peak region remains outside the current scope. Accurately describing the peak would require incorporating subleading power corrections, hadron-mass effects, and higher-order moments of the shape function, all of which are sensitive to the detailed dynamics of hadronization. These effects are expected to be non-universal and may significantly impact the distribution where $\tau_1^C \sim \Lambda_{\rm QCD}/Q_R$. 

A notable limitation appears in the far-tail region ($\tau_1^C \to 1$), particularly for small jet radii $R$. In this regime, the distributions are governed by hard, wide-angle radiation described by the fixed-order matching. We observe that the NLL, NNLL, and N$^3$LL predictions do not necessarily overlap in their uncertainty bands at the far-tail. This behavior is rooted in the discrete nature of the phase space at low perturbative orders; for instance, 
as verified in Eq.~\eqref{eq:tau1C-max-relation}, the maximum value of the observable scales as $\tau_{1,\max}^C \to n/(n+1)$, where $n+1$ denotes the number of final-state partons. Consequently, the LO ($n=1$) and NLO ($n=2$) calculations possess different kinematic endpoints ($\tau_{1,\max}^C = 1/2$ and $2/3$, respectively), leading to the observed deviations between the NLL+LO and N$^3$LL+NLO predictions in the far-tail.

Furthermore, the level of agreement with \textsc{Pythia} in the far-tail depends on the jet radius. While our predictions show excellent agreement for $R \sim 2$ even at large $\tau_1^C$, the deviation increases as $R$ decreases. This is because the sensitivity to hard radiation (and thus the importance of higher-order matching contributions) is significantly enhanced for smaller jet radii. 
While our current results are matched to NLO, a complete description of the far-tail for small $R$ would likely require NNLO matching, which would extend the kinematic endpoint to $\tau_{1,\max} = 3/4$.

In summary, this comparison validates the central claim of this work: the Centauric 1-jettiness observable admits a clean factorization and resummation structure, for $\tau_1^C$ values between peak and far-tail regions, with leading non-perturbative effects that are universal and controlled by a single non-perturbative parameter $\Omega_1$, with the corresponding shift scaling exactly as $1/R$.

\section{Conclusions}\label{sec:conc}
In this work we introduced the Centauric 1-jettiness, a global event-shape observable for deep inelastic scattering whose beam and jet regions are defined by the geometric boundary of the Centauro jet algorithm \cite{Arratia:2020ssx}. This construction introduces a continuously tunable jet-radius parameter $R$ while preserving the global, energy-flow nature of DIS 1-jettiness \cite{Kang:2013nha,Ee:2025scz}. We showed that, in the soft limit, the assignment of radiation depends only on the rapidity of the emission, so that the Centauric soft function maps onto a rescaled hemisphere soft function \cite{Kang:2013nha}. This places the observable in the same universality class as thrust-like event shapes and enables direct use of well-studied hemisphere soft-function ingredients \cite{Kelley:2011ng,Monni:2011gb,Hornig:2011iu,Kang:2015moa} in our N$^3$LL+NLO resummed predictions.

A key result of this work is the prediction of a simple form for the leading non-perturbative correction to the Centauric 1-jettiness distribution. The leading power correction enters as a shift of the spectrum governed by the universal soft matrix element $\Omega_1$ \cite{Lee:2006nr,Mateu:2012nk}, with a coefficient that scales as $1/R$. This scaling follows directly from the rapidity-based structure of the Centauro partition and the resulting factorization theorem; subleading corrections are suppressed by additional powers of $\Lambda_{\rm QCD}^2/Q^2$.

We also addressed a technical challenge in the perturbative implementation. Because the jet axis entering the measurement is determined dynamically, naive choices of measurement weights can lead to discontinuities in the fixed-order distribution when the identity of the primary jet changes. We showed that a hybrid prescription--using one set of weights for jet identification and a smoother set for the measurement--resolves these discontinuities while preserving the geometric Centauro jet boundary. This construction yields a stable observable suitable for high-precision perturbative calculations.

We tested the predicted $1/R$ scaling using hadron-level \textsc{Pythia} simulations across a range of $Q^2$ and Bjorken-$x_B$ values, finding agreement with the expected behavior at all nine kinematic points considered, with mild residual deviations discussed in Sec.~\ref{sec:omega1}. The grand global fit yields $\Omega_1 = 0.59 \pm 0.04$ GeV, comparable to the value $\Omega_1 \simeq 0.5$ GeV used in analyses of the standard DIS 1-jettiness observable $\tau_1^b$~\cite{Ee:2025scz,H1:2024aze}, supporting the universality of the underlying soft matrix element across DIS event shapes. As an independent cross-check, the SR analysis presented in Sec.~\ref{sec:SR} extracts $\Omega_1$ directly from the data without assuming the analytic form of the leading shift, yielding results consistent with the global fit value.

The explicit jet-radius dependence opens a new lever arm for disentangling perturbative and non-perturbative contributions in precision DIS event-shape analyses. While the perturbative radiation pattern is controlled by $\alpha_s$, the leading non-perturbative correction scales predictably as $1/R$. Measurements at multiple jet radii therefore offer a route to constrain the soft matrix element $\Omega_1$ independently of $\alpha_s$, with potential application to high-precision determinations of the strong coupling from DIS event shapes.

Overall, Centauric 1-jettiness provides a theoretically clean extension of the DIS event-shape framework in which the size of the jet region can be varied continuously. Because the observable shares the same perturbative ingredients and non-perturbative structure as hemisphere-based event shapes, it is readily amenable to high-order resummation and precision phenomenology. At the same time, the explicit $R$ dependence opens new possibilities for studying hadronization effects and testing QCD factorization in future DIS measurements. These properties make Centauric 1-jettiness a promising tool for precision QCD studies at the upcoming Electron-Ion Collider \cite{AbdulKhalek:2021gbh}, where its tunable jet radius offers direct experimental access to the interplay between perturbative resummation and universal non-perturbative effects.

\section*{Acknowledgments}
C.L. and Y.M. would like to thank L.~Cunqueiro, P.~Jacobs, H.~Klest, S.~Lee, and B.~Nachman for discussions and brainstorming which led to some of the early investigations that initially motivated this paper. C.L. and J.-H.E. would like to thank D. Kang and I. Stewart for their long and continuing collaboration with us on global DIS 1-jettiness. J.-H.E acknowledges helpful discussions with R.~Gupta on the statistical bootstrap method we used in \sec{omega1}.
The work of C.L., J.-H.E, and J.T.~was supported by the U.S.~Department of Energy, Office of Science, through the Office of Nuclear Physics, and the LDRD Program at Los Alamos National Laboratory under projects 20220715PRD1, 20230857PRD2, and 20250214ER. LANL is operated by Triad National Security, LLC, for the National Nuclear Security Administration of the U.S.~Department of Energy under Contract No. 89233218CNA000001. J.T. was also supported by the U.S. Department of Energy, Office of Science, Office of Nuclear Physics, contract No. DE-AC02-06CH11357.

\appendix

\section{Derivation of the Weights that Reproduce the Jet Boundary}
\label{sec:equivalence-derivation}
In Sec.~\ref{sec:Obs} we introduced the Centauric 1-jettiness observable and showed that, for an appropriate choice of reference-vector weights $\omega_{B,J}$, the minimization procedure reproduces the geometric jet boundary of the Centauro algorithm. The purpose of this appendix is to derive this result explicitly and to determine the form of the weights in Eq.~\eqref{eq:wJ-wB-cond} that lead to the Centauro partition of phase space. This establishes the equivalence between the Centauric 1-jettiness definition and the Centauro jet algorithm in the massless limit.

To derive Eq.~\eqref{eq:wJ-wB-cond}, we consider a single particle with index $i$ clustering into the jet, with azimuthal and polar angles $\phi_i$ and $\theta_i$, respectively. To determine the limits on the polar angle, we first consider the cases where the azimuthal angle is either $0$ or $\pi$, examining the distance measure $d_{iJ}$ in Eq.~\eqref{eq:diJ-i-to-Jet}. Setting the distance metric to $1$ with $\cos\phi_i=\pm1$, we find
\begin{align}
\label{eq:theta_min-max}
\theta_\textrm{max} = 2\tan^{-1}
\left[
\tan(\theta_J/2) + R/2
\right],
\quad
\theta_\textrm{min} = 2\tan^{-1}
\left[
\tan(\theta_J/2) - R/2
\right]\,.
\end{align}
The center of $\mathcal{H}_J$ corresponds to the unit vector
\begin{align}
\bm{n}_J'
&=
\left[
\cos\left(\frac{\theta_\textrm{max}-\theta_\textrm{min}}{2}\right)\right]^{-1}
\left(
\frac{\sin\theta_\textrm{max} + \sin\theta_\textrm{min}}{2}, 0,
\frac{\cos\theta_\textrm{max} + \cos\theta_\textrm{min}}{2}
\right)
\nonumber \\
&=
\left(
\sin\left(\frac{\theta_\textrm{max}+\theta_\textrm{min}}{2}\right), 0,
\cos\left(\frac{\theta_\textrm{max}+\theta_\textrm{min}}{2}\right)
\right),
\end{align}
which can be obtained as the normalized average of the boundary points $(\sin\theta_\textrm{max}, 0, \cos\theta_\textrm{max})$ and 
$(\sin\theta_\textrm{min}, 0, \cos\theta_\textrm{min})$.

Note that $\bm{n}_J'$ does not coincide
with the jet direction $\bm{n}_J$, which implies that $\mathcal{H}_J$ is not symmetric around the jet axis $\bm{n}_J$ identified by the jet clustering. 

In general, the condition for $\phi_i$ to make $d_{iJ}=1$ is given by
\begin{align}
\cos\phi_i(\theta_i)
=
\frac{\tan^2(\theta_i/2) + \tan^2(\theta_J/2) - R^2/4}{2\tan(\theta_i/2)\tan(\theta_J/2)}.
\end{align}
Then, with $\theta_i$ and $\phi_i(\theta_i)$, we can parametrize a trajectory of $\mathcal{H}_J$ on a unit sphere as follows:
\begin{align}
\label{eq:ni_thetai}
\bm{n}_i(\theta_i)
=
\left(
\cos\phi_i(\theta_i)\sin\theta_i,
\sin\phi_i(\theta_i)\sin\theta_i ,
\cos\theta_i
\right)\,.
\end{align}
We verify that $\bm{n}_i(\theta_i)$ lies on a cone around $\bm{n}_J'$ by 
computing $\bm{n}_i \cdot \bm{n}_J' = \cos((\theta_{\rm max}-\theta_{\rm min})/2)$, which is independent of $\theta_i$.
This confirms that the boundary of $\mathcal{H}_J$ determined by the Centauro algorithm is a cone around $\bm{n}_J'$ with the open-half angle being $(\theta_\textrm{max}-\theta_\textrm{min})/2$. 
This equally implies that if a particle with momentum $\bm{p}_i$ is in $\mathcal{H}_J$, it satisfies
\begin{align}
\frac{\bm{p}_i\cdot \bm{n}_J'}{|\bm{p}_i|}
>
\cos \frac{\theta_\textrm{max}-\theta_\textrm{min}}{2}.
\end{align}

In the previous discussion, the jet region $\mathcal{H}_J$ was identified by the Centauro jet algorithm. This time, let us consider the jet region using the standard 1-jettiness definition in Eq.~\eqref{eq:original-1-jettiness}. Here, we choose the beam and jet reference vectors as Eq.~\eqref{eq:reference-general} with $n_{B,J}$ being
\begin{align}
n_B = (1,0,0,-1),
\quad
n_J = (1, \sin\theta_J, 0, \cos\theta_J). 
\end{align}
The minimization operator in Eq.~\eqref{eq:original-1-jettiness} can group the final states into beam (when $q_B\cdot p_i<q_J\cdot p_i$) and jet (when $q_B\cdot p_i>q_J\cdot p_i$) regions. 
Let us assume that a four-momentum $p_i$ lies on the boundary $\mathcal{H}_J$ determined by Centauro jet algorithm. Then, the necessary  condition for the $\mathcal{H}_J$ to agree with the boundary determined by the minimization of 1-jettiness is 
\begin{align}
\label{eq:boundary_h_JB}
q_B\cdot n_i = q_J\cdot n_i,
\end{align}
for 
\begin{align}
n_i^\textrm{max} = (1, \sin\theta_\textrm{max}, 0, \cos\theta_\textrm{max}),
\end{align}
and
\begin{align}
n_i^\textrm{min} = (1, \sin\theta_\textrm{min}, 0, \cos\theta_\textrm{min}).
\end{align}
Then we obtain the two conditions,
\begin{align}
\frac{\omega_J}{\omega_B}\bigg|_\textrm{upper} = \frac{1+\cos\theta_\textrm{max}}{1-\cos(\theta_\textrm{max}-\theta_J)},
\end{align}
and
\begin{align}
\frac{\omega_J}{\omega_B}\bigg|_\textrm{lower} = \frac{1+\cos\theta_\textrm{min}}{1-\cos(\theta_\textrm{min}-\theta_J)}.
\end{align}
Using the explicit expressions for $\theta_\textrm{max,min}$ in Eq.~\eqref{eq:theta_min-max}, we can show that
\begin{align}
\label{eq:cond-1-jettiness-centauro}
\frac{\omega_J}{\omega_B}\bigg|_\textrm{upper}
=
\frac{\omega_J}{\omega_B}\bigg|_\textrm{lower}
=
\frac{4}{R^2\cos^2(\theta_J/2)}.
\end{align}
This confirms that the boundary $\mathcal{H}_J$ determined by the Centauro algorithm coincides with that of the 1-jettiness for $\phi_i=0,\pi$ when $\omega_J/\omega_B=4/(R^2\cos^2(\theta_J/2))$. 

We can check the condition in Eq.~\eqref{eq:cond-1-jettiness-centauro} for $n_i$ on the cone with general $\phi_i$ using Eq.~\eqref{eq:ni_thetai}:
\begin{align}
n_i(\theta_i)=  (1, \bm{n}_i(\theta_i)).
\end{align}
Then, we find 
\begin{align}
\frac{\omega_J}{\omega_B}
=
\frac{n_B\cdot n_i(\theta_i)}
{n_J\cdot n_i(\theta_i)}
=
\frac{1+\cos\theta_i}
{1-\sin\theta_J\sin\theta_i\cos\phi_i(\theta_i)
-\cos\theta_i \cos\theta_J}
=
\frac{4}{R^2\cos^2(\theta_J/2)}.
\end{align}
This confirms that once Eq.~\eqref{eq:wJ-wB-cond} is satisfied, then the boundary derived from the minimization in the definition of 1-jettiness in Eq.~\eqref{eq:original-1-jettiness} yields the same boundary $\mathcal{H}_J$ derived from the Centauro algorithm, and we can describe the clustered event in terms of the 1-jettiness.

\section{Additional Details on the Profile Functions}\label{sec:profile}
This appendix collects technical details on the profile functions introduced in Sec.~\ref{subsec:profile}. We first present the explicit forms of the variation profile $g$ and the spline function $\zeta$ used in Eq.~\eqref{eq:running-scale}, including a modified $\mu_\textrm{run}$ form that maintains monotonicity in regimes where the canonical scale ordering breaks down. We then describe the systematic scale variations used to 
estimate perturbative uncertainties, and the treatment of the non-singular scale $\mu_\textrm{ns}$.

The variation profile $g$ is defined by
\begin{align}
\label{eq:g_profile}
g({\tau_1^C}, \{t_0, t_3\})
=
\begin{cases}
1,
&\textrm{for~$0\le {\tau_1^C}< t_0$},
\\
\left(t_3-{\tau_1^C}\right)^2
/\left(t_3-t_0\right)^2,
&\textrm{for~$t_0\le {\tau_1^C}< t_3$},
\\
0,
&\textrm{for~${\tau_1^C}\ge t_3$}.
\end{cases}
\end{align}
This ensures that scale variations remain frozen in both the deep non-perturbative regime (${\tau_1^C} < t_0$) and the fixed-order-dominated region (${\tau_1^C} > t_3$).

The function $\zeta$ used in Eq.~\eqref{eq:running-scale} ensures a smooth transition between the canonical (tail) region ($t_1\le {\tau_1^C}\le t_2$) and the frozen/fixed regions. It is constructed as a piecewise quadratic spline:
\begin{align}
\label{eq:spline_profile}
\zeta(\tau,\{x_0,y_0,r_0\}, \{x_1,y_1,r_1\})
=
\begin{cases}
a\tau^2+b\tau+c,
&\textrm{for~$x_0\le \tau \le (x_0+x_1)/2$},
\\
d\tau^2+ e\tau+f,
&\textrm{for~$(x_0+x_1)/2\le \tau \le x_1$},
\end{cases}
\end{align}
where the coefficients $a$ through $f$ are determined by the continuity of $\zeta(\tau)$ and its first derivative at the boundaries $x_0, (x_0+x_1)/2,$ and $x_1$.
\begin{align}
\begin{split}
a
&=
\frac{(3r_0+r_1)(x_0-x_1) + 4(y_1 - y_0)}
{2(x_1-x_0)^2},
\\
b
&=
r_0 - 2a x_0,
\\
c
&=
y_0 - bx_0 -a x_0^2,
\\
d 
&=
\frac{(3r_1+r_0)(x_1-x_0) + 4(y_0 - y_1)}
{2(x_1-x_0)^2},
\\
e
&=
r_1 - 2d x_1,
\\
f
&=
y_1 - e x_1 -d x_1^2.
\end{split}
\end{align}

In Eq.~\eqref{eq:running-scale}, $\mu_\textrm{run}$ is designed to be monotonically increasing. However, for low hard scales ($\mu\sim 10~\textrm{GeV}$) or small jet radii ($R\lesssim 0.8$), the non-perturbative scale $\mu_0$ may exceed the canonical value at the transition point $t_1$ (i.e., $\mu_0/\mu > r t_1$), breaking monotonicity. In such cases, we adopt a modified form to restore monotonic behavior:
\begin{align}
\mu_\textrm{run}
({\tau_1^C}, \mu, \{\mu_0, r, t_0, t_1, t_2, t_3\})
=
\mu \times
\begin{cases}
\mu_0/\mu, 
&\textrm{for~$0\le {\tau_1^C} < t_0$},
\\
\zeta({\tau_1^C}, \{t_0,\mu_0, 0\}, \{t_3, 1, 0\}), 
&\textrm{for~$t_0\le {\tau_1^C} < t_3$},
\\
1,
&\textrm{for~${\tau_1^C} \ge t_3$}.
\end{cases}
\end{align}

To estimate perturbative uncertainties, we consider 16 scale variations around the central setting:
\begin{align}\label{eq:singular_scale_variations}
\begin{split}
\textrm{Variations 1--2}:~ & \mu_0 \to 1.1 \pm 0.2~\textrm{GeV},
\\
\textrm{Variations 3--4}:~ & t_1 \to (1\pm 0.2)t_1\big|_\textrm{central},
\\
\textrm{Variations 5--6}:~ & t_2 \to(1\pm 0.2)t_2\big|_\textrm{central},
\\
\textrm{Variations 7--8}:~ & \mu \to 2^{\pm 1} Q,
\\
\textrm{Variations 9--12}:~ & e_{B,J} \to \displaystyle\pm \frac{1}{3}, \pm\frac{1}{6},
\\
\textrm{Variations 13--16}:~ & e_{S} \to \displaystyle\pm \frac{1}{3}, \pm\frac{1}{6}.
\end{split}
\end{align}
Variations 1--6 probe the sensitivity to the profile shape and non-perturbative transition. Variations 7--8 estimate fixed-order uncertainties by shifting all scales. Variations 9--16 assess higher-order resummation uncertainties by varying the relations between $\mu_B, \mu_J,$ and $\mu_S$.

As discussed in Appendix~\ref{sec:power-corrections}, we make the renormalon subtraction scale $R_{\textrm{gap}}$ dependent on ${\tau_1^C}$ to avoid large logarithms in the subtraction terms. Following Ref.~\cite{Ee:2025scz}, we set
\begin{align}
\label{eq:R_profile_def}
R_{\textrm{gap}}\left(\tau_1^C\right)
=\mu_S\left(\tau_1^C\right)\big|_{\mu_0\to R_0},
\end{align}
with $R_0=0.85\mu_0$, which ensures the correct sign for the one-loop subtraction $\delta_1$ in the peak region. 

Finally, for the non-singular contributions, we define the central scale and its variations as:
\begin{align}\label{eq:non-singular_scale_variations}
\begin{split}
\textrm{Central}:~ & \mu_\textrm{ns} = Q,
\\
\textrm{Variations 1--2}:~ & \mu_\textrm{ns} = 2^{\pm 1}Q,
\\
\textrm{Variation 3}:~ & \mu_\textrm{ns} = \mu_J({\tau_1^C})\big|_\textrm{central},
\\
\textrm{Variation 4}:~ & \mu_\textrm{ns} = \frac{\mu_J({\tau_1^C}) + \mu_S({\tau_1^C})}{2}\bigg|_\textrm{central}.
\end{split}
\end{align}
To preserve correlations, when varying $\mu=2^{\pm 1}Q$ (Variations 7--8 in Eq.~\eqref{eq:singular_scale_variations}), we simultaneously apply Variations 1--2 for the non-singular contributions.

\section{Theoretical formulation}\label{app:Theory-Formalism}
This appendix collects the theoretical ingredients used to construct the singular cross section in Sec.~\ref{subsec:fac}, following our previous treatment for $\tau_1^b$ in Ref.~\cite{Ee:2025scz}, with the following Centauric modifications:
\begin{itemize}
\item The hard scale $Q$ is replaced by the $R$-dependent scale $Q_R$ (Eq.~\eqref{eq:1-jet-invariants}), and the rescaling variable $\xi$ in the resummed cross section is correspondingly normalized by $Q_R$.
\item The profile-function transition points carry explicit $R$-dependence (Eq.~\eqref{eq:profile_setting}), reflecting the $R$-dependent soft scale $\mu_S \sim (QR/2)\tau_1^C$ in the tail region.
\item The leading non-perturbative shift acquires a $1/R$ scaling (Eq.~\eqref{eq:R-dep-shift}), which is the central prediction studied in Sec.~\ref{subsec:soft}.
\end{itemize}
The remaining ingredients (RG evolution kernels, plus-distribution identities, $R$-gap subtraction, shape function model) are identical to those in Ref.~\cite{Ee:2025scz} and are referenced rather than reproduced. We summarize the Centauric-specific structure below.

\subsection{Resummation to \texorpdfstring{N$^3$LL}{N3LL}}\label{subsec:Resum}
The renormalization-group evolution of the hard, beam, jet, and soft functions takes the standard form
\begin{align}
\label{eq:RG-simple}
\begin{split}
H_q(y,Q^2,\mu)
&=
H_q(y,Q^2,\mu_H)
U_H(Q^2,\mu_H,\mu),
\\
{B}_q(t,x_B,\mu)
&=
\int dt'
{B}_q(t-t',x_B,\mu_B)
U_{B_q}(t',\mu_B,\mu),
\\
J_q(t,\mu)
&=
\int dt'
J_q(t-t',\mu_J)
U_{J_q}(t',\mu_J,\mu),
\\
S_\textrm{PT}(k,\mu)
&=
\int dk' 
S_\textrm{PT}(k-k',\mu_S)
U_S^2(k',\mu_S,\mu),
\end{split}
\end{align}
with the evolution factors $U_i$ and the corresponding kernels $K_i$, $\eta_i$ for $i=H,B,J,S$ given to N$^3$LL accuracy in Appendix~B of Ref.~\cite{Ee:2025scz}.
After rescaling the plus distributions in the fixed-order functions following Eq.~(3.26) of Ref.~\cite{Ee:2025scz} with the rescaling variable $\xi$ normalized by the Centauric $Q_R$, the resummed singular cross section takes the form
\begin{align}
\label{eq:singular-after-resummation}
\sigma^\textrm{s}_\textrm{PT}({\tau_1^C})
&=
\sigma_0^b
\frac{e^{\mathcal{K}-\gamma_\textrm{E}\Omega}}
{\Gamma(1+\Omega)}
\left(\frac{Q}{\mu_H}\right)^{\eta_H}
\left(\frac{\xi s_B}{Q_R\mu_B^2}\right)^{\eta_{B}}
\left(\frac{\xi s_J}{Q_R\mu_J^2}\right)^{\eta_{J}}
\left(\frac{\xi}{\mu_S}\right)^{2\eta_S}
\left(\frac{Q_R}{\xi}\right)
\nonumber \\
&
\times
\sum_q
\bigg[
H_{q}(y,Q^2,\mu_H)
\sum_{ \substack{ m_1,m_2, \\ m_3=-1 } }
{J}_{m_1}\left(\frac{\xi s_J}{Q_R\mu_J^2}\right)
B_{q,m_2}\left(x_B,\mu_B,\frac{\xi s_B}{Q_R\mu_B^2}\right)
S_{m_3}\left(\frac{\xi}{\mu_S}\right)
\nonumber \\
&
\times
\sum_{\ell_1=-1}^{m_1+m_2+1}
\sum_{\ell_2=-1}^{\ell_1+m_3+1}
\sum_{\ell_3=-1}^{\ell_2+1}
V_{\ell_1}^{m_1m_2} 
V_{\ell_2}^{\ell_1 m_3}
V_{\ell_3}^{\ell_2}(\Omega)
G_{\ell_3}\left(\frac{Q_R{\tau_1^C}}{\xi}\right)
+
(q\leftrightarrow \bar{q})
\bigg],
\end{align}
where the convolution coefficients $V_k^n(\Omega)$ and $V_k^{mn}$ are given in Sec.~E of Ref.~\cite{Kang:2013nha}, the kernels $\mathcal{K} = \sum_{i} K_i(\mu_i,\mu)$ and $\Omega = \eta_J + \eta_B + 2\eta_S$ are summed over the appropriate functions, and $G_{\ell_3}$ takes the form $\mathcal{L}_{\ell_3}^{\Omega}(Q_R\tau_1^C/\xi)$ for the differential cross section or $G_{\ell_3}^{\Omega}(Q_R\tau_1^C/\xi)$ for the cumulative. The modified plus distribution $\mathcal{L}_n^\Omega$ is defined as
\begin{align}
\mathcal{L}_n^\Omega(u)
=
\begin{cases}
\displaystyle
\delta(u),& \textrm{for $n=-1$,}
\\[2ex]
\displaystyle
\left[
\frac{\theta(u)\log^n u}{u^{1-\Omega}}
\right]_+,& \textrm{for $n>-1$.}
\end{cases}
\end{align}
For the cumulative case, the corresponding function is given by the lower incomplete gamma function $\gamma(s,t) = \int_0^t du\, u^{s-1} e^{-u}$ as
\begin{align}
\label{eq:def-of-G}
G^\Omega_{\ell_3}(u) = \frac{\gamma(1+\ell_3, -\Omega\log u)}{(-\Omega)^{1+\ell_3}}, \quad (\ell_3 > -1),
\qquad
G^\Omega_{-1}(u) = 1.
\end{align}
The cross section is evaluated with the $\tau_1^C$-dependent profile scales of Sec.~\ref{subsec:profile}.

\subsection{Renormalization-Group Evolution at \texorpdfstring{N$^3$LL}{N3LL}}\label{sec:N3LL-evolution}
The N$^3$LL evolution kernels appearing in Eq.~\eqref{eq:RG-simple} are given explicitly in Appendix~B of Ref.~\cite{Ee:2025scz}.
We refer the reader to that reference for explicit forms of:
\begin{itemize}
\item the evolution kernels $U_H$, $U_{B_q}$, $U_J$, and $U_S^2$;
\item the N$^3$LL evolution exponents $K_\Gamma$, $\eta_\Gamma$, and $K_\gamma$, written in terms of the integrated cusp and non-cusp anomalous dimensions and the QCD $\beta$ function;
\item the generalized plus distribution $\mathcal{L}^a(x)$ used in the kernels.
\end{itemize}
The perturbative inputs through three loops---the cusp and non-cusp anomalous dimensions $\Gamma_n^q$, $\gamma_n$ and the $\beta$-function coefficients $\beta_n$---are listed in Appendix~D of Ref.~\cite{Kang:2013nha}. To reach N$^3$LL accuracy we additionally use the four-loop cusp anomalous dimension $\Gamma_3^q$ from Ref.~\cite{Henn:2019swt} and the four-loop $\beta$-function coefficient $\beta_3$ from Ref.~\cite{vanRitbergen:1997va}.
 
For the running of the strong coupling, we employ the three-loop running of Ref.~\cite{DelDebbio:2007ee} for consistency with the NNPDF4.0 NNLO PDF sets used in this work, with the inputs $\alpha_s(m_Z) = 0.118$ and $m_Z = 91.1876~\textrm{GeV}$.

\subsection{Non-perturbative Power Corrections}\label{sec:power-corrections}
The treatment of non-perturbative power corrections follows Sec.~3.4 of Ref.~\cite{Ee:2025scz}, with the soft normalization $Q$ replaced by $Q_R$. We summarize the key steps here, emphasizing where the Centauric observable enters.

In the peak region $\tau_1^C \sim \Lambda_\textrm{QCD}/Q_R$, the soft measurement is sensitive to momenta of order $\Lambda_\textrm{QCD}$, and the perturbative soft function must be supplemented by a description of non-perturbative physics. In SCET, this is implemented by convolving $S_\textrm{PT}$ with a universal shape function $F(k)$ describing soft radiation at $k\sim\Lambda_\textrm{QCD}$ \cite{Korchemsky:1999kt,Korchemsky:2000kp,Hoang:2007vb},
\begin{align}
S_{\rm had}(k,\mu_S)
=
\int dk'\, S_{\rm PT}(k-k',\mu_S)\, F(k'),
\label{eq:shape-convolution}
\end{align}
with $F(k)$ normalized as $\int dk\, F(k) = 1$ and independent of $\mu_S$. Inserting Eq.~\eqref{eq:shape-convolution} into the factorization theorem in Eq.~\eqref{eq:jet-mass-before-simp-2} yields the standard convolution form for the cross section,
\begin{align}
\label{eq:power-correction}
\sigma({\tau_1^C})
=
\int dk\,
\sigma_\textrm{PT}
\left(
{\tau_1^C}
-
\frac{k}{Q_R}
\right)
F(k),
\end{align}
where the only Centauric modification relative to Ref.~\cite{Ee:2025scz} is the appearance of $Q_R$ in place of $Q$ in the rescaling of $k$. For $Q_R{\tau_1^C}\gg \Lambda_{\rm QCD}$, the convolution admits an OPE in which the only property of $F(k)$ needed at leading power is its first moment. Expanding the shape function about $k=0$ gives~\cite{Mateu:2012nk,Ee:2025scz}
\begin{align}
F(k) = \delta(k) - \delta'(k)\, \bar\Omega_1 
+ \mathcal{O}\!\left(\frac{\alpha_s\Lambda_\textrm{QCD}}{k^2}\right)
+ \mathcal{O}\!\left(\frac{\Lambda_\textrm{QCD}^2}{k^3}\right),
\end{align}
which yields the OPE for the cross section
\begin{align}
\label{eq:OPE_shape}
\sigma({\tau_1^C})
=
\left[
\sigma_\textrm{PT}({\tau_1^C})
-
\frac{2\bar\Omega_1}{Q_R}
\frac{d}{d{\tau_1^C}}\sigma({\tau_1^C})
\right]
\left[
1+\mathcal{O}\left(\frac{\alpha_s \Lambda_\textrm{QCD}}{Q_R{\tau_1^C}}\right)
+\mathcal{O}\left(\frac{\Lambda_\textrm{QCD}^2}{Q_R^2{{\tau_1^C}}^2}\right)
\right],
\end{align}
where the first moment of the shape function is defined as $2\bar{\Omega}_1
=
\int dk\, k F(k)$ \cite{Abbate:2010xh}.
This shows that the leading non-perturbative effect is a shift controlled by the $\overline{\textrm{MS}}$-scheme first moment $\bar\Omega_1$~\cite{Abbate:2010xh,Kang:2013nha}.

\paragraph{$R$-gap scheme.}
The $\overline{\textrm{MS}}$-scheme first moment $\bar\Omega_1$ and the perturbative soft function $S_\textrm{PT}$ each contain an $\mathcal{O}(\Lambda_\textrm{QCD})$ renormalon ambiguity that cancels in their combination~\cite{Hoang:2007vb}. To improve perturbative convergence, we adopt the $R$-gap scheme of Refs.~\cite{Hoang:2008fs,Jain:2008gb,Hoang:2008yj,Abbate:2010xh}, which introduces a gap parameter $\bar\Delta$ representing the minimum hadronic energy deposit and shifts the shape function as $F(k) \to F(k - 2\bar\Delta)$. Both $\bar\Delta$ and $\bar\Omega_1$ carry the same renormalon, which is removed by introducing renormalon-free parameters $\Delta(R_{\mathrm{gap}},\mu_S)$ and $\Omega_1(R_{\mathrm{gap}},\mu_S)$ via the perturbative subtraction series
[Eqs.~(3.47) and (3.49) of Ref.~\cite{Ee:2025scz}]
\begin{align}
\bar\Delta = \Delta(R_{\mathrm{gap}},\mu_S) + \delta(R_{\mathrm{gap}},\mu_S), \qquad
\bar\Omega_1 = \Omega_1(R_{\mathrm{gap}},\mu_S) + \delta(R_{\mathrm{gap}},\mu_S),
\end{align}
where 
\begin{align}
\delta(R_{\mathrm{gap}},\mu_S) = \frac{R_{\mathrm{gap}}}{2}\, e^{\gamma_\textrm{E}}
\frac{d}{d\log(ix)}\, [\log S_\textrm{PT}(x,\mu_S)]\big|_{x=(iR_{\textrm{gap}}e^{\gamma_\textrm{E}})^{-1}}.
\end{align}
$S_\textrm{PT}(x,\mu_S)$ is the position-space perturbative soft function, and $R_{\mathrm{gap}}$ is an infrared subtraction scale. With these definitions, the gap-subtracted shape function takes the form
\begin{align}
\label{eq:shape-function-after-R-gap}
S_{\rm had}(k,\mu_S) = \int dk'
\left[ e^{-2\delta(R_{\mathrm{gap}},\mu_S)(\partial/\partial k)} S_\textrm{PT}(k-k',\mu_S) \right]
F(k' - 2\Delta(R_{\mathrm{gap}},\mu_S)),
\end{align}
and the leading shift in Eq.~\eqref{eq:OPE_shape} becomes
\begin{align}
\label{eq:tau-shift}
{\tau_1^C} \;\longrightarrow\;
{\tau_1^C} - \frac{2\,[\Omega_1(R_{\mathrm{gap}},\mu_S) - \Delta(R_{\mathrm{gap}},\mu_S)]}{Q_R}\,.
\end{align}

\paragraph{The Centauric $1/R$ shift.}
Substituting the small-$\theta_J$ value $Q_R \simeq QR/2$ from Eq.~\eqref{eq:1-jet-invariants} into Eq.~\eqref{eq:tau-shift} yields
\begin{align}
{\tau_1^C} \;\longrightarrow\; {\tau_1^C} - \frac{4\,[\Omega_1 - \Delta]}{Q R},
\label{eq:R-dep-shift}
\end{align}
exhibiting the characteristic $1/R$ behavior at fixed $Q$. As in thrust in $e^+e^-$ collisions, hadronization does not distort the shape of the perturbative distribution but simply translates it; what is novel for the Centauric observable is that the magnitude of the translation is controlled by the jet radius $R$. This prediction is tested against \textsc{Pythia} hadron-level simulations in Sec.~\ref{sec:omega1}.

\paragraph{Shape-function model and input values.}
For numerical predictions we adopt the same one-parameter model used in Eq.~(3.38) of Ref.~\cite{Ee:2025scz},
\begin{align}
\label{eq:shape-definition}
F(k) = \frac{1}{\lambda} f_0\!\left(\frac{k}{\lambda}\right), \qquad
f_0(u) = 8\sqrt{\frac{2}{3}}e^{-2u}u^{3/2},
\qquad
\lambda = 2\left[\Omega_1(R_\Delta,\mu_\Delta) - \Delta(R_\Delta,\mu_\Delta)\right],
\end{align}
where the dependence on $(R_{\mathrm{gap}},\mu_S)$ cancels between $\Omega_1$ and $\Delta$. We specify input parameters at the reference scales $R_\Delta = \mu_\Delta = 2$~GeV: $\Delta(R_\Delta,\mu_\Delta) = 0.05$~GeV and $\Omega_1(R_\Delta,\mu_\Delta) = 0.59$~GeV (extracted from the \textsc{Pythia} fit in Eq.~\eqref{eq:omega_1_pythia_global}). Evolution to $(R_{\mathrm{gap}}, \mu_S)$ is performed using $R$-evolution~\cite{Hoang:2008yj,Hoang:2008fs} as implemented in Ref.~\cite{Ee:2025scz}.

\section{Non-perturbative Shift for Alternative Jet Algorithms}
\label{app:soft-universality}

In Sec.~\ref{subsec:soft} we showed that the Centauro jet algorithm leads to a soft function that belongs to the same universality class as hemisphere event shapes, implying that the leading non-perturbative correction is governed by the universal matrix element $\Omega_1$ with an $R$-dependent normalization. In this appendix we provide additional details supporting this result and compare the Centauro case to other commonly used jet algorithms. In Sec.~\ref{subsec:SI-soft} we analyze the spherically invariant anti-$k_T$ algorithm and show how universality can be recovered in the J-scheme. In Sec.~\ref{subsec:LI-soft} we consider the longitudinally invariant anti-$k_T$ algorithm and show that the leading power corrections acquire additional kinematic dependence, preventing a strict universality relation.

\subsection{Spherically Invariant Anti-\texorpdfstring{$k_T$}{kT} Algorithm}\label{subsec:SI-soft}
In this subsection we analyze the spherically invariant anti-$k_T$ algorithm in the Breit frame and show that the measurement functions introduce an additional dependence on the variable $r=p_T/m_T$. We demonstrate that within the J-scheme \cite{Mateu:2012nk} this dependence can be removed, restoring a mapping onto the same universal soft matrix element $\Omega_1$, but with a different $R$ dependence of the power corrections.

For the anti-$k_T$ algorithm defined in the Breit frame with spherical
invariance, the measurement functions retain a dependence on the variable
$r=p_T/m_T$. In this case universality is not manifest in the original
definition. However, within the J-scheme framework one can remove the
$r$-dependence by replacing $r\to1$ in the measurement functions, thereby
recovering a mapping onto the same universal soft matrix element.

The corresponding coefficients become
\begin{align}
C_{\text{J-SI}}^J = \tan\frac{R}{2},
\qquad
C_{\text{J-SI}}^B = \cot\frac{R}{2}.
\end{align}
Although the $R$ dependence differs from the Centauro case, the leading
non-perturbative correction remains controlled by the universal parameter
$\Omega_1$.

\subsection{Longitudinally Invariant Anti-\texorpdfstring{$k_T$}{kT} Algorithm}\label{subsec:LI-soft}
In this subsection we consider the longitudinally invariant anti-$k_T$ algorithm and show that, after transforming to the Breit frame, the leading non-perturbative corrections acquire an explicit dependence on the event inelasticity $y$. While the corrections remain governed by the same soft matrix element $\Omega_1$, this additional kinematic dependence prevents a strict universality relation of the type obtained for the Centauro algorithm.

For the longitudinally invariant (LI) anti-$k_T$ algorithm defined in the
lab or center-of-mass frame, the mapping to the Breit frame introduces an
additional dependence on the event inelasticity $y$. In the small-$R$ and
massless limit the coefficients governing the non-perturbative shifts take
the form
\begin{align}
C_{\text{J-LI}}^J = R\sqrt{1-y},
\qquad
C_{\text{J-LI}}^B = \frac{1}{R\sqrt{1-y}}.
\end{align}

In this case the leading non-perturbative corrections depend explicitly on the
event kinematics through $y$. While the corrections remain governed by the same
soft matrix element $\Omega_1$, the additional kinematic dependence prevents a
strict universality relation of the type obtained for the Centauro algorithm.

\bigskip

In summary, the Centauro jet algorithm is distinguished by the fact that its
soft measurement depends only on rapidity at leading power, allowing an exact
mapping onto the jet-mass universality class. Other jet algorithms can be
related to this class only after appropriate approximations or modifications,
such as the J-scheme treatment discussed above. For another generalized-$k_t$ algorithm appropriate for DIS, see the recent Ref.~\cite{vanBeekveld:2026rez}.

\section{Parametrization of the Two-body Final States}
\label{app:two-particle-kinematics}
In Sec.~\ref{subsec:instability}, we demonstrated that the perturbative stability of the Centauric 1-jettiness distribution depends on the choice of measurement weights and the jet axis selection in low-multiplicity final states.  In this appendix, we analyze the leading-order (LO) two-parton final state in detail to determine the phase-space regions where the jet axis aligns with $p_1$, $p_2$, or the clustered momentum $p_1+p_2$. We further derive the corresponding expressions for $\tau_1^C$ in each region. The kinematic variables and the phase-space decomposition presented here follow the general framework established in Refs.~\cite{Kang:2014qba, Chu:2022jgs}.

At LO partonic DIS, an initial parton carrying a momentum fraction $\xi$ of the proton ($x \le \xi \le 1$) scatters off a virtual photon with momentum $q$, producing two final-state partons with momenta $p_1$ and $p_2$:
\begin{align}
\label{eq:momentum-conservation}
\xi P^\mu + q^\mu = \left(p_1 + p_2\right)^\mu\,.
\end{align}
The proton and virtual photon momenta in the Breit frame are defined in Eq.~\eqref{eq:P-q-mu-in-breit}. The initial parton momentum is expressed as:
\begin{align}
\xi P^\mu = \frac{Q}{z} \frac{\bar{n}^\mu}{2}\,,
\end{align}
where $z \equiv x/\xi$ satisfies $x \le z \le 1$. We decompose the final-state momenta $p_1$ and $p_2$ in light-cone coordinates:
\begin{align}
p_1^\mu &= p_1^- \frac{n^\mu}{2} + p_1^+ \frac{\bar{n}^\mu}{2} -p_\perp^\mu,
\nonumber \\
p_2^\mu &= p_2^- \frac{n^\mu}{2} + p_2^+ \frac{\bar{n}^\mu}{2} +p_\perp^\mu,
\end{align}
where transverse momentum conservation is manifest. From Eq.~\eqref{eq:momentum-conservation}, the conservation of the light-cone components yields:
\begin{align}
p_1^- + p_2^- = Q,
\quad
p_1^+ + p_2^+ = \frac{1-z}{z} Q. 
\end{align}
Imposing the on-shell conditions $p_1^2 = p_2^2 = 0$, we arrive at the following parametrization in terms of $(z, v)$:
\begin{align}
\label{eq:p1p2-parametrization}
p_1^\mu &= Q(1-v) \frac{n^\mu}{2} + Q \frac{1-z}{z}v \frac{\bar{n}^\mu}{2} -
Q\sqrt{\frac{1-z}{z}\,(1-v)v}\, n_\perp^\mu,
\nonumber \\
p_2^\mu &= Qv \frac{n^\mu}{2} +  Q \frac{1-z}{z}(1-v) \frac{\bar{n}^\mu}{2} +
Q\sqrt{\frac{1-z}{z}\,(1-v)v}\, n_\perp^\mu,
\end{align}
where $v \in (0, 1)$ is a phase-space parameter and $n_\perp^\mu = (0, \cos\phi, \sin\phi, 0)$ is a unit spacelike vector. By exploiting the azimuthal symmetry, we set $\phi = 0$ without loss of generality, effectively confining the process to the $xz$-plane.

The behavior of $\tau_1^C$ depends on which axis is identified as the jet axis by the clustering algorithm.
When $d_{12}>1$, the jet axis is chosen along either $p_1$ or $p_2$.
If $p_1$ is selected as the jet axis ($n_J \propto p_1$), it lies within the jet cone, while $p_2$ remains outside. 
Since $p_1$ is massless and $n_J \propto p_1$, we have $n_J \cdot p_1 \propto p_1^2 = 0$, and the observable is determined solely by $p_2$:
\begin{align}
\tau_1^C = \frac{1}{Q^2} \omega_B n_B \cdot p_2 = \frac{v \omega_B}{Q}.
\end{align}
However, the algorithm selects $p_2$ as the axis if it has a smaller value of $\bar\eta$ ($\bar{\eta}_2 < \bar{\eta}_1$), which occurs when $v > 0.5$.
Consequently, when the jet axis is aligned with either $p_1$ or $p_2$, 
the value of $\tau_1^C$ is:
\begin{align}
\label{eq:tau1a-jet-p1orp2}
\tau_1^C\big|_{d_{12}>1} = \min(v, 1-v) \frac{\omega_B}{Q} \le \frac{\omega_B}{2Q},
\end{align}
with the kinematic maximum $\omega_B/(2Q)$ attained at $v=1/2$. For the choice $\omega_B = Q$ (corresponding to Choice I in Sec.~\ref{subsec:instability}), 
the maximum is $1/2$.

When the clustering condition $d_{12} < 1$ is met, both partons are combined into a single jet. The clustered momentum is:
\begin{align}
p_1^\mu + p_2^\mu = Q \frac{n^\mu}{2} + Q \frac{1-z}{z} \frac{\bar{n}^\mu}{2}.
\end{align}
The clustered axis is aligned with the $+z$ direction (i.e., the photon direction) 
only when $z > 1/2$, since this requires the coefficient of $n^\mu/2$ to exceed 
that of $\bar{n}^\mu/2$ in the total momentum.
In this case, both particles reside within the jet cone, and $\tau_1^C$ vanishes from the $n_B$ contribution, becoming:
\begin{align}
\label{eq:tau1a-p1p2-axis}
\tau_1^C = \frac{\omega_J}{Q^2} n_J \cdot (p_1 + p_2) = \frac{\omega_J}{Q} \frac{1-z}{z}.
\end{align}
Using Eq.~\eqref{eq:p1p2-parametrization}, the distance measure $d_{12}$ is computed as:
\begin{align}
d_{12} = \frac{(\bar{\eta}_1 + \bar{\eta}_2)^2}{R^2} = \frac{4}{R^2} \frac{1-z}{z} \frac{1}{v(1-v)}.
\end{align}
The condition $d_{12} < 1$ requires $z > z_c = 16/(R^2+16)$. 
At the boundary $z = z_c$ (where the clustered jet first becomes possible), 
the value of the observable is:
\begin{align}
\label{eq:tau1a-jet-p1andp2}
\tau_1^C\big|_{z=z_c} = \frac{R^2}{16} \frac{\omega_J}{Q},
\end{align}
which is the kinematic maximum within Region I.

Combining these results, the value of $\tau_1^C$ in the two-body final state 
takes the form:
\begin{align}
\tau_1^C = 
\begin{cases}
\min(v, 1-v) \frac{\omega_B}{Q}, & d_{12} > 1, \\
\frac{\omega_J}{Q} \frac{1-z}{z}, & d_{12} < 1,
\end{cases}
\end{align}
with maxima $\omega_B/(2Q)$ and $R^2 \omega_J/(16 Q)$ in the two regions, respectively.

The selection of the jet axis is partitioned into three distinct regions in $(z, v)$ space:
\begin{itemize}
\item \textbf{Region I ($p_1 + p_2$ jet):} $d_{12} < 1$, corresponding to $z > z_c$ and $v_- < v < v_+$, where $v_\pm = \frac{1}{2} \pm \frac{1}{2} \sqrt{1 - \frac{1-z}{z} \frac{16}{R^2}}$.
\item \textbf{Region II ($p_1$ jet):} $\bar{\eta}_1 < \bar{\eta}_2$ (i.e., $v < 1/2$), 
with the additional constraint $v < v_-$ when $z > z_c$.
\item \textbf{Region III ($p_2$ jet):} $\bar{\eta}_2 < \bar{\eta}_1$ (i.e., $v > 1/2$), 
with the additional constraint $v > v_+$ when $z > z_c$.
\end{itemize}

These analytical results provide the kinematic foundation for the phase-space partitioning and the resulting observable behavior discussed in the main text. Specifically, the boundaries defined by $z_c$ and $v_\pm$ delineate the domains in the $(z, v)$ plane shown in Fig.~\ref{fig:phase_space_LO}, where the jet algorithm transitions between different axis assignments. Mapping these kinematic regions onto the $\tau_1^C$ distribution yields the surface structures illustrated in Fig.~\ref{fig:LO-fixed-tau1a-two-vers} under the respective choices for the weights $\omega_{J,B}$.

\section{Final results at different $(Q, x_B)$}
\label{app:additional-kinematics}
\begin{figure}
    \centering
    \includegraphics[width=\linewidth]{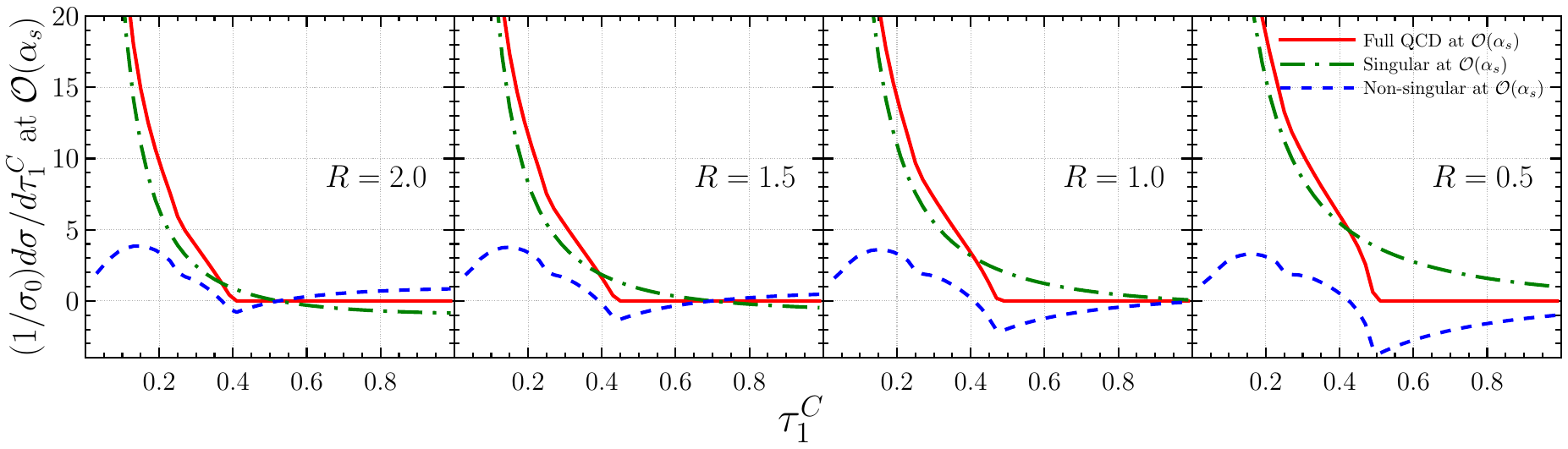}
    \includegraphics[width=\linewidth]{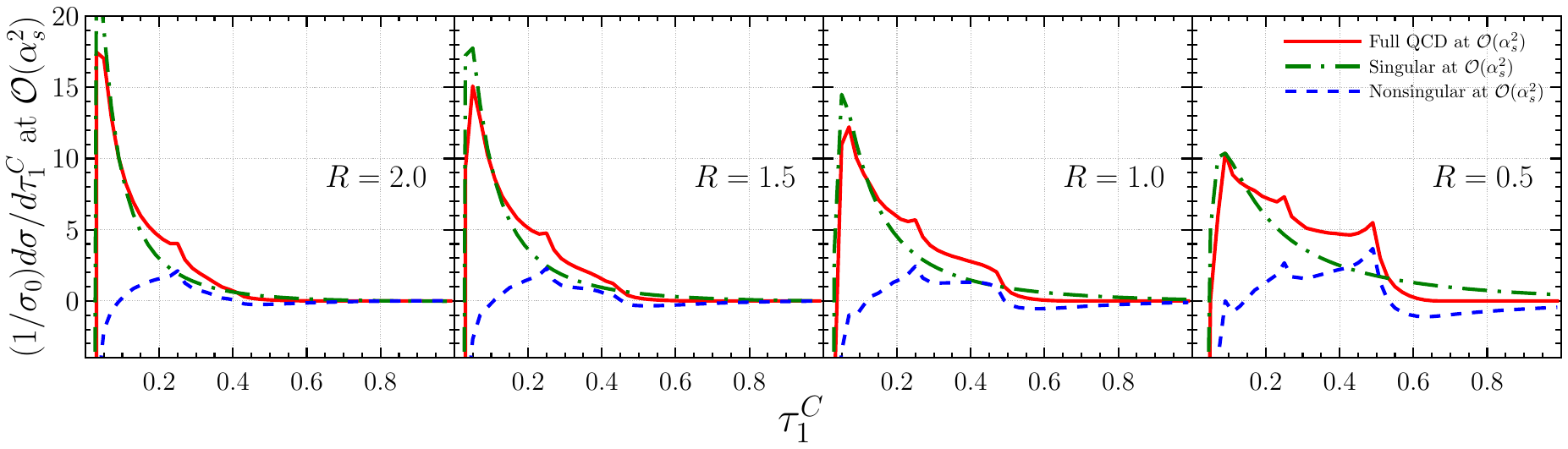}
    \caption{
    The same plots as Fig.~\ref{fig:ns_LO} at $Q=30$~GeV and $x_B=0.05$.
    }
    \label{fig:ns_LO_Q30_x005}
\end{figure}

\begin{figure}
    \centering
    \includegraphics[width=\linewidth]{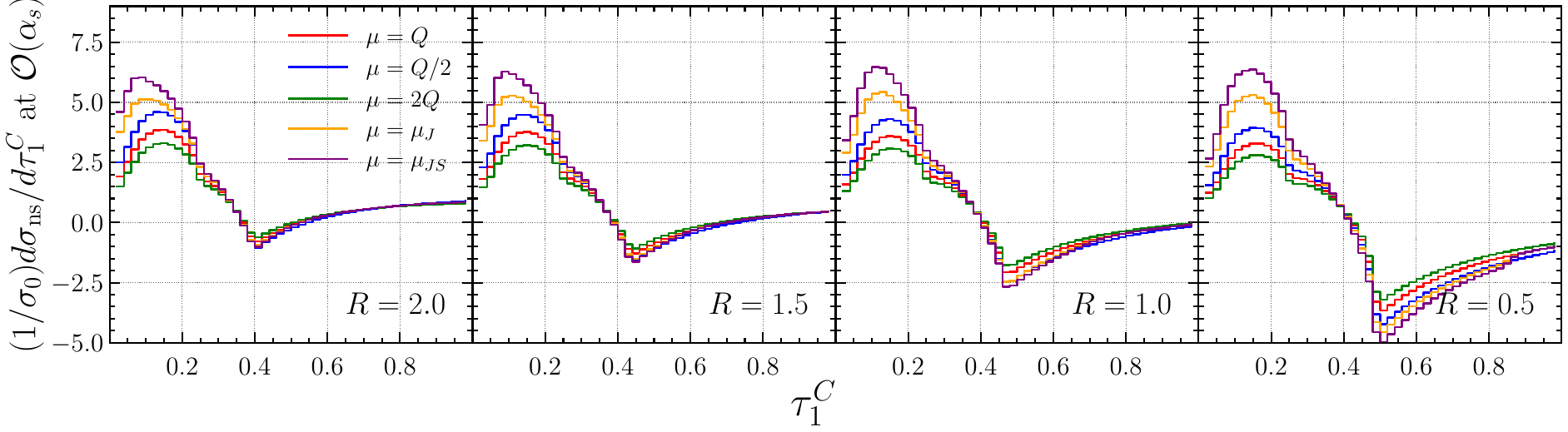}
    \includegraphics[width=\linewidth]{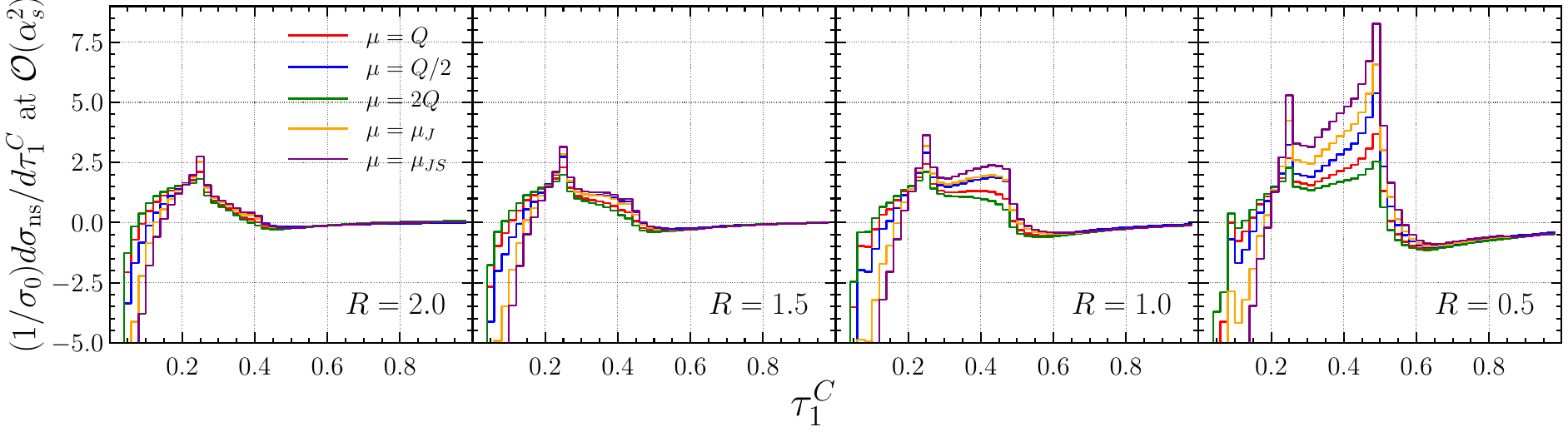}
    \caption{The same plots as Fig.~\ref{fig:ns_LO_NLO_scale_var} at $Q=30$~GeV and $x_B=0.05$.}
    \label{fig:ns_LO_NLO_scale_var_Q30_x005}
\end{figure}

\begin{figure}[t]
    \centering
    \includegraphics[width=0.40\linewidth]{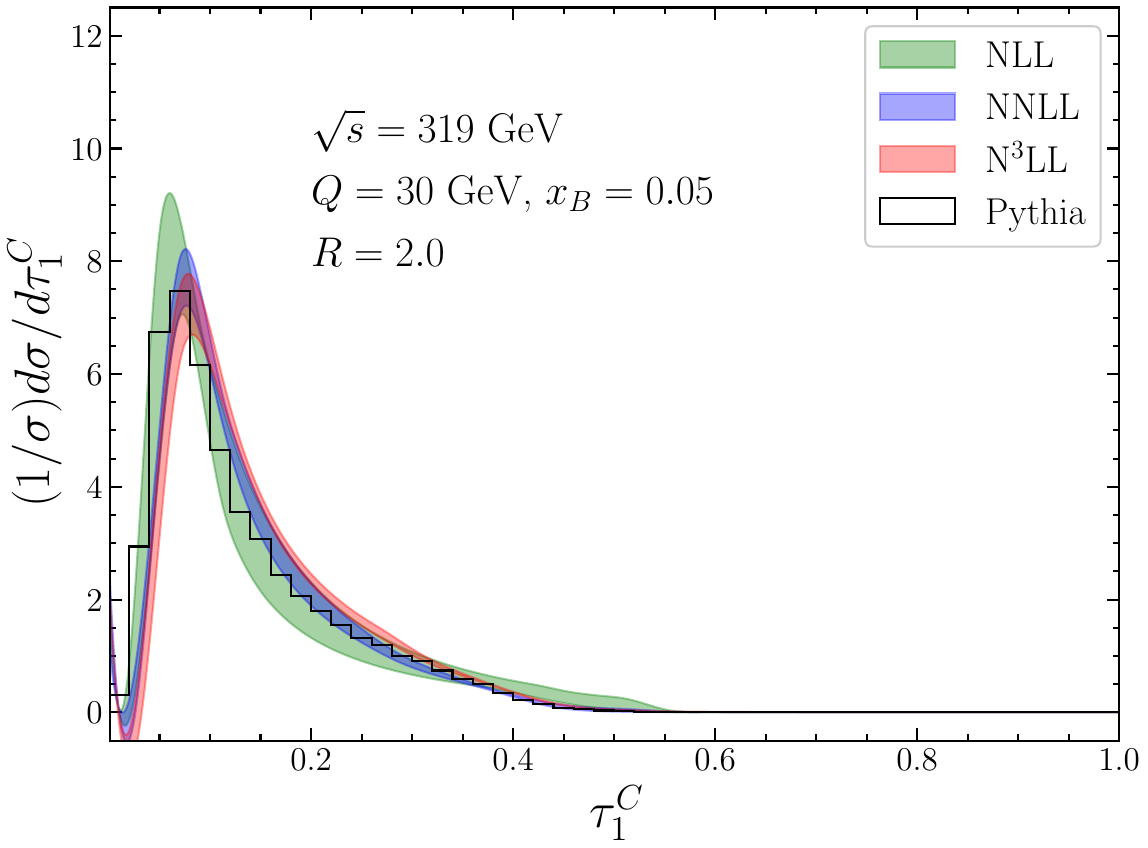}
    \includegraphics[width=0.40\linewidth]{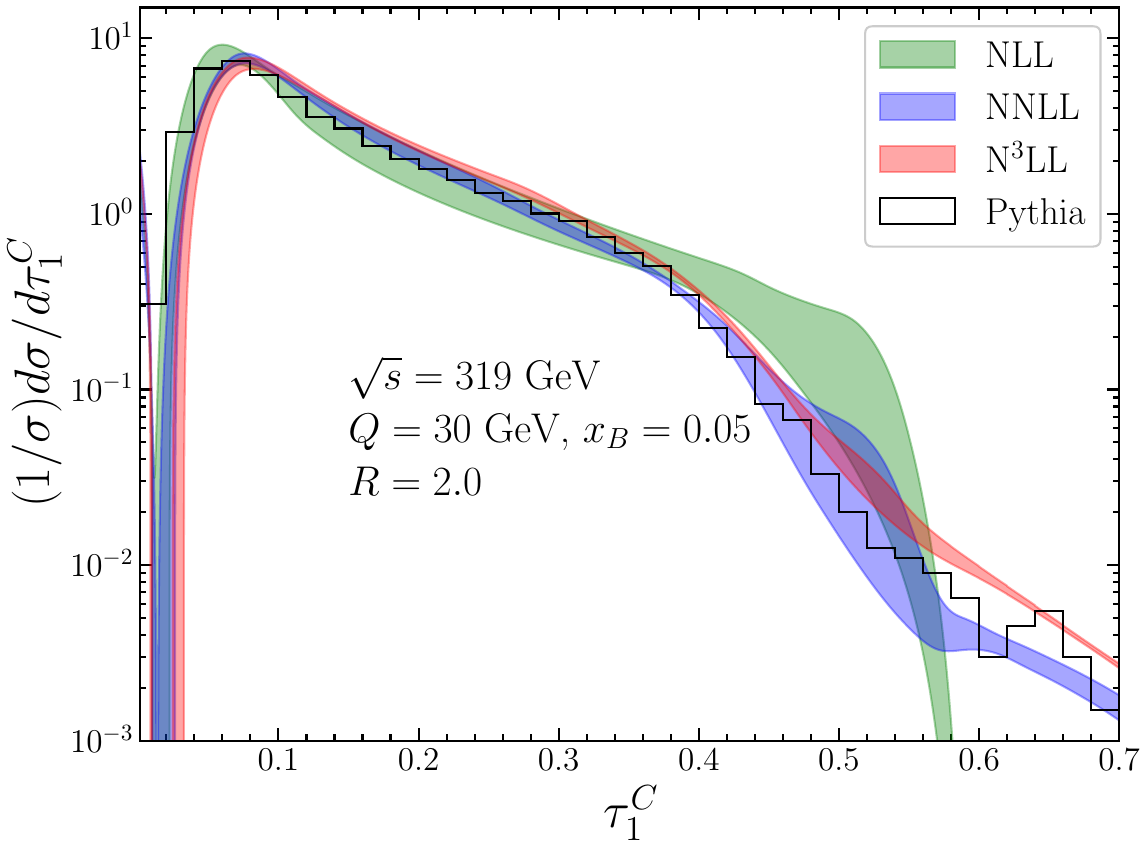}
    \includegraphics[width=0.40\linewidth]{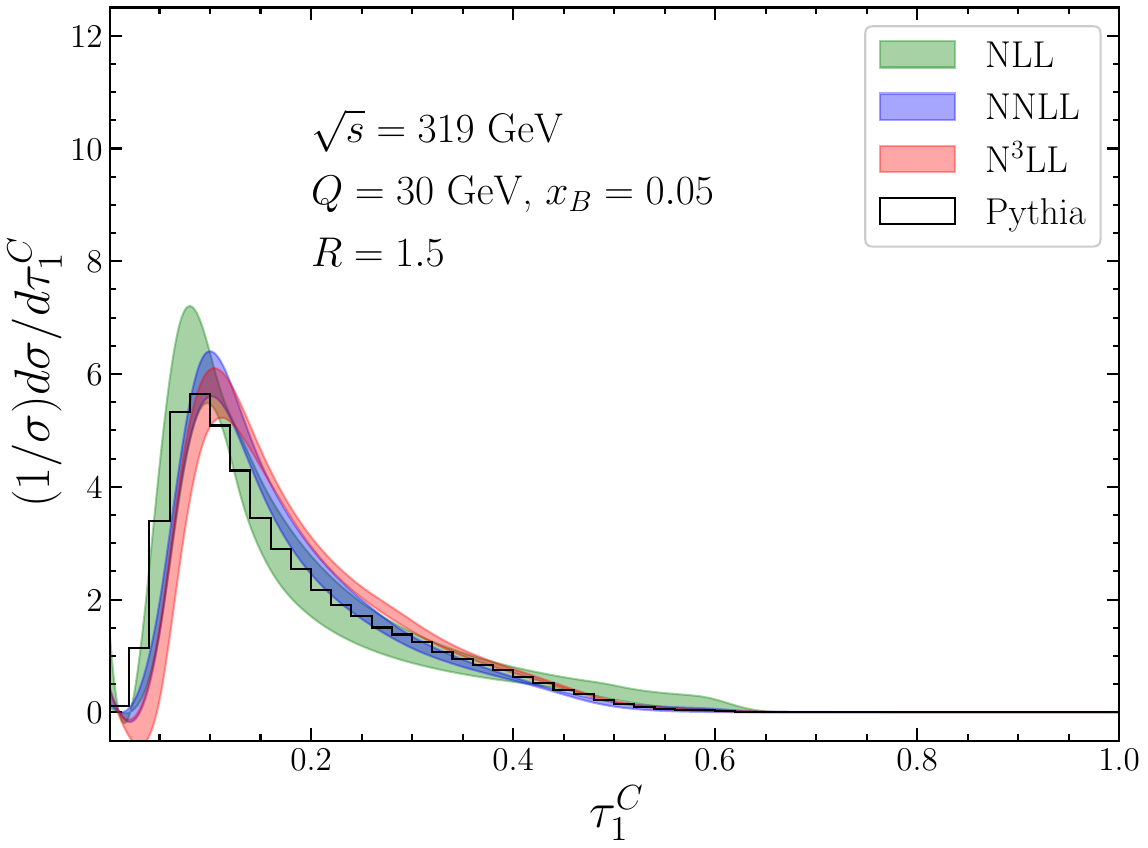}
    \includegraphics[width=0.40\linewidth]{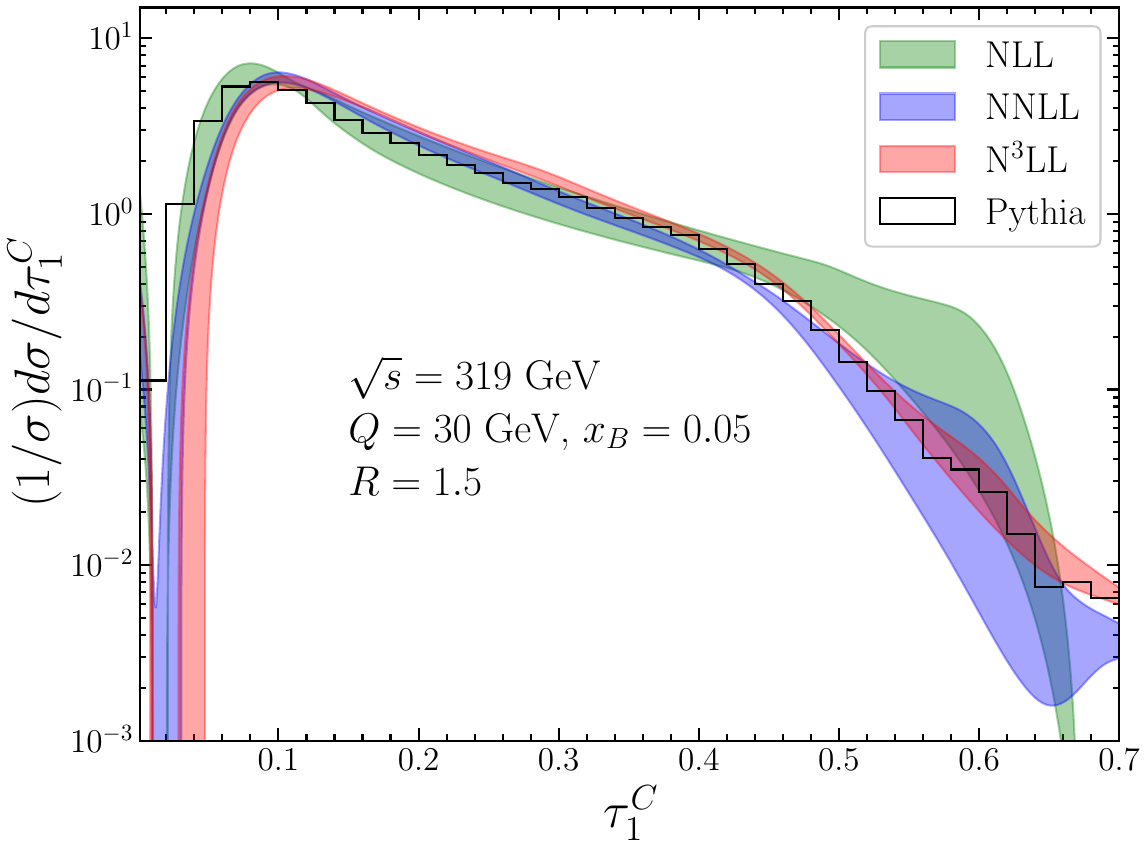}
    \includegraphics[width=0.40\linewidth]{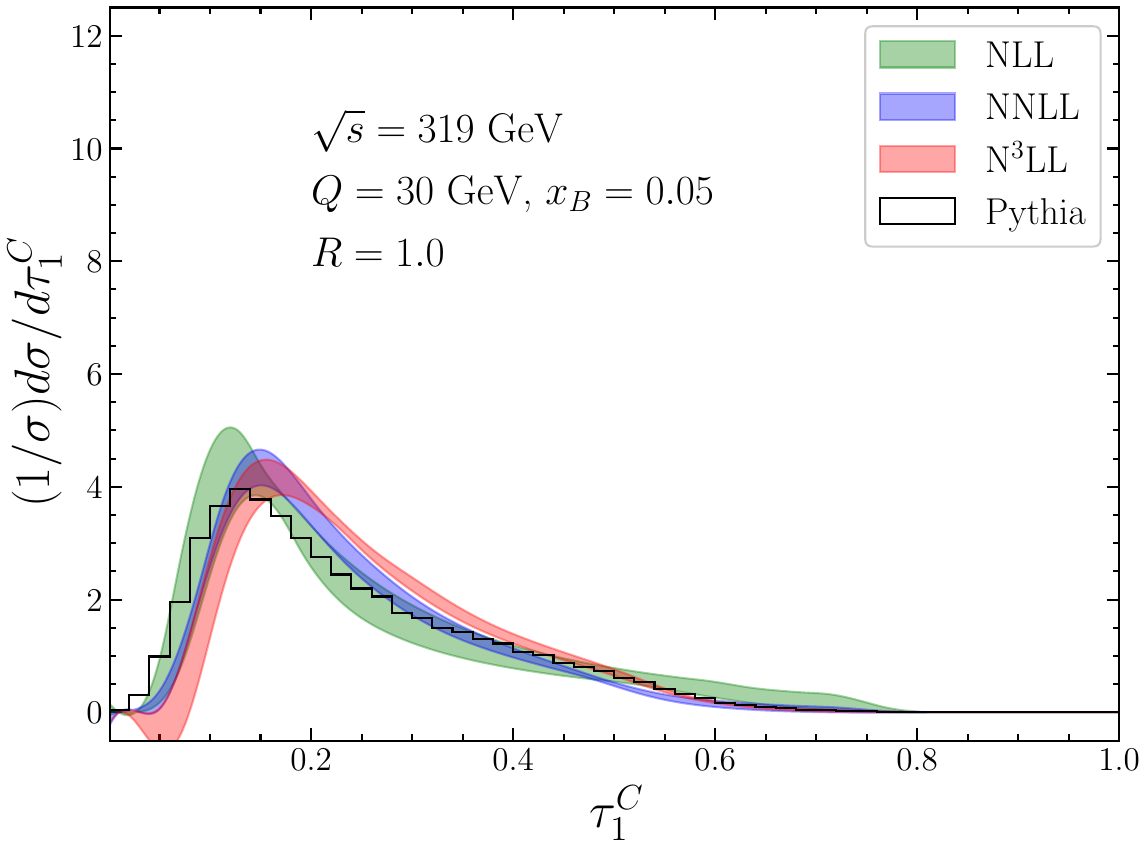}
    \includegraphics[width=0.40\linewidth]{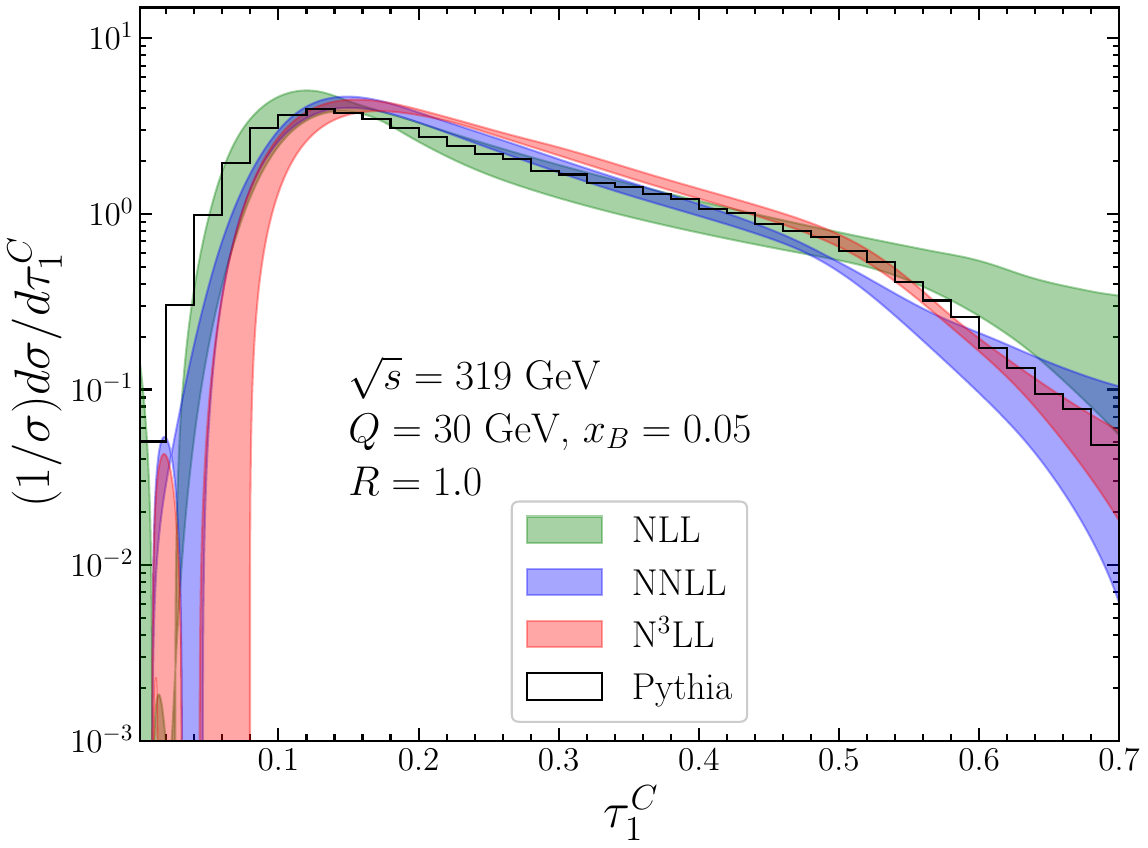}
    \includegraphics[width=0.40\linewidth]{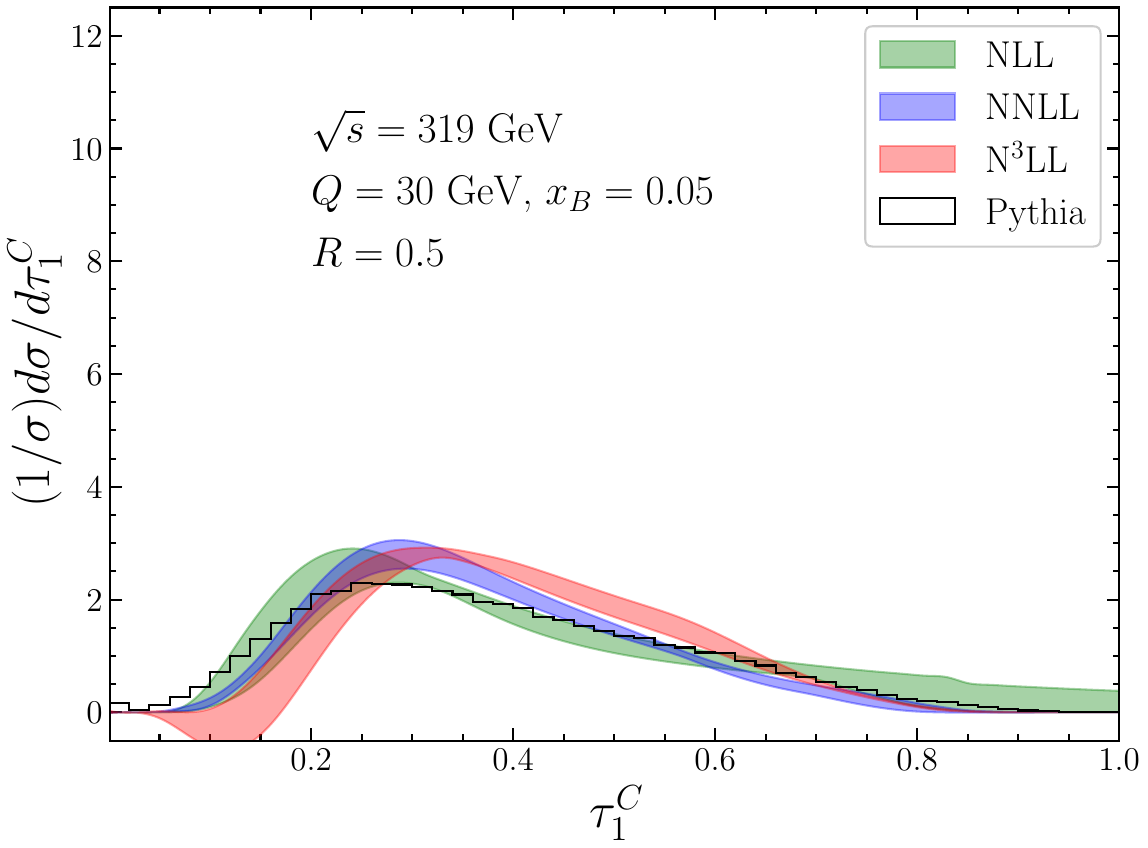}
    \includegraphics[width=0.40\linewidth]{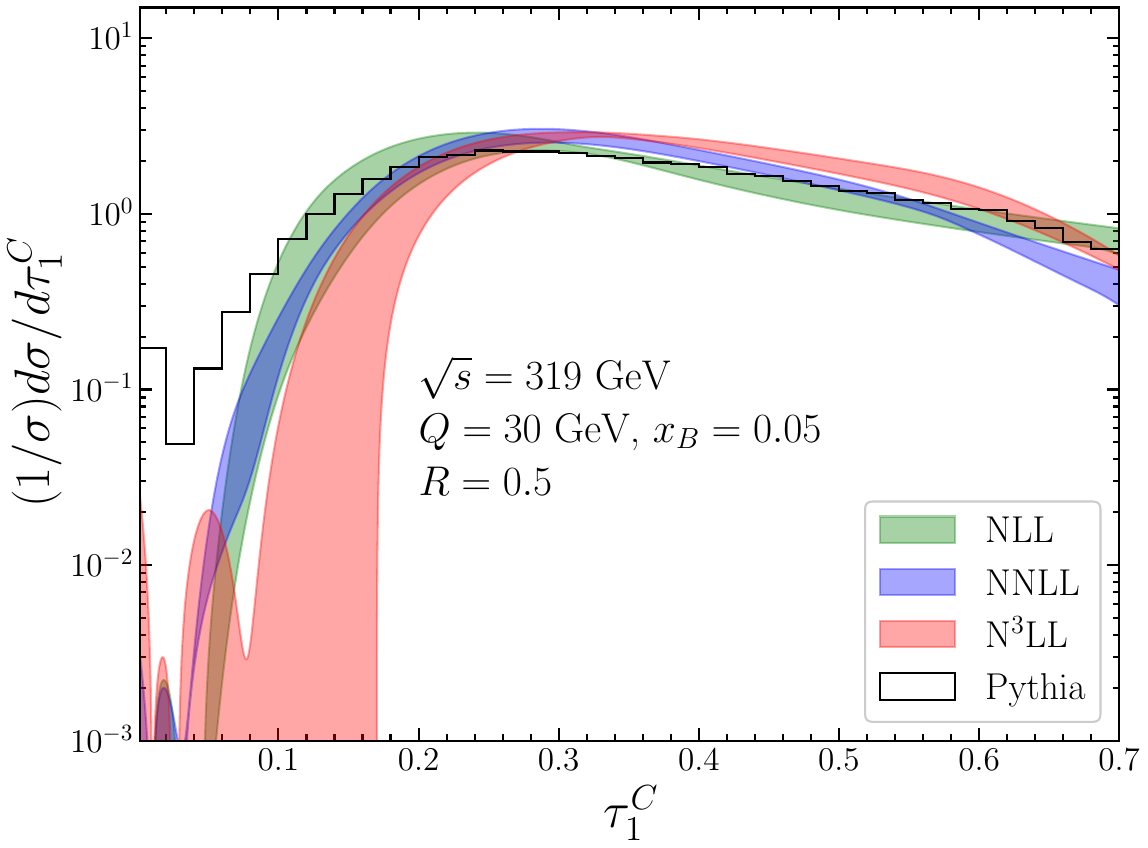}
    \vspace{-1em}
    \caption{The same plot as Fig.~\ref{fig:theory-final-hadron} at $Q=30$~GeV and $x_B=0.05$.}
    \label{fig:theory-final-hadron_Q30_x005}
\end{figure}

\begin{figure}
    \centering
    \includegraphics[width=\linewidth]{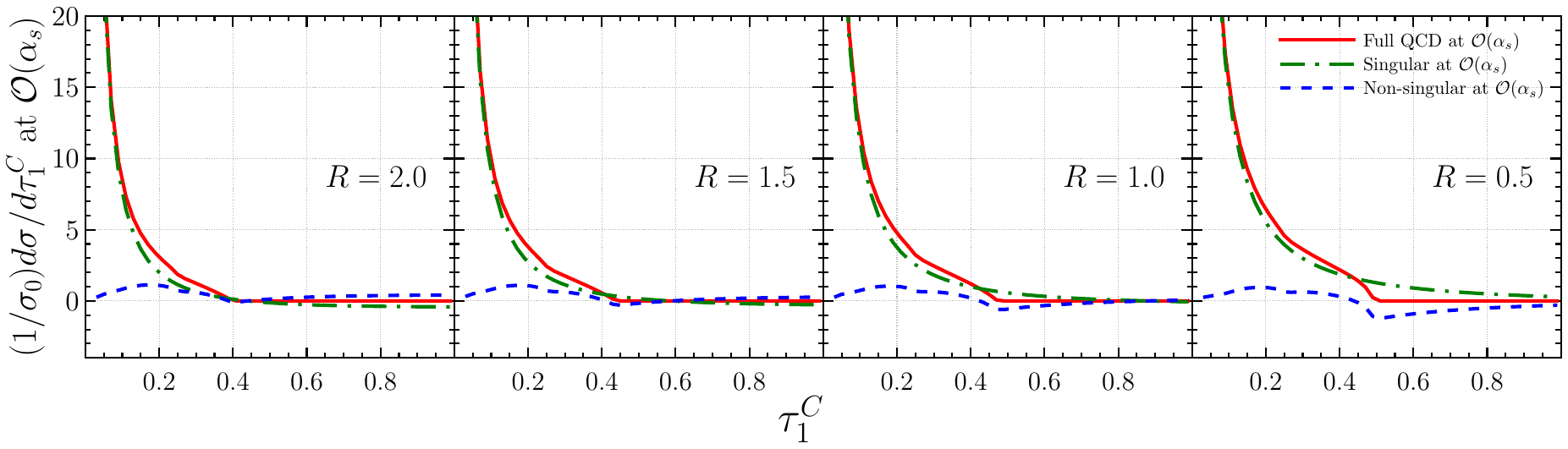}
    \includegraphics[width=\linewidth]{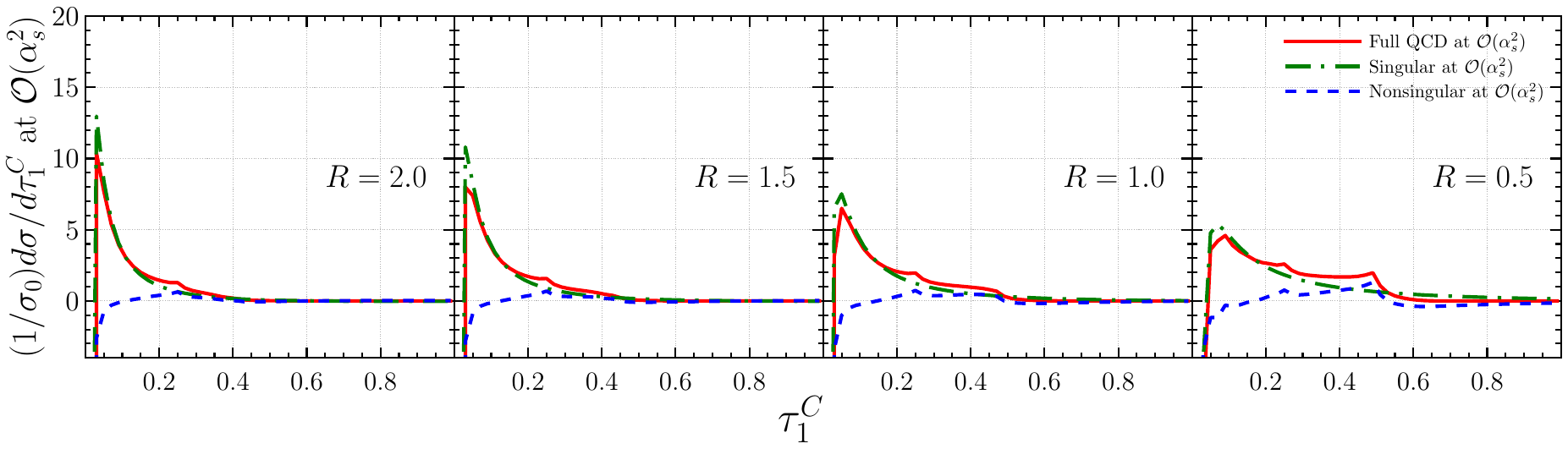}
    \caption{
    The same plots as Fig.~\ref{fig:ns_LO} at $Q=30$~GeV and $x_B=0.1$.
    }
    \label{fig:ns_LO_Q30_x010}
\end{figure}

\begin{figure}
    \centering
    \includegraphics[width=\linewidth]{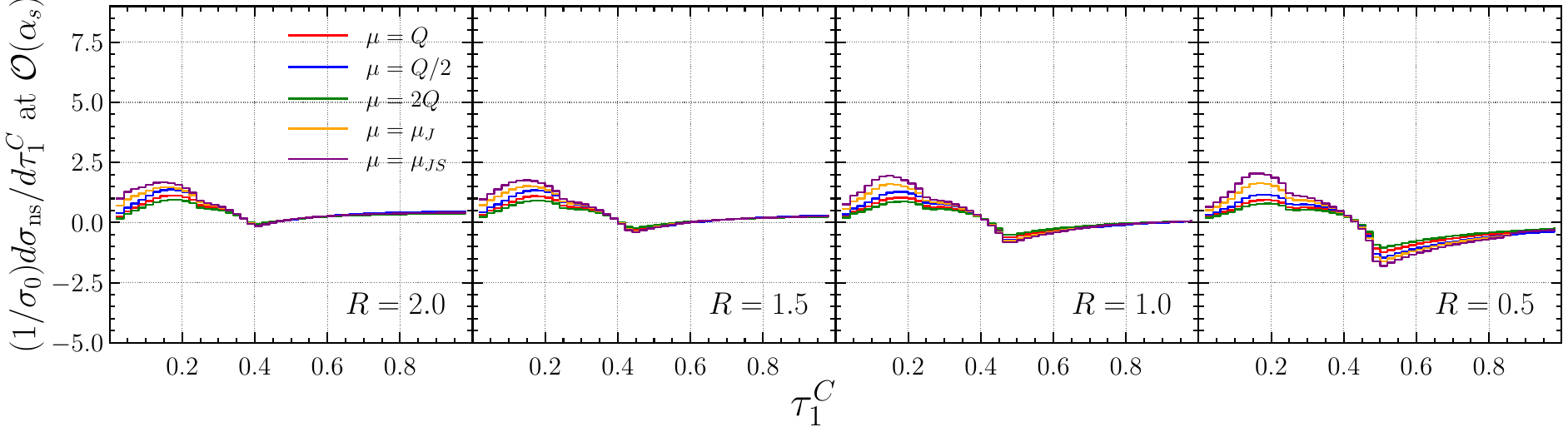}
    \includegraphics[width=\linewidth]{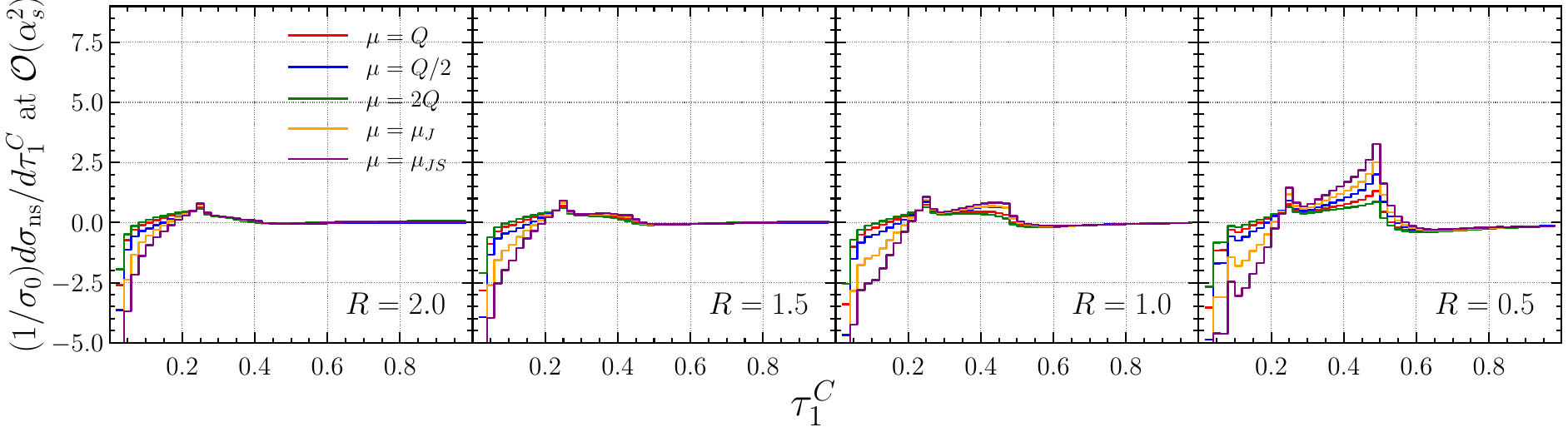}
    \caption{The same plots as Fig.~\ref{fig:ns_LO_NLO_scale_var} at $Q=30$~GeV and $x_B=0.1$.}
    \label{fig:ns_LO_NLO_scale_var_Q30_x010}
\end{figure}

\begin{figure}[t]
    \centering
    \includegraphics[width=0.40\linewidth]{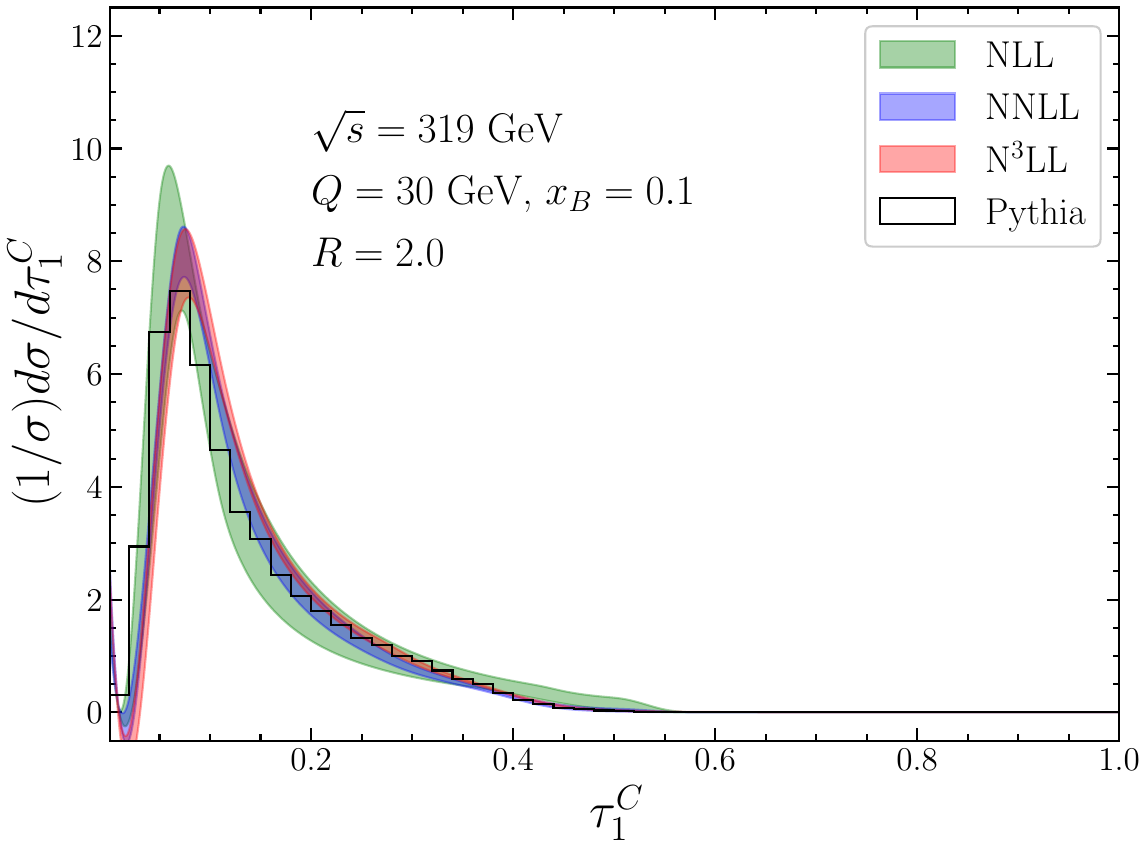}
    \includegraphics[width=0.40\linewidth]{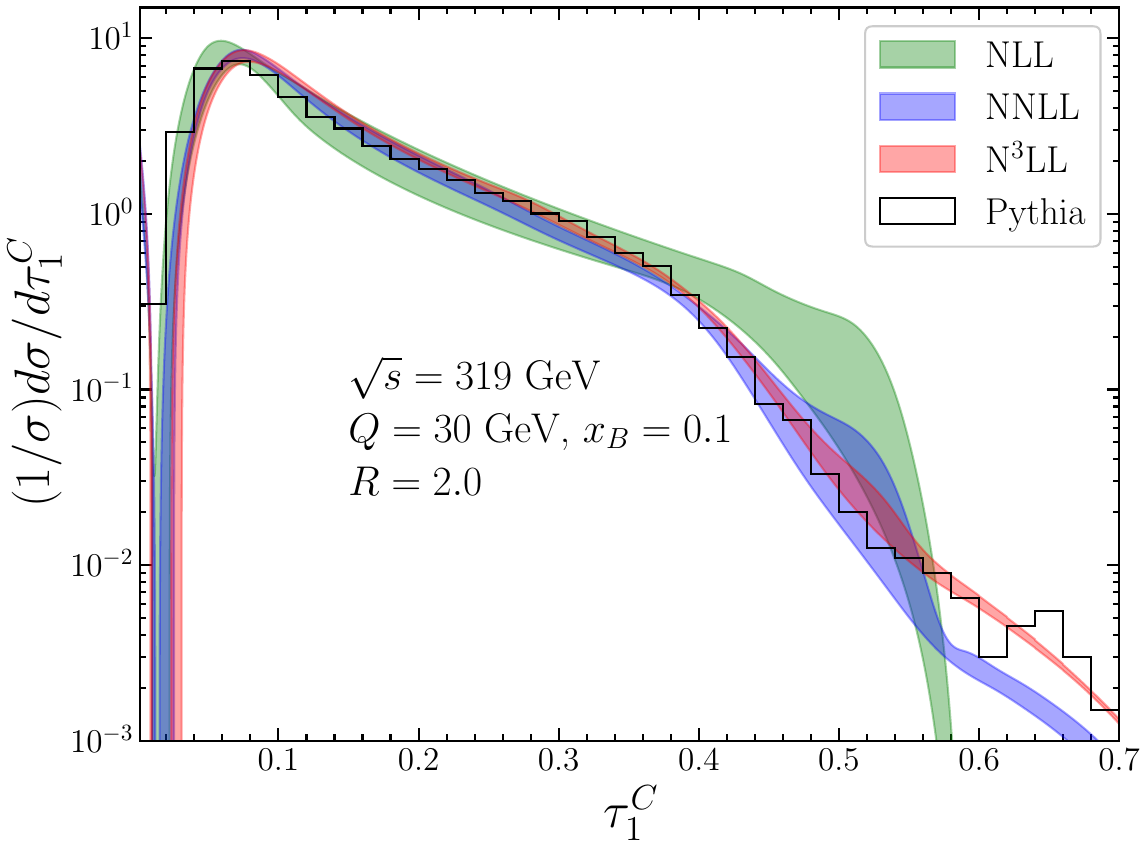}
    \includegraphics[width=0.40\linewidth]{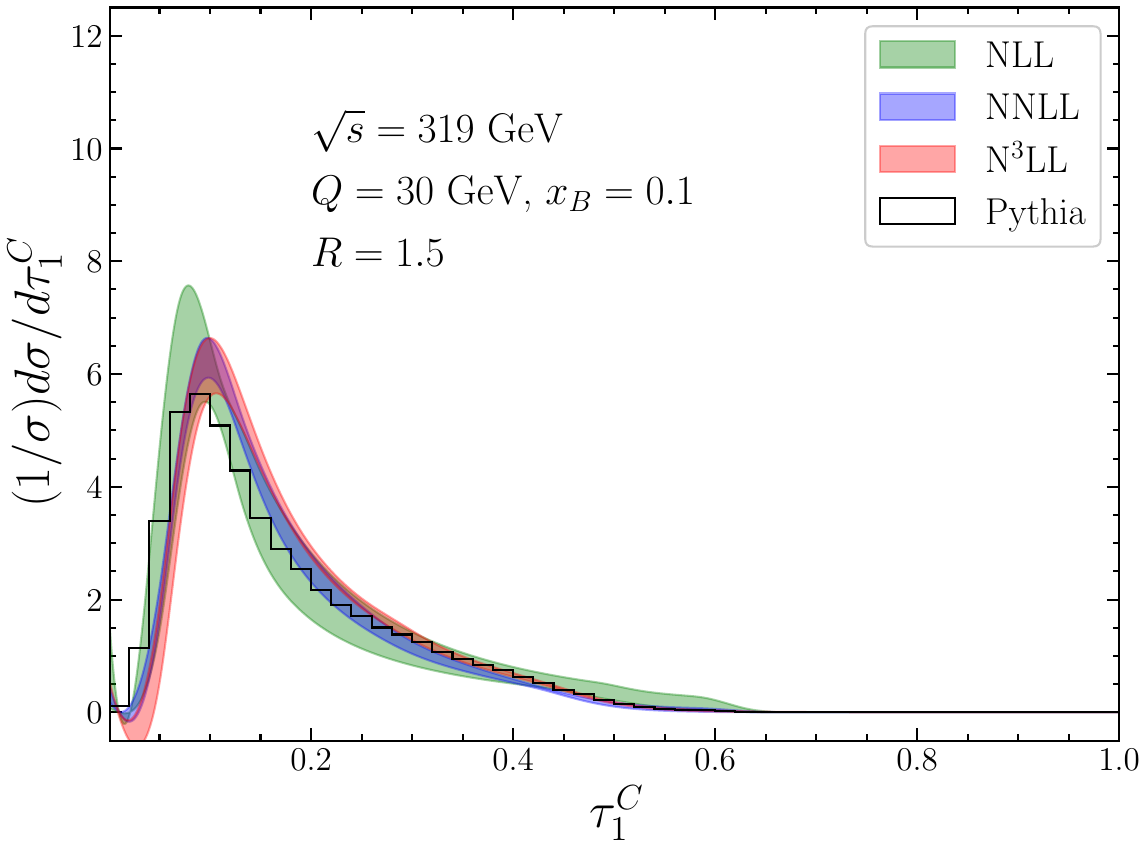}
    \includegraphics[width=0.40\linewidth]{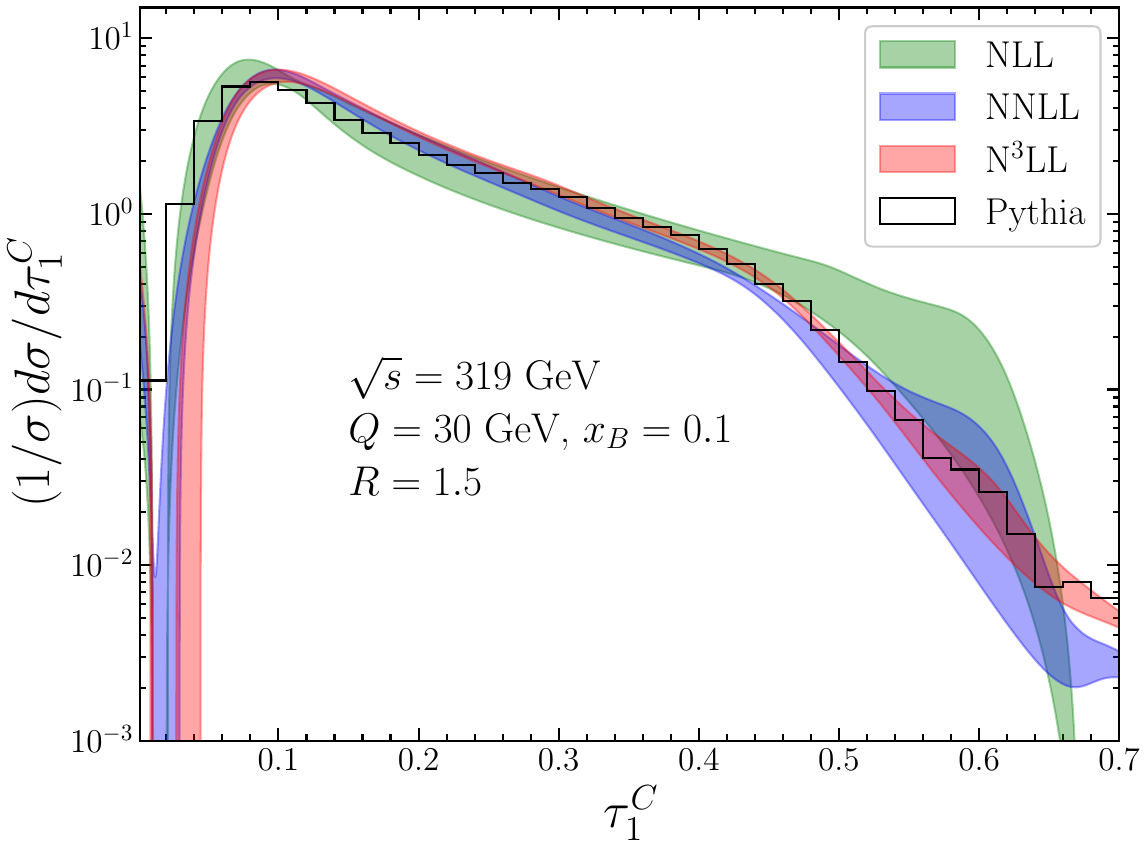}
    \includegraphics[width=0.40\linewidth]{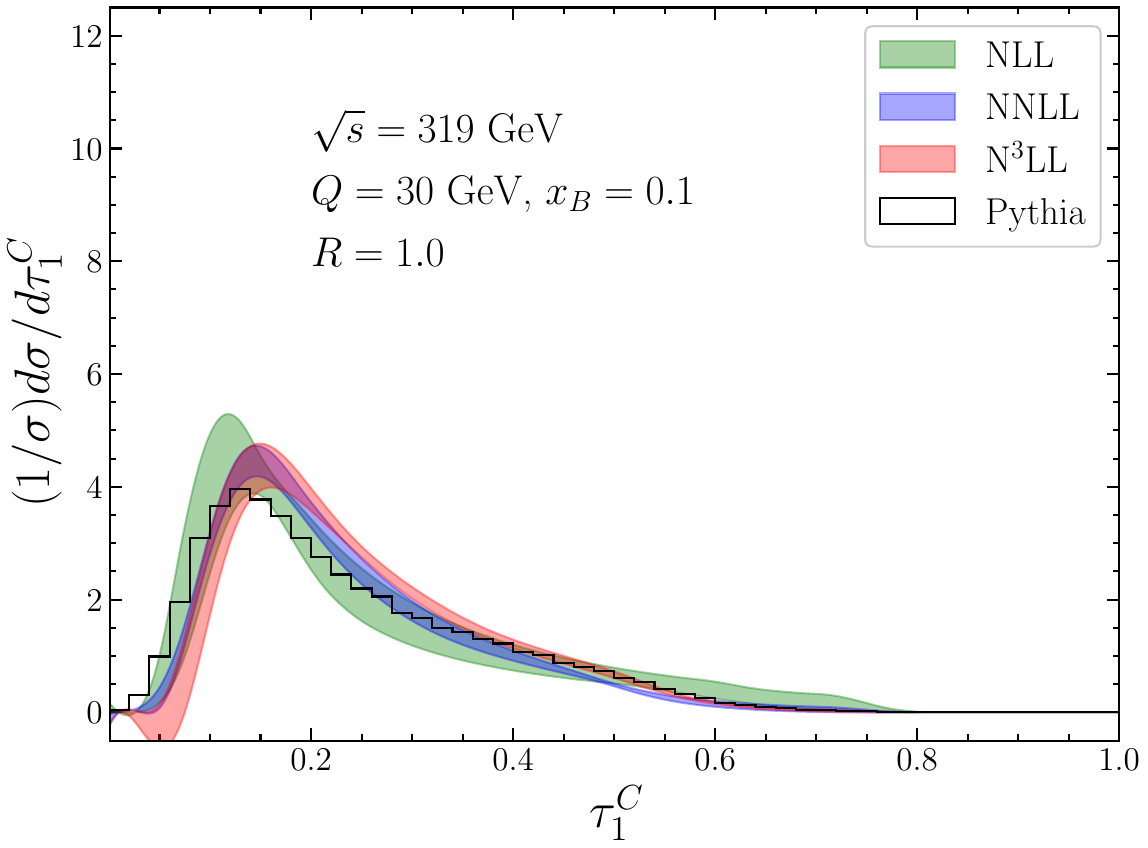}
    \includegraphics[width=0.40\linewidth]{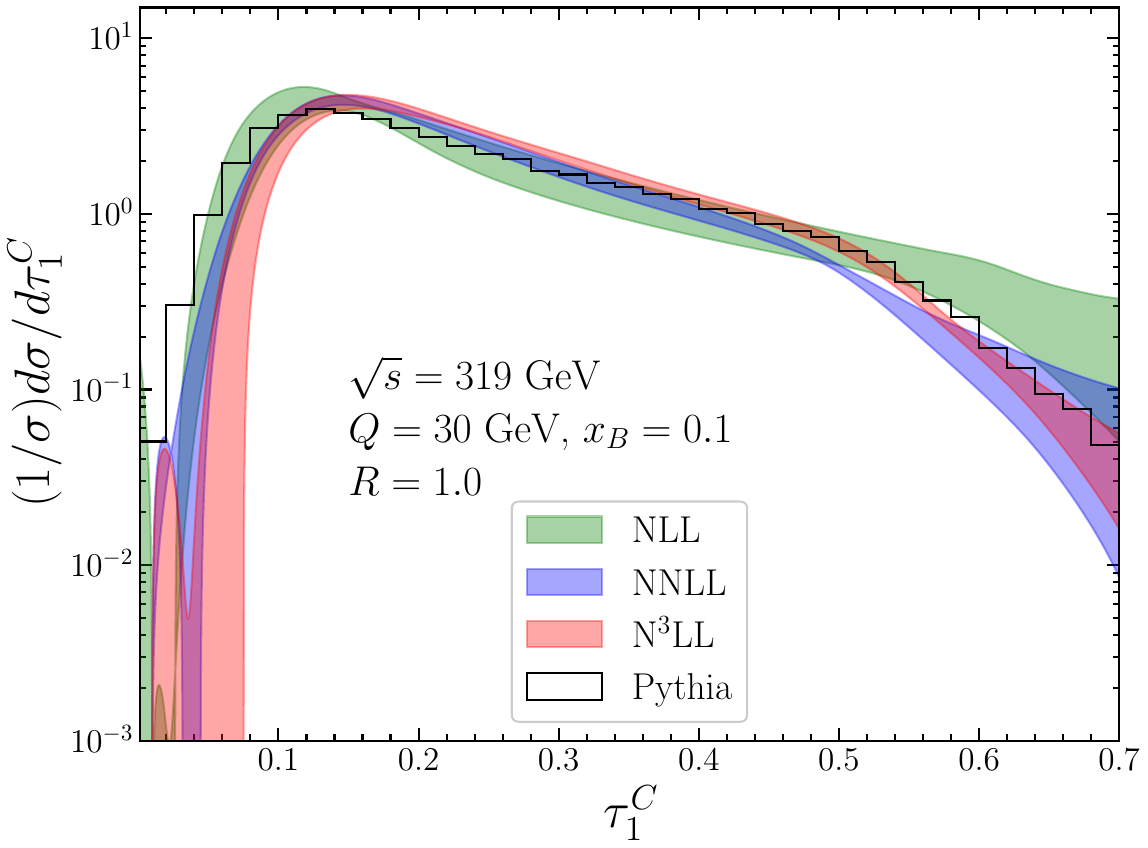}
    \includegraphics[width=0.40\linewidth]{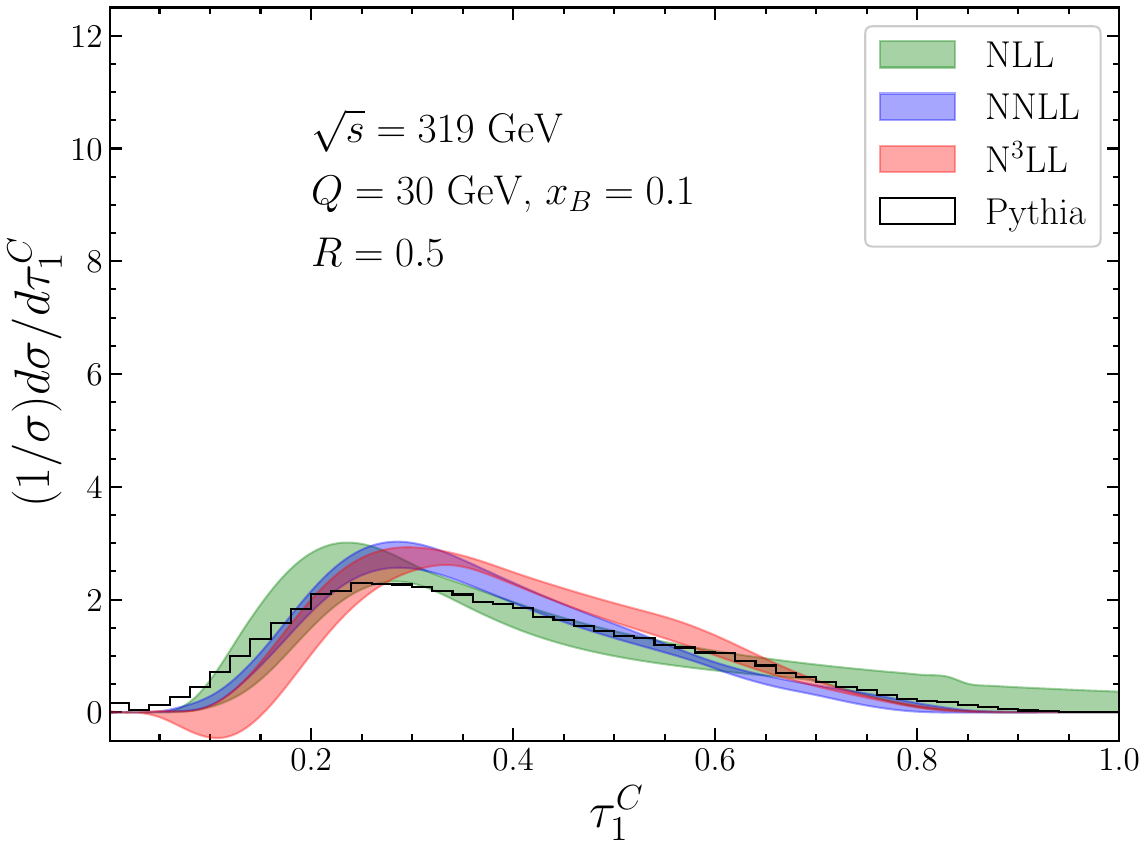}
    \includegraphics[width=0.40\linewidth]{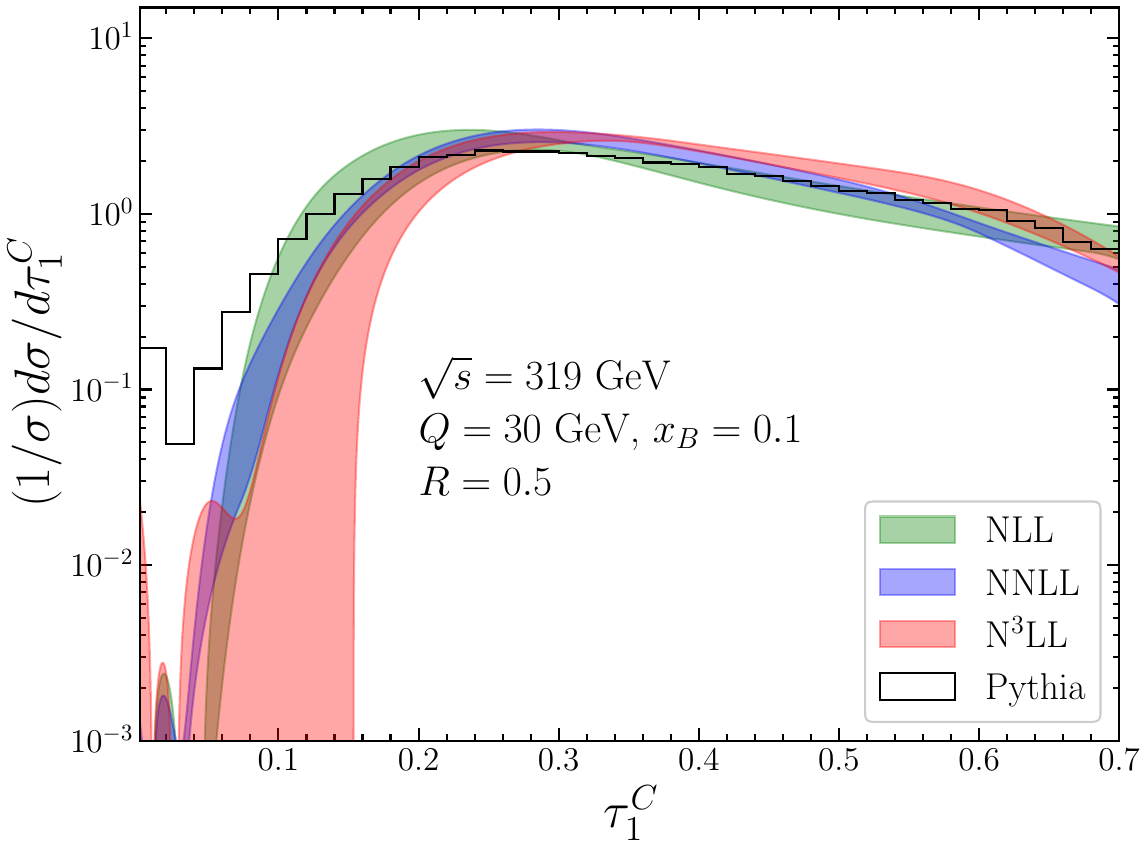}
    \vspace{-1em}
    \caption{The same plot as Fig.~\ref{fig:theory-final-hadron} at $Q=30$~GeV and $x_B=0.1$.}
    \label{fig:theory-final-hadron_Q30_x010}
\end{figure}

\begin{figure}
    \centering
    \includegraphics[width=\linewidth]{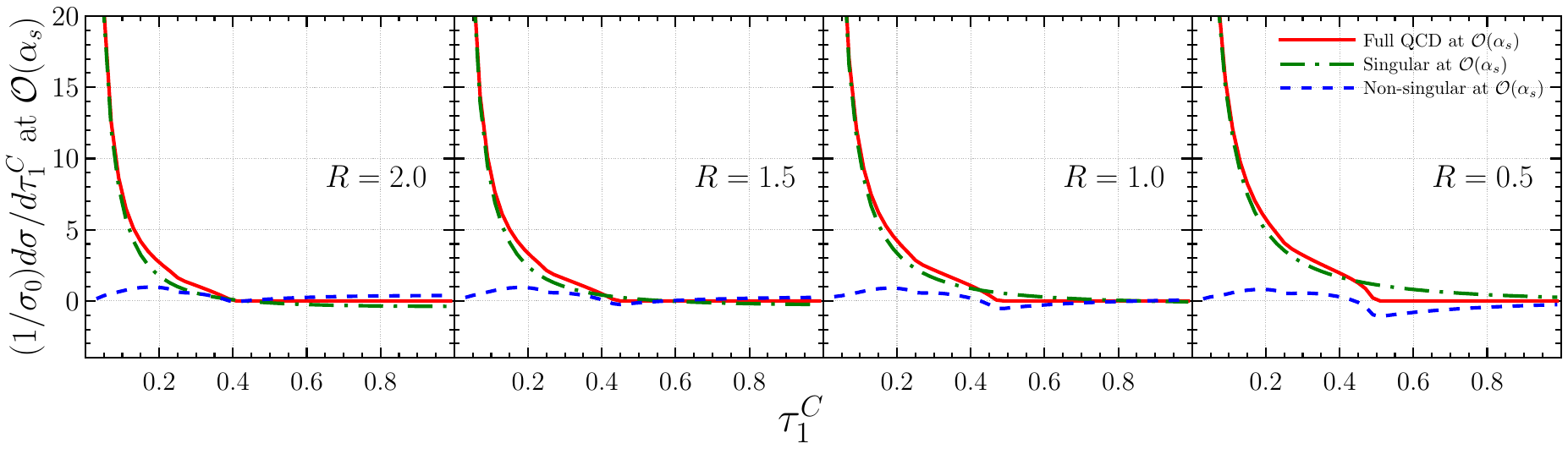}
    \includegraphics[width=\linewidth]{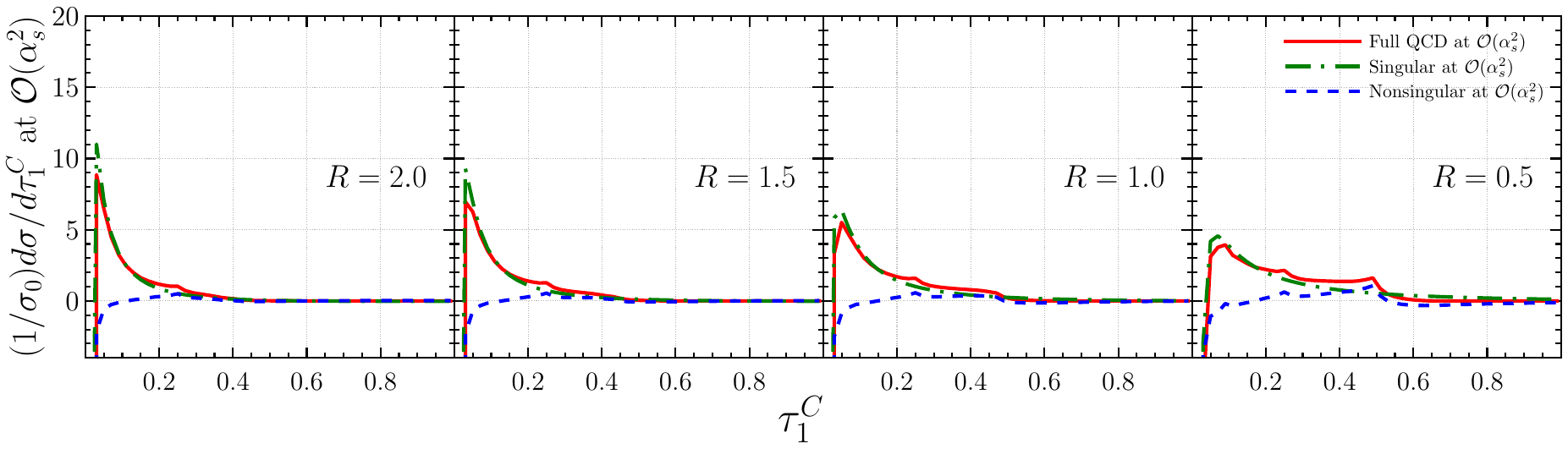}
    \caption{
    The same plots as Fig.~\ref{fig:ns_LO} at $Q=50$~GeV and $x_B=0.1$.
    }
    \label{fig:ns_LO_Q50_x010}
\end{figure}

\begin{figure}
    \centering
    \includegraphics[width=\linewidth]{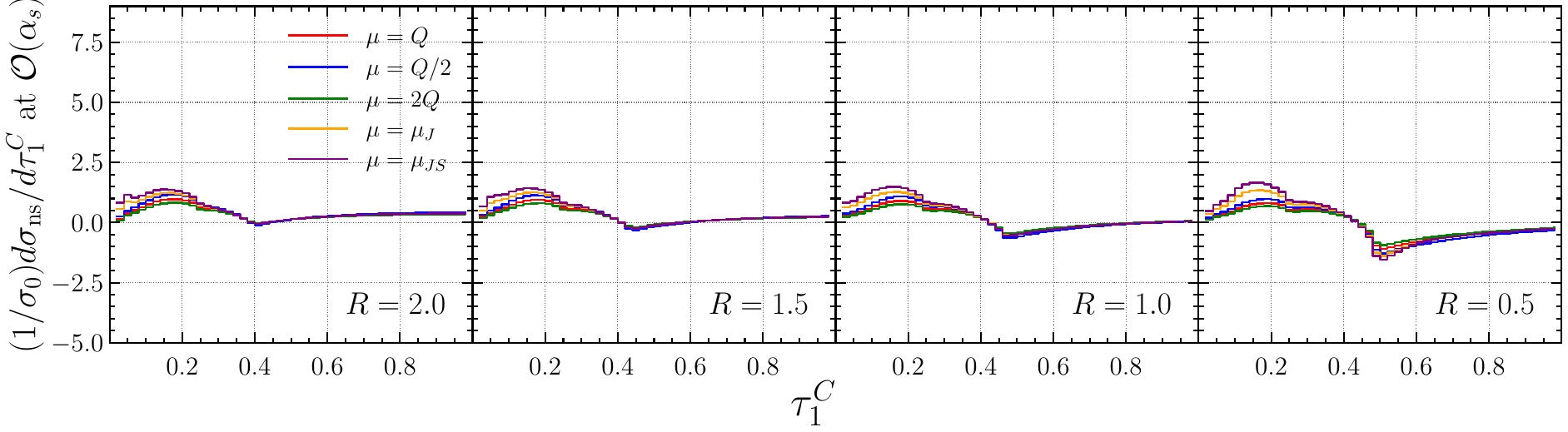}
    \includegraphics[width=\linewidth]{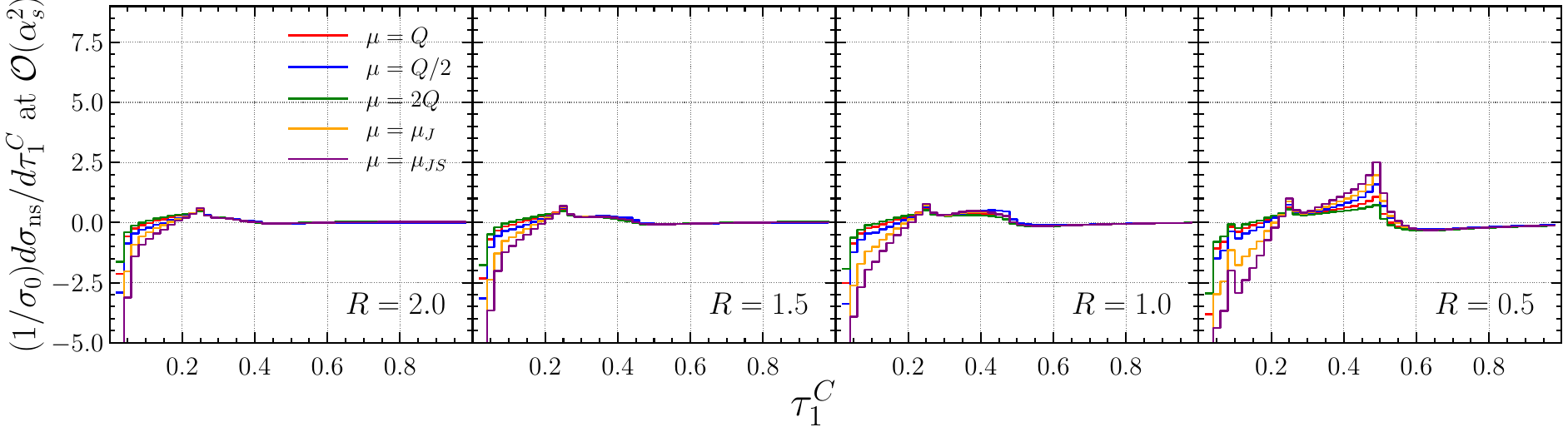}
    \caption{The same plots as Fig.~\ref{fig:ns_LO_NLO_scale_var} at $Q=50$~GeV and $x_B=0.1$.}
    \label{fig:ns_LO_NLO_scale_var_Q50_x010}
\end{figure}

\begin{figure}[t]
    \centering
    \includegraphics[width=0.40\linewidth]{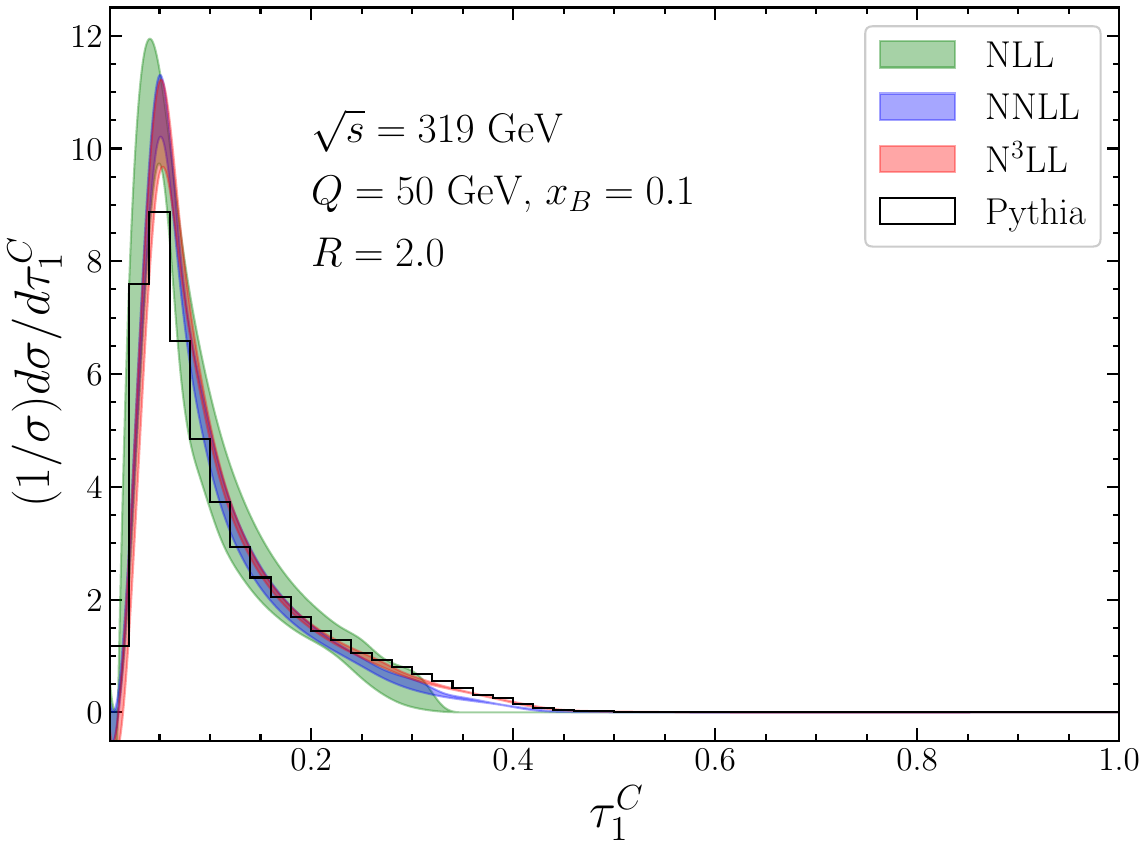}
    \includegraphics[width=0.40\linewidth]{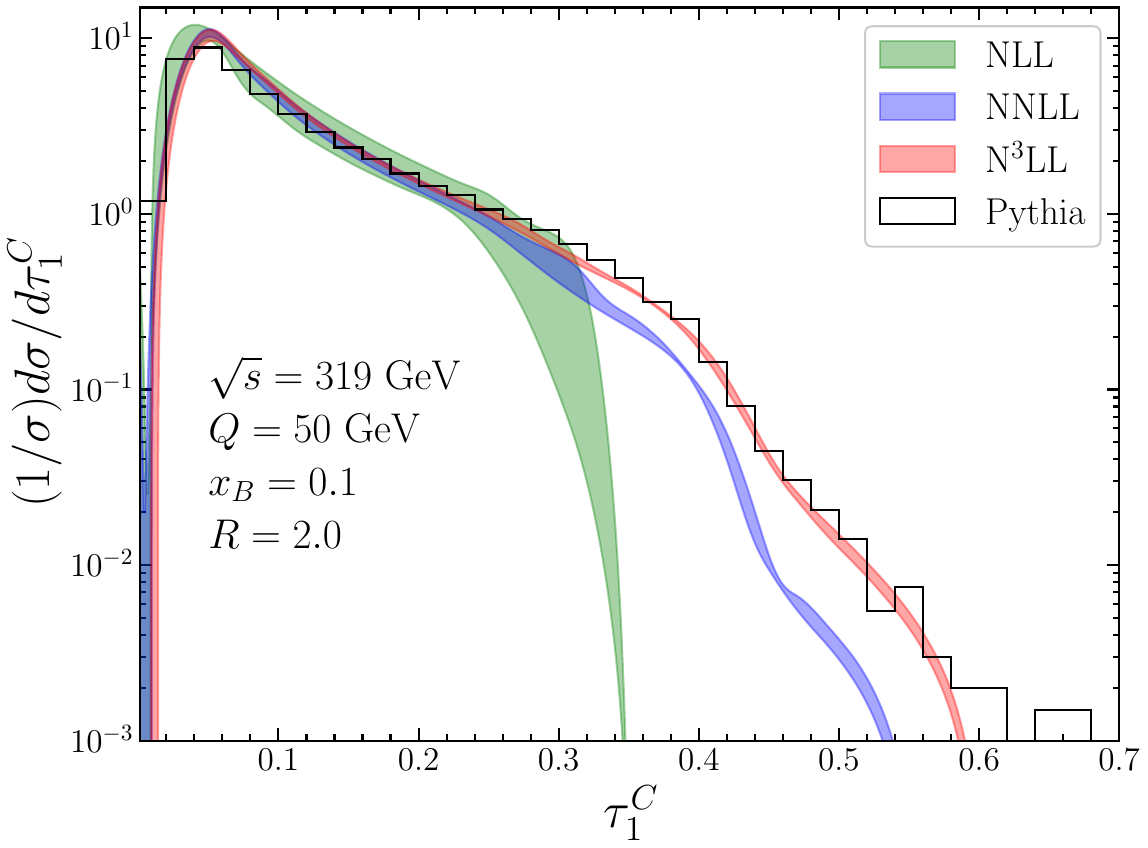}
    \includegraphics[width=0.40\linewidth]{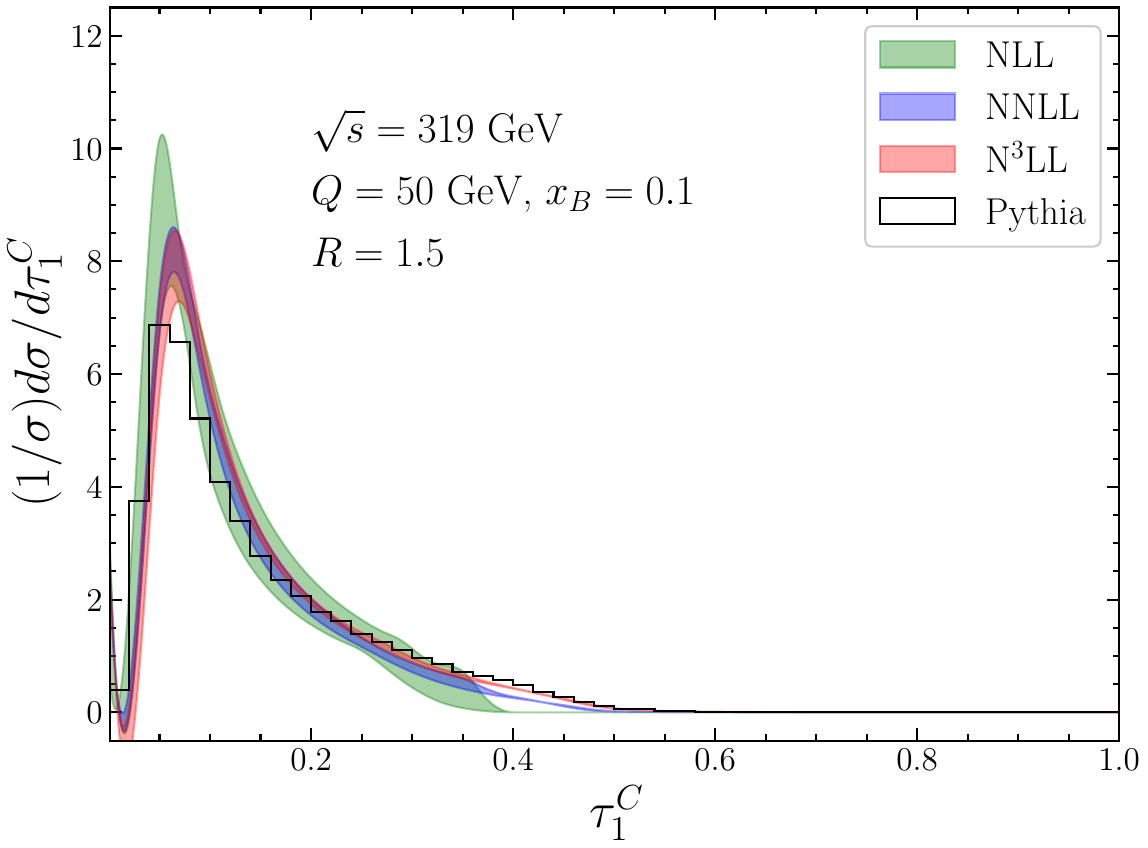}
    \includegraphics[width=0.40\linewidth]{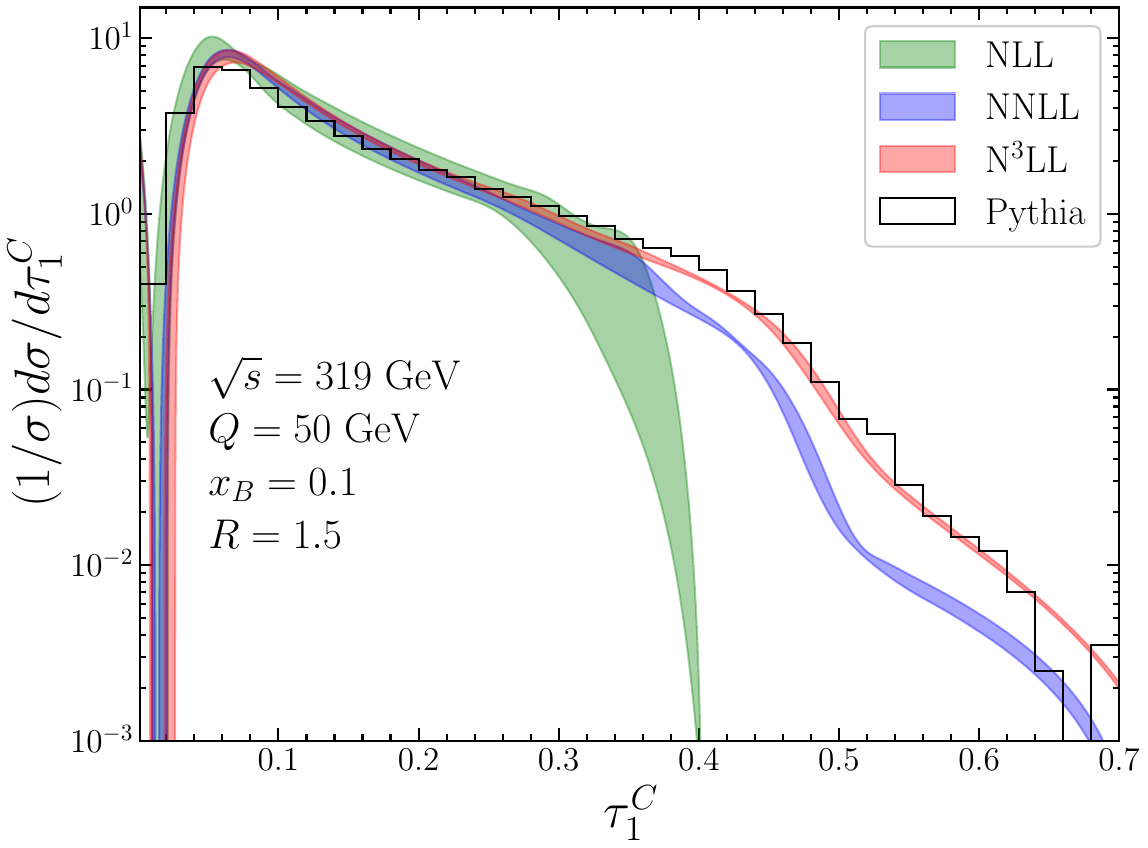}
    \includegraphics[width=0.40\linewidth]{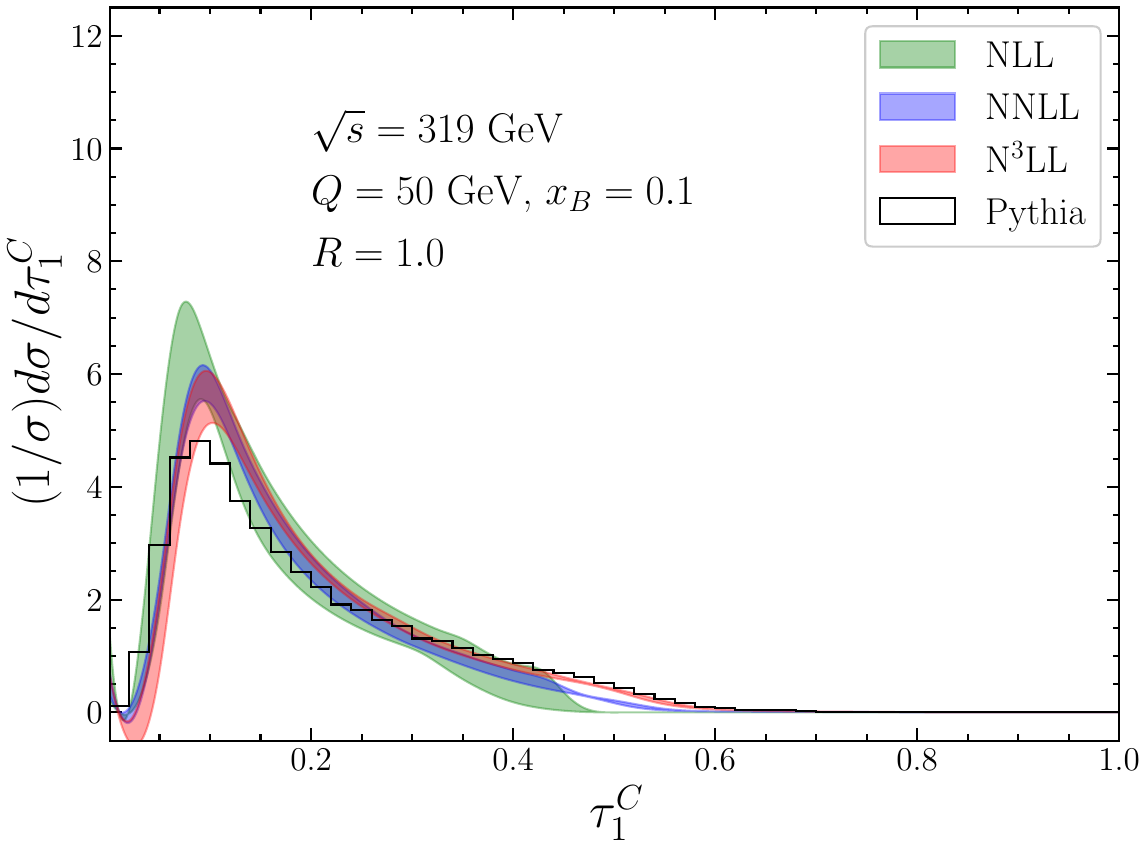}
    \includegraphics[width=0.40\linewidth]{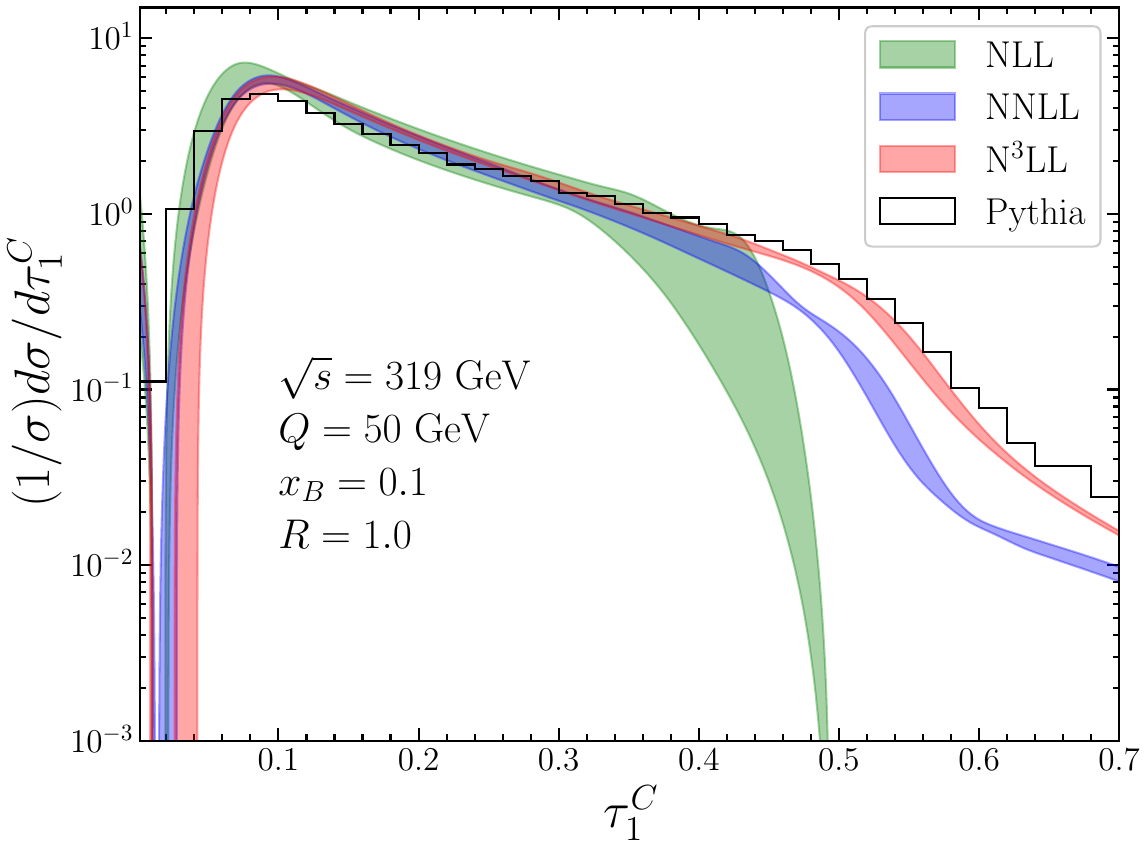}
    \includegraphics[width=0.40\linewidth]{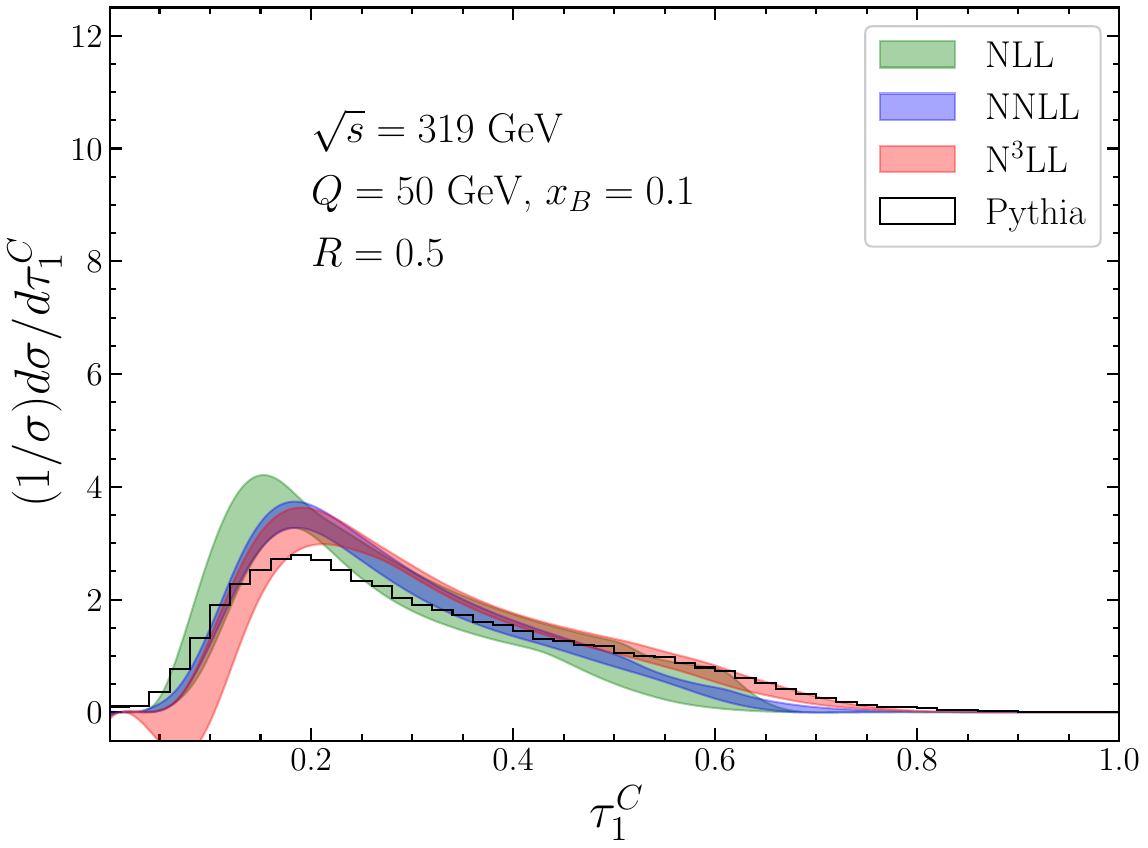}
    \includegraphics[width=0.40\linewidth]{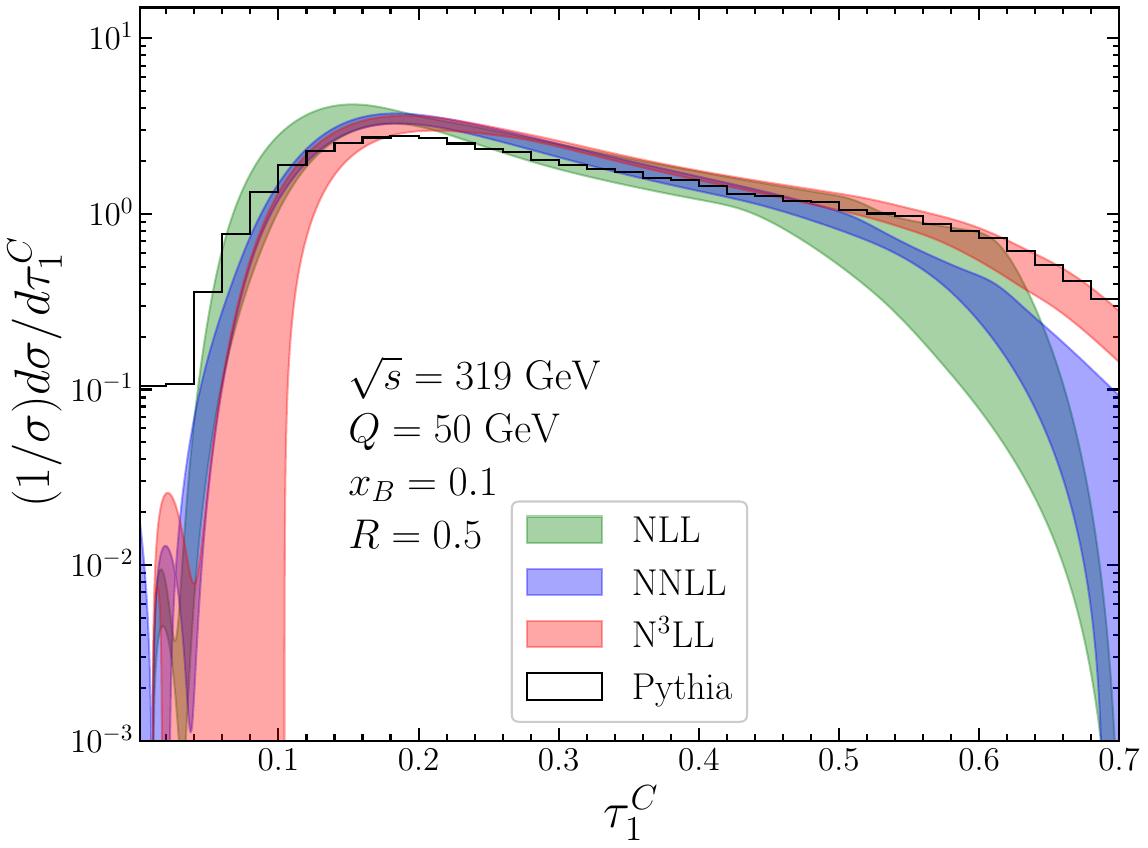}
    \vspace{-1em}
    \caption{The same plots as Fig.~\ref{fig:theory-final-hadron} at $Q=50$~GeV and $x_B=0.1$.}
    \label{fig:theory-final-hadron_Q50_x010}
\end{figure}

In this appendix, we present the same set of figures as those shown in the 
main text at three additional kinematic configurations: 
$(Q, x_B) = (30~\textrm{GeV}, 0.05)$, $(30~\textrm{GeV}, 0.1)$, and 
$(50~\textrm{GeV}, 0.1)$
(Figs.~\ref{fig:ns_LO_Q30_x005}–\ref{fig:theory-final-hadron_Q50_x010}). 
For each configuration, we show: (i) the LO and NLO fixed-order results, 
including the full QCD, singular, and non-singular contributions (analogous 
to Fig.~\ref{fig:ns_LO}); (ii) the corresponding non-singular scale 
variations (analogous to Fig.~\ref{fig:ns_LO_NLO_scale_var}); and (iii) 
the final theoretical predictions across four jet radii 
$R \in \{0.5, 1.0, 1.5, 2.0\}$ (analogous to Fig.~\ref{fig:theory-final-hadron}). 
The qualitative features observed in the main-text kinematics are preserved 
across all configurations: the LO non-singular contributions remain 
subdominant in the resummation region, and the final resummed predictions 
agree with the \textsc{Pythia} hadron-level distributions within their 
theoretical uncertainty in the tail region across all jet radii.

\clearpage

\bibliographystyle{JHEP}
\bibliography{Centauro.bib}

\end{document}